\documentclass[traditabstract]{aa} % for the letters 
\usepackage{natbib}
\usepackage{rotating}
\usepackage{graphicx}
\usepackage{amsmath}
\usepackage{lipsum}
\usepackage{txfonts}
\usepackage{amssymb}
\begin{document} 

   \title{Radio-emitting narrow-line Seyfert 1 \\ galaxies in the JVLA perspective}

   \author{M. Berton
	\inst{1,2}\thanks{marco.berton@unipd.it}
	\and E. Congiu\inst{1,2}
	\and E. J\"arvel\"a\inst{3,4} 
	\and R. Antonucci\inst{5}
	\and P. Kharb\inst{6}
	\and M.~L. Lister\inst{7}
	\and A. Tarchi\inst{8}
	\and \\
	A. Caccianiga\inst{2}
	\and S. Chen\inst{1,9,10}
	\and L. Foschini\inst{2}
	\and A. L\"ahteenm\"aki\inst{3,4,11}
	\and J.~L. Richards\inst{7}
	\and \\
	S. Ciroi\inst{1,12}
	\and V. Cracco\inst{1}
	\and M. Frezzato\inst{1}
	\and G. La Mura\inst{1}
	\and P. Rafanelli\inst{1}
         }

   \institute{$^{1}$ Dipartimento di Fisica e Astronomia "G. Galilei", Universit\`a di Padova, Vicolo dell'Osservatorio 3, 35122 Padova, Italy;\\
   $^{2}$ INAF - Osservatorio Astronomico di Brera, via E. Bianchi 46, 23807 Merate (LC), Italy;\\
 $^{3}$ Aalto University Mets{\"a}hovi Radio Observatory, Mets{\"a}hovintie 114, FIN-02540 Kylm{\"a}l{\"a}, Finland;\\
 $^4$ Aalto University Department of Electronics and Nanoengineering, P.O. Box 15500, FI-00076 AALTO, Finland;\\
 $^{5}$ Department of Physics, University of California, Santa Barbara, CA 93106-9530, USA; \\
 $^{6}$ National Centre for Radio Astrophysics - Tata Institute of Fundamental Research, Post Bag 3, Ganeshkhind, Pune 411007, India;\\
  $^{7}$ Department of Physics and Astronomy, Purdue University, 525 Northwestern Avenue, West Lafayette, IN 47907, USA;\\
 $^{8}$  INAF - Osservatorio Astronomico di Cagliari, Via della Scienza 5, 09047, Selargius (CA), Italy; \\
 $^{9}$ INFN - Sezione di Padova, Via Marzolo 8, 35131, Padova; \\
 $^{10}$ Center for Astrophysics, Guangzhou University, Guangzhou 510006, China; \\
 $^{11}$ Tartu Observatory, Observatooriumi 1, 61602 T\~{o}ravere, Estonia;\\ 
$^{12}$ INAF - Osservatorio Astronomico di Padova, Vicolo dell'Osservatorio 5, 35122 Padova, Italy.
            }

  % \date{}
\authorrunning{M. Berton et al.}
\titlerunning{NLS1s with JVLA}

\abstract{We report the first results of a survey on 74 narrow-line Seyfert 1 galaxies (NLS1s) carried out in 2015 with the Karl G. Jansky Very Large Array (JVLA) at 5 GHz in A-configuration. So far, this is the largest survey aimed to image the radio continuum of NLS1s. We produced radio maps in order to compare the general properties of three different samples of objects: radio-quiet NLS1s (RQNLS1s), steep-spectrum radio-loud NLS1s (S-NLS1s), and flat-spectrum radio-loud NLS1s (F-NLS1s). We find that the three classes correspond to different radio morphologies, with F-NLS1s being more compact, and RQNLS1s often showing diffuse emission on kpc scales. We also find that F-NLS1s might be low-luminosity and possibly young blazars, and that S-NLS1s are part of the parent population of F-NLS1s. Dedicated studies to RQNLS1s are needed in order to fully understand their role in the unification pictures.  }

\maketitle

\newcommand{\kms}{km s$^{-1}$}
\newcommand{\ergs}{erg s$^{-1}$}
\section{Introduction}

Narrow-line Seyfert 1 galaxies (NLS1s) are active galactic nuclei (AGN) characterized by optical emission spectra similar to those in ordinary Seyfert 1 galaxies (Sy1s), with ratio [O III]/H$\beta <$ 3 and strong Fe II multiplets, but unlike Sy1s they have narrower permitted lines \citep{Osterbrock85, Goodrich89}. 
These properties indicate that the high-density broad-line region (BLR) in NLS1s is visible, unlike in Seyfert 2s, but contains material moving much more slowly than in ordinary Sy1s. 
This, assuming a similar BLR size, implies that the central black hole masses should be smaller than in other Sy1s of similar luminosity. 
Indeed those measured in NLS1s range from 10$^5$-10$^8$ M$_\odot$, typically smaller than the 10$^7$-10$^9$ M$_\odot$ found in regular Sy1s, and the 10$^8$-10$^{10}$ M$_\odot$ typical of blazars \citep{Grupe00, Jarvela14, Cracco16}. 
NLS1s are also characterized by large accretion rates relative to the Eddington limit, often approaching that limit \citep{Boroson92, Sulentic00}. 
Many of their properties suggest they may represent young AGN in an early stage of evolution \citep[e.g.][]{Mathur00}. 
NLS1s are typically hosted by disk galaxies, whereas blazars and radio galaxies are normally found in ellipticals \citep{Crenshaw03, OrbandeXivry11}. \par

However, a disk-like BLR and a low inclination could also mimic the effect of a low-mass black hole. 
In this scenario, the narrowness of permitted lines would be due to the lack of Doppler broadening in a disk-BLR observed pole-on \citep{Decarli08}. 
Although this model is somewhat in contradiction with the current understanding of the physics behind the blazar sequence \citep{Ghisellini98, Foschini17}, the low-mass/low-inclination degeneracy is one of the most debated issues about NLS1s nature. \par

Most NLS1s are radio-quiet\footnote{\footnotesize According to \citet{Kellermann89}, radio-loudness is defined as the flux ratio RL = $S_{5 \\ GHz}$/S$_{B-band}$. If RL $>$ 10, the source is radio-loud, conversely it is radio-quiet. We define sources without radio detection as radio-silent.}, but not radio-silent \citep{Giroletti09, Doi13, Lahteenmaki17}. 
About 7\% are radio-loud, a lower incidence than among quasars \citep{Komossa06}; however this fraction is strongly dependent on the redshift, since it appears to be lower (3.5\%) in the nearby Universe \citep{Cracco16} and to be dependent on the large-scale environment \citep{Jarvela17}.  
The $\gamma$-ray emission detected by the Fermi Large Area Telescope (LAT) in 12 radio-loud NLS1s \citep{Foschini15, Liao15, Yao15, Dammando15, Berton17a} confirmed the presence of a relativistic jet. 
Indeed, images from Very Long Baseline Array (VLBA) observations of several NLS1s show blazar-like parsec-scale radio jets, with high brightness temperature, flat or inverted radio spectra, fast variability, and a significant degree of polarization \citep{Abdo09a, Ikejiri11, Lister13, Gu15}. 
In some cases, apparent superluminal motion is also present \citep{Lister16}. 
The production of relativistic jets among a population of AGN with small black hole masses and large Eddington accretion rates runs counter to the general trend of stronger jets in AGN with larger black holes and lower accretion rates \citep{Laor00, Boroson02, Boettcher02, Marscher09}. 
Furthermore, by analogy with stellar-mass black holes, AGN in high accretion states would be expected to correspond to the high/soft state, where jets are quenched \citep[e.g.][]{Maccarone03}. 
NLS1s thus present a puzzle with respect to the current understanding of jet production mechanisms. \par

It is not yet clear how to include radio-loud NLS1s in the prevailing orientation-driven AGN unification model \citep{Antonucci93, Urry95}. 
In this picture, the powerful blazars result when members of a jetted parent population $-$ the radio galaxies $-$ happen to be oriented with the jet nearly along our line of sight. 
In this preferred orientation, emission from the jet is relativistically beamed toward the observer, greatly enhancing its apparent luminosity. 
The presence of beamed emission in radio-loud NLS1s \citep{Abdo09c} suggests that these sources may be the low-mass and low-luminosity analogs of flat-spectrum radio quasars (FSRQ, \citealp{Foschini15, Berton16c}). 
If this is correct, a corresponding unbeamed parent population must exist. 
This population of “misaligned” NLS1s must be rather large: assuming a negligible opening angle of the jet cone, $\sim$2$\Gamma^2$ members of the parent population should be observable, where $\Gamma$ is the typical jet bulk Lorentz factor.   
For reasonable values $\Gamma \sim$ 10 \citep{Abdo09c}, the parent population should outnumber the radio-loud NLS1 population by a factor of about 200. 
The nature of this parent population is still unclear. 
Steep-spectrum\footnote{Typically evaluated with two frequencies around L band.} radio-loud NLS1s (S-NLS1s), disk-hosted radio galaxies and compact steep-spectrum sources (CSS)\footnote{CSS are here defined as jetted radio-loud sources with compact morphology and steep radio spectrum \citep[e.g.][]{Kunert10a}.} appear to be good candidates \citep{Berton15a, Berton16c, Berton17a}, but even among radio-quiet sources some parent objects could be hidden \citep{Tarchi11, Berton16b, Lahteenmaki17}. \par
Several studies have found NLS1s to be predominantly compact on kiloparsec scales; extended morphologies have been reported in some cases \citep{Gliozzi10, Doi12, Doi15, Richards15, Gu15, Caccianiga17, Congiu17}. 
However, the majority of observations on this scale come from the FIRST survey \citep{Becker95}, in which only sources larger than 2$^{\prime\prime}$ are resolved to the 5$^{\prime\prime}$ synthesized beam \citep{White97}. 
A wide gap between the 2$^{\prime\prime}$ scale and the scales of a few milliarcsecond explored by Very Large Baseline Interferometry (VLBI) observations remains largely unexplored. \par
In order to fill this gap, we observed a large sample of 74 NLS1s, both radio-loud and radio-quiet, with the Karl G. Jansky Very Large Array (JVLA) at 5 GHz in A-configuration. The latter has a longest baseline of 36.4 km, which gives an approximate beam size of $\sim$0.5$^{\prime\prime}$ with natural weighting. For many of these sources ($\sim40$), this is the first observation ever performed at this frequency, and will make it possible to get a spectral index measurement, which is essential to distinguish between flat- and steep-spectrum objects.
The aim of this study is also is to take advantage of the major sensitivy upgrade of the JVLA to search for kpc-scale jets which can contribute to increase the number of parent sources. \par
In this paper we will present the main results of this survey, focusing on the general properties of these sources. 
In an upcoming paper, we will focus in greater detail on the radio image properties as well as the multi-wavelength properties of the sample.
In Sect.~2 we will describe the sample, in Sect.~3 we will describe the observations and the data analysis, in Sect.~4 we will present our results, in Sect.~5 we will discuss them, and in Sect.~6 we will provide a brief summary of this work. 
Throughout this paper, we adopt a standard $\rm \Lambda CDM$ cosmology, with a Hubble constant $H_0 = 70$ \kms\ Mpc$^{-1}$, and $\Omega_\Lambda = 0.73$ \citep{Komatsu11}.
Spectral indexes are specified with flux density $S_{\nu} \propto \nu^{-\alpha}$ at frequency $\nu$.

\begin{table*}
\caption{The sample.}
\label{tab:summary1}
\centering
\scalebox{0.85}{
\footnotesize
\begin{tabular}{l c c c c c c c c c c} 
\hline\hline
Short name & NED Alias & R.A. & Dec. & z & Scale & Old & New & Map & Type \\
\hline\hline
J0006+2012 & Mrk 335 & 00 06 19.50 & +20 12 10.0 & 0.026 & 0.523 & Q & Q & Fig.~\ref{fig:J0006p2012} & C\\
J0100$-$0200 & FBQS J0100$-$0200 & 01 00 32.22 & $-$02 00 46.3 & 0.227 & 3.638 & F & S & Fig.~\ref{fig:J0100m0200} & C \\
J0146$-$0040 & 2MASX J01464481$-$0040426 & 01 46 44.80 & $-$00 40 43.0 & 0.083 & 1.561 & Q & Q & Fig.~\ref{fig:J0146m0040} & C \\
J0347+0105 & IRAS 03450+0055 & 03 47 40.20 & +01 05 14.0 & 0.031 & 0.620 & Q & Q & Fig.~\ref{fig:J0347p0105} & E \\ 
J0629$-$0545 & IRAS 06269$-$0543 & 06 29 24.70 & $-$05 45 30.0 & 0.117 & 2.116 & S & F & Fig.~\ref{fig:J0629m0545} & I \\
J0632+6340 & UGC 3478 & 06 32 47.20 & +63 40 25.0 & 0.013 & 0.266 & Q & Q & Fig.~\ref{fig:J0632p6340} & E \\
J0706+3901 & FBQS J0706+3901 & 07 06 25.15 & +39 01 51.6 & 0.086 & 1.612 & F & S & Fig.~\ref{fig:J0706p3901} & I \\
J0713+3820 & FBQS J0713+3820 & 07 13 40.29 & +38 20 40.1 & 0.123 & 2.210 & F & S & Fig.~\ref{fig:J0713p3820} & E \\
J0744+5149 & NVSS J074402+514917 & 07 44 02.24 & +51 49 17.5 & 0.460 & 5.831 & F & S & Fig.~\ref{fig:J0744p5149} & C \\
J0752+2617 & FBQS J0752+2617 & 07 52 45.60 & +26 17 36.0 & 0.082 & 1.544 & Q & Q & Fig.~\ref{fig:J0752p2617} & C \\
J0758+3920 & FBQS J075800.0+392029 & 07 58 00.05 & +39 20 29.1 & 0.096 & 1.779 & S & S & Fig.~\ref{fig:J0758p3920} & E \\
J0804+3853 & FBQS J0804+3853 & 08 04 09.24 & +38 53 48.7 & 0.212 & 3.453 & F & S & Fig.~\ref{fig:J0804p3853} & E \\ 
J0806+7248 & RGB J0806+728 & 08 06 38.96 & +72 48 20.4 & 0.098 & 1.812 & S & F & Fig.~\ref{fig:J0806p7248} & I \\ 
J0814+5609 & SDSS J081432.11+560956.6 & 08 14 32.13 & +56 09 56.6 & 0.510 & 6.168 & F & F & Fig.~\ref{fig:J0814p5609} & I \\
J0849+5108* & SBS 0846+513 & 08 49 57.99 & +51 08 28.8 & 0.585 & 6.606 & G & F & Fig.~\ref{fig:J0849p5108} & C \\
J0850+4626 & SDSS J085001.17+462600.5 & 08 50 01.17 & +46 26 00.5 & 0.524 & 6.256 & S & F & Fig.~\ref{fig:J0850p4626} & C \\
J0902+0443 & SDSS J090227.16+044309.5 & 09 02 27.15 & +04 43 09.4 & 0.533 & 6.311 & F & F & Fig.~\ref{fig:J0902p0443} & C \\
J0913+3658 & RX J0913.2+3658 & 09 13 13.70 & +36 58 17.0 & 0.107 & 1.958 & Q & Q & Fig.~\ref{fig:J0913p3658} & I \\
J0925+5217 & Mrk 110 & 09 25 12.90 & +52 17 11.0 & 0.035 & 0.697 & Q & Q & Fig.~\ref{fig:J0925p5217} & E \\
J0926+1244 & Mrk 705 & 09 26 03.30 & +12 44 04.0 & 0.029 & 0.581 & Q & Q & Fig.~\ref{fig:J0926p1244} & I \\
J0937+3615 & SDSS J093703.03+361537.2 & 09 37 03.01 & +36 15 37.3 & 0.180 & 3.035 & F & S & Fig.~\ref{fig:J0937p3615} & E \\
J0945+1915 & SDSS J094529.23+191548.8 & 09 45 29.21 & +19 15 48.9 & 0.284 & 4.287 & F & S & Fig.~\ref{fig:J0945p1915} & C \\
J0948+5029 & Mrk 124 & 09 48 42.60 & +50 29 31.0 & 0.056 & 1.087 & Q & Q & Fig.~\ref{fig:J0948p5029} & C \\
J0952$-$0136 & Mrk 1239 & 09 52 19.10 & $-$01 36 43.0 & 0.020 & 0.405 & Q & Q & Fig.~\ref{fig:J0952m0136} & I \\
J0957+2433 & RX J0957.1+2433 & 09 57 07.20 & +24 33 16.0 & 0.082 & 1.544 & Q & Q & Fig.~\ref{fig:J0957p2433} & C \\
J1031+4234 & SDSS J103123.73+423439.3 & 10 31 23.73 & +42 34 39.4 & 0.377 & 5.180 & F & S & Fig.~\ref{fig:J1031p4234} & C \\
J1034+3938 & KUG 1031+398 & 10 34 38.60 & +39 38 28.0 & 0.042 & 0.829 & S & F & Fig.~\ref{fig:J1034p3938} & I \\
J1037+0036 & SDSS J103727.45+003635.6 & 10 37 27.45 & +00 36 35.8 & 0.595 & 6.659 & F & F & Fig.~\ref{fig:J1037p0036} & C \\
J1038+4227 & SDSS J103859.58+422742.3 & 10 38 59.59 & +42 27 42.0 & 0.220 & 3.552 & F & S & Fig.~\ref{fig:J1038p4227} & E \\
J1047+4725 & SDSS J104732.68+472532.0 & 10 47 32.65 & +47 25 32.2 & 0.799 & 7.505 & F & F & Fig.~\ref{fig:J1047p4725} & E \\
J1048+2222 & SDSS J104816.57+222238.9 & 10 48 16.56 & +22 22 40.1 & 0.330 & 4.752 & F & S & Fig.~\ref{fig:J1048p2222} & I \\
J1102+2239* & SDSS J110223.38+223920.7 & 11 02 23.36 & +22 39 20.7 & 0.453 & 5.781 & G & S & Fig.~\ref{fig:J1102p2239} & I \\
J1110+3653 & SDSS J111005.03+365336.3 & 11 10 05.03 & +36 53 36.1 & 0.629 & 6.830 & F & F & Fig.~\ref{fig:J1110p3653} & I \\
J1114+3241 & B2 1111+32 & 11 14 38.89 & +32 41 33.4 & 0.189 & 3.156 & S & F & Fig.~\ref{fig:J1114p3241} & C \\
J1121+5351 & SBS 1118+541 & 11 21 08.60 & +53 51 21.0 & 0.103 & 1.893 & Q & Q & Fig.~\ref{fig:J1121p5351} & I \\
J1138+3653 & SDSS J113824.54+365327.1 & 11 38 24.54 & +36 53 27.1 & 0.356 & 4.994 & F & F & Fig.~\ref{fig:J1138p3653} & I \\
J1146+3236 & SDSS J114654.28+323652.3 & 11 46 54.30 & +32 36 52.2 & 0.465 & 5.867 & F & F & Fig.~\ref{fig:J1146p3236} & C \\
J1159+2838 & SDSS J115917.32+283814.5 & 11 59 17.31 & +28 38 14.8 & 0.210 & 3.427 & F & S & Fig.~\ref{fig:J1159p2838} & I \\
J1203+4431 & NGC 4051 & 12 03 09.60 & +44 31 53.0 & 0.002 & 0.041 & Q & Q & Fig.~\ref{fig:J1203p4431} & E \\
J1209+3217 & RX J1209.7+3217 & 12 09 45.20 & +32 17 01.0 & 0.144 & 2.527 & Q & Q & Fig.~\ref{fig:J1209p3217} & I \\
J1215+5442 & SBS 1213+549A & 12 15 49.40 & +54 42 24.0 & 0.150 & 2.614 & Q & Q & Fig.~\ref{fig:J1215p5442} & I \\
J1218+2948 & Mrk 766 & 12 18 26.50 & +29 48 45.8 & 0.013 & 0.266 & Q & Q & Fig.~\ref{fig:J1218p2948} & E \\
J1227+3214 & SDSS J122749.14+321458.9 & 12 27 49.15 & +32 14 59.0 & 0.136 & 2.408 & F & F & Fig.~\ref{fig:J1227p3214} & I \\
J1238+3942 & SDSS J123852.12+394227.8 & 12 27 49.15 & +39 42 27.6 & 0.622 & 6.796 & F & F & Fig.~\ref{fig:J1238p3942} & C \\
J1242+3317 & WAS 61 & 12 42 10.60 & +33 17 03.0 & 0.044 & 0.866 & Q & Q & Fig.~\ref{fig:J1242p3317} & E \\
J1246+0222 & PG 1244+026 & 12 46 35.20 & +02 22 09.0 & 0.048 & 0.941 & Q & Q & Fig.~\ref{fig:J1246p0222} & E \\
J1246+0238* & SDSS J124634.65+023809.0 & 12 46 34.68 & +02 38 09.0 & 0.367 & 5.092 & G & F & Fig.~\ref{fig:J1246p0238} & C \\
J1302+1624 & Mrk 783 & 13 02 58.80 & +16 24 27.0 & 0.067 & 1.284 & S & S & Fig.~\ref{fig:J1302p1624} & E \\
J1305+5116* & SDSS J130522.74+511640.2 & 13 05 22.75 & +51 16 40.3 & 0.788 & 7.469 & S & F & Fig.~\ref{fig:J1305p5116} & E \\
J1317+6010 & SBS 1315+604 & 13 17 50.30 & +60 10 41.0 & 0.137 & 2.423 & Q & Q & Fig.~\ref{fig:J1317p6010} & E \\
J1333+4141 & SDSS J133345.47+414127.7 & 13 33 45.47 & +41 41 28.2 & 0.225 & 3.614 & F & S & Fig.~\ref{fig:J1333p4141} & C \\
J1337+2423 & IRAS 13349+2438 & 13 37 18.70 & +24 23 03.0 & 0.108 & 1.974 & Q & Q & Fig.~\ref{fig:J1337p2423} & I \\
J1346+3121 & SDSS J134634.97+312133.7 & 13 46 35.07 & +31 21 33.9 & 0.246 & 3.864 & F & F & Fig.~\ref{fig:J1346p3121} & C \\
J1348+2622 & SDSS J134834.28+262205.9 & 13 48 34.25 & +26 22 05.9 & 0.917 & 7.832 & F & S & Fig.~\ref{fig:J1348p2622} & I \\
J1355+5612 & SBS 1353+564 & 13 55 16.50 & +56 12 45.0 & 0.122 & 2.194 & Q & Q & Fig.~\ref{fig:J1355p5612} & C \\
J1358+2658 & SDSS J135845.38+265808.4 & 13 58 45.40 & +26 58 08.3 & 0.331 & 4.762 & F & S & Fig.~\ref{fig:J1358p2658} & I \\
J1402+2159 & RX J1402.5+2159 & 14 02 34.40 & +21 59 52.0 & 0.066 & 1.266 & Q & Q & Fig.~\ref{fig:J1402p2159} & I \\
J1421+2824 & SDSS J142114.05+282452.8 & 14 21 14.07 & +28 24 52.2 & 0.539 & 6.347 & F & F & Fig.~\ref{fig:J1421p2824} & C \\
J1443+4725* & SDSS J144318.56+472556.7 & 14 43 18.56 & +47 25 56.7 & 0.705 & 7.166 & S & F & Fig.~\ref{fig:J1443p4725} & C \\
J1505+0326* & SDSS J150506.47+032630.8 & 15 05 06.47 & +03 26 30.8 & 0.408 & 5.438 & G & F & Fig.~\ref{fig:J1505p0326} & C \\
J1536+5433 & Mrk 486 & 15 36 38.30 & +54 33 33.0 & 0.039 & 0.772 & Q & Q & Fig.~\ref{fig:J1536p5433} & I \\
J1537+4942 & SDSS J153732.61+494247.5 & 15 37 32.60 & +49 42 48.0 & 0.280 & 4.244 & Q & Q & Fig.~\ref{fig:J1537p4942} & C \\
J1548+3511 & SDSS J154817.92+351128.0 & 15 48 17.92 & +35 11 28.4 & 0.479 & 5.964 & F & F & Fig.~\ref{fig:J1548p3511} & C \\
J1555+1911 & Mrk 291 & 15 55 07.90 & +19 11 33.0 & 0.035 & 0.697 & Q & Q & Fig.~\ref{fig:J1555p1911} & E \\
J1559+3501 & Mrk 493 & 15 59 09.60 & +35 01 47.0 & 0.031 & 0.620 & Q & Q & Fig.~\ref{fig:J1559p3501} & E \\
J1612+4219 & SDSS J161259.83+421940.3 & 16 12 59.83 & +42 19 40.0 & 0.233 & 3.710 & F & F & Fig.~\ref{fig:J1612p4219} & E \\
J1629+4007 & SDSS J162901.30+400759.9 & 16 29 01.31 & +40 07 59.6 & 0.272 & 4.157 & F & F & Fig.~\ref{fig:J1629p4007} & C \\
J1633+4718 & SDSS J163323.58+471858.9 & 16 33 23.58 & +47 18 59.0 & 0.116 & 2.101 & F & F & Fig.~\ref{fig:J1633p4718} & I \\
J1634+4809 & SDSS J163401.94+480940.2 & 16 34 01.94 & +48 09 40.1 & 0.495 & 6.071 & F & F & Fig.~\ref{fig:J1634p4809} & C \\
J1703+4540 & SDSS J170330.38+454047.1 & 17 03 30.38 & +45 40 47.2 & 0.060 & 1.159 & S & F & Fig.~\ref{fig:J1703p4540} & I \\
J1709+2348 & SDSS J170907.80+234837.7 & 17 09 07.82 & +23 48 38.2 & 0.254 & 3.956 & F & S & Fig.~\ref{fig:J1709p2348} & C \\
J1713+3523 & FBQS J1713+3523 & 17 13 04.48 & +35 23 33.4 & 0.083 & 1.561 & S & S & Fig.~\ref{fig:J1713p3523} & I \\
J2242+2943 & Ark 564 & 22 42 39.30 & +29 43 31.0 & 0.025 & 0.504 & Q & Q & Fig.~\ref{fig:J2242p2943} & E \\
J2314+2243 & RX J2314.9+2243 & 23 14 55.70 & +22 43 25.0 & 0.169 & 2.884 & S & F & Fig.~\ref{fig:J2314p2243} & I \\
\hline
\end{tabular}
}
\tablefoot{Columns: (1) short name; (2) NED alias; (3) right ascension (J2000); (4) declination (J2000); (5) redshift; (6) scale (kpc arcsec$^{-1}$); (7) old classification: Q indicates radio-quiet, S is steep-spectrum radio-loud, F is flat-spectrum radio-loud; G is gamma-ray emitter; (8) updated classification (G sources are included in the F sample); (9) figure showing the radio map; (10) morphology type: C indicates compact, I is intermediate morphology, and E indicates the presence of diffuse emission. The asterisk indicates that the source has been detected at $\gamma$-rays by the \textit{Fermi} satellite.}
\end{table*}

\section{Sources selection}

Our sample is mainly based on the samples of FIRST detected sources (flux density limit $\sim$1 mJy at 1.4 GHz) analyzed by \citet{Foschini15} and \citet{Berton15a}. 
From those samples, we excluded sources already observed in detail at 5 GHz (e.g. PMN J0948+0022, observed repeatedly), and also three already known to have extended emission \citep{Richards15}.
We have however included in our sample a few more objects found in the literature that were claimed to harbor a relativistic jet.
These new sources were Mrk 335 (e.g. \citealp{Gallo13}), IRAS 03450+0055 \citep{Tarchi11}, IRAS 06269-0543 \citep{Moran00}, B2 1111+32 \citep{Komossa06}, SBS 1315+604 \citep{Wadadekar04}, and Ark 564 \citep{Moran00}. 
Among these, only IRAS 03450+0055 was included in the FIRST.
The others were instead detected in the NVSS survey \citep{Condon98}. 
Five more sources were also originally observed, but we did not include them in the sample we will analyze in the following for different reasons. 
SDSS J164200.55+533950.6 and SBS 0748+499 were classified in the literature as NLS1s, but after a more careful analysis of their optical spectra we concluded that they do not belong to this class: they are in fact a regular Seyfert 1 \citep{Puchnarewicz92} and an intermediate-type Seyfert, respectively.
Mrk 739W was observed in the normal way, but resulted in a non-detection. 
Finally, both SDSS J143221.43+322318.5 and RX J1016.7+4210 were not correctly pointed. 
All of our sources have a FWHM(H$\beta$) $<$ 2000 \kms, a ratio [O III]/H$\beta$ $<$ 3, and Fe II multiplets in their optical spectrum. 
The total number of sources is then 74, listed in Tab.~\ref{tab:summary1}. \par

\section{Data analysis}
\label{sec:data}
The data we present in this paper were obtained in the A configuration of the JVLA in several sessions between 2015 July and September. 
The observations were carried out at 5 GHz with a bandwidth of 2 GHz, for a total observing time of approximately 24 hours. 
The project code is 15A-283 (P.I. Richards). 
The list of sources with their main parameters and the exposure time for each object is reported in Tab.~\ref{tab:source}. \par
\begin{figure*}[t!]
\centering
\includegraphics[width=\hsize]{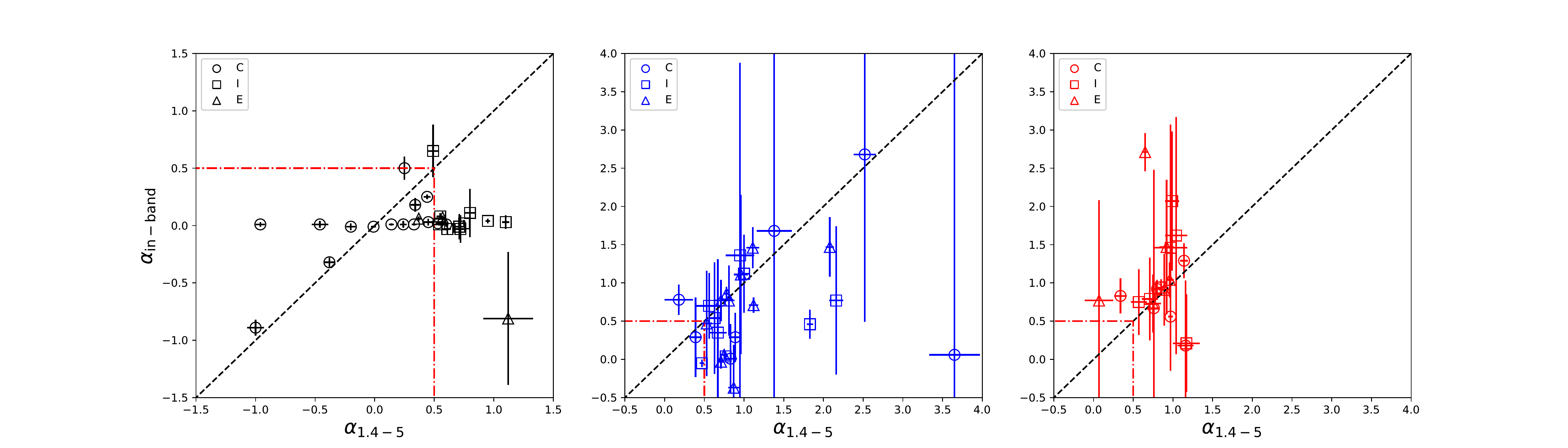} 
\caption{Spectral index between 1.4 and 5 GHz vs in-band spectral index at 5 GHz. From the left, F sample, Q sample, and S sample. The colors indicate the classification, from left to right: black is for F objects,  blue for the Q sample, and red for the S sources. Circles indicate compact sources, squares intermediate sources and triangles extended sources. The dashed black line indicates the 1:1 ratio, the dashed-dotted red lines separate the region of flat-spectrum sources ($\alpha < 0.5$) from that of steep-spectrum objects.}
\label{fig:consistency}
\end{figure*}
We reduced and analyzed the data using the Common Astronomy Software Applications (CASA) version 4.7.2 and the standard Expanded VLA (EVLA) data reduction pipeline version 4.7.2. 
For each dataset, a different flux calibrator was used. 
We point out here that one of the main calibrators, 3C 286, is a S-NLS1 partially resolved on VLA scales \citep{An17, Berton17a}. \par
We split off the measurement set of each object from the main datasets, averaging over the $64$ channels of each one of the 16 spectral windows (centered on 4.271, 4.399, 4.527, 4.655, 4.783, 4.911, 5.039, 5.167, 5.301, 5.429, 5.557, 5.685, 5.813, 5.941, 6.069, 6.197 GHz) and over 10 seconds of exposure time. 
To produce the maps, we used a pixel size of 0.05 arcsec. 
First we analyzed a region of 8192$\times$8192 px centered on the source coordinates that covers most of the primary beam, checking for the presence of nearby sources to avoid contamination from their sidelobes. 
When nearby sources were present we modelled them along with the main target. 
In a few cases, the non perfect modeling of nearby sources severely affected the quality of the maps (e.g. J1612+4219, affected by CRATES J1613+4223). 
Each region was CLEANed using all the available spectral windows and a natural weighting to create a first tentative image.
In three bright sources (J0849+5108, J1505+0326, J1047+4725), we applied a uniform weighting to reduce the sidelobes contribution. 
In J1209+3217, instead, we applied a Briggs weighting intermediate between natural and uniform, to reduce the sidelobe amplitudes without an excessive loss of dynamic range. \par
After the first cleaning, when the noise level was close to the predicted thermal noise of $\sim$10 $\mu$Jy beam$^{-1}$, we did not proceed any further. 
Conversely, for higher noise levels, we proceeded with iterative cycles of phase only self-calibration on the visibilities, in order to improve the dynamic range. 
When possible, we applied to the visibilities a final self-calibration both in amplitude and phase. 
For each source we modelled the beam with a Gaussian in the image plane and we deconvolved it from the core.
It was therefore possible to recover the core deconvolved size, its position angle, the peak and the integrated (i.e. total) flux density.
When the source had multiple components, to measure the total extension of the emission we applied to the visibilities a Gaussian taper with a radius of 120 k$\lambda$, and we derived the total flux of the object. 
We then derived the integrated and peak luminosity (in \ergs) following as usual $L = 4\pi d_L^2 \nu S_\nu (1 + z)^{1 - \alpha_\nu}$, where $\alpha_\nu$ is the spectral index associated to each flux density (in erg s$^{-1}$ cm$^{-2}$ Hz$^{-1}$) and $d_L$ is the luminosity distance.
In particular, we estimated the in-band spectral index measuring the flux of the source in two bands obtained by splitting the data in two separate frequency windows, centered on 4.7 and 5.7 GHz respectively.
To minimize the difference in beam size as a function of frequency, in each source we selected only the uv range common to all spectral windows. 
Finally, to estimate the diffuse luminosity, we calculated the extended flux by subtracting the peak flux from the total flux, and we converted it in luminosity assuming a spectral index of $\alpha_\nu = -1$. 
For sources where the two measurements are in agreement given the errors, we regarded their diffuse luminosity as upper limits.
A dedicated analysis of the extended sources and their tapered images will be presented in a subsequent paper. 
The errors associated with each measurement are those produced by CASA in all cases but the extended emission, where it was estimated as the product of the rms per beam as measured in an empty region of the image times the square root of the object size expressed in beams. \par
We also measured the brightness temperature T$_b$ for each source, which was calculated following \citet{Doi13} and using 
\begin{equation}
T_b = 1.8\times 10^9 (1 + z) \frac{S_{\nu,p}}{\nu^2 \theta_{maj} \theta_{min}} [\mathrm{K}] \; , 
\end{equation}
where $S_{\nu,p}$ is the peak flux density in mJy beam$^{-1}$, $\nu$ is the observing frequency, and $\theta_{maj}$, $\theta_{min}$ are the major and minor axis of the source core in milliarcsec, deconvolved from beam. 
When the latter measurement was not available, we used as beam size the average deconvolved beam size calculated within each sample. \par

\begin{figure}[t!]
\centering
\includegraphics[width=\hsize]{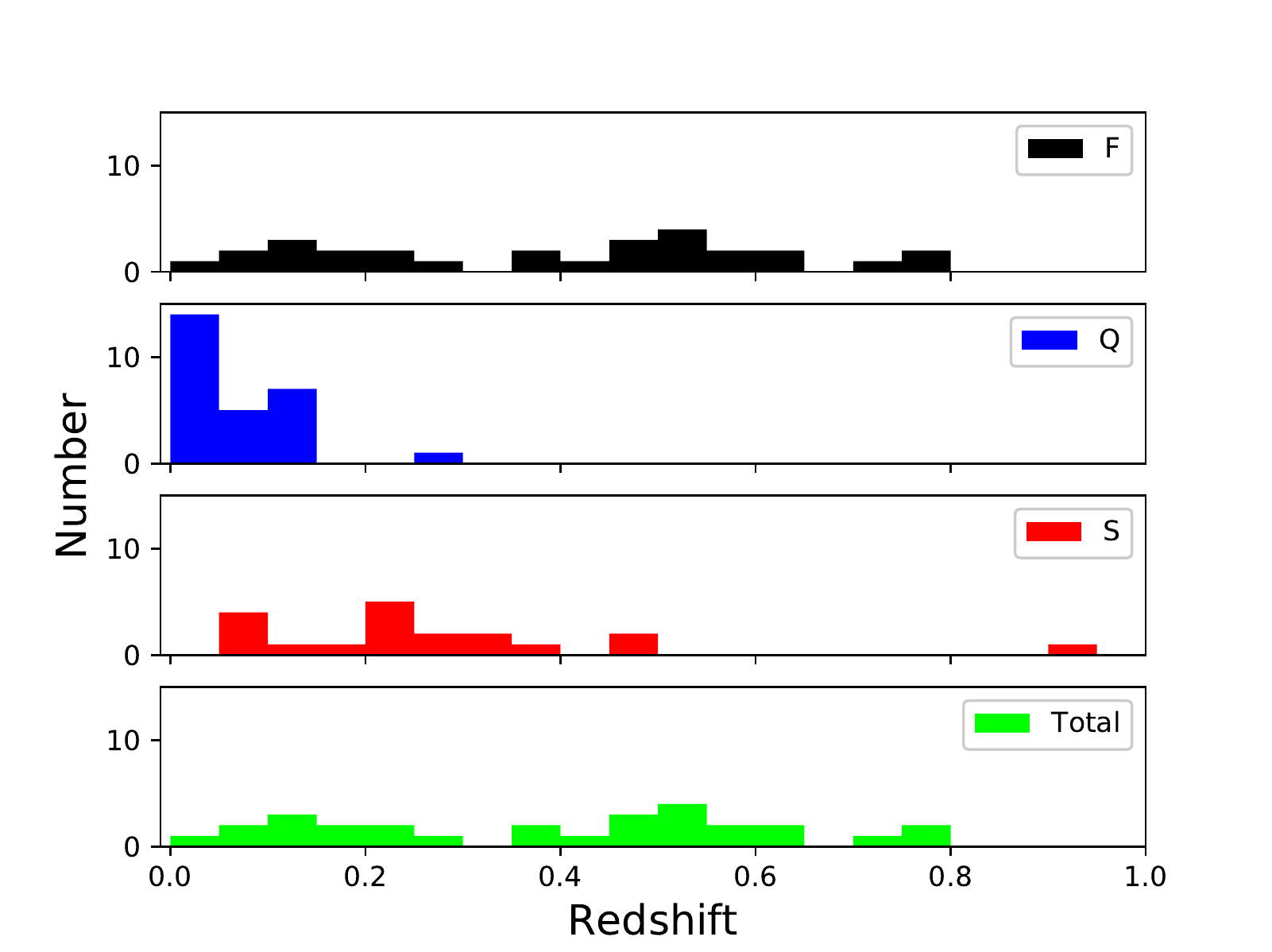} 
\caption{The redshift distribution of the three subsamples and of the entire sample. From top to bottom, F, Q, S subsamples, and the entire sample.}
\label{fig:redshift_distribution}
\end{figure}

\section{Spectral indexes and samples}
\label{sec:spindexes}
During the observations, our sample was initially divided in 4 subsamples, according to their radio and $\gamma$-ray properties, that is $\gamma$-ray emitters, flat-spectrum radio-loud, steep-spectrum radio-loud, and radio-quiet.
However, this classification has to be updated in the light of the results of the present survey. 
We decided to regroup the sources as a function of their radio properties only, regardless of the $\gamma$-ray emission. 
At the time of the original sample creation, indeed, only four of our sources were detected at $\gamma$-rays (marked with G in Col.~7 of Tab.~\ref{tab:summary1}), while to date at least two more have been found \citep{Liao15}. 
Furthermore, since radio emission can change in time because of source variability, we tried to avoid a classification based on non-simultaneous observations, hence relying primarily on in-band spectral indexes (Tab.~\ref{tab:spind}, see Sect.~\ref{sec:spindexes}). \par
The spectral indexes between two frequencies were derived modelling the spectrum as a power law following the usual definition:
\begin{equation}
\alpha_{1,2} = -\frac{\log_{10} \left(\frac{S_1}{S_2}\right)}{\log_{10} \left(\frac{\nu_1}{\nu_2}\right)}
\end{equation}
For all of our sources we derived the in-band spectral index of the peak flux $\alpha_{p}$, the in-band spectral index $\alpha_{in-band}$ of the integrated flux, and the spectral index $\alpha_{1.4-5}$ between 1.4 and 5 GHz, using the integrated flux we measured and the FIRST or NVSS survey flux. The results are shown in Tab.~\ref{tab:spind}. \par
Using what we found, the three spectral classes we created are the following.
\begin{itemize}
\item \textbf{F sources:} flat-spectrum radio-loud NLS1s (F-NLS1s) have a flat or inverted radio spectrum around 5 GHz ($\alpha < 0.5$). Initially we examined the in-band spectral index of the total flux $\alpha_{in-band}$. When, including the error bars, the latter satisfies the required criterion $<$0.5, we included the source in this sample. When this was not the case, we compared $\alpha_{in-band}$ with the broad-band spectral index between 1.4 and 5 GHz, $\alpha_{1.4-5}$. We classified the source based on the measurement with the lowest relative error. The total number of objects in this sample is 28, including all but one of the $\gamma$-ray emitters. 
\item \textbf{S sources:} S-NLS1s must have a steep radio spectrum ($\alpha > 0.5$) in our measurements. To be included directly in this sample, $\alpha_{in-band} > 0.5$ within the error bars. Otherwise, in a similar fashion to F objects, we compared it with $\alpha_{1.4-5}$ and chose according to the lowest relative error. With this selection, the number of S objects is 19, including the $\gamma$-ray source J1102+2239 \citep{Foschini15}. Both F-NLS1s and S-NLS1s belong to the class of radio-loud NLS1s (RLNLS1s). 
\item \textbf{Q sources:} radio-quiet NLS1s (RQNLS1s) finally are radio-quiet (R $<10$), but not radio-silent, that is they all have a previous radio detection in FIRST or NVSS. This classification did not change with respect to previous works published in the literature \citep[e.g.][]{Berton15a}. This subsample contains 27 sources. 
\end{itemize}
The redshift distributions of the four subsamples and the whole sample are shown in Fig.~\ref{fig:redshift_distribution}. The median values are 0.437, 0.048, and 0.225 for the F, Q, and S sample, respectively. For the whole sample, the median redshift value is 0.160. \par
As a consistency check, we compared our in-band indexes, calculated over the total flux, with the $\alpha_{1.4-5}$. 
The results of our check are shown in Fig.~\ref{fig:consistency}. 
In the F sample the sources show a widespread distribution of broad-band spectral indexes, while the in-band index is instead often close to 0. 
However, as we already mentioned, the large variability observed in some of our sources might affect this result, since beamed relativistic jet can vary on very short time scales \citep{Foschini15, Kshama17}. 
Our result seems to confirm this impression: the broad-band index derived from non-simultaneous observations is often unreliable, and very different from the (flat) in-band index, because of this issue. \par
Also radio-quiet NLS1s, at least in the optical, are known to be variable \citep[e.g.][]{Shapovalova12}, but the effect should be less extreme \citep{Rakshit17b}. 
Indeed, in the Q sample, the majority of our sources have in- and broad-band indexes consistent with each other within the errors. 
It is worth noting that error bars, especially those of $\alpha_{in-band}$, are very large. 
This is due to the low luminosity of Q objects, which inevitably affected our measurements. We also highlight that the majority of Q objects lie below the 1:1 ratio. 
The broad-band index seems then to be steeper with respect to the in-band one. 
Finally, the S sample also has very large errors, but on average the two measurements are consistent. \par

\section{Physical properties}
The results of our measurements are shown in Tab.~\ref{tab:dimensioni}, and Tab.~\ref{tab:flussi} . \par
\subsection{Morphology}
\label{sec:morphology}

\begin{figure}[t!]
\centering
\includegraphics[width=\hsize]{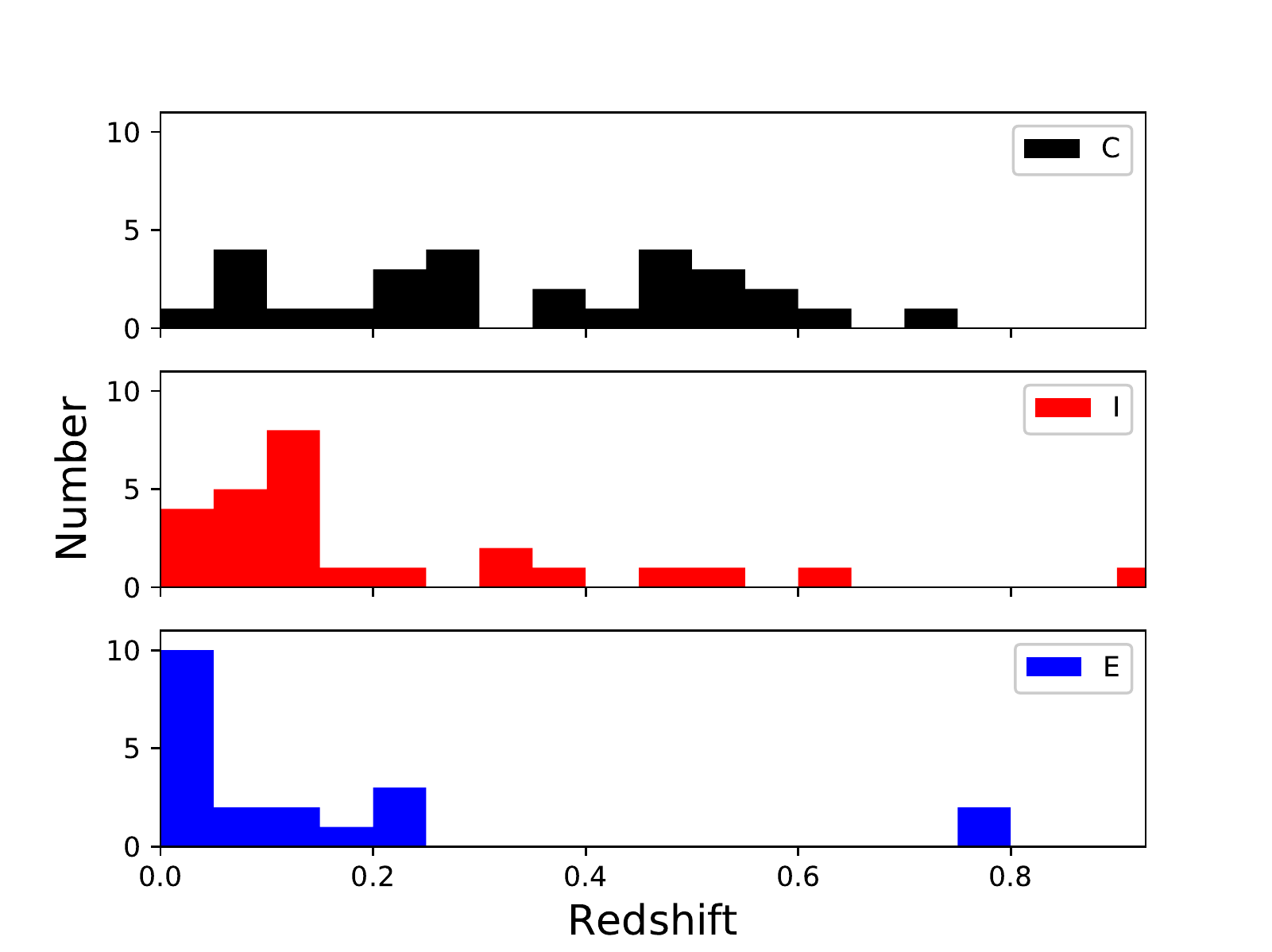} 
\caption{The redshift distribution of the three morphological samples. From top to bottom, C, I, and E sample.}
\label{fig:redshift_cumulative}
\end{figure}

Another result that can be derived from maps is a morphological classification of the sources. We divided them into three classes, according to the ratio between the peak and the total flux density of each source, $\mathcal{R}$ = S$_p$/S$_{int}$:

\begin{itemize}
\item \textbf{Compact:}  labeled as C in the last column of Tab.~\ref{tab:summary1}, are those sources with $\mathcal{R} \geq 0.95$. A typical object is shown in Fig.~\ref{fig:J1114p3241}; 
\item \textbf{Intermediate:} all sources with $\mathcal{R} < 0.95$ and $\mathcal{R} \geq 0.75$, they are labeled as I in Tab.~\ref{tab:summary1}. An example is shown in Fig.~\ref{fig:J1159p2838};
\item \textbf{Extended:} all the remaining sources ($\mathcal{R} < 0.75$). They are labeled as E in Tab.~\ref{tab:summary1}. An example is shown in Fig.~\ref{fig:J1203p4431}.
\end{itemize}

The thresholds were selected in such a way that the classification reflects as much as possible the properties of the sources as seen in the maps.
The morphological distributions of the samples are shown in Tab.~\ref{tab:morphology}. The three morphological classes are almost equally represented in the sample. The plurality is made of compact sources (36\%), followed by a 35\% of I sources, and a 28\% of extended sources, 12 of which are below the nominal equality between compact and extended emission (i.e. 0.5).\footnote{The numbers do not add to 100\% because of rounding.} 
However, the single subsamples significantly differ from each other. Among F sources, more than half are compact, while $\sim$1/3 have intermediate morphology and only 11\% has an extended emission. In the S sample, sources are distributed almost uniformly among the three morphologies, with a numerical predominance of intermediate morphologies. Finally, radio-quiet objects exhibit an opposite behavior with respect to F objects, often having a significant extended emission, and less frequently compact (22\%). \par
\begin{table}
\caption{Morphological distribution of the sources. }
\label{tab:morphology}
\centering
\begin{tabular}{l c c c}
\hline
Sample & C & I & E \\
\hline\hline
Total & 27 (36\%) & 26 (35\%) & 21 (28\%) \\
F & 15 (54\%) & 10 (36\%) & 3 (11\%) \\
Q & 6 (22\%) & 9 (33\%) & 12 (44\%) \\
S & 6 (32\%) & 7 (37\%) & 6 (32\%) \\
\hline\hline
\end{tabular}
\tablefoot{Columns: (1) sample; (2) compact sources; (3) intermediate sources; (4) extended emission.}
\end{table}
The morphological classification seems to have a dependence on the redshift. This result is somewhat expected, since distant sources are less likely to be resolved. As shown in Fig.~\ref{fig:redshift_cumulative}, sources with extended emission are typically concentrated at low z, and the same is true for I sources, although their distribution is slightly more uniform. The number of compact objects instead increases almost linearly with z. However, there are objects which show extended emission even at high z (e.g. the I source J0814+5609, Fig.~\ref{fig:J0814p5609}, z = 0.510), indicating a brightness of the diffuse emission much larger than the average. Objects with such large-scale emission would not be visible at smaller distances because of instrumental limitations. The JVLA A-configuration in fact resolves-out all those structures with a size larger than $\sim$18$^{\prime\prime}$. We cannot exclude then the presence of similar objects at low $z$.
In the following, along with the original classification used to create the sample and based on the spectral indexes, the radio-loudness, and the $\gamma$-ray detection, we will analyze the properties of the three morphological classes which, in principle, are based on a less variable property of the sources.
In particular, the radio-loudness parameter is often considered to be somewhat ambiguous \citep[e.g.][]{Ho01, Kharb14, Foschini17, Padovani17}. 
The spectral index measurements, particularly broad-band ones but also in-band, can be affected by the strong variability of these sources, especially the beamed ones \citep{Foschini15}. 
Therefore, the spectral classification of a source can change in time.
For instance, the flat-spectrum $\gamma$-ray emitter PKS 2004-447, was classified in the past as a compact steep-spectrum source likely because of a different state of activity \citep{Oshlack01, Gallo06}.  
The morphological classification hence is an helpful way around this variability issue. \par

\begin{figure}[t!]
\centering
\includegraphics[width=\hsize]{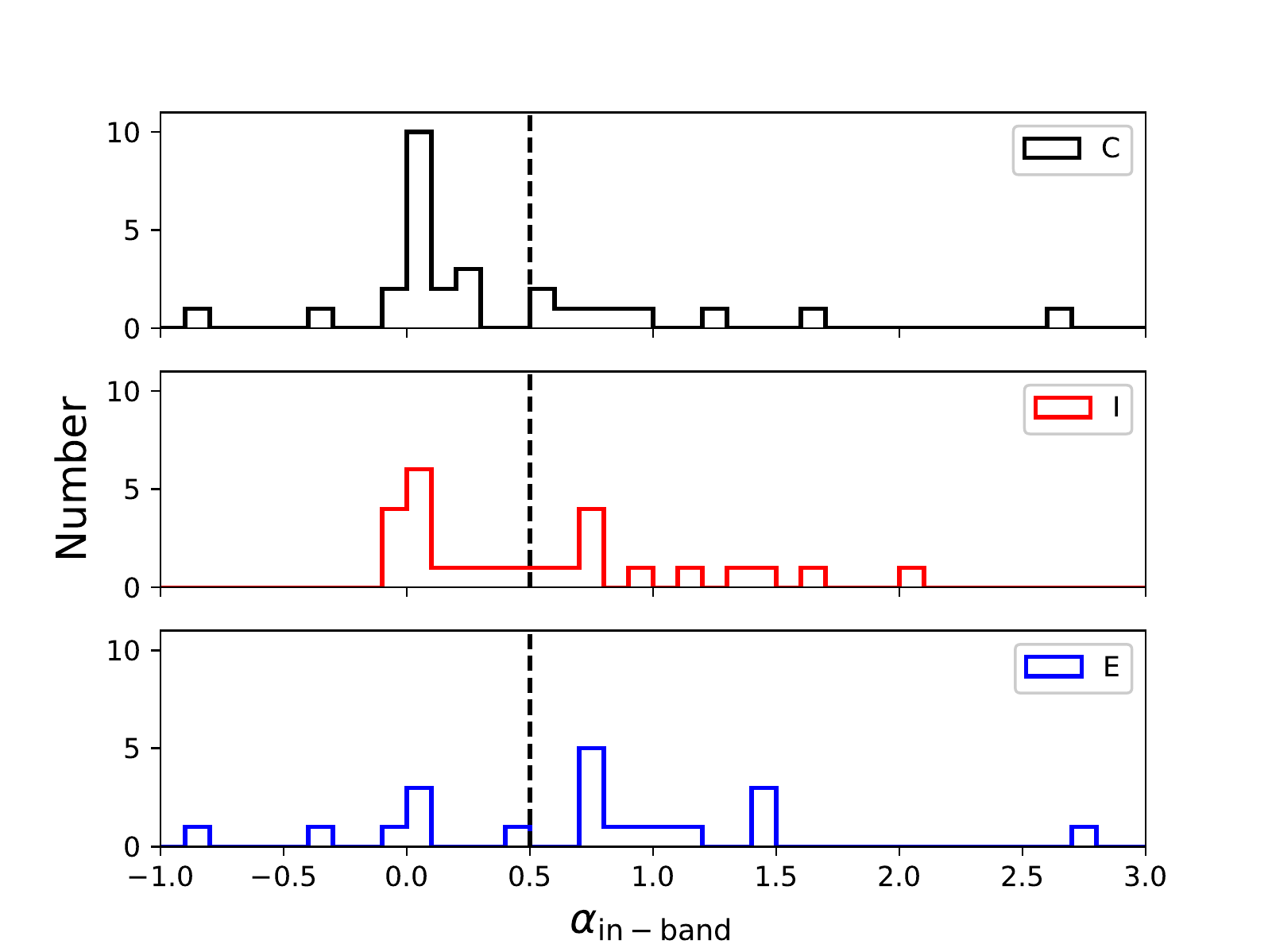} 
\caption{Histogram of the in-band spectral index, calculated on the integrated flux, in the three different morphological classes. From top to bottom, C, I, and E. The vertical dashed line indicates the position of the 0.5 value, which separates flat- from steep-spectrum sources.}
\label{fig:spind_morphology}
\end{figure}

We determined the distribution of spectral indexes as a function of the morphological classification.  
The results are shown in Fig.~\ref{fig:spind_morphology}, and briefly summarized in Tab.~\ref{tab:spind_summary}. 
The data show that the fraction of sources with a steep spectrum is higher in sources with extended emission, while the compact one mainly show a flat spectrum.
Indeed, while compact sources have an average in-band index of 0.36, both intermediate and extended sources are on average above the $\alpha_\nu = 0.5$ threshold.
It is also worth mentioning that the interquartile range (IQR) of the three distribution increases from C to E objects, with the largest spread found in the E sample. \par
To statistically verify the differences between the samples, we decided to use the Kolmogorov-Smirnov (K-S) test, that measures the distance between the cumulative distributions.
The null hypothesis is that two brightness temperature distributions are originated from the same population of sources. 
To reject the null hypothesis, we require a confidence level of 5\%, meaning a p-value lower than 0.05. 
Indeed, the K-S test confirms this. 
The only two samples for which it is possible to reject the null hypothesis are the compact and extended sample (p-value 0.02). 
This therefore indicates that the two distributions have not the same origin, and that, not surprisingly, the extended sources have intrinsically steeper spectra with respect to compact objects. \par

\begin{figure}[t!]
\centering
\includegraphics[width=\hsize]{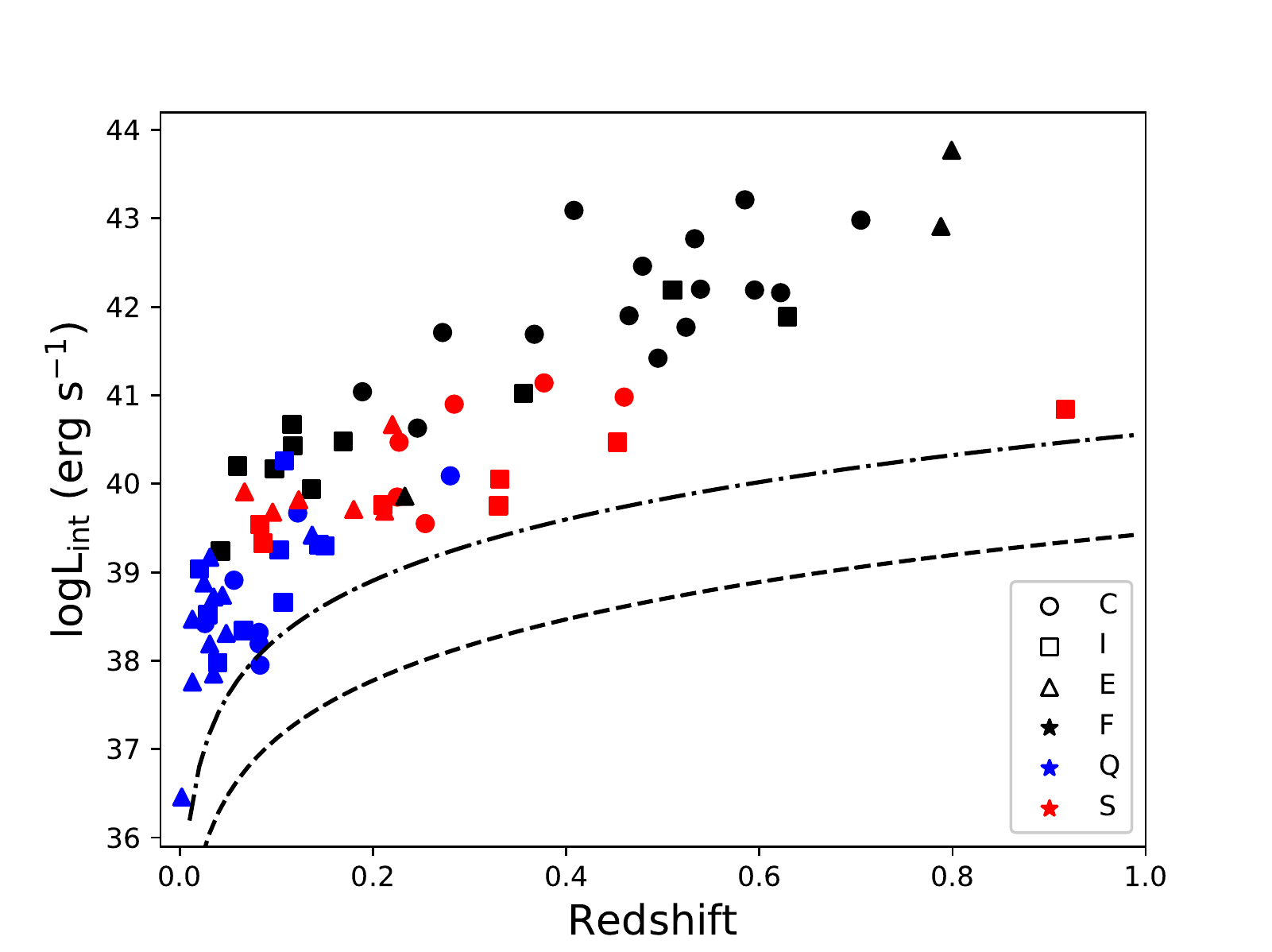} 
\caption{The logarithm of integrated radio luminosity (5 GHz) as a function of redshift. Colors and symbols are the same as in Fig.\,\ref{fig:consistency}. The dashed black line represents the JVLA detection limit for our survey parameters (10 $\mu$Jy, assuming $\alpha_\nu = -1$), the dashed-dotted line represents the FIRST survey limit (1 mJy, same spectral index as before).}
\label{fig:redshift_luminosity}
\end{figure}

\begin{table}
\caption{Statistics of the in-band spectral index for each morphological class.}
\label{tab:spind_summary}
\centering
\begin{tabular}{l c c c c c c c}
\hline
Sample & Mean & Median & Std. dev. & IQR & N & N (\%) \\
\hline\hline
C & 0.36 & 0.12 & 0.67 & 0.58 & 8 & 29\% \\
I & 0.54 & 0.41 & 0.58 & 0.75 & 12 & 46\% \\
E & 0.71 & 0.77 & 0.75 & 0.99 & 13 & 65\% \\
\hline\hline
\end{tabular}
\tablefoot{Columns: (1) sample; (2) mean in-band spectral index; (3) median spectral index; (4) standard deviation; (5) interquartile range; (6) number of objects with $\alpha_{in-band} > 0.5$; (7) percentage of steep-spectrum sources.}
\end{table}

\subsection{Luminosity}
\label{sec:luminosity}
The integrated luminosity of our sources (expressed as $\nu$L$_\nu$) ranges between 10$^{36}$ and 10$^{44}$ \ergs, increasing from radio-quiet to radio-loud objects as shown in Fig.~\ref{fig:redshift_luminosity}. As expected, radio-quiet sources are concentrated at low redshift and lower luminosities, while radio-loud sources are predominant at larger distances. The logarithm of the mean total luminosity for Q sources and the standard deviation of the distribution are (38.67$\pm$0.77) \ergs, while it is (40.11$\pm$0.55) \ergs\ and (41.57$\pm$1.15) \ergs\ for S and F sources, respectively. 
\begin{figure}[t!]
\centering
\includegraphics[width=\hsize]{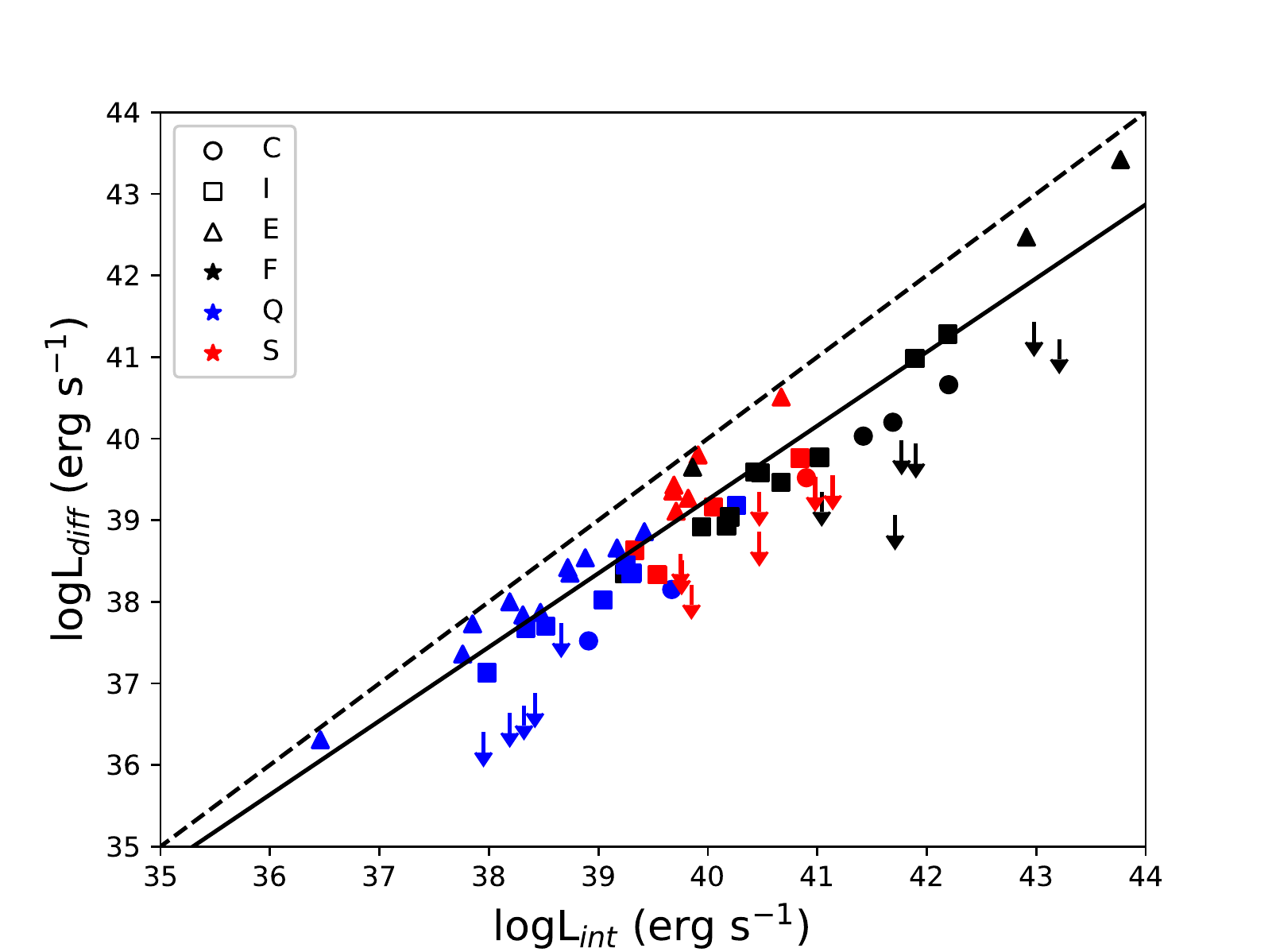} 
\caption{Logarithm of the diffuse luminosity as a function of the logarithm of the total integrated luminosity. The black dashed line is the 1:1 ratio, the black solid line is the linear best-fit of the points. Arrows indicate upper limits. Colors and symbols as in Fig.~\ref{fig:consistency}.}
\label{fig:integrated_diffuse}
\end{figure}

In terms of peak luminosity, the results are very similar for radio-loud sources, with mean values (41.52$\pm$1.16) \ergs, and (39.99$\pm$0.60) \ergs\ for F and S samples. The largest difference is observed in Q sources, in which the mean peak luminosity is (38.53$\pm$0.87) \ergs, which is $\sim$0.2 dex smaller than the integrated luminosity. This is in agreement with the observed large number of sources with extended emission in the Q sample. \par
If we limit our analysis only to the 6 $\gamma$-ray sources, the luminosity appears to be the highest, with a mean value of (42.39$\pm$1.00) \ergs. Their peak luminosity, given their typically compact structure, is very similar, 42.34$\pm$0.99 \ergs, and is also the highest of the sample. It is worth noting that the four out of the five most luminous sources in the samples are all $\gamma$-ray emitters. \par
A relatively large fraction of F and S sources also shows extended emission. The best-fit line shown in Fig.~\ref{fig:integrated_diffuse}, calculated excluding the upper limits, has a slope of 0.90$\pm$0.01, meaning that the diffuse luminosity, i.e. the difference between integrated and peak luminosity, is larger in brighter (radio-loud) sources than in fainter (radio-quiet) sources. Q sources in fact lie close to the 1:1 ratio between total and diffuse emission while, in our sample, the highest is the integrated luminosity, the smallest becomes the contribution of the extended emission with respect to the total luminosity.

\subsection{Brightness temperature}
\begin{figure}[t!]
\centering
\includegraphics[width=\hsize]{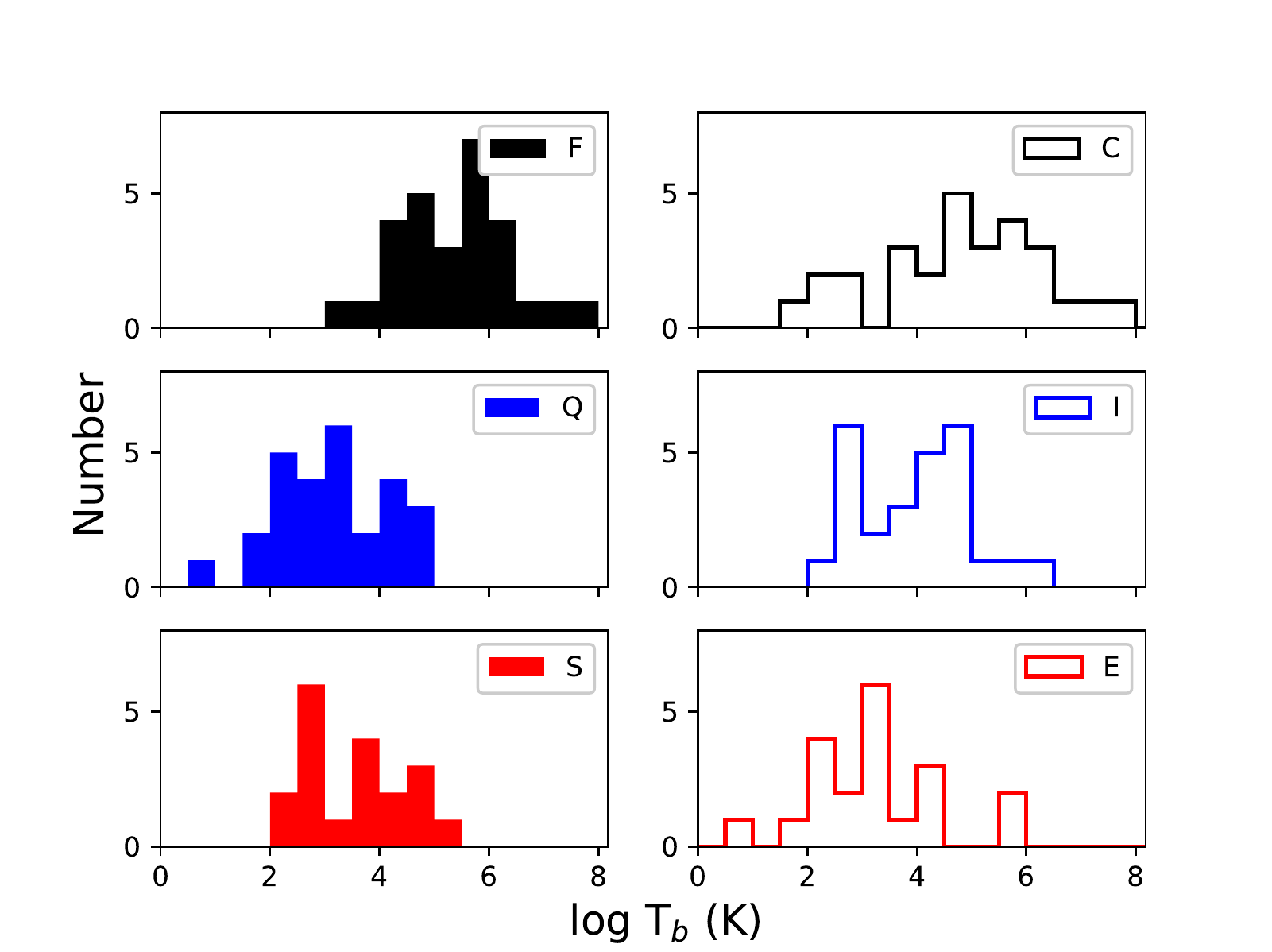} 
\caption{Brightness temperature distributions in the different samples. Left column, from top to bottom, F, Q, and S samples. Right column, from top to bottom, C, I, and E morphology. We highlight that, as mentioned in the text, the T$_b$ values are only lower limits.}
\label{fig:tb_distribution}
\end{figure}

\begin{table}
\caption{Statistics of the brightness temperature in the different samples.}
\label{tab:tb_summary}
\centering
\begin{tabular}{l c c c c c c c}
\hline
Sample & Mean & Median & Std. dev. & IQR \\
\hline\hline
F & 5.33 & 5.43 & 1.00 & 1.27 \\
Q & 3.15 & 3.32 & 0.98 & 1.40 \\
S & 3.51 & 3.57 & 0.88 & 1.37 \\
C & 4.78 & 4.77 & 1.49 & 2.20 \\
I & 3.96 & 4.08 & 1.01 & 1.65 \\
E & 3.21 & 3.18 & 1.10 & 1.38 \\
\hline\hline
\end{tabular}
\tablefoot{Columns: (1) sample; (2) mean brightness temperature; (3) median brightness temperature; (4) standard deviation; (5) interquartile range.}
\end{table}
The brightness temperature is usually calculated using very high resolution observations.
Therefore, our brightness temperature measurements are all lower limits, because the sizes measured from the deconvolution are only upper limits to the real size of the core. 
However, our measurements are comparable with each other, therefore this parameter can provide some useful insights even in our case. 
We show the distribution of brightness temperatures as a function of the sample and the morphology in Fig.~\ref{fig:tb_distribution}, and we summarize the statistics of each sample in Tab.~\ref{tab:tb_summary}. 
In those sources where the core size could not be determined, we assumed a deconvolved size of the core equal to the average size of each sample. 
The mean logarithmic value and standard deviation for each one of our samples are 5.33$\pm$1.00, 3.15$\pm$0.98, and 3.51$\pm$0.98, for F, Q, and S samples, respectively. 
The value for F sources is significantly different from those of the other two samples.
The different morphological groups instead have mean values 4.78$\pm$1.49,3.96$\pm$1.01, and 3.21$\pm$1.10 for compact, intermediate, and extended sources, respectively. 
The differences between these values seem to be less pronounced that those found using the spectral classification. \par
We tested the possibility that the difference in brightness temperature is connected with redshift. The Spearman rank correlation coefficient between these two quantities is 0.52, with a p-value of 2.0$\times$10$^{-6}$. This indicates only a trend between these two quantities, and suggests that even if the redshift has some impact on the brightness temperature, it is not its main driver.\par
In analogy with spectral indexes, we used again the K-S test to verify whether the samples are actually different, although in the case of lower limits the K-S test is suggestive at best.
The null hypothesis, as before, is that the two samples are originated from the same population of sources, and the p-value threshold is still 0.05. 
The K-S test indicates that F-NLS1s are well distinct in terms of brightness temperature from both S-NLS1s and radio-quiet objects. 
The p-values are indeed 2.7$\times$10$^{-5}$ and 3.5$\times$10$^{-8}$, respectively. 
Interestingly, the brightness temperature distributions of S and Q objects might instead be originated from the same population of sources (p-value = 0.46). \par
In terms of morphology, the results are similar, with the K-S test convincingly distinguishing between compact and extended sources (p-value 1$\times$10$^{-4}$). 
Also extended and intermediate objects seem to be different (p-value 0.02), and the same is true for compact and intermediate objects (p-value = 0.04). \par
Finally, it is worth mentioning that the six $\gamma$-ray sources included in the sample have a mean brightness temperature of 5.67$\pm$1.52, which is higher than the other samples (although still consistent within the distribution widths). 
In particular, J0849+5108 and J1505+0326 are the only two sources with a brightness temperature above 10$^7$ K. 

\begin{table}
\caption{Properties of the optical spectra used to calculate the new black hole masses.}
\label{tab:mass_spectra}
\centering
\begin{tabular}{l c c c c}
\hline
Name & Spectrum & Date & Exp. time (s) & $\log{M}$ \\
\hline\hline
J0347+0105 & Asiago & 2013-11-30 & 4800 & 7.33 \\
J0629$-$0545 & Asiago & 2015-12-10 & 6000 & 7.80 \\
J1114+3241 & SDSS & 2005-09-05 & 900 & 6.56 \\
J1317+6010 & SDSS & 2002-05-16 & 900 & 6.59 \\
J2242+2943 & Asiago & 2015-08-06 & 18000 & 6.51 \\
\hline\hline
\end{tabular}
\tablefoot{Columns: (1) source name; (2) observatory; (3) observing date; (4) exposure time (seconds); (5) logarithm of the black hole mass.}
\end{table}

\subsection{Black hole mass}

Another fundamental property characterizing NLS1s is their black hole mass. Almost all the values we used in this analysis are those obtained in the papers by \citet{Foschini15}, \citet{Berton15a}, and \citet{Berton16b}. For J0006+2012 we used the value derived from reverberation mapping calculated by \citet{Grier12}. In five more cases, J0347+0105, J0629$-$0545, J1114+3241, J1317+6010, and J2242+2943, we calculated the black hole mass using either SDSS or new spectra obtained with the Asiago 1.22m telescope (grism 300 mm$^{-1}$, R$\sim$700, summary in Tab.~\ref{tab:mass_spectra}), and following the same procedure already described in detail by \citet{Foschini15}. We briefly summarize it here. We corrected the spectra for redshift and Galactic absorption ($N_H$ values from \citealp{Kalberla05}), and we subtracted the continuum and the Fe II multiplets \citep{Kovacevic10, Shapovalova12}. We reproduced the H$\beta$ profile using three Gaussians, two to represent the broad component and one to represent the narrow. We estimated then the second-order moment of the whole broad H$\beta$, obtained by adding up the two broad Gaussians, as a proxy for gas velocity. The BLR radius was calculated from the total H$\beta$ flux following \citet{Greene10}. Assuming that the gas is virialized and that the scaling factor is equal to 3.85 \citep{Peterson04, Collin06}, we derived the black hole mass. \par
\begin{figure}[t!]
\centering
\includegraphics[width=\hsize]{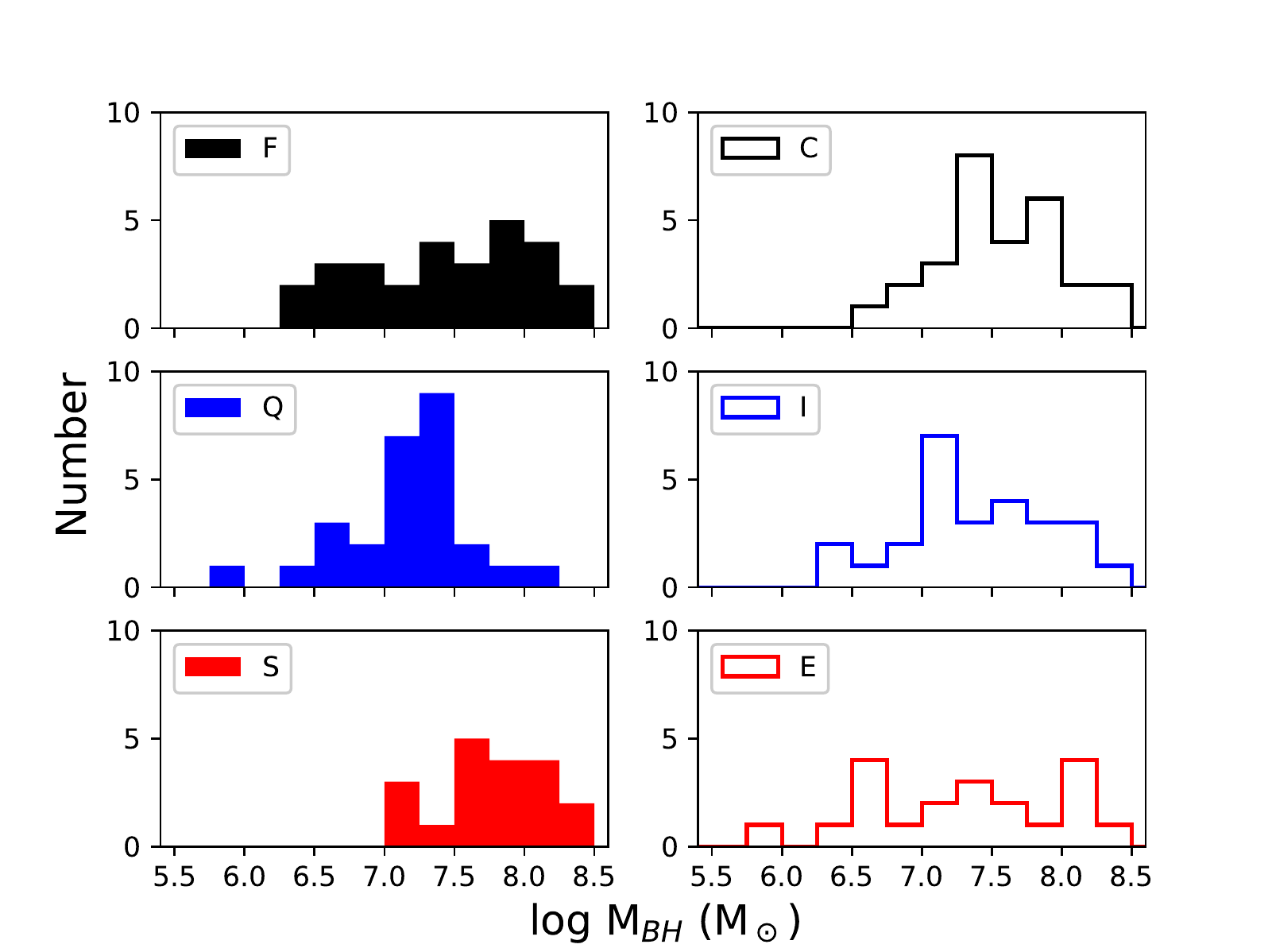} 
\caption{Black hole mass distributions in the different samples. Left column, from top to bottom, F, Q, and S sample. Right column, from top to bottom, C, I, and E morphology.}
\label{fig:mass_distribution}
\end{figure}
\begin{table}
\caption{Statistics of the brightness temperature in the different samples.}
\label{tab:mass_summary}
\centering
\begin{tabular}{l c c c c c c c}
\hline
Sample & Mean & Median & Std. dev. & IQR \\
\hline\hline
F & 7.46 & 7.56 & 0.61 & 1.06 \\
Q & 7.13 & 7.19 & 0.46 & 0.44 \\
S & 7.76 & 7.84 & 0.41 & 0.52 \\
C & 7.56 & 7.53 & 0.45 & 0.62 \\
I & 7.38 & 7.35 & 0.53 & 0.69 \\
E & 7.28 & 7.35 & 0.7 & 1.24 \\
\hline\hline
\end{tabular}
\tablefoot{Columns: (1) sample; (2) mean black hole mass; (3) median black hole mass; (4) standard deviation; (5) interquartile range.}
\end{table}

The resulting distribution is showed in Fig.~\ref{fig:mass_distribution} and summarized in Tab.~\ref{tab:mass_summary}. In terms of spectral classification, the mean logarithmic black hole mass values (in M$_\odot$ units) and the standard deviation in each sample are 7.46$\pm$0.61, 7.13$\pm$0.46, and 7.76$\pm$0.41 for F, Q, and S sample, respectively. As done before, we tested the different distributions by means of the K-S test. The null hypothesis in this case can be rejected when the Q sample is compared with both the F (p-value = 0.02) and the S sample (p-value = 9$\times$10$^{-5}$). Conversely, it cannot be rejected when F and S samples are compared (p-value = 0.25). This therefore suggests that, while F and S mass distributions might be drawn from the same population, Q sources could be different in terms of black hole mass.  \par
From a morphological point of view, we did not find any significant difference between the samples. The mean logarithmic values of the black hole masses in this case are 7.55$\pm$0.45, 7.38$\pm$0.53, and 7.28$\pm$0.70 for C, I, and E sources, respectively. Unlike before, these values are very close to each other. The K-S test indeed does not allows us to reject the null hypothesis (p-values 0.19 C-I, 0.16 C-E, 0.50 E-I), suggesting that the black hole mass does not have a strong impact on the source morphology. \par
Lastly, the six $\gamma$-ray sources have a median logarithmic mass of 7.72, with a standard deviation of 0.40 inside the sample. This value is not significantly different neither from that of F sources, nor from that of S sources, as confirmed by the K-S test (p-values 0.46 and 0.76, respectively).

\section{Discussion}
\label{sec:discussion}

\subsection{Origin of the radio emission}

The first aspect that we highlight is that each one of the three spectral classes seem to have their own, particular, physical properties. F-NLS1s are typically compact, very bright in the radio and, as a consequence, they typically have a fairly high brightness temperature. 
In particular, the $\gamma$-ray sources, which are almost all concentrated in this class, are the most extreme objects in terms of the brightness temperatures. S-NLS1s are instead somewhat intermediate sources, which typically neither are compact nor show an extended emission. 
Their luminosity is on average lower than that of F-NLS1s, and the same is true for their brightness temperatures. 
Both radio-loud classes can be found up to high $z$. Obviously, given their higher luminosity, F sources are most common at larger distances, but the farthest source, J1348+2622 (z = 0.966), belongs to the S class. RQNLS1s instead seem to be rather different from both the other classes. Firstly, they are located at lower $z$, but this is very likely a selection effect. Faint NLS1s were not included in the FIRST because of their weak emission below the detection limit of the survey, and consequently they were excluded from our survey. Furthermore, radio-quiet NLS1s often exhibit large scale structures and they have a relatively low brightness temperature. \par
This rather clear separation between the samples might be interpreted in terms of different origin of the radio emission and incidence of relativistic beaming. F-NLS1s, in analogy with the other classes of blazars, are observed inside their beaming cone, and their radio emission is due to optically thick synchrotron radiation. The relativistic beaming enhances greatly both their brightness temperature and their luminosity, making them visible at large distances. S-NLS1s are instead observed at larger angles, therefore the effect of beaming is less evident, and the radio spectrum becomes dominated by the optically thin sinchrotron radiation of the jet. This scenario is nothing new, but it confirms what we already knew of these objects. \par
In radio-quiet objects the situation is instead less clear. Their brightness temperature distribution and their spectral indexes are similar to those of S-NLS1s, thus suggesting that the presence of a misaligned relativistic jet cannot be ruled out. Some examples are already known. J0952-0136 harbors a pc-scale relativistic jets observed with the VLBA \citep{Doi15}, and our map reveals indeed a faint diffuse emission around its core. Its brightness temperature is the highest in the Q sample ($\sim$8.3$\times$10$^4$ K), a value which is also above the median of the S sample. There is at least another source with similar brightness temperature and morphology, J1337+2423, which in our map (see Fig.~\ref{fig:J1337p2423}) shows a marginally resolved structure extended for $\sim$7 kpc. No high-resolution images are available for this source, but it could represent another example of a jetted RQNLS1. \par
In several cases instead the low radio luminosity of RQNLS1s is associated with a very low brightness temperature. This might indicate that in such sources the origin of the radio emission is not a misaligned relativistic jet, but instead star formation. NLS1s are well-known to be objects with a very high circumnuclear star formation rate \citep{Sani10}, possibly due to the fact that these sources are young or rejuvenated by galaxy mergers \citep{Mathur00, Mathur12, Leontavares14}. The same appears to be true in many RLNLS1s: \citet{Caccianiga15} estimated the q22 parameter, defined as
\begin{equation}
q22 = \log\frac{S_{22 \; \mu m}}{S_{1.4 \; GHz}}
\end{equation}

where S$_{22\mu m}$ is the flux density at 22$\mu$m derived from the WISE colors, and S$_{1.4\;GHz}$ is the flux density at 1.4 GHz. A high q22 (e.g. $\sim$1), when associated with a high level of star formation rate, might suggest that a large fraction of the radio emission could be attributed to the ongoing star formation. Many of the radio-loud sources common to their and our sample that have a high q22, also have a low brightness temperature, seemingly confirming this hypothesis.

\subsection{Comparison to other surveys}

Our survey is the largest survey aimed at imaging the radio continuum emission in NLS1 carried out at this frequency so far.
\citet{Richards15} performed a similar study at 7.6 GHz on three objects. They randomly selected three sources in a sample of objects with positive declination and centimeter-band flux density above 30 mJy, and found kpc-scale emission in all of them. In our study we found several sources showing extended emission, and indeed their number is fairly high. As discussed in Sect.~\ref{sec:morphology}, among our 47 RLNLS1s 26 show an intermediate or extended morphology ($\sim$55\% of the total). Unlike them, though, we do not find many objects clearly showing a fully developed, large scale one-sided relativistic jet. Only a few objects, like J0814+5609 or J1110+3653 have similar morphology and projected size to those sources. Both of them anyway show a flat radio spectrum, likely because of the overwhelming core contribution. Many of the other objects instead have smaller projected sizes, possibly suggesting a low inclination, as expected from type 1 (and beamed) AGN. \par
At 5 GHz \citet{Gu15} investigated a small sample of 14 RLNLS1 with the VLBA. With the only exception of SDSS J095317.09+283601.5, their sample is completely included in ours. In general, it appears that those sources resolved on VLBA scales show a similar structure on the larger VLA scales. Only one exception can be found, that is J1548+3511. As seen in Fig.~\ref{fig:J1548p3511}, on the VLA scales it is compact, while its VLBA image reveals a well resolved core-jet structure which, given its size ($\sim$350 pc), is entirely inside our unresolved core ($\sim$700$\times$300pc). In other cases, the extended emission they detect is clearly seen also in sources which are strongly compact. An interesting difference is found in J0902+0443, which in their study shows a jet pointing toward North-East, while in our case we found signs of diffuse emission toward South-East (see Sect.~\ref{sec:J0902+0443}), possibly suggesting a change in jet direction. \par
Another small NLS1 survey at 5 GHz was performed by \citet{Ulvestad95} to complete the radio characterization of the sample objects with spectropolarimetric observations from \citet{Goodrich89}.
Only two of the 7 NLS1s observed in this paper are included in our sample: Mrk 766 and Mrk 1239.
The objects were observed with the same VLA configuration we are using but only Mrk 766 was ``partially" resolved.
We observe a similar morphology in Mrk 766 (Fig.~\ref{fig:J1218p2948}), with more diffuse emission, and a well defined extended structure on both sides of the nucleus of Mrk 1239 (Fig~\ref{fig:J0952m0136}), at the same position angle of the emission detected by \citet{Doi13} at 1.7 GHz with the VLBA. \par
Our radio-quiet objects often show significant faint extended emission, located around an unresolved core. \citet{Doi13} investigated seven of these sources, all included in our sample, at 1.7 GHz using VLBA. All of those sources revealed a high brightness temperature, suggesting a non-thermal origin of the radio emission. In our sample, all these sources belong to the I or E class, showing a significant extended emission on kpc scale. On VLBA scale, instead, only three of them show signs of extended emission on 10-pc scale, while the others are compact or not-detected. This indicates that the extended emission is resolved-out by VLBI observations. 

\subsection{F-NLS1s and other blazars}

One of the first observations of blazars with the VLA in A-configuration were carried out at 1.4 GHz by \citet{Antonucci85a}. Despite the much lower sensitivity with respect to that of present-day JVLA, they found that those sources had a powerful kpc-scale extended emission in some cases comparable to that of FR II doubles, thus confirming the unification between blazars and radio galaxies. Subsequent observations with the VLA confirmed this result, suggesting that both BL Lacs and FSRQs can be associated with an FR II-like extended emission \citep{Murphy93, Cooper07, Kharb10}, even though different associations between these classes of objects can be found in the literature \citep{Urry95}. Our maps of F-NLS1s do not show anything comparable. Although some extended emission in some cases is present, only 3 out of 28 F sources clearly show diffuse emission with a luminosity comparable to that of the core, and two of them are in the top-5 of our most luminous sources. Ten of them instead show some diffuse emission, which is however not comparable to that observed in blazars. Similar results, although at different scales, were already found in previous studies \citep{Doi06, Yuan08, Foschini11c, Giroletti11}. \par
This difference in terms of extended emission suggests that F-NLS1s and FSRQs have different physical properties. As already mentioned, one of the possible interpretations regarding the nature of NLS1s is that they are a projection effect due to the lack of Doppler broadening in a disk-like BLR viewed pole-on \citep{Decarli08}. Several authors suggested that this is true for F-NLS1s as well, which would have a black hole mass comparable to that of typical FSRQs \citep{Calderone13}. If this were true, F-NLS1s should show powerful extended emission, just like other FSRQs. Since they do not, we conclude that they are not the same class of sources. \par
Given that the typical luminosity of F-NLS1s is lower than that of FSRQs, it is possible that the former are the low-luminosity version of the latter \citep{Foschini15, Berton16c}. Additionally, since the jet power is theoretically expected to scale with the black hole mass \citep{Heinz03, Foschini14} F-NLS1s might also be low-mass, possibly young, FSRQs that, unlike the latter, are not usually capable of producing radio lobes. In fact, if NLS1s are young objects, their relativistic jets might not have had enough time to develop the extended emission yet \citep{Berton15a}. Conversely, sources such as J0814+5609, J1110+3653, or J1633+4718, might be instead slightly older objects which already managed to form the radio lobes. This young age hypothesis could account for the relative rarity of radio lobes among RLNLS1s, and give RQNLS1s a role to play in the parent population problem (see Sect.~\ref{sec:parent_pop}). \par

\subsection{Parent population}
\label{sec:parent_pop}
Several authors have already suggested a possible link between CSS sources and RLNLS1s \citep{Oshlack01, Gallo06, Komossa06, Gu15, Caccianiga14, Caccianiga17, Berton16c}. A possibility is that CSS are the misaligned counterpart of F-NLS1s (their parent population), with S-NLS1s, in particular those with high luminosity, being a population of type 1 CSS sources \citep{Berton17a}. 
Typical CSS sources indeed have a double-sided morphology \citep[e.g.][]{Saikia03, Dallacasa13, Orienti16}, an indication that their inclination is large like in type 2 AGN.
CSS sources by definition have powerful extended emission, with $\log{P}_{1.4} \geq$ 25 W Hz$^{-1}$. This value corresponds at 1.4 GHz to $\log{(\nu F_\nu)} \geq $ 41.15 \ergs, and using $\alpha_\nu = 1$, at 5 GHz it is $\log{(\nu F_\nu)} \geq $ 40.60 \ergs. 
Furthermore, CSS have a small size (between 1 and 20 kpc) and a convex radio spectrum that peaks in the MHz range and steepens toward higher frequencies \citep{Odea98}. \par
Whilst the small size and convex spectrum at radio wavelengths can be true for NLS1s as well \citep{Caccianiga17}, the condition on luminosity is not always respected. Several RLNLS1s are indeed below that threshold (see Fig.~\ref{fig:redshift_luminosity}), especially because, unlike CSS, they often do not have powerful extended emission. CSS sources indeed on VLBI scales very often exhibit a complex morphology \citep{Dallacasa13}, while RLNLS1s have a simpler core-jet or compact structure \citep{Gu10}. On VLA scale the situation is not different. Around $\sim$33\% of CSS sources are compact, while the others show double or triple morphology, or a core-jet structure \citep{Fanti01}. Among our RLNLS1s the percentage of compact sources is slightly higher ($\sim$44\%), while only a few doubles and core-jet structures are present. \par
RLNLS1s seem also to share several aspects with the low-power CSS studied by \citet{Kunert10a}, particularly those classified as high-excitation radio galaxies (HERGs), which typically show less disturbed morphology with respect to higher-luminosity CSS \citep{Kunert06}. These objects might be short-lived sources, whose activity is switched on and off by disk instabilities \citep{Czerny09}. RLNLS1s might also show intermittent activity due to the same physical mechanism. An aborted jet has often been invoked to explain their radio emission \citep{Ghisellini04}, and the occurrence of this phenomenon was maybe observed in X-rays \citep{Wilkins15}. One of our sources, J1302+1624, is also a possible example of intermittent activity, since its extended emission might be associated with a relic \citep{Congiu17}. \par
A more proper unification might then be sought between these low-luminosity CSS and RLNLS1s. However, another option exists. The flux density of beamed objects is S$_{beam} = \delta^{3-\alpha_\nu}$ S$_{unb}$, where S$_{beam}$ is the observed beamed flux density, S$_{unb}$ is the unbeamed flux density, $\delta$ is the kinematic Doppler factor of the jet, and $\alpha_\nu$ is the intrinsic slope of the jet power-law (that we assume to be $\alpha_\nu = 1$). Knowing that $\delta = [\Gamma(1 - \beta \cos \theta)]^{-1}$, where $\theta$ is the observing angle ($\sim$0$^\circ$ for a blazar), $\beta = v/c$, and $\Gamma$ is the bulk Lorentz factor of the jet. A flux density of $\sim$50 mJy, similar to that of the $\gamma$-ray source J1305+5116, if $\Gamma \sim 10$ \citep{Abdo09c} and $\beta = 0.99$, corresponds to an unbeamed flux of 0.5 mJy, which is below the detection limit of FIRST. If observed at large angles then a bright source like J1305+5116 would appear as radio-silent, hence it could not be included in our sample. Radio-silent objects might therefore harbor hidden relativistic jets, and be part of the parent population. This aspect was already investigated in \citet{Berton16b} by means of the [O III] lines. They found that the number of blue outliers, that is sources with [O III] blue-shifted with respect to their restframe position, is much larger in RLNLS1s than in radio-quiet and radio-silent objects. This suggests interaction between the relativistic jet and the narrow-line region occurring in radio-loud objects, and the lack of it in radio-quiet/silent sources. However, another interpretation could account for this phenomenon. Some authors \citep[e.g.][]{Boroson11, Risaliti11, Bisogni17} affirm that the [O III] emission line can be used as an inclination indicator. In particular \citet{Boroson11} focused on the blueshifted component. If sources with a large blueshift are those observed pole-on, this could mean that the lack of blue outliers among radio-quiet NLS1s could be due to their large viewing angle, instead of the lack of a relativistic jet interacting with the interstellar medium. \par
In this scenario, radio-quiet NLS1s are part of the parent population of F-NLS1s. However, an important property seems to indicate that RQNLS1s and S-NLS1s are very different objects. Their black hole mass distributions, indeed, are significantly different, as confirmed by the K-S test. The possible impact of redshift on this difference is not clear. Low redshift might in fact enable us to see faint sources that, because of the mass scaling, harbor low-mass black holes. However, it is also worth noting that all S sources of our samples have a black hole mass larger than $10^7$ M$_\odot$ regardless of z, while a significant fraction of Q objects lie below this value. To test this difference, we carried out the K-S test limiting the two samples (Q and S) at z $<$ 0.3, and the result does not change: the null hypothesis is rejected again (p-value = 8$\times$10$^{-3}$). \par
Curiously, several F sources also fall below the 10$^7$ M$_\odot$ threshold. A possible interpretation for this might be the presence of a flattened BLR in some objects, that leads to an underestimate of the black hole mass even when using the second-order moment of the permitted line, as it was done in our case. The "real" masses would then be closer to those found for S sources, which on average are 7$\times$10$^7$ M$_\odot$, in the high-mass tail of the NLS1s distribution (but below the typical mass of FSRQs). \par
Finally, the morphology of Q objects appears to be different with respect to radio-loud NLS1s. In Q sources a faint, diffuse emission is present in many cases, their unresolved cores are much fainter, and a larger contribution of star formation to the total emission appears plausible. We cannot exclude, though, that our objects are different only because of a selection effect due to their low redshift. In order to obtain an unbiased comparison, one should study radio-quiet/silent objects at higher redshift, to test whether faint relativistic jets are present and they belong to the parent population of F-NLS1s. \par

\section{Summary}
We have presented the first results of a survey on 74 NLS1s carried out with the JVLA. The sources were divided according to their radio properties into radio-quiet (Q), flat-spectrum radio-loud (F), and steep-spectrum radio-loud (S). Additionally, we classified the sources based on their morphology into compact (C), intermediate (I), and extended (E). We found that the majority of F sources has a compact morphology, a high luminosity, high redshift, and high brightness temperature, signs that these objects are basically low-luminosity blazars, that is objects observed inside their relativistic jet. S objects instead show often intermediate or extended morphology, and lower luminosity and brightness temperature with respect to F objects, although they remain brighter than Q sources. The latter instead often show a faint diffuse emission on kpc scale which surrounds the unresolved core. Their brightness temperature is comparable to that of S sources, but given their different morphology its origin might be found in star formation instead of jet activity. The black hole mass distributions confirms that S sources are part of the F parent population, while the connection between F and Q objects is unclear. We also notice how several shared characteristics seem to indicate that low-luminosity CSS sources, and some classical CSS as well, might also be part of F-NLS1s parent population. To understand the role of Q sources, a dedicated survey at higher redshift is needed, in order to find out whether faint misaligned relativistic jets are present also in this class of objects. An upcoming paper with analyze in detail the morphology and the properties of extended sources, both radio-quiet and radio-loud, in order to compare them with CSS and other AGN classes. 

\begin{acknowledgements}
The authors are deeply grateful to Dr. M. Giroletti for the helpful discussion about the JVLA data reduction. The National Radio Astronomy Observatory is a facility of the National Science Foundation operated under cooperative agreement by Associated Universities, Inc. This paper is based on observations collected with the 1.22m \textit{Galileo} telescope of the Asiago Astrophysical Observatory, operated by the Department of Physics and Astronomy "G. Galilei" of the University of Padova. This work has been partially supported by PRIN INAF 2014 ``Jet and astro-particle physics of gamma-ray blazars'' (P.I. F. Tavecchio). This research has made use of the NASA/IPAC Extragalactic Database (NED) which is operated by the Jet Propulsion Laboratory, California Institute of Technology, under contract with the National Aeronautics and  Space Admistration. 
\end{acknowledgements}

\bibliographystyle{aa}
\bibliography{./biblio}

\begin{thebibliography}{135}
\expandafter\ifx\csname natexlab\endcsname\relax\def\natexlab#1{#1}\fi

\bibitem[{{Abdo} {et~al.}(2009{\natexlab{a}}){Abdo}, {Ackermann}, {Ajello},
  {Axelsson}, {Baldini}, {Ballet}, {Barbiellini}, {Bastieri}, {Battelino}, \&
  {Baughman}}]{Abdo09a}
{Abdo}, A.~A., {Ackermann}, M., {Ajello}, M., {et~al.} 2009{\natexlab{a}},
  \apj, 699, 976

\bibitem[{{Abdo} {et~al.}(2009{\natexlab{b}}){Abdo}, {Ackermann}, {Ajello},
  {Baldini}, {Ballet}, {Barbiellini}, {Bastieri}, {Bechtol}, {Bellazzini}, \&
  {Berenji}}]{Abdo09c}
{Abdo}, A.~A., {Ackermann}, M., {Ajello}, M., {et~al.} 2009{\natexlab{b}},
  \apjl, 707, L142

\bibitem[{{Ackermann} {et~al.}(2015){Ackermann}, {Ajello}, {Atwood}, {Baldini},
  {Ballet}, {Barbiellini}, {Bastieri}, {Becerra Gonzalez}, {Bellazzini},
  {Bissaldi}, {Blandford}, {Bloom}, {Bonino}, {Bottacini}, {Brandt}, {Bregeon},
  {Britto}, {Bruel}, {Buehler}, {Buson}, {Caliandro}, {Cameron}, {Caragiulo},
  {Caraveo}, {Carpenter}, {Casandjian}, {Cavazzuti}, {Cecchi}, {Charles},
  {Chekhtman}, {Cheung}, {Chiang}, {Chiaro}, {Ciprini}, {Claus},
  {Cohen-Tanugi}, {Cominsky}, {Conrad}, {Cutini}, {D'Abrusco}, {D'Ammando}, {de
  Angelis}, {Desiante}, {Digel}, {Di Venere}, {Drell}, {Favuzzi}, {Fegan},
  {Ferrara}, {Finke}, {Focke}, {Franckowiak}, {Fuhrmann}, {Fukazawa},
  {Furniss}, {Fusco}, {Gargano}, {Gasparrini}, {Giglietto}, {Giommi},
  {Giordano}, {Giroletti}, {Glanzman}, {Godfrey}, {Grenier}, {Grove},
  {Guiriec}, {Hewitt}, {Hill}, {Horan}, {Itoh}, {J{\'o}hannesson}, {Johnson},
  {Johnson}, {Kataoka}, {Kawano}, {Krauss}, {Kuss}, {La Mura}, {Larsson},
  {Latronico}, {Leto}, {Li}, {Li}, {Longo}, {Loparco}, {Lott}, {Lovellette},
  {Lubrano}, {Madejski}, {Mayer}, {Mazziotta}, {McEnery}, {Michelson},
  {Mizuno}, {Moiseev}, {Monzani}, {Morselli}, {Moskalenko}, {Murgia}, {Nuss},
  {Ohno}, {Ohsugi}, {Ojha}, {Omodei}, {Orienti}, {Orlando}, {Paggi}, {Paneque},
  {Perkins}, {Pesce-Rollins}, {Piron}, {Pivato}, {Porter}, {Rain{\`o}},
  {Rando}, {Razzano}, {Razzaque}, {Reimer}, {Reimer}, {Romani}, {Salvetti},
  {Schaal}, {Schinzel}, {Schulz}, {Sgr{\`o}}, {Siskind}, {Sokolovsky}, {Spada},
  {Spandre}, {Spinelli}, {Stawarz}, {Suson}, {Takahashi}, {Takahashi},
  {Tanaka}, {Thayer}, {Thayer}, {Tibaldo}, {Torres}, {Torresi}, {Tosti},
  {Troja}, {Uchiyama}, {Vianello}, {Winer}, {Wood}, \& {Zimmer}}]{Ackermann15}
{Ackermann}, M., {Ajello}, M., {Atwood}, W.~B., {et~al.} 2015, \apj, 810, 14

\bibitem[{{An} {et~al.}(2017){An}, {Lao}, {Zhao}, {Mohan}, {Cheng}, {Cui}, \&
  {Zhang}}]{An17}
{An}, T., {Lao}, B.-Q., {Zhao}, W., {et~al.} 2017, \mnras, 466, 952

\bibitem[{{Angelakis} {et~al.}(2015){Angelakis}, {Fuhrmann}, {Marchili},
  {Foschini}, {Myserlis}, {Karamanavis}, {Komossa}, {Blinov}, {Krichbaum},
  {Sievers}, {Ungerechts}, \& {Zensus}}]{Angelakis15}
{Angelakis}, E., {Fuhrmann}, L., {Marchili}, N., {et~al.} 2015, \aap, 575, A55

\bibitem[{{Antonucci}(1993)}]{Antonucci93}
{Antonucci}, R. 1993, ARA\&A, 31, 473

\bibitem[{{Antonucci} \& {Ulvestad}(1985)}]{Antonucci85a}
{Antonucci}, R.~R.~J. \& {Ulvestad}, J.~S. 1985, \apj, 294, 158

\bibitem[{{Becker} {et~al.}(1991){Becker}, {White}, \& {Edwards}}]{Becker91}
{Becker}, R.~H., {White}, R.~L., \& {Edwards}, A.~L. 1991, \apjs, 75, 1

\bibitem[{{Becker} {et~al.}(1995){Becker}, {White}, \& {Helfand}}]{Becker95}
{Becker}, R.~H., {White}, R.~L., \& {Helfand}, D.~J. 1995, \apj, 450, 559

\bibitem[{{Berton} {et~al.}(2016{\natexlab{a}}){Berton}, {Caccianiga},
  {Foschini}, {Peterson}, {Mathur}, {Terreran}, {Ciroi}, {Congiu}, {Cracco},
  {Frezzato}, {La Mura}, \& {Rafanelli}}]{Berton16c}
{Berton}, M., {Caccianiga}, A., {Foschini}, L., {et~al.} 2016{\natexlab{a}},
  \aap, 591, A98

\bibitem[{{Berton} {et~al.}(2017){Berton}, {Foschini}, {Caccianiga}, {Ciroi},
  {Congiu}, {Cracco}, {Frezzato}, {La Mura}, \& {Rafanelli}}]{Berton17a}
{Berton}, M., {Foschini}, L., {Caccianiga}, A., {et~al.} 2017, Frontiers in
  Astronomy and Space Sciences, 4, 8

\bibitem[{{Berton} {et~al.}(2016{\natexlab{b}}){Berton}, {Foschini}, {Ciroi},
  {Cracco}, {La Mura}, {Di Mille}, \& {Rafanelli}}]{Berton16b}
{Berton}, M., {Foschini}, L., {Ciroi}, S., {et~al.} 2016{\natexlab{b}}, \aap,
  591, A88

\bibitem[{{Berton} {et~al.}(2015){Berton}, {Foschini}, {Ciroi}, {Cracco}, {La
  Mura}, {Lister}, {Mathur}, {Peterson}, {Richards}, \&
  {Rafanelli}}]{Berton15a}
{Berton}, M., {Foschini}, L., {Ciroi}, S., {et~al.} 2015, \aap, 578, A28

\bibitem[{{Bisogni} {et~al.}(2017){Bisogni}, {Marconi}, \&
  {Risaliti}}]{Bisogni17}
{Bisogni}, S., {Marconi}, A., \& {Risaliti}, G. 2017, \mnras, 464, 385

\bibitem[{{Boroson}(2002)}]{Boroson02}
{Boroson}, T.~A. 2002, \apj, 565, 78

\bibitem[{{Boroson}(2011)}]{Boroson11}
{Boroson}, T.~A. 2011, \apjl, 735, L14

\bibitem[{{Boroson} \& {Green}(1992)}]{Boroson92}
{Boroson}, T.~A. \& {Green}, R.~F. 1992, \apjs, 80, 109

\bibitem[{{B{\"o}ttcher} \& {Dermer}(2002)}]{Boettcher02}
{B{\"o}ttcher}, M. \& {Dermer}, C.~D. 2002, \apj, 564, 86

\bibitem[{{Caccianiga} {et~al.}(2014){Caccianiga}, {Ant{\'o}n}, {Ballo},
  {Dallacasa}, {Della Ceca}, {Fanali}, {Foschini}, {Hamilton}, {Kraus},
  {Maccacaro}, {Mack}, {March{\~a}}, {Paulino-Afonso}, {Sani}, \&
  {Severgnini}}]{Caccianiga14}
{Caccianiga}, A., {Ant{\'o}n}, S., {Ballo}, L., {et~al.} 2014, \mnras, 441, 172

\bibitem[{{Caccianiga} {et~al.}(2015){Caccianiga}, {Ant{\'o}n}, {Ballo},
  {Foschini}, {Maccacaro}, {Della Ceca}, {Severgnini}, {March{\~a}}, {Mateos},
  \& {Sani}}]{Caccianiga15}
{Caccianiga}, A., {Ant{\'o}n}, S., {Ballo}, L., {et~al.} 2015, \mnras, 451,
  1795

\bibitem[{{Caccianiga} {et~al.}(2017){Caccianiga}, {Dallacasa}, {Ant{\'o}n},
  {Ballo}, {Berton}, {Mack}, \& {Paulino-Afonso}}]{Caccianiga17}
{Caccianiga}, A., {Dallacasa}, D., {Ant{\'o}n}, S., {et~al.} 2017, \mnras, 464,
  1474

\bibitem[{{Calderone} {et~al.}(2013){Calderone}, {Ghisellini}, {Colpi}, \&
  {Dotti}}]{Calderone13}
{Calderone}, G., {Ghisellini}, G., {Colpi}, M., \& {Dotti}, M. 2013, \mnras,
  431, 210

\bibitem[{{Collin} {et~al.}(2006){Collin}, {Kawaguchi}, {Peterson}, \&
  {Vestergaard}}]{Collin06}
{Collin}, S., {Kawaguchi}, T., {Peterson}, B.~M., \& {Vestergaard}, M. 2006,
  \aap, 456, 75

\bibitem[{{Condon} {et~al.}(1998){Condon}, {Cotton}, {Greisen}, {Yin},
  {Perley}, {Taylor}, \& {Broderick}}]{Condon98}
{Condon}, J.~J., {Cotton}, W.~D., {Greisen}, E.~W., {et~al.} 1998, \aj, 115,
  1693

\bibitem[{{Congiu} {et~al.}(2017{\natexlab{a}}){Congiu}, {Berton}, {Giroletti},
  {Antonucci}, {Caccianiga}, {Kharb}, {Lister}, {Foschini}, {Ciroi}, {Cracco},
  {Frezzato}, {J{\"a}rvel{\"a}}, {La Mura}, {Richards}, \&
  {Rafanelli}}]{Congiu17}
{Congiu}, E., {Berton}, M., {Giroletti}, M., {et~al.} 2017{\natexlab{a}}, \aap,
  603, A32

\bibitem[{{Congiu} {et~al.}(2017{\natexlab{b}}){Congiu}, {Contini}, {Ciroi},
  {Cracco}, {Di Mille}, {Berton}, {Frezzato}, {La Mura}, {Rafanelli}, \&
  {.}}]{Congiu17b}
{Congiu}, E., {Contini}, M., {Ciroi}, S., {et~al.} 2017{\natexlab{b}}, ArXiv
  e-prints [\eprint[arXiv]{1710.01173}]

\bibitem[{{Cooper} {et~al.}(2007){Cooper}, {Lister}, \&
  {Kochanczyk}}]{Cooper07}
{Cooper}, N.~J., {Lister}, M.~L., \& {Kochanczyk}, M.~D. 2007, \apjs, 171, 376

\bibitem[{{Cracco} {et~al.}(2016){Cracco}, {Ciroi}, {Berton}, {Di Mille},
  {Foschini}, {La Mura}, \& {Rafanelli}}]{Cracco16}
{Cracco}, V., {Ciroi}, S., {Berton}, M., {et~al.} 2016, \mnras, 462, 1256

\bibitem[{{Crenshaw} {et~al.}(2003){Crenshaw}, {Kraemer}, \&
  {Gabel}}]{Crenshaw03}
{Crenshaw}, D.~M., {Kraemer}, S.~B., \& {Gabel}, J.~R. 2003, \aj, 126, 1690

\bibitem[{{Czerny} {et~al.}(2009){Czerny}, {Siemiginowska}, {Janiuk},
  {Nikiel-Wroczy{\'n}ski}, \& {Stawarz}}]{Czerny09}
{Czerny}, B., {Siemiginowska}, A., {Janiuk}, A., {Nikiel-Wroczy{\'n}ski}, B.,
  \& {Stawarz}, {\L}. 2009, \apj, 698, 840

\bibitem[{{Dallacasa} {et~al.}(1998){Dallacasa}, {Bondi}, {Alef}, \&
  {Mantovani}}]{Dallacasa98}
{Dallacasa}, D., {Bondi}, M., {Alef}, W., \& {Mantovani}, F. 1998, \aaps, 129,
  219

\bibitem[{{Dallacasa} {et~al.}(2013){Dallacasa}, {Orienti}, {Fanti}, {Fanti},
  \& {Stanghellini}}]{Dallacasa13}
{Dallacasa}, D., {Orienti}, M., {Fanti}, C., {Fanti}, R., \& {Stanghellini}, C.
  2013, \mnras, 433, 147

\bibitem[{{Dallacasa} {et~al.}(2000){Dallacasa}, {Stanghellini}, {Centonza}, \&
  {Fanti}}]{Dallacasa00}
{Dallacasa}, D., {Stanghellini}, C., {Centonza}, M., \& {Fanti}, R. 2000, \aap,
  363, 887

\bibitem[{{D'Ammando} {et~al.}(2016){D'Ammando}, {Orienti}, {Finke}, {Hovatta},
  {Giroletti}, {Max-Moerbeck}, {Pearson}, {Readhead}, {Reeves}, \&
  {Richards}}]{Dammando16}
{D'Ammando}, F., {Orienti}, M., {Finke}, J., {et~al.} 2016, \mnras, 463, 4469

\bibitem[{{D'Ammando} {et~al.}(2012){D'Ammando}, {Orienti}, {Finke}, {Raiteri},
  {Angelakis}, {Fuhrmann}, {Giroletti}, {Hovatta}, {Max-Moerbeck}, {Perkins},
  {Readhead}, {Richards}, {Stawarz}, \& {Donato}}]{Dammando12}
{D'Ammando}, F., {Orienti}, M., {Finke}, J., {et~al.} 2012, \mnras, 426, 317

\bibitem[{{D'Ammando} {et~al.}(2015){D'Ammando}, {Orienti}, {Larsson}, \&
  {Giroletti}}]{Dammando15}
{D'Ammando}, F., {Orienti}, M., {Larsson}, J., \& {Giroletti}, M. 2015, \mnras,
  452, 520

\bibitem[{{Decarli} {et~al.}(2008){Decarli}, {Dotti}, {Fontana}, \&
  {Haardt}}]{Decarli08}
{Decarli}, R., {Dotti}, M., {Fontana}, M., \& {Haardt}, F. 2008, \mnras, 386,
  L15

\bibitem[{{Doi} {et~al.}(2013){Doi}, {Asada}, {Fujisawa}, {Nagai}, {Hagiwara},
  {Wajima}, \& {Inoue}}]{Doi13}
{Doi}, A., {Asada}, K., {Fujisawa}, K., {et~al.} 2013, \apj, 765, 69

\bibitem[{{Doi} {et~al.}(2011){Doi}, {Asada}, \& {Nagai}}]{Doi11}
{Doi}, A., {Asada}, K., \& {Nagai}, H. 2011, \apj, 738, 126

\bibitem[{{Doi} {et~al.}(2007){Doi}, {Fujisawa}, {Inoue}, {Wajima}, {Nagai},
  {Harada}, {Suematsu}, {Habe}, {Honma}, {Kawaguchi}, {Kawai}, {Kobayashi},
  {Koyama}, {Kuboki}, {Murata}, {Omodaka}, {Sorai}, {Sudou}, {Takaba},
  {Takashima}, {Takeda}, {Tamura}, \& {Wakamatsu}}]{Doi07}
{Doi}, A., {Fujisawa}, K., {Inoue}, M., {et~al.} 2007, \pasj, 59, 703

\bibitem[{{Doi} {et~al.}(2006){Doi}, {Nagai}, {Asada}, {Kameno}, {Wajima}, \&
  {Inoue}}]{Doi06}
{Doi}, A., {Nagai}, H., {Asada}, K., {et~al.} 2006, \pasj, 58, 829

\bibitem[{{Doi} {et~al.}(2012){Doi}, {Nagira}, {Kawakatu}, {Kino}, {Nagai}, \&
  {Asada}}]{Doi12}
{Doi}, A., {Nagira}, H., {Kawakatu}, N., {et~al.} 2012, \apj, 760, 41

\bibitem[{{Doi} {et~al.}(2016){Doi}, {Oyama}, {Kono}, {Yamauchi}, {Suzuki},
  {Matsumoto}, \& {Tazaki}}]{Doi16}
{Doi}, A., {Oyama}, T., {Kono}, Y., {et~al.} 2016, \pasj, 68, 73

\bibitem[{{Doi} {et~al.}(2015){Doi}, {Wajima}, {Hagiwara}, \& {Inoue}}]{Doi15}
{Doi}, A., {Wajima}, K., {Hagiwara}, Y., \& {Inoue}, M. 2015, \apjl, 798, L30

\bibitem[{{Douglas} {et~al.}(1996){Douglas}, {Bash}, {Bozyan}, {Torrence}, \&
  {Wolfe}}]{Douglas96}
{Douglas}, J.~N., {Bash}, F.~N., {Bozyan}, F.~A., {Torrence}, G.~W., \&
  {Wolfe}, C. 1996, \aj, 111, 1945

\bibitem[{{Edelson}(1987)}]{Edelson87}
{Edelson}, R.~A. 1987, \apj, 313, 651

\bibitem[{{Fanti} {et~al.}(2001){Fanti}, {Pozzi}, {Dallacasa}, {Fanti},
  {Gregorini}, {Stanghellini}, \& {Vigotti}}]{Fanti01}
{Fanti}, C., {Pozzi}, F., {Dallacasa}, D., {et~al.} 2001, \aap, 369, 380

\bibitem[{{Foschini}(2011)}]{Foschini11}
{Foschini}, L. 2011, in Narrow-Line Seyfert 1 Galaxies and their Place in the
  Universe, Proc. of Science, Vol. NLS1, id. 24

\bibitem[{{Foschini}(2014)}]{Foschini14}
{Foschini}, L. 2014, International Journal of Modern Physics Conference Series,
  28, 1460188

\bibitem[{{Foschini}(2017)}]{Foschini17}
{Foschini}, L. 2017, Frontiers in Astronomy and Space Sciences, 4, 6

\bibitem[{{Foschini} {et~al.}(2015){Foschini}, {Berton}, {Caccianiga}, {Ciroi},
  {Cracco}, {Peterson}, {Angelakis}, {Braito}, {Fuhrmann}, {Gallo}, {Grupe},
  {J{\"a}rvel{\"a}}, {Kaufmann}, {Komossa}, {Kovalev}, {L{\"a}hteenm{\"a}ki},
  {Lisakov}, {Lister}, {Mathur}, {Richards}, {Romano}, {Sievers},
  {Tagliaferri}, {Tammi}, {Tibolla}, {Tornikoski}, {Vercellone}, {La Mura},
  {Maraschi}, \& {Rafanelli}}]{Foschini15}
{Foschini}, L., {Berton}, M., {Caccianiga}, A., {et~al.} 2015, \aap, 575, A13

\bibitem[{{Foschini} {et~al.}(2011){Foschini}, {Ghisellini}, {Kovalev},
  {Lister}, {D'Ammando}, {Thompson}, {Tramacere}, {Angelakis}, {Donato},
  {Falcone}, {Fuhrmann}, {Hauser}, {Kovalev}, {Mannheim}, {Maraschi},
  {Max-Moerbeck}, {Nestoras}, {Pavlidou}, {Pearson}, {Pushkarev}, {Readhead},
  {Richards}, {Stevenson}, {Tagliaferri}, {Tibolla}, {Tavecchio}, \&
  {Wagner}}]{Foschini11c}
{Foschini}, L., {Ghisellini}, G., {Kovalev}, Y.~Y., {et~al.} 2011, \mnras, 413,
  1671

\bibitem[{{Gallo}(2006)}]{Gallo06}
{Gallo}, L.~C. 2006, \mnras, 368, 479

\bibitem[{{Gallo} {et~al.}(2013){Gallo}, {Fabian}, {Grupe}, {Bonson},
  {Komossa}, {Longinotti}, {Miniutti}, {Walton}, {Zoghbi}, \&
  {Mathur}}]{Gallo13}
{Gallo}, L.~C., {Fabian}, A.~C., {Grupe}, D., {et~al.} 2013, \mnras, 428, 1191

\bibitem[{{Ghisellini} {et~al.}(1998){Ghisellini}, {Celotti}, {Fossati},
  {Maraschi}, \& {Comastri}}]{Ghisellini98}
{Ghisellini}, G., {Celotti}, A., {Fossati}, G., {Maraschi}, L., \& {Comastri},
  A. 1998, \mnras, 301, 451

\bibitem[{{Ghisellini} {et~al.}(2004){Ghisellini}, {Haardt}, \&
  {Matt}}]{Ghisellini04}
{Ghisellini}, G., {Haardt}, F., \& {Matt}, G. 2004, \aap, 413, 535

\bibitem[{{Giroletti} \& {Panessa}(2009)}]{Giroletti09}
{Giroletti}, M. \& {Panessa}, F. 2009, \apjl, 706, L260

\bibitem[{{Giroletti} {et~al.}(2011){Giroletti}, {Paragi}, {Bignall}, {Doi},
  {Foschini}, {Gab{\'a}nyi}, {Reynolds}, {Blanchard}, {Campbell}, {Colomer},
  {Hong}, {Kadler}, {Kino}, {van Langevelde}, {Nagai}, {Phillips}, {Sekido},
  {Szomoru}, \& {Tzioumis}}]{Giroletti11}
{Giroletti}, M., {Paragi}, Z., {Bignall}, H., {et~al.} 2011, \aap, 528, L11

\bibitem[{{Gliozzi} {et~al.}(2010){Gliozzi}, {Papadakis}, {Grupe}, {Brinkmann},
  {Raeth}, \& {Kedziora-Chudczer}}]{Gliozzi10}
{Gliozzi}, M., {Papadakis}, I.~E., {Grupe}, D., {et~al.} 2010, \apj, 717, 1243

\bibitem[{{Goodrich}(1989)}]{Goodrich89}
{Goodrich}, R.~W. 1989, \apj, 342, 224

\bibitem[{{Greene} {et~al.}(2010){Greene}, {Hood}, {Barth}, {Bennert}, {Bentz},
  {Filippenko}, {Gates}, {Malkan}, {Treu}, {Walsh}, \& {Woo}}]{Greene10}
{Greene}, J.~E., {Hood}, C.~E., {Barth}, A.~J., {et~al.} 2010, \apj, 723, 409

\bibitem[{{Gregory} \& {Condon}(1991)}]{Gregory91}
{Gregory}, P.~C. \& {Condon}, J.~J. 1991, \apjs, 75, 1011

\bibitem[{{Grier} {et~al.}(2012){Grier}, {Peterson}, {Pogge}, {Denney},
  {Bentz}, {Martini}, {Sergeev}, {Kaspi}, {Minezaki}, {Zu}, {Kochanek},
  {Siverd}, {Shappee}, {Stanek}, {Araya Salvo}, {Beatty}, {Bird}, {Bord},
  {Borman}, {Che}, {Chen}, {Cohen}, {Dietrich}, {Doroshenko}, {Drake},
  {Efimov}, {Free}, {Ginsburg}, {Henderson}, {King}, {Koshida}, {Mogren},
  {Molina}, {Mosquera}, {Nazarov}, {Okhmat}, {Pejcha}, {Rafter}, {Shields},
  {Skowron}, {Szczygiel}, {Valluri}, \& {van Saders}}]{Grier12}
{Grier}, C.~J., {Peterson}, B.~M., {Pogge}, R.~W., {et~al.} 2012, \apj, 755, 60

\bibitem[{{Griffith} {et~al.}(1995){Griffith}, {Wright}, {Burke}, \&
  {Ekers}}]{Griffith95}
{Griffith}, M.~R., {Wright}, A.~E., {Burke}, B.~F., \& {Ekers}, R.~D. 1995,
  \apjs, 97, 347

\bibitem[{{Grupe}(2000)}]{Grupe00}
{Grupe}, D. 2000, New Astron. Rev., 44, 455

\bibitem[{{Gu} \& {Chen}(2010)}]{Gu10}
{Gu}, M. \& {Chen}, Y. 2010, \aj, 139, 2612

\bibitem[{{Gu} {et~al.}(2015){Gu}, {Chen}, {Komossa}, {Yuan}, {Shen}, {Wajima},
  {Zhou}, \& {Zensus}}]{Gu15}
{Gu}, M., {Chen}, Y., {Komossa}, S., {et~al.} 2015, \apjs, 221, 3

\bibitem[{{Gupta} {et~al.}(2013){Gupta}, {Mathur}, {Krongold}, \&
  {Nicastro}}]{Gupta13}
{Gupta}, A., {Mathur}, S., {Krongold}, Y., \& {Nicastro}, F. 2013, \apj, 772,
  66

\bibitem[{{Heinz} \& {Sunyaev}(2003)}]{Heinz03}
{Heinz}, S. \& {Sunyaev}, R.~A. 2003, \mnras, 343, L59

\bibitem[{{Ho} \& {Peng}(2001)}]{Ho01}
{Ho}, L.~C. \& {Peng}, C.~Y. 2001, \apj, 555, 650

\bibitem[{{Ikejiri} {et~al.}(2011){Ikejiri}, {Uemura}, {Sasada}, {Ito},
  {Yamanaka}, {Sakimoto}, {Arai}, {Fukazawa}, {Ohsugi}, {Kawabata}, {Yoshida},
  {Sato}, \& {Kino}}]{Ikejiri11}
{Ikejiri}, Y., {Uemura}, M., {Sasada}, M., {et~al.} 2011, \pasj, 63, 639

\bibitem[{{J{\"a}rvel{\"a}} {et~al.}(2015){J{\"a}rvel{\"a}},
  {L{\"a}hteenm{\"a}ki}, \& {Le{\'o}n-Tavares}}]{Jarvela14}
{J{\"a}rvel{\"a}}, E., {L{\"a}hteenm{\"a}ki}, A., \& {Le{\'o}n-Tavares}, J.
  2015, \aap, 573, A76

\bibitem[{{J{\"a}rvel{\"a}} {et~al.}(2017){J{\"a}rvel{\"a}},
  {L{\"a}hteenm{\"a}ki}, {Lietzen}, {Poudel}, {Hein{\"a}m{\"a}ki}, \&
  {Einasto}}]{Jarvela17}
{J{\"a}rvel{\"a}}, E., {L{\"a}hteenm{\"a}ki}, A., {Lietzen}, H., {et~al.} 2017,
  \aap, 606, A9

\bibitem[{{Kalberla} {et~al.}(2005){Kalberla}, {Burton}, {Hartmann}, {Arnal},
  {Bajaja}, {Morras}, \& {P{\"o}ppel}}]{Kalberla05}
{Kalberla}, P.~M.~W., {Burton}, W.~B., {Hartmann}, D., {et~al.} 2005, \aap,
  440, 775

\bibitem[{{Kataoka} {et~al.}(2011){Kataoka}, {Stawarz}, {Takahashi}, {Cheung},
  {Hayashida}, {Grandi}, {Burnett}, {Celotti}, {Fegan}, {Fortin}, {Maeda},
  {Nakamori}, {Taylor}, {Tosti}, {Digel}, {McConville}, {Finke}, \&
  {D'Ammando}}]{Kataoka11}
{Kataoka}, J., {Stawarz}, {\L}., {Takahashi}, Y., {et~al.} 2011, \apj, 740, 29

\bibitem[{{Kellermann} {et~al.}(1989){Kellermann}, {Sramek}, {Schmidt},
  {Shaffer}, \& {Green}}]{Kellermann89}
{Kellermann}, K.~I., {Sramek}, R., {Schmidt}, M., {Shaffer}, D.~B., \& {Green},
  R. 1989, \aj, 98, 1195

\bibitem[{{Kharb} {et~al.}(2010){Kharb}, {Lister}, \& {Cooper}}]{Kharb10}
{Kharb}, P., {Lister}, M.~L., \& {Cooper}, N.~J. 2010, \apj, 710, 764

\bibitem[{{Kharb} {et~al.}(2014){Kharb}, {O'Dea}, {Baum}, {Hardcastle},
  {Dicken}, {Croston}, {Mingo}, \& {Noel-Storr}}]{Kharb14}
{Kharb}, P., {O'Dea}, C.~P., {Baum}, S.~A., {et~al.} 2014, \mnras, 440, 2976

\bibitem[{{Kinney} {et~al.}(2000){Kinney}, {Schmitt}, {Clarke}, {Pringle},
  {Ulvestad}, \& {Antonucci}}]{Kinney00}
{Kinney}, A.~L., {Schmitt}, H.~R., {Clarke}, C.~J., {et~al.} 2000, \apj, 537,
  152

\bibitem[{{Komatsu} {et~al.}(2011){Komatsu}, {Smith}, {Dunkley}, {Bennett},
  {Gold}, {Hinshaw}, {Jarosik}, {Larson}, {Nolta}, {Page}, {Spergel},
  {Halpern}, {Hill}, {Kogut}, {Limon}, {Meyer}, {Odegard}, {Tucker}, {Weiland},
  {Wollack}, \& {Wright}}]{Komatsu11}
{Komatsu}, E., {Smith}, K.~M., {Dunkley}, J., {et~al.} 2011, \apjs, 192, 18

\bibitem[{{Komossa} {et~al.}(2006){Komossa}, {Voges}, {Xu}, {Mathur}, {Adorf},
  {Lemson}, {Duschl}, \& {Grupe}}]{Komossa06}
{Komossa}, S., {Voges}, W., {Xu}, D., {et~al.} 2006, \aj, 132, 531

\bibitem[{{Komossa} {et~al.}(2015){Komossa}, {Xu}, {Fuhrmann}, {Grupe}, {Yao},
  {Fan}, {Myserlis}, {Angelakis}, {Karamanavis}, {Yuan}, \&
  {Zensus}}]{Komossa15}
{Komossa}, S., {Xu}, D., {Fuhrmann}, L., {et~al.} 2015, \aap, 574, A121

\bibitem[{{Kova{\v c}evi{\'c}} {et~al.}(2010){Kova{\v c}evi{\'c}},
  {Popovi{\'c}}, \& {Dimitrijevi{\'c}}}]{Kovacevic10}
{Kova{\v c}evi{\'c}}, J., {Popovi{\'c}}, L.~{\v C}., \& {Dimitrijevi{\'c}},
  M.~S. 2010, \apjs, 189, 15

\bibitem[{{Kshama} {et~al.}(2017){Kshama}, {Paliya}, \& {Stalin}}]{Kshama17}
{Kshama}, S.~K., {Paliya}, V.~S., \& {Stalin}, C.~S. 2017, \mnras, 466, 2679

\bibitem[{{Kukula} {et~al.}(1998){Kukula}, {Dunlop}, {Hughes}, \&
  {Rawlings}}]{Kukula98}
{Kukula}, M.~J., {Dunlop}, J.~S., {Hughes}, D.~H., \& {Rawlings}, S. 1998,
  \mnras, 297, 366

\bibitem[{{Kunert-Bajraszewska} {et~al.}(2010){Kunert-Bajraszewska},
  {Gawro{\'n}ski}, {Labiano}, \& {Siemiginowska}}]{Kunert10a}
{Kunert-Bajraszewska}, M., {Gawro{\'n}ski}, M.~P., {Labiano}, A., \&
  {Siemiginowska}, A. 2010, \mnras, 408, 2261

\bibitem[{{Kunert-Bajraszewska} {et~al.}(2006){Kunert-Bajraszewska}, {Marecki},
  \& {Thomasson}}]{Kunert06}
{Kunert-Bajraszewska}, M., {Marecki}, A., \& {Thomasson}, P. 2006, \aap, 450,
  945

\bibitem[{{L{\"a}hteenm{\"a}ki} {et~al.}(2017){L{\"a}hteenm{\"a}ki},
  {J{\"a}rvel{\"a}}, {Hovatta}, {Tornikoski}, {Harrison}, {L{\'o}pez-Caniego},
  {Max-Moerbeck}, {Mingaliev}, {Pearson}, {Ramakrishnan}, {Readhead}, {Reeves},
  {Richards}, {Sotnikova}, \& {Tammi}}]{Lahteenmaki17}
{L{\"a}hteenm{\"a}ki}, A., {J{\"a}rvel{\"a}}, E., {Hovatta}, T., {et~al.} 2017,
  \aap, 603, A100

\bibitem[{{Lal} {et~al.}(2011){Lal}, {Shastri}, \& {Gabuzda}}]{Lal11}
{Lal}, D.~V., {Shastri}, P., \& {Gabuzda}, D.~C. 2011, \apj, 731, 68

\bibitem[{{Laor}(2000)}]{Laor00}
{Laor}, A. 2000, \apjl, 543, L111

\bibitem[{{Laurent-Muehleisen} {et~al.}(1997){Laurent-Muehleisen}, {Kollgaard},
  {Ryan}, {Feigelson}, {Brinkmann}, \& {Siebert}}]{Laurent97}
{Laurent-Muehleisen}, S.~A., {Kollgaard}, R.~I., {Ryan}, P.~J., {et~al.} 1997,
  \aaps, 122 [\eprint{astro-ph/9607058}]

\bibitem[{{Leipski} {et~al.}(2006){Leipski}, {Falcke}, {Bennert}, \&
  {H{\"u}ttemeister}}]{Leipski06}
{Leipski}, C., {Falcke}, H., {Bennert}, N., \& {H{\"u}ttemeister}, S. 2006,
  \aap, 455, 161

\bibitem[{{Le{\'o}n Tavares} {et~al.}(2014){Le{\'o}n Tavares}, {Kotilainen},
  {Chavushyan}, {A{\~n}orve}, {Puerari}, {Cruz-Gonz{\'a}lez},
  {Pati{\~n}o-Alvarez}, {Ant{\'o}n}, {Carrami{\~n}ana}, {Carrasco}, {Guichard},
  {Karhunen}, {Olgu{\'{\i}}n-Iglesias}, {Sanghvi}, \& {Valdes}}]{Leontavares14}
{Le{\'o}n Tavares}, J., {Kotilainen}, J., {Chavushyan}, V., {et~al.} 2014,
  \apj, 795, 58

\bibitem[{{Liao} {et~al.}(2015){Liao}, {Liang}, {Weng}, {Berton}, {Gu}, \&
  {Fan}}]{Liao15}
{Liao}, N.-H., {Liang}, Y.-F., {Weng}, S.-S., {et~al.} 2015, ArXiv e-prints
  [\eprint[arXiv]{1510.05584}]

\bibitem[{{Lister} {et~al.}(2013){Lister}, {Aller}, {Aller}, {Homan},
  {Kellermann}, {Kovalev}, {Pushkarev}, {Richards}, {Ros}, \&
  {Savolainen}}]{Lister13}
{Lister}, M.~L., {Aller}, M.~F., {Aller}, H.~D., {et~al.} 2013, \aj, 146, 120

\bibitem[{{Lister} {et~al.}(2016){Lister}, {Aller}, {Aller}, {Homan},
  {Kellermann}, {Kovalev}, {Pushkarev}, {Richards}, {Ros}, \&
  {Savolainen}}]{Lister16}
{Lister}, M.~L., {Aller}, M.~F., {Aller}, H.~D., {et~al.} 2016, \aj, 152, 12

\bibitem[{{Maccarone} {et~al.}(2003){Maccarone}, {Gallo}, \&
  {Fender}}]{Maccarone03}
{Maccarone}, T.~J., {Gallo}, E., \& {Fender}, R. 2003, \mnras, 345, L19

\bibitem[{{Marscher}(2009)}]{Marscher09}
{Marscher}, A.~P. 2009, ArXiv e-prints [\eprint[arXiv]{0909.2576}]

\bibitem[{{Mathur}(2000)}]{Mathur00}
{Mathur}, S. 2000, \mnras, 314, L17

\bibitem[{{Mathur} {et~al.}(2012){Mathur}, {Fields}, {Peterson}, \&
  {Grupe}}]{Mathur12}
{Mathur}, S., {Fields}, D., {Peterson}, B.~M., \& {Grupe}, D. 2012, \apj, 754,
  146

\bibitem[{{Maune} {et~al.}(2014){Maune}, {Eggen}, {Miller}, {Marshall},
  {Readhead}, {Hovatta}, \& {King}}]{Maune14}
{Maune}, J.~D., {Eggen}, J.~R., {Miller}, H.~R., {et~al.} 2014, \apj, 794, 93

\bibitem[{{Moran}(2000)}]{Moran00}
{Moran}, E.~C. 2000, New Astron. Rev., 44, 527

\bibitem[{{Murphy} {et~al.}(1993){Murphy}, {Browne}, \& {Perley}}]{Murphy93}
{Murphy}, D.~W., {Browne}, I.~W.~A., \& {Perley}, R.~A. 1993, \mnras, 264, 298

\bibitem[{{Nelson}(2000)}]{Nelson00}
{Nelson}, C.~H. 2000, \apjl, 544, L91

\bibitem[{{O'Dea}(1998)}]{Odea98}
{O'Dea}, C.~P. 1998, \pasp, 110, 493

\bibitem[{{Orban de Xivry} {et~al.}(2011){Orban de Xivry}, {Davies},
  {Schartmann}, {Komossa}, {Marconi}, {Hicks}, {Engel}, \&
  {Tacconi}}]{OrbandeXivry11}
{Orban de Xivry}, G., {Davies}, R., {Schartmann}, M., {et~al.} 2011, \mnras,
  417, 2721

\bibitem[{{Orienti}(2016)}]{Orienti16}
{Orienti}, M. 2016, Astronomische Nachrichten, 337, 9

\bibitem[{{Orienti} \& {Prieto}(2010)}]{Orienti10}
{Orienti}, M. \& {Prieto}, M.~A. 2010, \mnras, 401, 2599

\bibitem[{{Oshlack} {et~al.}(2001){Oshlack}, {Webster}, \&
  {Whiting}}]{Oshlack01}
{Oshlack}, A.~Y.~K.~N., {Webster}, R.~L., \& {Whiting}, M.~T. 2001, \apj, 558,
  578

\bibitem[{{Osterbrock} \& {Pogge}(1985)}]{Osterbrock85}
{Osterbrock}, D.~E. \& {Pogge}, R.~W. 1985, \apj, 297, 166

\bibitem[{{Padovani}(2017)}]{Padovani17}
{Padovani}, P. 2017, Nature Astronomy, 1, 0194

\bibitem[{{Peterson} {et~al.}(2004){Peterson}, {Ferrarese}, {Gilbert}, {Kaspi},
  {Malkan}, {Maoz}, {Merritt}, {Netzer}, {Onken}, {Pogge}, {Vestergaard}, \&
  {Wandel}}]{Peterson04}
{Peterson}, B.~M., {Ferrarese}, L., {Gilbert}, K.~M., {et~al.} 2004, \apj, 613,
  682

\bibitem[{{Puchnarewicz} {et~al.}(1992){Puchnarewicz}, {Mason}, {Cordova},
  {Kartje}, {Brabduardi-A A Puchnarewicz}, {Mason}, {Cordova}, {Kartje},
  {Branduardi-Raymont}, {Mittaz}, {Murdin}, \&
  {Allington-Smith}}]{Puchnarewicz92}
{Puchnarewicz}, E.~M., {Mason}, K.~O., {Cordova}, F.~A., {et~al.} 1992, \mnras,
  256, 589

\bibitem[{{Rakshit} \& {Stalin}(2017)}]{Rakshit17b}
{Rakshit}, S. \& {Stalin}, C.~S. 2017, \apj, 842, 96

\bibitem[{{Richards} \& {Lister}(2015)}]{Richards15}
{Richards}, J.~L. \& {Lister}, M.~L. 2015, \apjl, 800, L8

\bibitem[{{Risaliti} {et~al.}(2011){Risaliti}, {Salvati}, \&
  {Marconi}}]{Risaliti11}
{Risaliti}, G., {Salvati}, M., \& {Marconi}, A. 2011, \mnras, 411, 2223

\bibitem[{{Saikia} {et~al.}(2003){Saikia}, {Jeyakumar}, {Mantovani}, {Salter},
  {Spencer}, {Thomasson}, \& {Wiita}}]{Saikia03}
{Saikia}, D.~J., {Jeyakumar}, S., {Mantovani}, F., {et~al.} 2003, \pasa, 20, 50

\bibitem[{{Sani} {et~al.}(2010){Sani}, {Lutz}, {Risaliti}, {Netzer}, {Gallo},
  {Trakhtenbrot}, {Sturm}, \& {Boller}}]{Sani10}
{Sani}, E., {Lutz}, D., {Risaliti}, G., {et~al.} 2010, \mnras, 403, 1246

\bibitem[{{Schmitt} {et~al.}(2001){Schmitt}, {Ulvestad}, {Antonucci}, \&
  {Kinney}}]{Schmitt01}
{Schmitt}, H.~R., {Ulvestad}, J.~S., {Antonucci}, R.~R.~J., \& {Kinney}, A.~L.
  2001, \apjs, 132, 199

\bibitem[{{Sergeev} {et~al.}(2007){Sergeev}, {Klimanov}, {Chesnok}, \&
  {Pronik}}]{Sergeev07}
{Sergeev}, S.~G., {Klimanov}, S.~A., {Chesnok}, N.~G., \& {Pronik}, V.~I. 2007,
  Astronomy Letters, 33, 429

\bibitem[{{Shapovalova} {et~al.}(2012){Shapovalova}, {Popovi{\'c}}, {Burenkov},
  {Chavushyan}, {Ili{\'c}}, {Kova{\v c}evi{\'c}}, {Kollatschny}, {Kova{\v
  c}evi{\'c}}, {Bochkarev}, {Valdes}, {Torrealba}, {Le{\'o}n-Tavares},
  {Mercado}, {Ben{\'{\i}}tez}, {Carrasco}, {Dultzin}, \& {de la
  Fuente}}]{Shapovalova12}
{Shapovalova}, A.~I., {Popovi{\'c}}, L.~{\v C}., {Burenkov}, A.~N., {et~al.}
  2012, \apjs, 202, 10

\bibitem[{{Sikora} {et~al.}(2007){Sikora}, {Stawarz}, \& {Lasota}}]{Sikora07}
{Sikora}, M., {Stawarz}, {\L}., \& {Lasota}, J.-P. 2007, \apj, 658, 815

\bibitem[{{Snellen} {et~al.}(2004){Snellen}, {Mack}, {Schilizzi}, \&
  {Tschager}}]{Snellen04}
{Snellen}, I.~A.~G., {Mack}, K.-H., {Schilizzi}, R.~T., \& {Tschager}, W. 2004,
  \mnras, 348, 227

\bibitem[{{Sulentic} {et~al.}(2000){Sulentic}, {Zwitter}, {Marziani}, \&
  {Dultzin-Hacyan}}]{Sulentic00}
{Sulentic}, J.~W., {Zwitter}, T., {Marziani}, P., \& {Dultzin-Hacyan}, D. 2000,
  \apjl, 536, L5

\bibitem[{{Tarchi} {et~al.}(2011){Tarchi}, {Castangia}, {Columbano}, {Panessa},
  \& {Braatz}}]{Tarchi11}
{Tarchi}, A., {Castangia}, P., {Columbano}, A., {Panessa}, F., \& {Braatz},
  J.~A. 2011, \aap, 532, A125

\bibitem[{{Thean} {et~al.}(2000){Thean}, {Pedlar}, {Kukula}, {Baum}, \&
  {O'Dea}}]{Thean00}
{Thean}, A., {Pedlar}, A., {Kukula}, M.~J., {Baum}, S.~A., \& {O'Dea}, C.~P.
  2000, \mnras, 314, 573

\bibitem[{{Tombesi} {et~al.}(2015){Tombesi}, {Mel{\'e}ndez}, {Veilleux},
  {Reeves}, {Gonz{\'a}lez-Alfonso}, \& {Reynolds}}]{Tombesi15}
{Tombesi}, F., {Mel{\'e}ndez}, M., {Veilleux}, S., {et~al.} 2015, \nat, 519,
  436

\bibitem[{{Ulvestad} {et~al.}(1995){Ulvestad}, {Antonucci}, \&
  {Goodrich}}]{Ulvestad95}
{Ulvestad}, J.~S., {Antonucci}, R.~R.~J., \& {Goodrich}, R.~W. 1995, \aj, 109,
  81

\bibitem[{{Urry} \& {Padovani}(1995)}]{Urry95}
{Urry}, C.~M. \& {Padovani}, P. 1995, \pasp, 107, 803

\bibitem[{{Wadadekar}(2004)}]{Wadadekar04}
{Wadadekar}, Y. 2004, \aap, 416, 35

\bibitem[{{White} {et~al.}(1997){White}, {Becker}, {Helfand}, \&
  {Gregg}}]{White97}
{White}, R.~L., {Becker}, R.~H., {Helfand}, D.~J., \& {Gregg}, M.~D. 1997,
  \apj, 475, 479

\bibitem[{{Wilkins} {et~al.}(2015){Wilkins}, {Gallo}, {Grupe}, {Bonson},
  {Komossa}, \& {Fabian}}]{Wilkins15}
{Wilkins}, D.~R., {Gallo}, L.~C., {Grupe}, D., {et~al.} 2015, \mnras, 454, 4440

\bibitem[{{Wu} \& {Cao}(2005)}]{Wu05}
{Wu}, Q. \& {Cao}, X. 2005, \apj, 621, 130

\bibitem[{{Yao} {et~al.}(2015){Yao}, {Yuan}, {Zhou}, {Komossa}, {Zhang},
  {Qiao}, \& {Liu}}]{Yao15}
{Yao}, S., {Yuan}, W., {Zhou}, H., {et~al.} 2015, \mnras, 454, L16

\bibitem[{{Yuan} {et~al.}(2008){Yuan}, {Zhou}, {Komossa}, {Dong}, {Wang}, {Lu},
  \& {Bai}}]{Yuan08}
{Yuan}, W., {Zhou}, H.~Y., {Komossa}, S., {et~al.} 2008, \apj, 685, 801

\end{thebibliography}

\clearpage
\begin{appendix}

\section{Tables}
\begin{table}[h!]
\caption{Observational details. }
\label{tab:source}
\centering
\scalebox{0.8}{
\footnotesize
\begin{tabular}{l c c c c c} 
\hline\hline
Short name & Date & Exposure & Short name & Date & Exposure \\
\hline\hline
J0006+2012 & 2015-07-05 & 538 & J1159+2838 & 2015-08-14 & 506 \\
J0100-0200 & 2015-07-05 & 538 & J1203+4431 & 2015-09-05 & 566 \\
J0146-0040 & 2015-07-05 & 538 & J1209+3217 & 2015-09-05 & 568 \\
J0347+0150 & 2015-07-05 & 538 & J1215+5442 & 2015-09-05 & 568 \\
J0629-0545 & 2015-07-10 & 506 & J1218+2948 & 2015-08-14 & 506 \\
J0632+6340 & 2015-07-10 & 536 & J1227+3214 & 2015-08-14 & 506 \\
J0706+3901 & 2015-07-10 & 536 & J1238+3942 & 2015-09-05 & 568 \\
J0713+3820 & 2015-07-10 & 536 & J1242+3317 & 2015-09-06 & 552 \\
J0744+5149 & 2015-07-14 & 536 & J1246+0222 & 2015-09-06 & 552 \\
J0752+2617 & 2015-08-14 & 506 & J1246+0238 & 2015-09-06 & 552 \\
J0754+3920 & 2015-07-10 & 506 & J1302+1624 & 2015-09-06 & 552 \\
J0804+3853 & 2015-07-10 & 508 & J1305+5116 & 2015-08-30 & 566 \\
J0806+7248 & 2015-07-14 & 536 & J1317+6010 & 2015-08-30 & 552 \\
J0814+5609 & 2015-07-14 & 536 & J1333+4141 & 2015-08-30 & 566 \\
J0849+5108 & 2015-07-10 & 506 & J1337+2423 & 2015-09-04 & 536 \\
J0850+4626 & 2015-08-01 & 508 & J1346+3121 & 2015-09-04 & 538 \\
J0902+0443 & 2015-07-28 & 492 & J1348+2622 & 2015-09-04 & 538 \\
J0913+3658 & 2015-08-15 & 716 & J1355+5612 & 2015-08-30 & 550 \\
J0925+5217 & 2015-08-01 & 506 & J1358+2658 & 2015-09-03 & 596 \\
J0926+1244 & 2015-07-28 & 492 & J1402+2159 & 2015-09-03 & 596 \\
J0937+3615 & 2015-08-15 & 714 & J1421+2824 & 2015-09-03 & 596 \\
J0945+1915 & 2015-08-14 & 506 & J1443+4725 & 2015-08-29 & 652 \\
J0948+5029 & 2015-08-01 & 506 & J1505+0326 & 2015-09-01 & 654\\
J0952-0136 & 2015-07-28 & 492 & J1536+5433 & 2015-08-29 & 656 \\
J0957+2444 & 2015-08-14 & 506 & J1537+4942 & 2015-08-29 & 654 \\
J1031+4234 & 2015-08-29 & 594 & J1548+3511 & 2015-07-23 & 522 \\
J1034+3938 & 2015-08-29 & 594 & J1555+1911 & 2015-09-01 & 654 \\
J1037+0036 & 2015-07-28 & 492 & J1559+3501 & 2015-07-23 & 522 \\
J1038+4227 & 2015-08-29 & 594 & J1612+4219 & 2015-07-23 & 522 \\
J1047+4725 & 2015-08-29 & 594 & J1629+4007 & 2015-07-23 & 522 \\
J1048+2222 & 2015-08-29 & 596 & J1633+4718 & 2015-07-22 & 536 \\
J1102+2239 & 2015-08-29 & 596 & J1634+4809 & 2015-07-22 & 536 \\
J1110+3653 & 2015-08-14 & 598 & J1703+4540 & 2015-07-22 & 538 \\
J1114+3241 & 2015-08-15 & 596 & J1709+2348 & 2015-09-01 & 638 \\
J1121+5351 & 2015-08-29 & 596 & J1713+3523 & 2015-09-01 & 640 \\
J1138+3653 & 2015-08-15 & 598 & J2242+2943 & 2015-07-02 & 596 \\
J1146+3236 & 2015-08-15 & 596 & J2314+2243 & 2015-07-02 & 596 \\
\hline
\end{tabular}
}
\tablefoot{Columns: (1) short name; (2) observation date; (3) exposure time (s); (4) short name; (5) observation date; (6) exposure time (s).}
\end{table}
\begin{table}
\caption{Spectral indexes of the sources. The 1.4 GHz data are derived from FIRST or NVSS.}
\label{tab:spind}
\centering
\scalebox{0.85}{
\footnotesize
\begin{tabular}{l c c c} 
\hline\hline
Short name & $\alpha_p$ & $\alpha_{in-band}$ & $\alpha_{1.4-5}$ \\
\hline\hline
  J0006+2012 &    0.77  $\pm$ 0.06 &    0.78  $\pm$ 0.20 &     0.18  $\pm$ 0.18  \\
  J0100$-$0200 &  0.77  $\pm$ 0.07 &    0.83  $\pm$ 0.23 &     0.34  $\pm$ 0.08  \\
  J0146$-$0040 & -3.31  $\pm$ 1.36 &    0.06  $\pm$ 4.05 &     3.65  $\pm$ 0.32  \\
  J0347+0105 &    0.34  $\pm$ 0.07 &    -0.03 $\pm$ 0.09 &     0.71  $\pm$ 0.03  \\
  J0629$-$0545 &  0.14  $\pm$ 0.01 &    -0.03 $\pm$ 0.05 &     0.61  $\pm$ 0.03  \\
  J0632+6340 &    0.83  $\pm$ 0.09 &    0.71  $\pm$ 0.10 &     1.12  $\pm$ 0.06  \\
  J0706+3901 &    0.84  $\pm$ 0.11 &    0.75  $\pm$ 0.43 &     0.57  $\pm$ 0.10  \\
  J0713+3820 &    0.89  $\pm$ 0.11 &    1.04  $\pm$ 0.23 &     0.96  $\pm$ 0.05  \\
  J0744+5149 &    1.25  $\pm$ 0.05 &    1.29  $\pm$ 0.23 &     1.14  $\pm$ 0.05  \\
  J0752+2617 &    3.23  $\pm$ 1.11 &    1.68  $\pm$ 3.66 &     1.38  $\pm$ 0.22  \\
  J0758+3920 &    0.65  $\pm$ 0.20 &    0.73  $\pm$ 0.38 &     0.75  $\pm$ 0.10  \\
  J0804+3853 &    1.79  $\pm$ 0.46 &    1.47  $\pm$ 0.88 &     0.92  $\pm$ 0.09  \\
  J0806+7248 &    0.06  $\pm$ 0.01 &    0.03  $\pm$ 0.06 &     1.10  $\pm$ 0.03  \\
  J0814+5609 &    0.02  $\pm$ 0.01 &    0.08  $\pm$ 0.03 &     0.55  $\pm$ 0.03  \\
  J0849+5108 &    0.02  $\pm$ 0.01 &    0.01  $\pm$ 0.01 &     0.14  $\pm$ 0.02  \\
  J0850+4626 &    0.20  $\pm$ 0.02 &    0.18  $\pm$ 0.06 &     0.34  $\pm$ 0.04  \\
  J0902+0443 &    0.01  $\pm$ 0.01 &    0.01  $\pm$ 0.01 &     0.33  $\pm$ 0.02  \\
  J0913+3658 &    1.40  $\pm$ 2.02 &    1.36  $\pm$ 2.52 &     0.95  $\pm$ 0.18  \\
  J0925+5217 &    0.25  $\pm$ 0.16 &    0.77  $\pm$ 0.46 &     0.81  $\pm$ 0.07  \\
  J0926+1244 &    0.74  $\pm$ 0.06 &    0.46  $\pm$ 0.19 &     1.83  $\pm$ 0.04  \\
  J0937+3615 &    1.39  $\pm$ 0.19 &    0.91  $\pm$ 0.47 &     0.89  $\pm$ 0.07  \\
  J0945+1915 &    0.92  $\pm$ 0.05 &    0.92  $\pm$ 0.12 &     0.80  $\pm$ 0.04  \\
  J0948+5029 &    0.71  $\pm$ 0.10 &    0.29  $\pm$ 0.32 &     0.89  $\pm$ 0.08  \\
  J0952-0136 &    0.08  $\pm$ 0.03 &    0.04  $\pm$ 0.05 &     0.77  $\pm$ 0.03  \\
  J0957+2433 &    0.49  $\pm$ 0.46 &    2.68  $\pm$ 2.19 &     2.52  $\pm$ 0.14  \\
  J1031+4234 &    0.59  $\pm$ 0.02 &    0.56  $\pm$ 0.10 &     0.97  $\pm$ 0.03  \\
  J1034+3928 &    0.06  $\pm$ 0.02 &    0.11  $\pm$ 0.21 &     0.80  $\pm$ 0.05  \\
  J1037+0036 &    0.01  $\pm$ 0.01 &    0.01  $\pm$ 0.03 &     0.24  $\pm$ 0.03  \\
  J1038+4227 &    0.56  $\pm$ 0.10 &    0.77  $\pm$ 1.31 &     0.07  $\pm$ 0.18  \\
  J1047+4725 &    0.47  $\pm$ 0.16 &    0.07  $\pm$ 0.01 &     0.57  $\pm$ 0.02  \\
  J1048+2222 &    0.34  $\pm$ 0.45 &    1.62  $\pm$ 1.55 &     1.04  $\pm$ 0.14  \\
  J1102+2239 &    1.23  $\pm$ 0.26 &    1.46  $\pm$ 1.61 &     0.97  $\pm$ 0.21  \\
  J1110+3653 &    0.01  $\pm$ 0.04 &    0.08  $\pm$ 0.10 &     0.54  $\pm$ 0.05  \\
  J1114+3241 &    0.03  $\pm$ 0.01 &    0.01  $\pm$ 0.03 &     -0.46 $\pm$ 0.07  \\
  J1121+5351 &    0.91  $\pm$ 0.23 &    0.70  $\pm$ 0.43 &     0.56  $\pm$ 0.17  \\
  J1138+3653 &    0.06  $\pm$ 0.03 &    -0.03 $\pm$ 0.12 &     0.72  $\pm$ 0.04  \\
  J1146+3236 &    0.02  $\pm$ 0.01 &    -0.01 $\pm$ 0.03 &     -0.20 $\pm$ 0.04  \\
  J1159+2838 &    1.22  $\pm$ 0.18 &    0.79  $\pm$ 0.54 &     0.71  $\pm$ 0.10  \\
  J1203+4431 &    0.71  $\pm$ 0.08 &    1.47  $\pm$ 0.39 &     2.08  $\pm$ 0.05  \\
  J1209+3217 &    0.93  $\pm$ 0.33 &    0.54  $\pm$ 0.73 &     0.63  $\pm$ 0.08  \\
  J1215+5442 &    1.08  $\pm$ 0.25 &    1.12  $\pm$ 0.51 &     1.00  $\pm$ 0.06  \\
  J1218+2948 &    0.24  $\pm$ 0.09 &    0.06  $\pm$ 0.06 &     0.75  $\pm$ 0.03  \\
  J1227+3214 &    0.66  $\pm$ 0.13 &    0.65  $\pm$ 0.23 &     0.49  $\pm$ 0.05  \\
  J1238+3942 &    -0.30 $\pm$ 0.02 &    -0.32 $\pm$ 0.04 &     -0.38 $\pm$ 0.04  \\
  J1242+3317 &    1.29  $\pm$ 0.12 &    0.77  $\pm$ 0.27 &     0.71  $\pm$ 0.08  \\
  J1246+0222 &    0.38  $\pm$ 0.34 &    -0.37 $\pm$ 0.56 &     0.87  $\pm$ 0.07  \\
  J1246+0238 &    0.29  $\pm$ 0.01 &    0.25  $\pm$ 0.02 &     0.44  $\pm$ 0.03  \\
  J1302+1624 &    0.70  $\pm$ 0.15 &    2.71  $\pm$ 0.25 &     0.65  $\pm$ 0.04  \\
  J1305+5116 &    -0.12 $\pm$ 0.01 &    0.06  $\pm$ 0.02 &     0.37  $\pm$ 0.02  \\
  J1317+6010 &    0.05  $\pm$ 0.29 &    0.47  $\pm$ 0.69 &     0.53  $\pm$ 0.07  \\
  J1333+4141 &    0.67  $\pm$ 0.32 &    0.67  $\pm$ 1.81 &     0.76  $\pm$ 0.07  \\
  J1337+2423 &    0.02  $\pm$ 0.02 &    -0.05 $\pm$ 0.05 &     0.47  $\pm$ 0.03  \\
  J1346+3121 &    -0.79 $\pm$ 0.03 &    -0.89 $\pm$ 0.07 &     -1.00 $\pm$ 0.07  \\
  J1348+2622 &    1.48  $\pm$ 0.45 &    2.07  $\pm$ 0.91 &     0.99  $\pm$ 0.09  \\
  J1355+5612 &    0.02  $\pm$ 0.11 &    0.01  $\pm$ 0.45 &     0.83  $\pm$ 0.08  \\
  J1358+2658 &    0.72  $\pm$ 0.30 &    0.21  $\pm$ 0.64 &     1.17  $\pm$ 0.17  \\
  J1402+2159 &    1.02  $\pm$ 0.40 &    0.77  $\pm$ 0.97 &     2.16  $\pm$ 0.09  \\
  J1421+2824 &    0.03  $\pm$ 0.01 &    0.03  $\pm$ 0.03 &     0.45  $\pm$ 0.04  \\
  J1443+4725 &    0.04  $\pm$ 0.01 &    0.02  $\pm$ 0.01 &     0.53  $\pm$ 0.02  \\
  J1505+0326 &    0.01  $\pm$ 0.01 &    -0.01 $\pm$ 0.01 &     -0.01 $\pm$ 0.02  \\
  J1536+5433 &    0.65  $\pm$ 0.34 &    0.35  $\pm$ 0.96 &     0.67  $\pm$ 0.11  \\
  J1537+4942 &    0.41  $\pm$ 0.19 &    0.29  $\pm$ 0.52 &     0.39  $\pm$ 0.08  \\
  J1548+3511 &    0.02  $\pm$ 0.01 &    0.01  $\pm$ 0.02 &     0.60  $\pm$ 0.02  \\
  J1555+1911 &    2.29  $\pm$ 1.06 &    1.11  $\pm$ 1.04 &     0.96  $\pm$ 0.09  \\
  J1559+3501 &    1.26  $\pm$ 0.52 &    1.46  $\pm$ 0.27 &     1.11  $\pm$ 0.08  \\
  J1612+4219 &    -0.24 $\pm$ 1.36 &    -0.81 $\pm$ 0.58 &     1.12  $\pm$ 0.21  \\
  J1629+4007 &    0.01  $\pm$ 0.01 &    0.01  $\pm$ 0.02 &     -0.96 $\pm$ 0.04  \\
  J1633+4718 &    0.03  $\pm$ 0.01 &    0.02  $\pm$ 0.03 &     0.75  $\pm$ 0.02  \\
  J1634+4809 &    0.52  $\pm$ 0.04 &    0.50  $\pm$ 0.10 &     0.25  $\pm$ 0.02  \\
  J1703+4540 &    0.13  $\pm$ 0.02 &    0.04  $\pm$ 0.02 &     0.95  $\pm$ 0.02  \\
  J1709+2348 &    1.51  $\pm$ 0.43 &    0.18  $\pm$ 0.85 &     1.16  $\pm$ 0.10  \\
  J1713+3523 &    1.03  $\pm$ 0.04 &    0.94  $\pm$ 0.11 &     0.85  $\pm$ 0.04  \\
  J2242+2943 &    1.37  $\pm$ 0.04 &    0.86  $\pm$ 0.09 &     0.78  $\pm$ 0.03  \\
  J2314+2243 &    0.03  $\pm$ 0.03 &    -0.01 $\pm$ 0.11 &     0.71  $\pm$ 0.05  \\
\hline\hline
\end{tabular}
}
\tablefoot{Columns: (1) short name; (2) in-band spectral index of the peak flux; (3) in-band spectral index of the integral flux; (4) spectral index between 1.4 and 5 GHz. }
\end{table}

\begin{table*}
\caption{Geometrical properties of the sources. Missing values indicate an unresolved core.}
\label{tab:dimensioni}
\centering
\scalebox{0.85}{
\footnotesize
\begin{tabular}{l c c c c c c } 
\hline\hline
Short name & Beam maj & Beam min & Beam P.A. & Core maj & Core min & Core P.A. \\
\hline\hline
J0006+2012 & 0.46 & 0.43 & $+$38.40 & 0.095 $\pm$ 0.016 & 0.054 $\pm$ 0.041 & 14 $\pm$ 21 \\
J0100$-$0200 & 0.55 & 0.40 & $+$17.93 & 0.084 $\pm$ 0.020 & 0.065 $\pm$ 0.022 & 19 $\pm$ 65 \\
J0146$-$0040 & 0.54 & 0.38 & $-$174.08 & $<$0.39 & $<$0.18 & {} \\
J0347+0105 & 0.52 & 0.38 & $-$9.64 & 0.330 $\pm$ 0.011 & 0.185 $\pm$ 6.900 & 169 $\pm$ 2 \\
J0629$-$0545 & 1.49 & 0.44 & $-$54.31 & 0.297 $\pm$ 0.014 & 0.183 $\pm$ 0.004 & 111 $\pm$ 3 \\
J0632+6340 & 0.80 & 0.40 & $-$84.06 & 0.251 $\pm$ 0.065 & 0.219 $\pm$ 0.068 & 83 $\pm$ 89 \\
J0706+3901 & 1.09 & 0.58 & $-$57.69 & 0.508 $\pm$ 0.033 & 0.257 $\pm$ 0.039 & 154 $\pm$ 6 \\
J0713+3820 & 1.09 & 0.51 & $-$59.99 & 0.526 $\pm$ 0.029 & 0.266 $\pm$ 0.060 & 169 $\pm$ 6 \\
J0744+5149 & 0.82 & 0.38 & $-$73.18 & 0.122 $\pm$ 0.027 & 0.058 $\pm$ 0.029 & 64 $\pm$ 22 \\
J0752+2617 & 1.17 & 0.39 & $+$61.90 & {} & {} & {} \\
J0758+3920 & 1.15 & 0.53 & $-$54.23 & 1.218 $\pm$ 0.053 & 0.416 $\pm$ 0.025 & 143 $\pm$ 2 \\
J0804+3853 & 0.45 & 0.41 & +51.27 & 0.590 $\pm$ 0.061 & 0.363 $\pm$ 0.057 & 128 $\pm$ 10 \\
J0806+7248 & 0.73 & 0.40 & $-$85.95 & 0.141 $\pm$ 0.007 & 0.117 $\pm$ 0.002 & 89 $\pm$ 7 \\
J0814+5609 & 0.82 & 0.37 & $-$72.62 & 0.075 $\pm$ 0.007 & 0.031 $\pm$ 0.004 & 122 $\pm$ 5 \\
J0849+5108 & 0.43 & 0.41 & +14.12 & 0.048 $\pm$ 0.001 & 0.038 $\pm$ 0.002 & 28  $\pm$ 7 \\
J0850+4626 & 0.48 & 0.42 & $-$53.08 & 0.060 $\pm$ 0.007 & 0.056 $\pm$ 0.010 & 31 $\pm$ 75 \\
J0902+0443 & 1.35 & 0.41 & +55.57 & 0.058 $\pm$ 0.017 & 0.036 $\pm$ 0.013 & 81 $\pm$ 42 \\
J0913+3658 & 1.18 & 0.41 & $-$62.63 & 0.346 $\pm$ 0.192 & 0.178 $\pm$ 0.030 & 113 $\pm$ 7 \\
J0925+5217 & 0.54 & 0.46 & $-$88.78 & 0.283 $\pm$ 0.021 & 0.186 $\pm$ 0.022 & 101 $\pm$ 11 \\
J0926+1244 & 0.94 & 0.41 & $+$58.26 & 0.166 $\pm$ 0.056 & 0.103 $\pm$ 0.026 & 90 $\pm$ 24 \\
J0937+3615 & 1.29 & 0.41 & $-$61.46 & 0.463 $\pm$ 0.095 & 0.314 $\pm$ 0.015 & 122 $\pm$ 19 \\
J0945+1915 & 0.72 & 0.42 & +66.87 & 0.114 $\pm$ 0.013 & 0.077 $\pm$ 0.034 & 165 $\pm$ 70 \\
J0948+5029 & 0.52 & 0.45 & $-$83.83 & 0.145 $\pm$ 0.018 & 0.044 $\pm$ 0.026 & 97 $\pm$ 12 \\
J0952$-$0136 & 0.94 & 0.45 & $+$45.40 & 0.207 $\pm$ 0.018 & 0.082 $\pm$ 0.005 & 46 $\pm$ 2 \\
J0957+2433 & 0.69 & 0.41 & $+$69.47 & {} & {} & {} \\
J1031+4234 & 0.50 & 0.43 & +81.12 & $<$ 0.11 & $<$ 0.02 & {} \\
J1034+3938 & 0.56 & 0.45 & $+$52.83 & 0.118 $\pm$ 0.008 & 0.080 $\pm$ 0.010 & 84 $\pm$ 9 \\
J1037+0036 & 0.65 & 0.41 & +42.82 & 0.056 $\pm$ 0.006 & 0.014 $\pm$ 0.012 & 49 $\pm$ 6 \\
J1038+4227 & 0.54 & 0.45 & +56.01 & 0.129 $\pm$ 0.037 & 0.116 $\pm$ 0.062 & 98 $\pm$ 77 \\
J1047+4725 & 0.40 & 0.31 & $-$74.46 & 0.474 $\pm$ 0.016 & 0.165 $\pm$ 0.013 & 127 $\pm$ 2 \\
J1048+2222 & 0.66 & 0.42 & +58.93 & $<$ 0.25 & $<$ 0.18 & {} \\
J1102+2239 & 0.65 & 0.41 & +58.60 & 0.201 $\pm$ 0.070 & 0.090 $\pm$ 0.053 & 85 $\pm$ 34 \\
J1110+3653 & 0.50 & 0.41 & +84.14 & 0.072 $\pm$ 0.012 & 0.057 $\pm$ 0.020 & 134 $\pm$ 20 \\
J1114+3241 & 0.55 & 0.44 & $+$66.94 & 0.074 $\pm$ 0.004 & 0.065 $\pm$ 0.004 & 103 $\pm$ 20 \\
J1121+5351 & 0.51 & 0.45 & $-$76.04 & 0.261 $\pm$ 0.035 & 0.141 $\pm$ 0.072 & 156 $\pm$ 16 \\
J1138+3653 & 0.53 & 0.46 & +64.17 & 0.158 $\pm$ 0.005 & 0.049 $\pm$ 0.021 & 127 $\pm$ 3 \\
J1146+3236 & 0.53 & 0.43 & +60.65 & 0.061 $\pm$ 0.005 & 0.040 $\pm$ 0.006 & 37 $\pm$ 12 \\
J1159+2838 & 0.52 & 0.42 & +85.32 & 0.126 $\pm$ 0.039 & 0.103 $\pm$ 0.075 & 62 $\pm$ 77 \\
J1203+4431 & 0.43 & 0.41 & $+$42.41 & 0.805 $\pm$ 0.055 & 0.260 $\pm$ 0.037 & 67 $\pm$ 2 \\
J1209+3217 & 0.33 & 0.30 & $+$44.69 & 0.139 $\pm$ 0.038 & 0.068 $\pm$ 0.050 & 122 $\pm$ 67 \\
J1215+5442 & 0.46 & 0.41 & $+$45.18 & 0.184 $\pm$ 0.075 & 0.101 $\pm$ 0.068 & 88 $\pm$ 42 \\
J1218+2948 & 0.52 & 0.43 & $+$85.15 & 0.302 $\pm$ 0.011 & 0.234 $\pm$ 0.017 & 8 $\pm$ 10 \\
J1227+3214 & 0.51 & 0.42 & +87.18 & 0.188 $\pm$ 0.025 & 0.094 $\pm$ 0.042 & 30 $\pm$ 15 \\
J1238+3942 & 0.43 & 0.41 & +70.94 & $<$ 0.06 & $<$ 0.03 & {} \\
J1242+3317 & 0.50 & 0.38 & $+$83.17 & 0.505 $\pm$ 0.019 & 0.161 $\pm$ 0.019 & 65 $\pm$ 4 \\
J1246+0222 & 0.56 & 0.45 & $-$44.26 & 0.464 $\pm$ 0.063 & 0.251 $\pm$ 0.086 & 177 $\pm$ 15 \\
J1246+0238 & 0.55 & 0.41 & $-$35.07 & {} & {} & {} \\
J1302+1624 & 0.45 & 0.40 & $-$54.67 & 0.216 $\pm$ 0.021 & 0.173 $\pm$ 0.024 & 153 $\pm$ 22 \\
J1305+5116 & 1.70 & 0.44 & +51.86 & 0.146 $\pm$ 0.799 & 0.058 $\pm$ 0.110 & 52 $\pm$ 1 \\
J1317+6010 & 1.13 & 0.49 & $+$58.10 & 0.462 $\pm$ 0.119 & 0.380 $\pm$ 0.134 & 82 $\pm$ 66 \\
J1333+4141 & 2.30 & 1.25 & +56.87 & 0.399 $\pm$ 0.306 & 0.041 $\pm$ 0.264 & 27 $\pm$ 30 \\
J1337+2423 & 1.24 & 0.47 & $+$62.17 & 0.155 $\pm$ 0.044 & 0.065 $\pm$ 0.011 & 67 $\pm$ 14 \\
J1346+3121 & 1.06 & 0.43 & +63.78 & 0.187 $\pm$ 0.154 & 0.068 $\pm$ 0.034 & 55 $\pm$ 11 \\
J1348+2622 & 1.31 & 0.42 & +60.44 & 0.438 $\pm$ 0.221 & 0.164 $\pm$ 0.035 & 58 $\pm$ 5 \\
J1355+5612 & 1.18 & 0.58 & $+$64.91 & $<$ 0.27 & $<$ 0.11 & {} \\
J1358+2658 & 1.31 & 0.41 & +60.51 & $<$ 0.75 & $<$ 0.11 & {} \\
J1402+2159 & 1.74 & 0.42 & $+$58.57 & {} & {} & {} \\
J1421+2824 & 1.31 & 0.42 & +60.47 & 0.239 $\pm$ 0.019 & 0.068 $\pm$ 0.003 & 63 $\pm$ 1 \\
J1443+4725 & 0.81 & 0.42 & +68.47 & 0.102 $\pm$ 0.008 & 0.072 $\pm$ 0.015 & 120 $\pm$ 15 \\
J1505+0326 & 0.42 & 0.33 & $-$58.68 & 0.043 $\pm$ 0.001 & 0.018 $\pm$ 0.005 & 179 $\pm$ 4 \\
J1536+5433 & 0.69 & 0.43 & $+$81.60 & 0.350 $\pm$ 0.069 & 0.116 $\pm$ 0.073 & 86 $\pm$ 11 \\
J1537+4942 & 0.72 & 0.43 & $+$75.89 & {} & {} & {} \\
J1548+3511 & 1.56 & 0.43 & +61.87 & 0.128 $\pm$ 0.008 & 0.054 $\pm$ 0.001 & 54 $\pm$ 2 \\
J1555+1911 & 0.78 & 0.43 & $-$70.92 & 1.040 $\pm$ 0.230 & 0.960 $\pm$ 0.270 & 178 $\pm$ 71 \\
J1559+3501 & 1.47 & 0.41 & $+$59.78 & 0.987 $\pm$ 0.362 & 0.543 $\pm$ 0.137 & 68 $\pm$ 35 \\
J1612+4219 & 1.00 & 0.42 & +56.70 & $<$ 1.4 & $<$ 0.38 & {} \\
J1629+4007 & 1.20 & 0.40 & +62.18 & 0.063 $\pm$ 0.010 & 0.017 $\pm$ 0.002 & 60 $\pm$ 3 \\
J1633+4718 & 0.76 & 0.40 & +73.97 & 0.091 $\pm$ 0.005 & 0.056 $\pm$ 0.001 & 76 $\pm$ 3 \\
J1634+4809 & 0.79 & 0.40 & +72.76 & 0.149 $\pm$ 0.022 & 0.056 $\pm$ 0.026 & 108 $\pm$ 13 \\
J1703+4540 & 0.77 & 0.41 & $+$80.66 & 0.228 $\pm$ 0.006 & 0.103 $\pm$ 0.002 & 82 $\pm$ 1 \\
J1709+2348 & 0.90 & 0.40 & $-$65.79 & {} & {} & {} \\
J1713+3523 & 0.74 & 0.37 & -73.62 & 0.151 $\pm$ 0.022 & 0.106 $\pm$ 0.006 & 107 $\pm$ 15 \\
J2242+2943 & 0.53 & 0.46 & $+$50.45 & 0.655 $\pm$ 0.048 & 0.213 $\pm$ 0.067 & 174 $\pm$ 4 \\
J2314+2243 & 0.68 & 0.44 & $+$43.11 & 0.327 $\pm$ 0.005 & 0.098 $\pm$ 0.005 & 31 $\pm$ 1 \\
\hline\hline
\end{tabular}
}
\tablefoot{Columns: (1) short name; (2) beam major axis (arcsec); (3) beam minor axis (arcsec); (4) beam position angle (degrees); (5) core major axis deconvolved from beam (arcsec); (6) core minor axis deconvolved from beam (arcsec); (7) core position angle (degrees). }
\end{table*}

\begin{table*}
\caption{Flux densities, luminosities and brightness temperature of the sources.}
\label{tab:flussi}
\centering
\scalebox{0.85}{
\footnotesize
\begin{tabular}{l c c c c c c c  } 
\hline\hline
Short name & rms & S$_{int}$ & S$_p$ & $\log$L$_{int}$ & $\log$L$_p$ & $\log$L$_{diff}$ & $\log$T$_b$ \\
\hline\hline
J0006+2012 & 10 & 3.25 $\pm$ 0.04 & 3.16 $\pm$ 0.01 & 38.42 $\pm$ 0.01 & 38.41 $\pm$ 0.01 & 36.88 $\pm$ 0.26 & 4.62 \\
J0100-0200 & 10 & 3.58 $\pm$ 0.04 & 3.49 $\pm$ 0.02 & 40.47 $\pm$ 0.01 & 40.46 $\pm$ 0.01 & 38.86 $\pm$ 0.32 & 4.72 \\
J0146-0040 & 10 & 0.07 $\pm$ 0.02 & 0.07 $\pm$ 0.01 & 37.95 $\pm$ 0.12 & 37.94 $\pm$ 0.06 & 36.40 $\pm$ 0.89 & 1.90 \\
J0347+0105 & 15 & 12.67 $\pm$ 0.08 & 8.80 $\pm$ 0.07 & 39.17 $\pm$ 0.01 & 39.01 $\pm$ 0.01 & 38.66 $\pm$ 0.02 & 4.00 \\
J0629-0545 & 12 & 14.59 $\pm$ 0.04 & 12.47 $\pm$ 0.01 & 40.43 $\pm$ 0.01 & 40.36 $\pm$ 0.01 & 39.59 $\pm$ 0.01 & 4.23 \\
J0632+6340 & 10 & 2.94 $\pm$ 0.02 & 1.77 $\pm$ 0.03 & 37.76 $\pm$ 0.01 & 37.54 $\pm$ 0.01 & 37.36 $\pm$ 0.02 & 3.34 \\
J0706+3901 & 13 & 2.22 $\pm$ 0.05 & 1.77 $\pm$ 0.02 & 39.33 $\pm$ 0.01 & 39.23 $\pm$ 0.01 & 38.63 $\pm$ 0.07 & 2.99 \\
J0713+3820 & 12 & 3.18 $\pm$ 0.05 & 2.29 $\pm$ 0.02 & 39.82 $\pm$ 0.01 & 39.68 $\pm$ 0.01 & 39.27 $\pm$ 0.04 & 3.09 \\
J0744+5149 & 10 & 2.58 $\pm$ 0.04 & 2.49 $\pm$ 0.01 & 40.98 $\pm$ 0.01 & 40.97 $\pm$ 0.01 & 39.53 $\pm$ 0.22 & 4.53 \\
J0752+2617 & 12 & 0.21 $\pm$ 0.04 & 0.21 $\pm$ 0.02 & 38.19 $\pm$ 0.09 & 38.17 $\pm$ 0.03 & 36.64 $\pm$ 0.36 & 2.03 \\
J0758+3920 & 18 & 3.92 $\pm$ 0.10 & 2.04 $\pm$ 0.03 & 39.68 $\pm$ 0.01 & 39.39 $\pm$ 0.01 & 39.36 $\pm$ 0.03 & 2.47 \\
J0804+3853 & 11 & 0.83 $\pm$ 0.04 & 0.37 $\pm$ 0.02 & 39.69 $\pm$ 0.02 & 39.33 $\pm$ 0.03 & 39.43 $\pm$ 0.05 & 2.14 \\
J0806+7248 & 10 & 11.79 $\pm$ 0.05 & 11.12 $\pm$ 0.01 & 40.17 $\pm$ 0.01 & 40.15 $\pm$ 0.01 & 38.93 $\pm$ 0.04 & 4.69 \\
J0814+5609 & 10 & 29.59 $\pm$ 0.06 & 25.94 $\pm$ 0.01 & 42.19 $\pm$ 0.01 & 42.13 $\pm$ 0.01 & 41.28 $\pm$ 0.01 & 6.05 \\
J0849+5108 & 35 & 222.09 $\pm$ 0.17 & 219.82 $\pm$ 0.06 & 43.21 $\pm$ 0.01 & 43.20 $\pm$ 0.01 & 41.22 $\pm$ 0.04 & 7.11 \\
J0850+4626 & 11 & 10.44 $\pm$ 0.04 & 10.27 $\pm$ 0.02 & 41.77 $\pm$ 0.01 & 41.76 $\pm$ 0.01 & 39.98 $\pm$ 0.15 & 5.50 \\
J0902+0443 & 13 & 102.09 $\pm$ 0.04 & 101.51 $\pm$ 0.04 & 42.77 $\pm$ 0.01 & 42.77 $\pm$ 0.01 & {} & 6.70 \\
J0913+3658 & 11 & 0.31 $\pm$ 0.04 & 0.28 $\pm$ 0.09 & 38.66 $\pm$ 0.06 & 38.61 $\pm$ 0.15 & 37.74 $\pm$ 1.51 & 2.52 \\
J0925+5217 & 11 & 3.48 $\pm$ 0.11 & 1.72 $\pm$ 0.02 & 38.72 $\pm$ 0.01 & 38.42 $\pm$ 0.01 & 38.42 $\pm$ 0.03 & 3.35 \\
J0926+1244 & 12 & 3.31 $\pm$ 0.04 & 2.82 $\pm$ 0.02 & 38.52 $\pm$ 0.01 & 38.45 $\pm$ 0.01 & 37.70 $\pm$ 0.06 & 4.05 \\
J0937+3658 & 12 & 1.08 $\pm$ 0.04 & 0.80 $\pm$ 0.01 & 39.71 $\pm$ 0.02 & 39.58 $\pm$ 0.01 & 39.11 $\pm$ 0.08 & 2.64 \\
J0945+1915 & 11 & 5.91 $\pm$ 0.04 & 5.67 $\pm$ 0.03 & 40.90 $\pm$ 0.01 & 40.88 $\pm$ 0.01 & 39.52 $\pm$ 0.13 & 4.74 \\
J0948+5029 & 15 & 2.05 $\pm$ 0.05 & 1.97 $\pm$ 0.02 & 38.91 $\pm$ 0.01 & 38.89 $\pm$ 0.01 & 37.52 $\pm$ 0.36 & 4.34 \\
J0952-0136 & 11 & 22.77 $\pm$ 0.08 & 20.59 $\pm$ 0.06 & 39.04 $\pm$ 0.01 & 38.99 $\pm$ 0.01 & 38.02 $\pm$ 0.03 & 4.92 \\
J0957+2433 & 10 & 0.23 $\pm$ 0.03 & 0.23 $\pm$ 0.01 & 38.32 $\pm$ 0.06 & 38.31 $\pm$ 0.01 & 36.73 $\pm$ 2.34 & 2.28 \\
J1031+4234 & 10 & 5.43 $\pm$ 0.04 & 5.29 $\pm$ 0.01 & 41.14 $\pm$ 0.01 & 41.13 $\pm$ 0.01 & 39.55 $\pm$ 0.15 & 5.34 \\
J1034+3928 & 10 & 8.15 $\pm$ 0.10 & 7.13 $\pm$ 0.01 & 39.24 $\pm$ 0.01 & 39.18 $\pm$ 0.01 & 38.34 $\pm$ 0.05 & 4.72 \\
J1037+0036 & 11 & 20.41 $\pm$ 0.04 & 20.32 $\pm$ 0.01 & 42.19 $\pm$ 0.01 & 42.19 $\pm$ 0.01 & {} & 6.43 \\
J1038+4227 & 10 & 6.35 $\pm$ 0.15 & 1.96 $\pm$ 0.01 & 40.67 $\pm$ 0.01 & 40.16 $\pm$ 0.01 & 40.51 $\pm$ 0.02 & 4.03 \\
J1047+4725 & 33 & 374.50 $\pm$ 0.25 & 207.8 $\pm$ 3.30 & 43.77 $\pm$ 0.01 & 43.51 $\pm$ 0.01 & 43.42 $\pm$ 0.01 & 5.50 \\
J1048+2222 & 10 & 0.30 $\pm$ 0.03 & 0.28 $\pm$ 0.01 & 39.75 $\pm$ 0.05 & 39.72 $\pm$ 0.02 & 38.59 $\pm$ 0.87 & 2.76 \\
J1102+2239 & 15 & 0.75 $\pm$ 0.07 & 0.70 $\pm$ 0.01 & 40.47 $\pm$ 0.04 & 40.44 $\pm$ 0.01 & 39.35 $\pm$ 0.60 & 3.57 \\
J1110+3653 & 10 & 8.95 $\pm$ 0.05 & 7.85 $\pm$ 0.02 & 41.89 $\pm$ 0.01 & 41.83 $\pm$ 0.01 & 40.98 $\pm$ 0.03 & 5.32 \\
J1114+3241 & 10 & 20.98 $\pm$ 0.04 & 20.55 $\pm$ 0.01 & 41.04 $\pm$ 0.01 & 41.03 $\pm$ 0.01 & 39.35 $\pm$ 0.05 & 5.53 \\
J1121+5351 & 11 & 1.24 $\pm$ 0.04 & 1.04 $\pm$ 0.02 & 39.25 $\pm$ 0.01 & 39.17 $\pm$ 0.01 & 38.45 $\pm$ 0.12 & 3.32 \\
J1138+3653 & 10 & 4.68 $\pm$ 0.03 & 4.42 $\pm$ 0.01 & 41.02 $\pm$ 0.01 & 41.00 $\pm$ 0.01 & 39.77 $\pm$ 0.07 & 4.71 \\
J1146+3236 & 11 & 18.79 $\pm$ 0.04 & 18.59 $\pm$ 0.01 & 41.90 $\pm$ 0.01 & 41.89 $\pm$ 0.01 & 39.94 $\pm$ 0.11 & 5.87 \\
J1159+2838 & 10 & 0.86 $\pm$ 0.03 & 0.81 $\pm$ 0.01 & 39.76 $\pm$ 0.02 & 39.74 $\pm$ 0.01 & 38.51 $\pm$ 0.36 & 3.70 \\
J1203+4431 & 12 & 6.35 $\pm$ 0.17 & 1.85 $\pm$ 0.02 & 36.46 $\pm$ 0.01 & 35.93 $\pm$ 0.01 & 36.31 $\pm$ 0.02 & 2.77 \\
J1209+3217 & 13 & 0.70 $\pm$ 0.03 & 0.62 $\pm$ 0.02 & 39.31 $\pm$ 0.02 & 39.26 $\pm$ 0.01 & 38.35 $\pm$ 0.29 & 3.70 \\
J1215+5442 & 10 & 0.64 $\pm$ 0.02 & 0.57 $\pm$ 0.02 & 39.30 $\pm$ 0.01 & 39.25 $\pm$ 0.01 & 38.34 $\pm$ 0.25 & 3.37 \\
J1218+2948 & 9 & 14.78 $\pm$ 0.05 & 11.01 $\pm$ 0.09 & 38.47 $\pm$ 0.01 & 38.34 $\pm$ 0.01 & 37.87 $\pm$ 0.02 & 4.02 \\
J1227+3214 & 12 & 3.41 $\pm$ 0.05 & 3.08 $\pm$ 0.04 & 39.94 $\pm$ 0.01 & 39.90 $\pm$ 0.01 & 38.92 $\pm$ 0.11 & 4.12 \\
J1238+3942 & 13 & 17.06 $\pm$ 0.06 & 17.04 $\pm$ 0.03 & 42.16 $\pm$ 0.01 & 42.16 $\pm$ 0.01 & {} & 6.04 \\
J1242+3317 & 12 & 2.35 $\pm$ 0.04 & 1.40 $\pm$ 0.01 & 38.74 $\pm$ 0.01 & 38.51 $\pm$ 0.01 & 38.35 $\pm$ 0.02 & 3.08 \\
J1246+0222 & 11 & 0.70 $\pm$ 0.03 & 0.46 $\pm$ 0.01 & 38.31 $\pm$ 0.02 & 38.12 $\pm$ 0.01 & 37.84 $\pm$ 0.08 & 2.44 \\
J1246+0238 & 12 & 20.33 $\pm$ 0.04 & 19.66 $\pm$ 0.01 & 41.69 $\pm$ 0.01 & 41.67 $\pm$ 0.01 & 40.20 $\pm$ 0.03 & 5.36 \\
J1302+1624 & 11 & 14.18 $\pm$ 0.19 & 3.32 $\pm$ 0.04 & 39.91 $\pm$ 0.01 & 39.28 $\pm$ 0.01 & 39.80 $\pm$ 0.01 & 3.80 \\
J1305+5116 & 13 & 53.80 $\pm$ 0.07 & 34.20 $\pm$ 0.01 & 42.91 $\pm$ 0.01 & 42.71 $\pm$ 0.01 & 42.47 $\pm$ 0.01 & 5.68 \\
J1317+6010 & 11 & 0.89 $\pm$ 0.04 & 0.65 $\pm$ 0.02 & 39.42 $\pm$ 0.02 & 39.28 $\pm$ 0.01 & 38.86 $\pm$ 0.11 & 2.45 \\
J1333+4141 & 14 & 0.92 $\pm$ 0.05 & 0.90 $\pm$ 0.02 & 39.85 $\pm$ 0.02 & 39.84 $\pm$ 0.01 & 38.21 $\pm$ 1.41 & 3.65 \\
J1337+2423 & 13 & 10.71 $\pm$ 0.04 & 9.83 $\pm$ 0.01 & 40.26 $\pm$ 0.01 & 40.23 $\pm$ 0.01 & 39.18 $\pm$ 0.03 & 4.86 \\
J1346+3121 & 12 & 4.48 $\pm$ 0.02 & 4.48 $\pm$ 0.01 & 40.63 $\pm$ 0.01 & 40.63 $\pm$ 0.01 & {} & 4.47 \\
J1348+2622 & 12 & 0.43 $\pm$ 0.02 & 0.40 $\pm$ 0.01 & 40.84 $\pm$ 0.02 & 40.80 $\pm$ 0.02 & 39.76 $\pm$ 0.34 & 2.85 \\
J1355+5612 & 12 & 2.06 $\pm$ 0.05 & 1.99 $\pm$ 0.02 & 39.67 $\pm$ 0.01 & 39.65 $\pm$ 0.01 & 38.15 $\pm$ 0.49 & 3.70 \\
J1358+2658 & 12 & 0.60 $\pm$ 0.03 & 0.52 $\pm$ 0.01 & 40.05 $\pm$ 0.02 & 39.99 $\pm$ 0.01 & 39.16 $\pm$ 0.22 & 2.75 \\
J1402+2159 & 12 & 0.40 $\pm$ 0.02 & 0.32 $\pm$ 0.01 & 38.34 $\pm$ 0.02 & 38.24 $\pm$ 0.01 & 37.67 $\pm$ 0.15 & 2.00 \\
J1421+2824 & 11 & 26.35 $\pm$ 0.05 & 25.59 $\pm$ 0.04 & 42.20 $\pm$ 0.01 & 42.18 $\pm$ 0.01 & 40.66 $\pm$ 0.05 & 5.21 \\
J1443+4725 & 13 & 83.24 $\pm$ 0.08 & 80.91 $\pm$ 0.09 & 42.98 $\pm$ 0.01 & 42.97 $\pm$ 0.01 & 41.43 $\pm$ 0.03 & 6.10 \\
J1505+0326 & 37 & 403.28 $\pm$ 0.14 & 399.46 $\pm$ 0.11 & 43.09 $\pm$ 0.01 & 43.09 $\pm$ 0.01 & {} & 7.68 \\
J1536+5433 & 11 & 0.52 $\pm$ 0.03 & 0.44 $\pm$ 0.01 & 37.98 $\pm$ 0.03 & 37.92 $\pm$ 0.01 & 37.13 $\pm$ 0.29 & 2.88 \\
J1537+4942 & 10 & 0.84 $\pm$ 0.03 & 0.84 $\pm$ 0.01 & 40.09 $\pm$ 0.02 & 40.09 $\pm$ 0.01 & {} & 2.89 \\
J1555+1911 & 11 & 0.50 $\pm$ 0.03 & 0.12 $\pm$ 0.01 & 37.85 $\pm$ 0.03 & 37.23 $\pm$ 0.05 & 37.73 $\pm$ 0.05 & 0.92 \\
J1559+3501 & 12 & 1.36 $\pm$ 0.03 & 0.49 $\pm$ 0.03 & 38.19 $\pm$ 0.01 & 37.75 $\pm$ 0.02 & 38.00 $\pm$ 0.03 & 1.80 \\
J1548+3511 & 12 & 64.22 $\pm$ 0.08 & 63.54 $\pm$ 0.01 & 42.46 $\pm$ 0.01 & 42.46 $\pm$ 0.01 & {} & 5.96 \\
J1612+4219 & 39 & 0.87 $\pm$ 0.13 & 0.34 $\pm$ 0.05 & 39.86 $\pm$ 0.06 & 39.45 $\pm$ 0.06 & 39.65 $\pm$ 0.14 & 3.28 \\
J1629+4007 & 10 & 42.39 $\pm$ 0.05 & 42.29 $\pm$ 0.01 & 41.71 $\pm$ 0.01 & 41.70 $\pm$ 0.01 & 39.06 $\pm$ 0.29 & 3.53 \\
J1633+4718 & 9 & 25.80 $\pm$ 0.06 & 24.20 $\pm$ 0.01 & 40.67 $\pm$ 0.01 & 40.64 $\pm$ 0.01 & 39.46 $\pm$ 0.02 & 5.55 \\
J1634+4809 & 10 & 5.42 $\pm$ 0.04 & 5.20 $\pm$ 0.02 & 41.42 $\pm$ 0.01 & 41.40 $\pm$ 0.01 & 40.03 $\pm$ 0.12 & 4.79 \\
J1703+4540 & 11 & 34.9 $\pm$ 0.05 & 32.46 $\pm$ 0.05 & 40.20 $\pm$ 0.01 & 40.16 $\pm$ 0.01 & 39.04 $\pm$ 0.02 & 4.99 \\
J1713+3523 & 13 & 3.91 $\pm$ 0.03 & 3.67 $\pm$ 0.01 & 39.54 $\pm$ 0.01 & 39.51 $\pm$ 0.01 & 38.33 $\pm$ 0.09 & 4.22 \\
J1709+2348 & 12 & 0.35 $\pm$ 0.02 & 0.35 $\pm$ 0.01 & 39.55 $\pm$ 0.03 & 39.55 $\pm$ 0.01 & {} & 2.76 \\
J2242+2943 & 13 & 10.40 $\pm$ 0.06 & 5.68 $\pm$ 0.02 & 38.88 $\pm$ 0.01 & 38.62 $\pm$ 0.01 & 38.54 $\pm$ 0.01 & 3.44 \\
J2314+2243 & 11 & 7.38 $\pm$ 0.05 & 6.46 $\pm$ 0.01 & 40.48 $\pm$ 0.01 & 40.42 $\pm$ 0.01 & 39.58 $\pm$ 0.03 & 4.20 \\
\hline\hline
\end{tabular}
}
\tablefoot{Columns: (1) short name; (2) map rms ($\mu$Jy beam$^{-1}$); (3) integrated flux (mJy); (4) peak flux (mJy beam$^{-1}$); (5) logarithm of the integrated luminosity (\ergs); (6) logarithm of the peak luminosity (\ergs); (7) logarithm of the diffuse luminosity (\ergs); (8) logarithm of the brightness temperature (K). }
\end{table*}

\section{Notes on single objects}
In the following, unless otherwise specified, when they are present the reference for 22 GHz VLBI measurement is \citet{Doi16}, that for 37 GHz with the Mets\"ahovi 14m radio telescope is \citet{Lahteenmaki17}, and that for the q22 parameter and star formation rate is \citet{Caccianiga15}. 
\subsection{J0006+2012}
\label{sec:J0006+2012}
Mrk 335, a well-known radio-quiet NLS1, was already observed at 5 GHz by \citet{Edelson87} with the Owen Valley Radio Telescope (OVRO), finding a flux of 3.30$\pm$0.18, which is essentially identical to our result. Another observation was carried out more recently with the VLA in A configuration \citep{Leipski06}, finding a point source and a total flux of 3.58$\pm$0.05 mJy, slightly larger than our measurement. Our map, shown in Fig.~\ref{fig:J0006p2012} shows an unresolved core, with a source size deconvolved from beam of 50$\times$28 pc.  
\subsection{J0100$-$0200}
\label{sec:J0100-0200}
\begin{figure*}
\centering
\includegraphics[trim={0cm 2cm 0cm 0cm},clip,width=7cm]{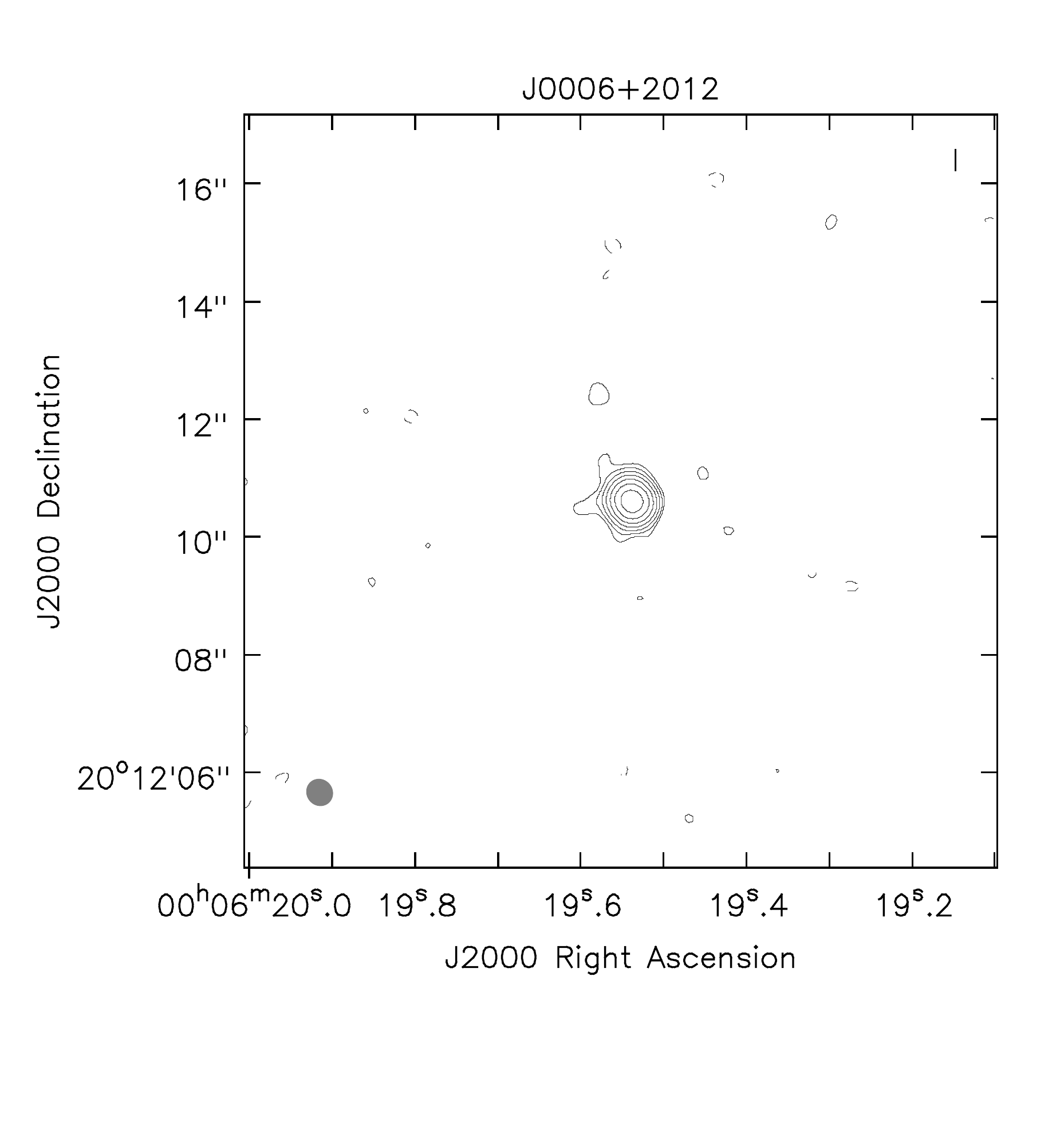} 
\includegraphics[trim={0cm 2cm 0cm 0cm},clip,width=7cm]{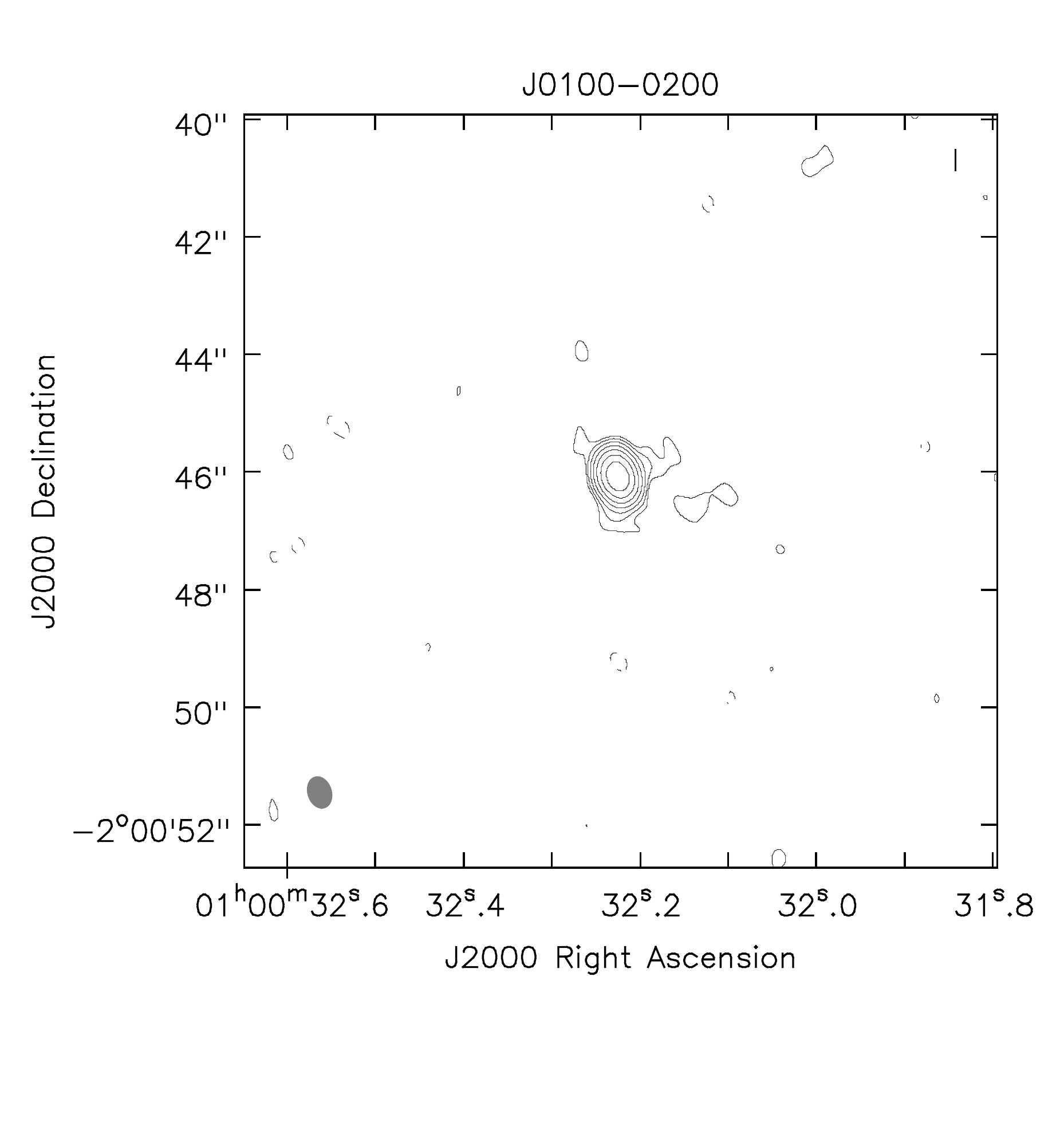} 
\caption{\textbf{Left panel:} J0006+2012, rms = 10 $\mu$Jy, contour levels at -3, 3$\times$2$^n$, n $\in$ [0,7], beam size 0.24$\times$0.22 kpc. \textbf{Right panel:} J0100-0200, rms = 11 $\mu$Jy, contour levels at -3, 3$\times$2$^n$, n $\in$ [0,6], beam size 2.00$\times$1.46 kpc. }
\label{fig:J0006p2012}
\label{fig:J0100m0200}
\end{figure*}

FBQS J0100-0200 was one of the sources included \textit{bona-fide} (i.e. without a measured spectral index) in the F sample by \citet{Foschini15}. While the broad-band index is consistent with a flat-spectrum object, the in-band spectral indexes reveal that it might be a steep-spectrum source. The map shown in Fig.~\ref{fig:J0100m0200} seems to show an unresolved core with a size of 307$\times$238 pc. It is not clear whether the faint emission located on the west side of the core is real or just noise. Given its uncertain nature, this source would likely require a follow-up at higher resolution. Because of its compact size and steep spectrum it might be classified as a low-power CSS. Its luminosity is indeed lower than classical CSS sources \citep{Odea98}, but close to another example of NLS1 classified as CSS, J1432+3014 \citep{Caccianiga14, Caccianiga17}. 
\subsection{J0146$-$0040}
\label{sec:J0146-0040}
2MASX J01464481-0040426 is a radio-quiet NLS1, that in our map (Fig.~\ref{fig:J0146m0040}) shows only an unresolved core whose size we could not determine. Its broad-band spectral index is very steep, since the source is clearly visible in the NVSS survey with a flux of 8.7 mJy, while it is barely detected at 6$\sigma$ in our survey (total flux 0.07 mJy). Moreover, J0146$-$0040 is not detected at 4.7 GHz, while it is clearly visible at 5.7 GHz. This could be a sign of a minimum in the spectrum around 5 GHz, and of an inverted spectrum going to higher frequencies. Despite the inverted spectrum, the source has not detection at high frequency (37 GHz).
\subsection{J0347+0105}
\begin{figure*}
\centering
\includegraphics[trim={0cm 2cm 0cm 0cm},clip,width=7cm]{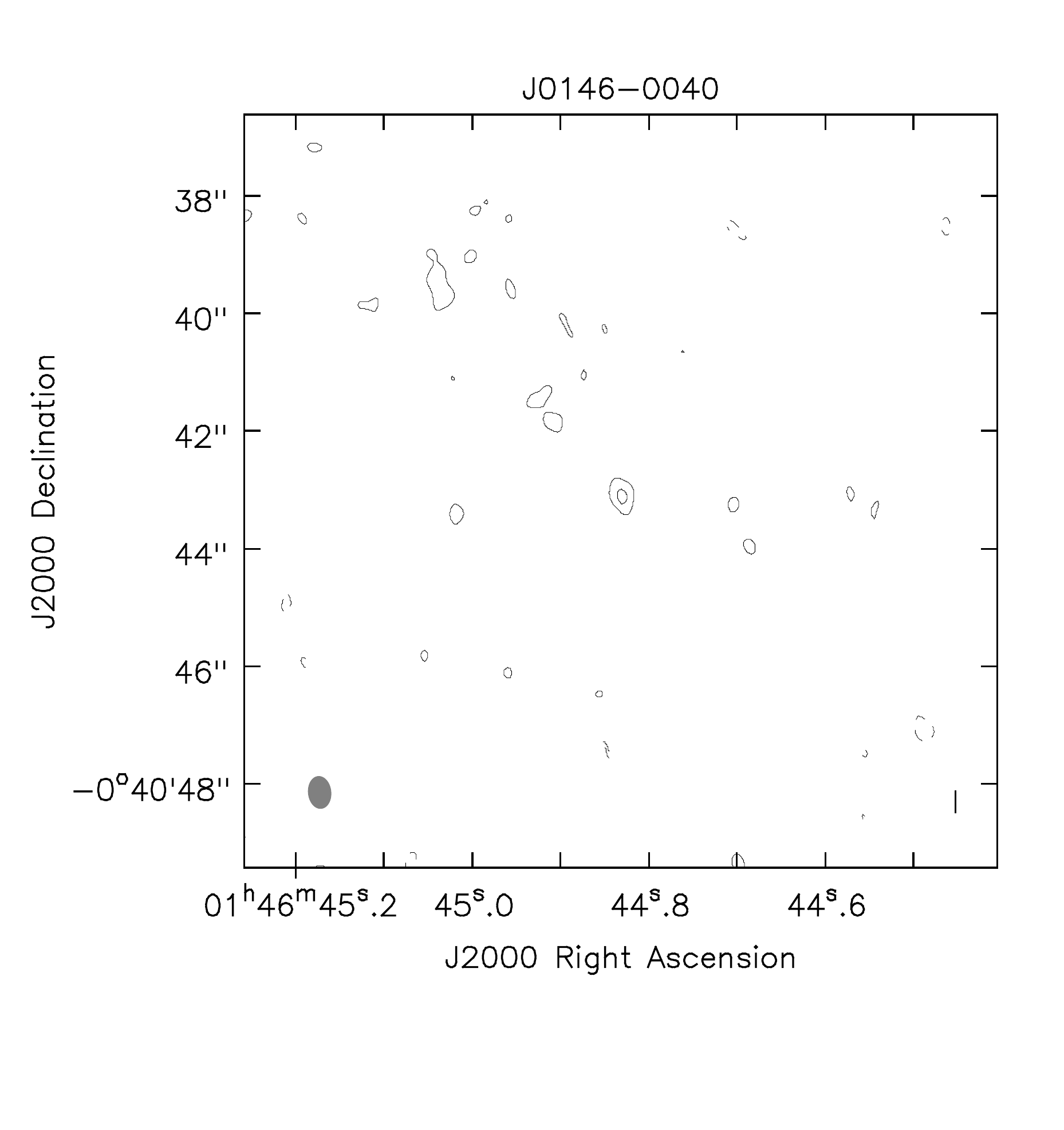} 
\includegraphics[trim={0cm 2cm 0cm 0cm},clip,width=7cm]{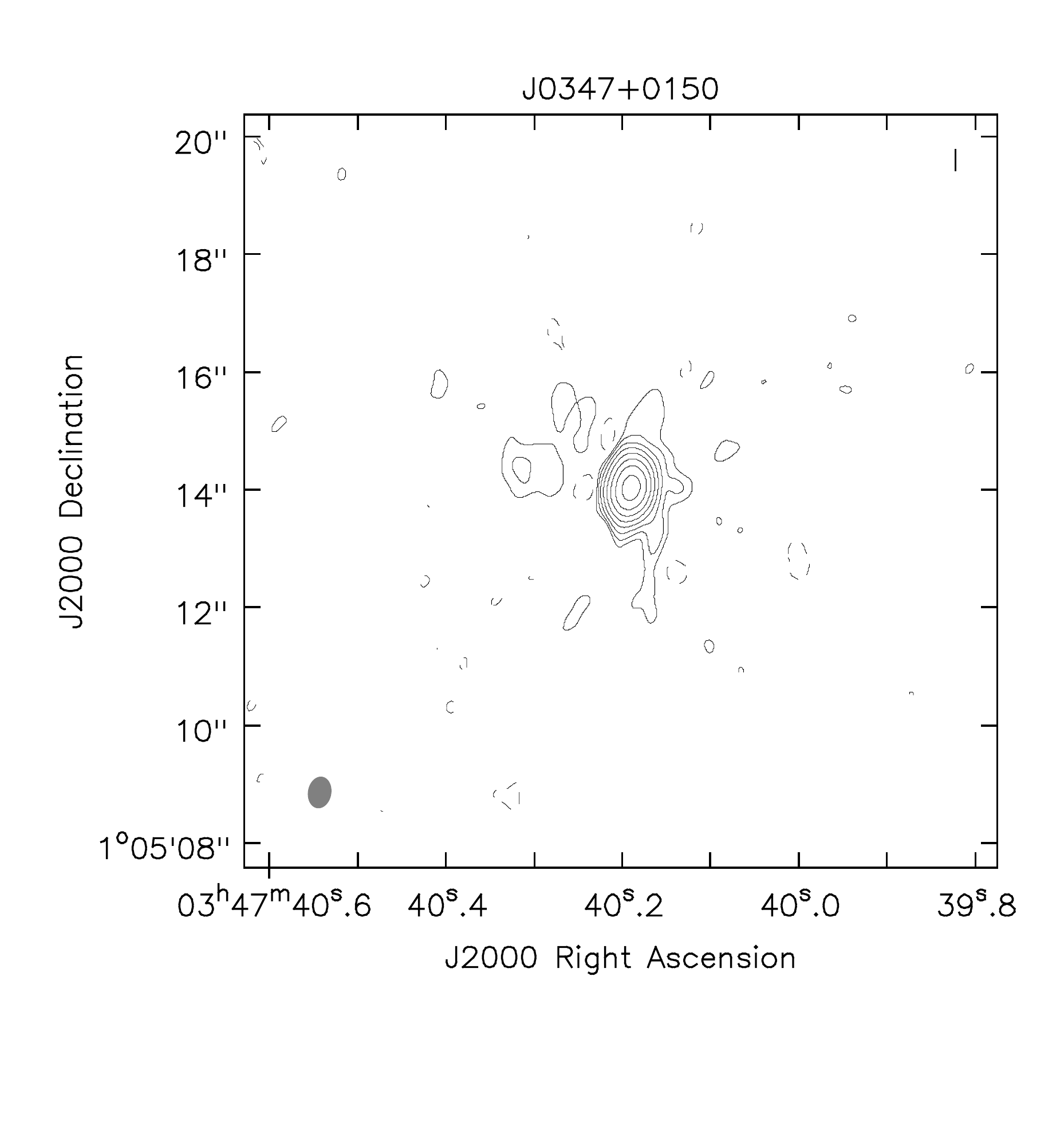} 
\caption{\textbf{Left panel:} J0146-0040, rms = 11 $\mu$Jy, contour levels at -3, 3$\times$2$^n$, n $\in$ [0,1], beam size 0.84$\times$0.59 kpc. \textbf{Right panel:} J0347+0150, rms = 19 $\mu$Jy, contour levels at -3, 3$\times$2$^n$, n $\in$ [0,7], beam size 0.32$\times$0.24 kpc.}
\label{fig:J0146m0040}
\label{fig:J0347p0105}
\end{figure*}

IRAS 03450+0055 is a radio-quiet NLS1 associated with a water maser, often observed in NLS1s \citet{Tarchi11}. The source was already studied at 8.4 GHz with the VLA-A by \citet{Thean00}, who found a flux of 6.8 mJy and an unresolved core, while it was not detected neither through VLBI observations at 22 GHz nor at 37 GHz. In our map (Fig.~\ref{fig:J0347p0105}) the source seems to show some diffuse emission, particularly extended $\sim$2$^{\prime\prime}$ (1.2 kpc) both South (PA$\sim$188$^\circ$) and East (PA$\sim$81$^\circ$) of the core. The map noise is higher with respect to the rest of the sample, possibly because of the lack of four antennas during the observations. The core size deconvolved from the beam is 204$\times$115 pc. It is worth noting that the broad- and in-band spectral indexes are not in agreement, possibly indicating some variability in this object.  
\subsection{J0629$-$0545}
IRAS 06269-0543 is classified as a steep-spectrum radio-loud NLS1, and was observed with the VLA-A at 8.4 GHz by \citet{Moran00}. This source was detected in the NVSS, but it is not present in the FIRST survey. The source is not centered on the NVSS coordinates, but it is found within the error bars. The map (Fig.~\ref{fig:J0629m0545}), indeed, shows two evident components, separated by $\sim$2.4 kpc, in perfect agreement with \citet{Moran00} results. The south-western component is the brightest and the closest to the NVSS coordinates, with a peak flux of 12.47 mJy and a spectral index of 0.14$\pm$0.01. The north-eastern component (PA$\sim$73$^\circ$ with respect to the other component) has instead a flux of 0.42 mJy, and an inverted spectral index (-0.03 $\pm$ 0.04). In \citet{Moran00} the north-eastern component is instead brighter than the other, with a peak of 0.81 mJy/beam. This might indicate that the north-eastern source has a strongly inverted spectrum, while the south-western component has a steeper slope. It is not clear which one of these two components is the source core. 
\subsection{J0632+6340}
\begin{figure*}
\centering
\includegraphics[trim={0cm 2cm 0cm 0cm},clip,width=7cm]{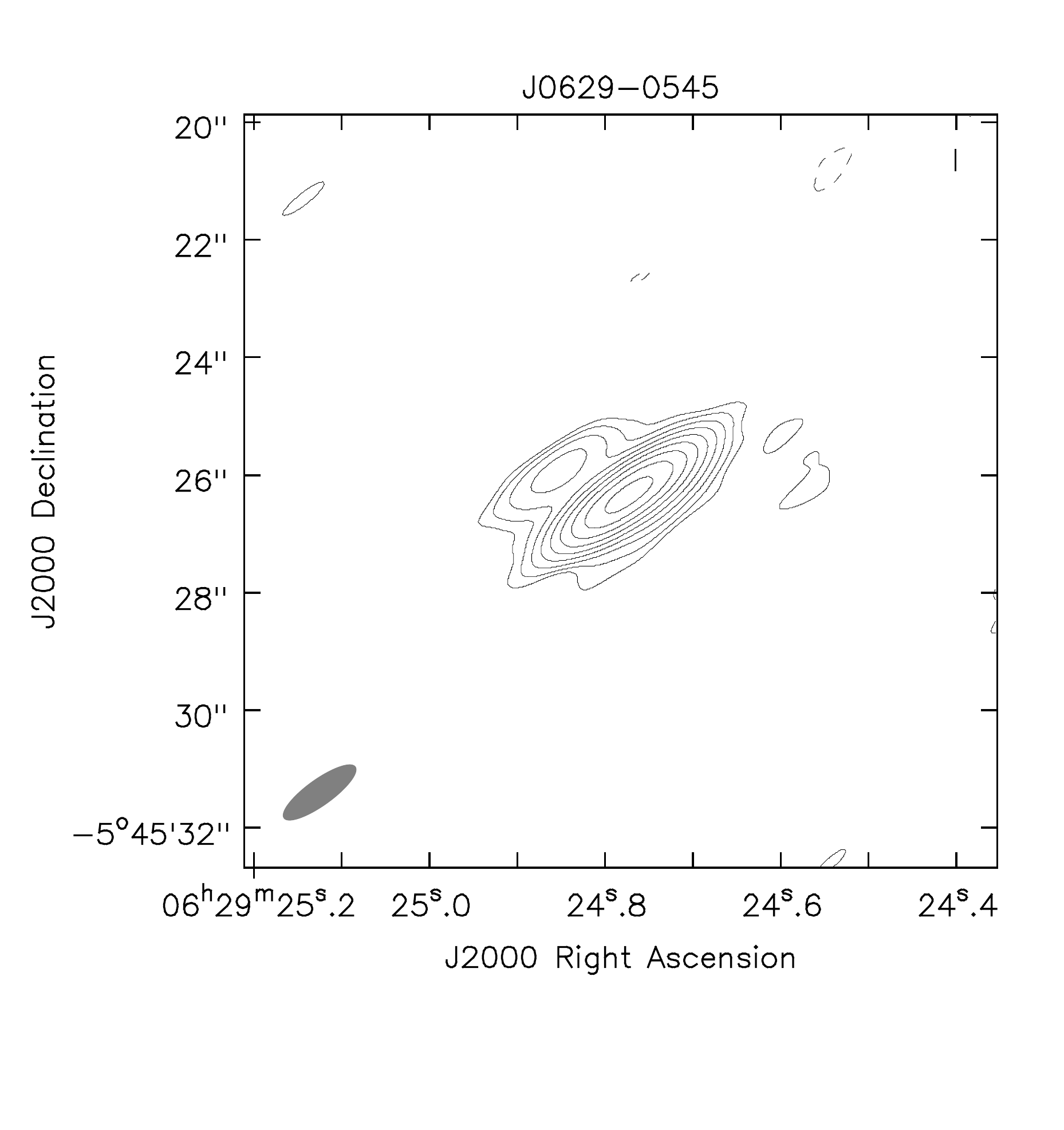} 
\includegraphics[trim={0cm 2cm 0cm 0cm},clip,width=7cm]{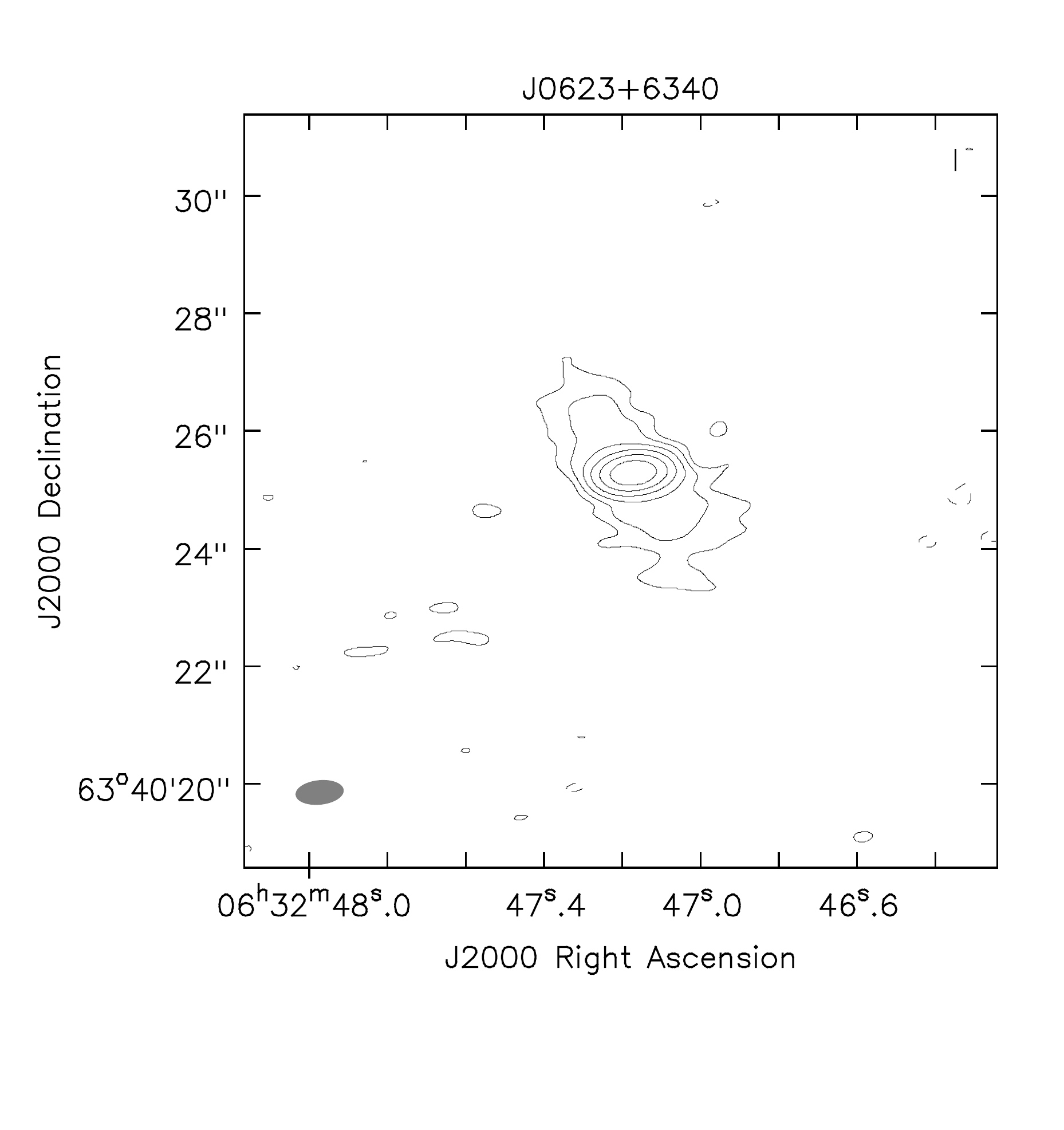} 
\caption{\textbf{Left panel:} J0629$-$0545, rms = 12 $\mu$Jy, contour levels at -3, 3$\times$2$^n$, n $\in$ [0,8], beam size 3.15$\times$0.93 kpc. \textbf{Right panel:} J0623+6340, rms = 10 $\mu$Jy, contour levels at -3, 3$\times$2$^n$, n $\in$ [0,5], beam size 0.21$\times$0.11 kpc.}
\label{fig:J0629m0545}
\label{fig:J0632p6340}
\end{figure*}

UGC 3478 is a radio-quiet NLS1. It was previously studied by \citet{Kinney00} and \citet{Schmitt01} at 8.4 GHz, finding an unresolved radio structure with a scale of 25 pc and a flux of 1.4 mJy. Our map (Fig.~\ref{fig:J0632p6340}) reveal an extended structure on larger scales, with emission diffuse for $\sim$0.6 kpc both north and south of the unresolved core, roughly aligned along PA$\sim$30$^\circ$. The core has a deconvolved size of 66$\times$58 pc. Both the latter and the integrated emission have a steep spectrum, and the broad-band index is even steeper. 
\subsection{J0706+3901}
FBQS J0706+3901 is a \textit{bona-fide} F-NLS1s by \citet{Foschini15}. Our measurements instead indicate that this object is a steep-spectrum radio-loud NLS1, since all of its spectral indexes are larger than 0.5. The map (Fig.~\ref{fig:J0706p3901}) shows an unresolved core, with an approximate size of 800$\times$400 pc, and an intermediate morphology, with a significant diffuse emission around the core. Since it is compact and it has a steep radio spectrum, but a relatively low luminosity, it is classifiable as a low-luminosity CSS.
\subsection{J0713+3820}
\begin{figure*}
\centering
\includegraphics[trim={0cm 2cm 0cm 0cm},clip,width=7cm]{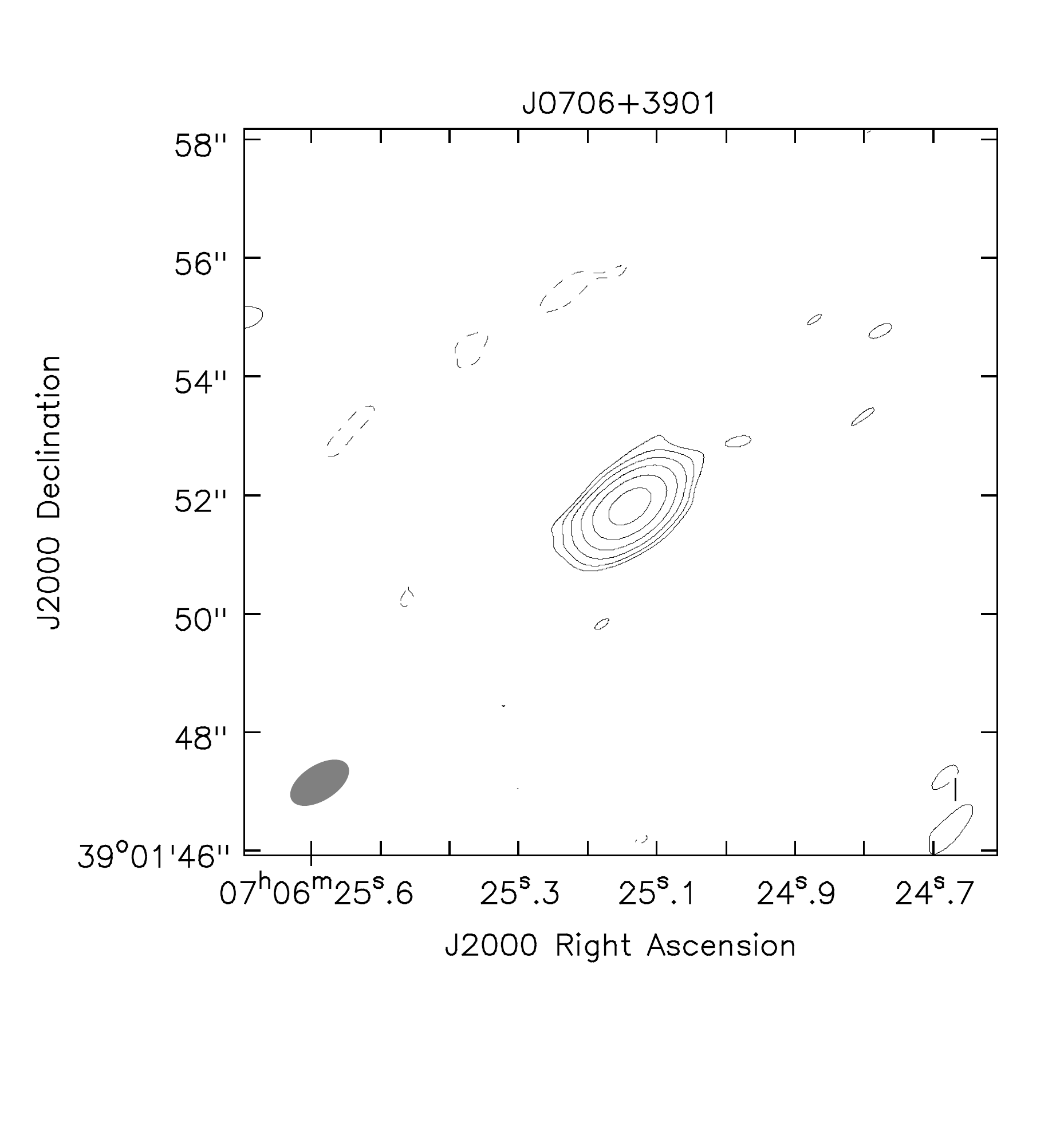} 
\includegraphics[trim={0cm 2cm 0cm 0cm},clip,width=7cm]{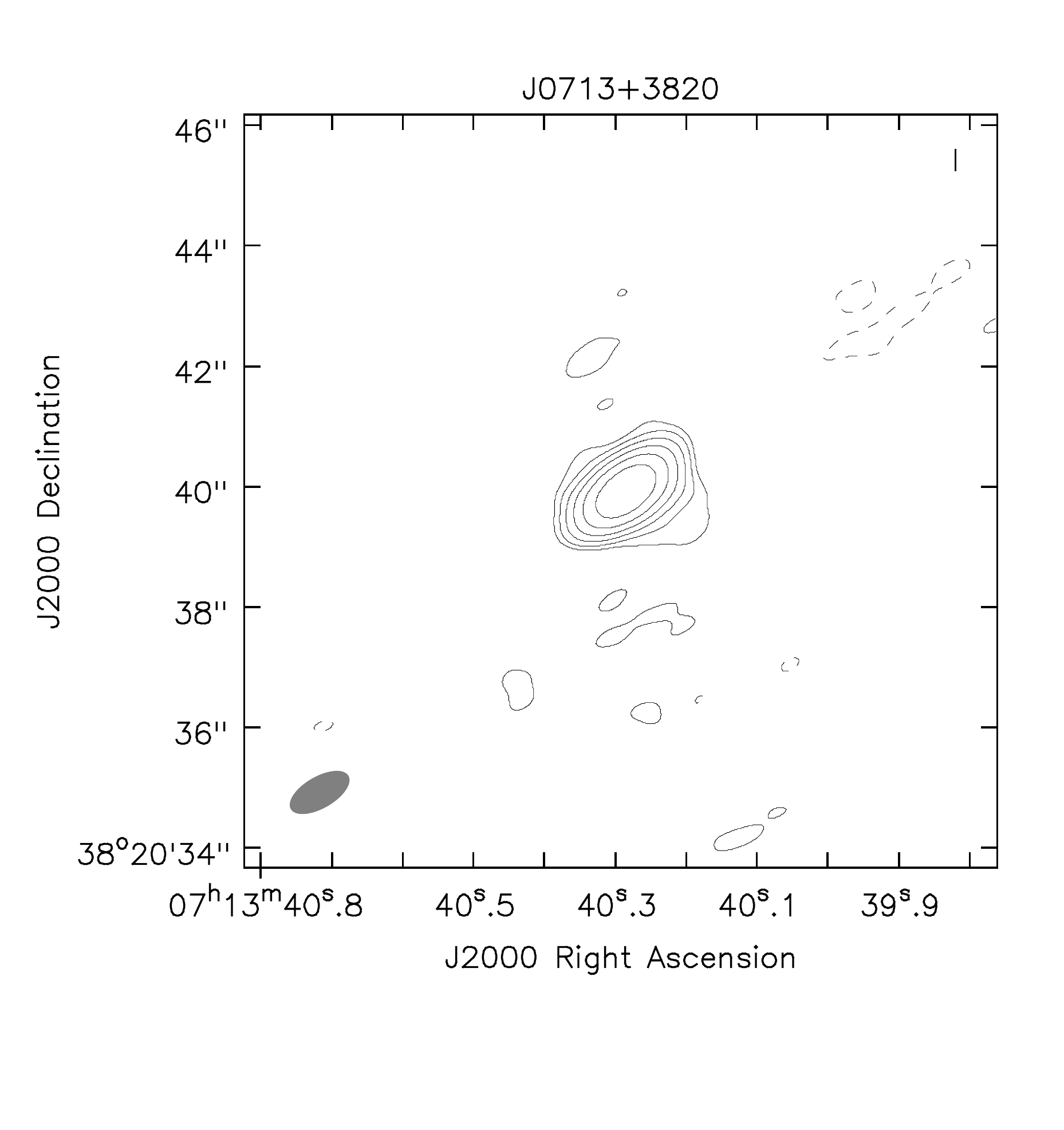} 
\caption{\textbf{Left panel:} J0706+3901, rms = 13 $\mu$Jy, contour levels at -3, 3$\times$2$^n$, n $\in$ [0,5], beam size 1.76$\times$0.93 kpc. \textbf{Right panel:} J0713+3820, rms = 12 $\mu$Jy, contour levels at -3, 3$\times$2$^n$, n $\in$ [0,6], beam size 2.41$\times$1.13 kpc.}
\label{fig:J0706p3901}
\label{fig:J0713p3820}
\end{figure*}

FBQS J0713+3820 was classified as F source by \citet{Foschini15}, with a spectral index of 0.58$\pm$0.12 measured at frequencies below 1.4 GHz. Our observations seem instead to indicate that the index becomes steeper at higher frequencies, in a similar fashion to CSS objects. The source indeed was not detected neither at 22 nor at 37 GHz. The map (Fig.~\ref{fig:J0713p3820}) indicates that the radio emission is compact, with a core size of 1.16$\times$0.59 kpc. As other RLNLS1s, J0713+3820 likely belongs to the class of low-luminosity CSS. 
\subsection{J0744+5149}
Little is known about NVSS J074402+514917, which was included \textit{bona-fide} in \citet{Foschini15} F sample. However, all of its spectral index are in agreement within the error bars, and indicate a rather steep radio spectrum, $\alpha_\nu \sim$1.2, consistent with the non-detection at 22 GHz. The map (Fig.~\ref{fig:J0744p5149}) shows only an unresolved core with a size of 0.7$\times$0.3 kpc. Its total luminosity is approximately 10$^{41}$ \ergs, which is higher with respect to other radio-loud NLS1s, and comparable to that of the CSS-NLS1 J1432+3014 \citep{Caccianiga15}. A strong ongoing star formation (300 M$_\odot$ yr$^{-1}$) was found investigating the WISE colors of this object \citep{Caccianiga15}. The q22 ratio in this source is 0.47, possibly indicating that the contribution of star formation is not the main component of the radio emission. 
\subsection{J0752+2617}
\begin{figure*}
\centering
\includegraphics[trim={0cm 2cm 0cm 0cm},clip,width=7cm]{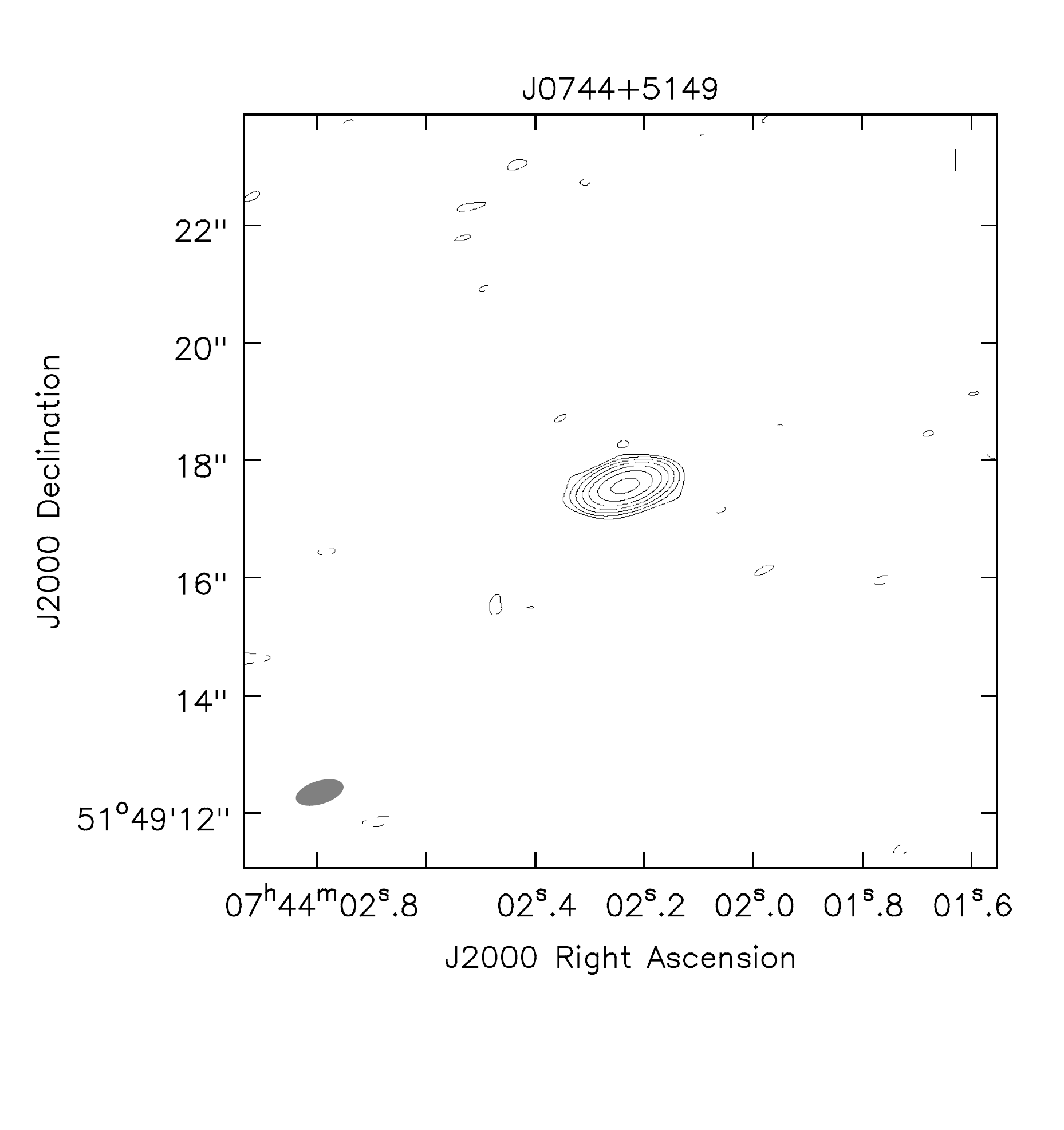} 
\includegraphics[trim={0cm 2cm 0cm 0cm},clip,width=7cm]{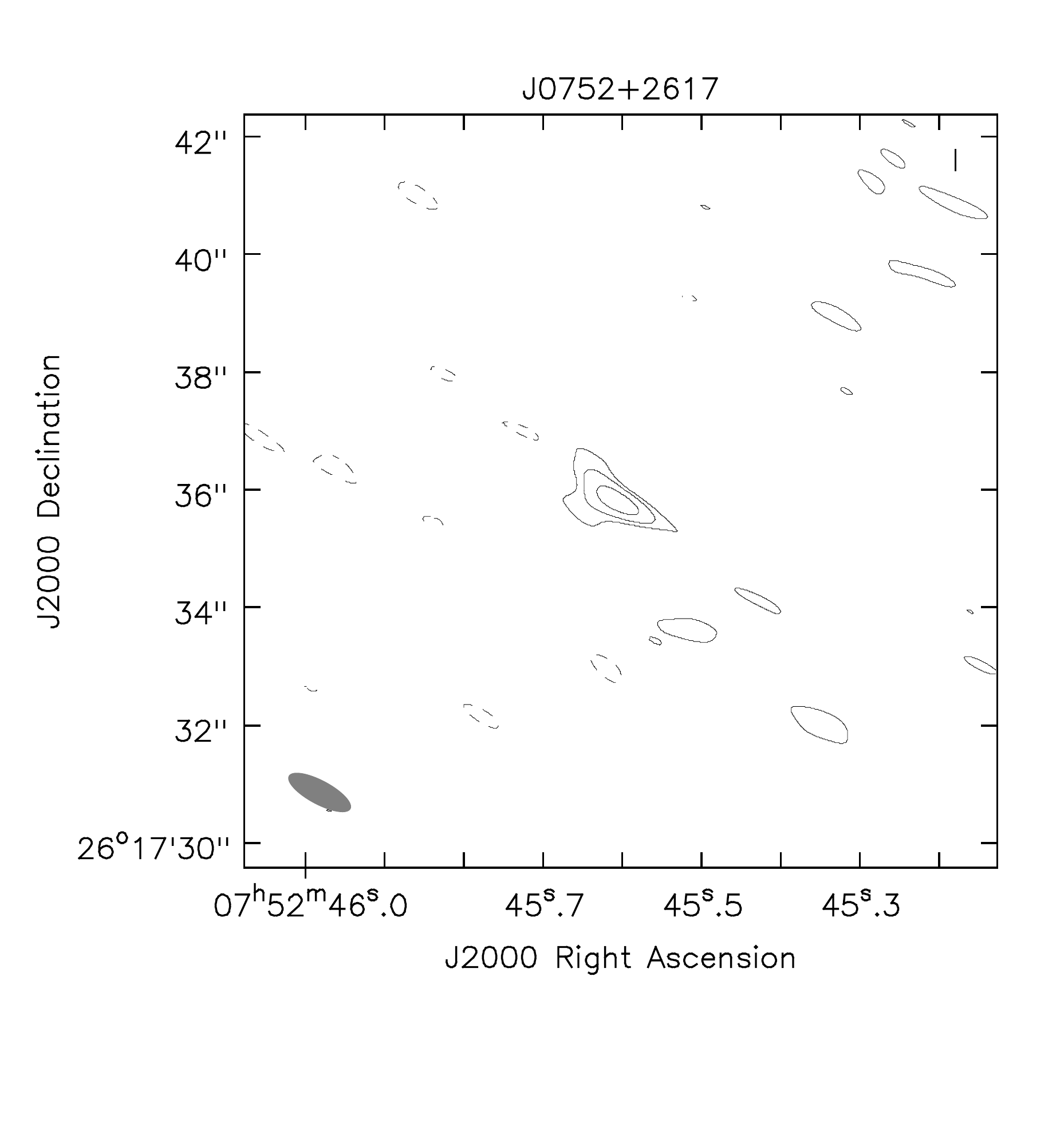} 
\caption{\textbf{Left panel:} J0744+5149, rms = 10 $\mu$Jy, contour levels at -3, 3$\times$2$^n$, n $\in$ [0,6], beam size 4.78$\times$2.21 kpc. \textbf{Right panel:} J0752+2617, rms = 12 $\mu$Jy, contour levels at -3, 3$\times$2$^n$, n $\in$ [0,2], beam size 1.81$\times$0.60 kpc.}
\label{fig:J0744p5149}
\label{fig:J0752p2617}
\end{figure*}

Also known as FBQS J0752+2617, it is classified as a radio-quiet NLS1 and detected only at a 12$\sigma$ level. The map of Fig.~\ref{fig:J0752p2617} shows only an unresolved core whose size could not be determined. The spectral indexes seem all to point out that the source has a fairly steep spectrum, with $\alpha_\nu$ larger than 1.
\subsection{J0758+3920} 
B3 0754+394 is a radio-loud NLS1 which showed a steep-spectrum below 1.4 GHz \citep{Berton15a}. All of our measurements are consistent with this classification. The lack of strong variability found by \citet{Sergeev07} in a reverberation mapping campaign is also supportive of its unbeamed emission. In the map shown in Fig.~\ref{fig:J0758p3920} the source do not show large scale structure, with a core size of 2.2$\times$0.7 kpc, but a large fraction of its flux from diffuse emission outside the core. At higher frequencies the source was detected neither at 22 GHz, nor at 37 GHz. 
\subsection{J0804+3853}
\begin{figure*}
\centering
\includegraphics[trim={0cm 2cm 0cm 0cm},clip,width=7cm]{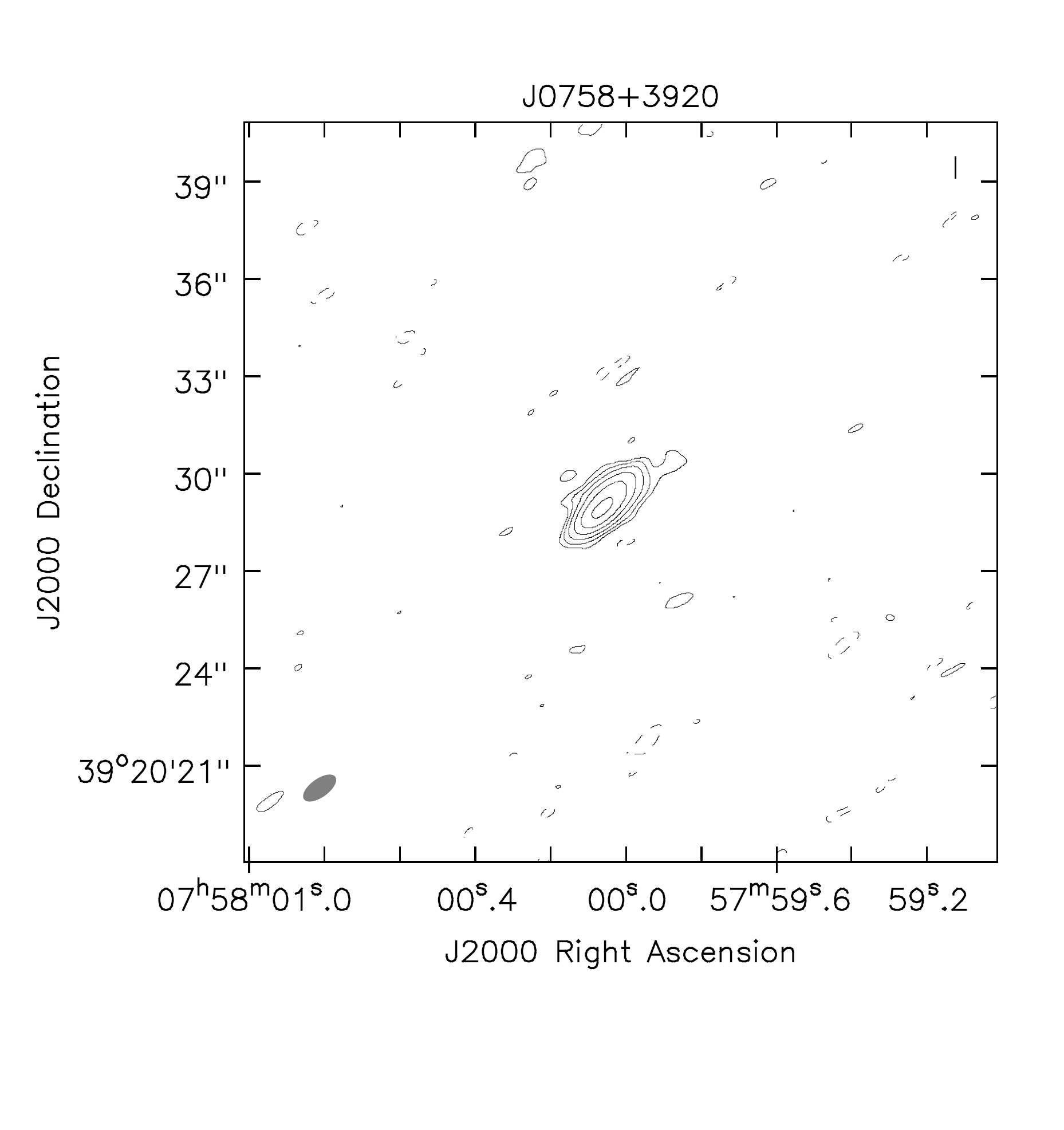} 
\includegraphics[trim={0cm 2cm 0cm 0cm},clip,width=7cm]{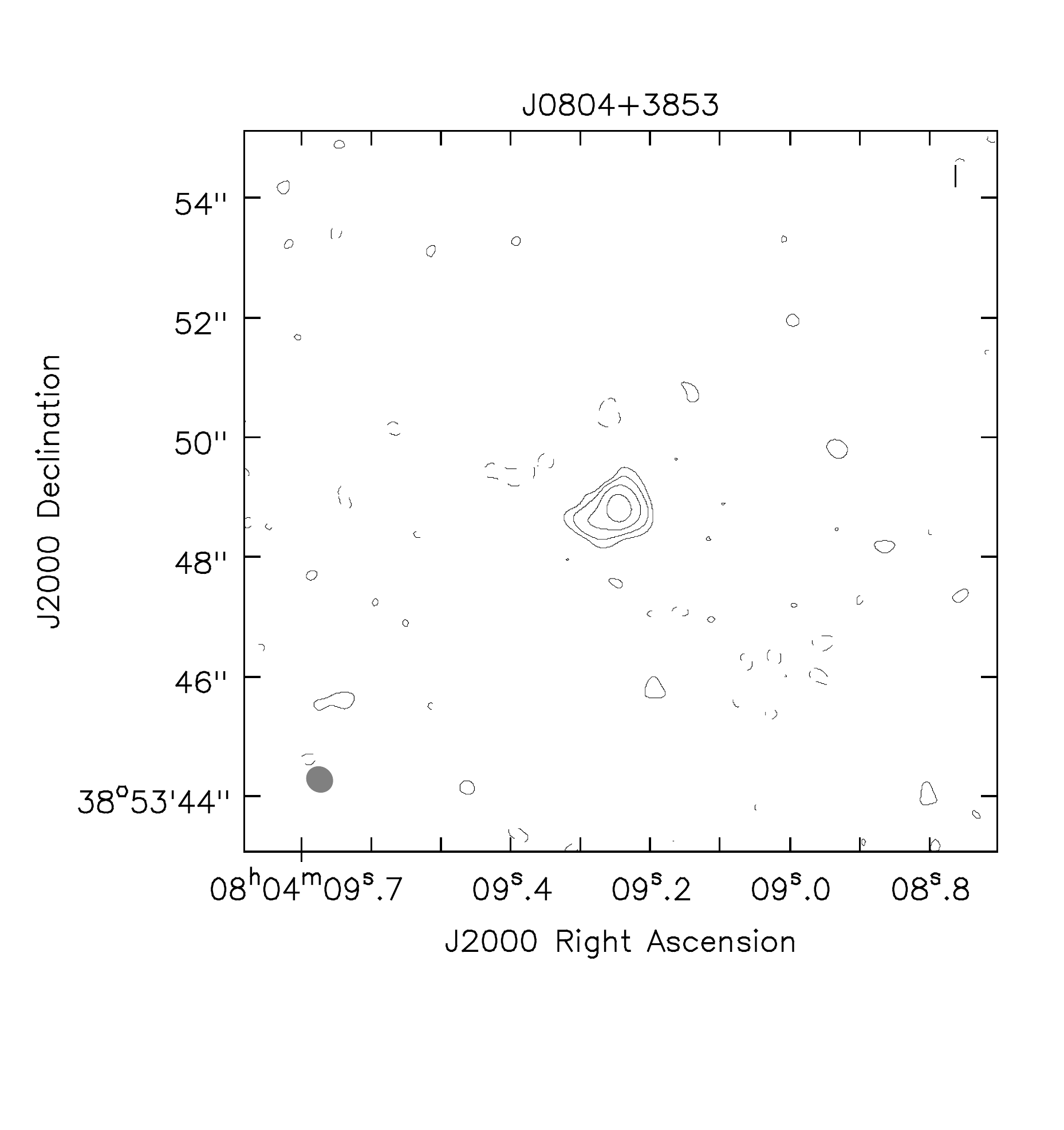} 
\caption{\textbf{Left panel:} J0758+3920, rms = 18 $\mu$Jy, contour levels at -3, 3$\times$2$^n$, n $\in$ [0,5], beam size 2.05$\times$0.94 kpc. \textbf{Right panel:} J0804+3853, rms = 11 $\mu$Jy, contour levels at -3, 3$\times$2$^n$, n $\in$ [0,3], beam size 1.55$\times$1.42 kpc.}
\label{fig:J0758p3920}
\label{fig:J0804p3853}
\end{figure*}

FBQS J0804+3853 if one of the \textit{bona-fide} F sources. Its broad- and in-band spectral indexes however indicate that it should be classified as a S source. The source, as expected from its spectrum, is not detected at 37 GHz. The map (Fig.~\ref{fig:J0804p3853}) shows a moderate extension of the contours in the East direction (PA$\sim$113$^\circ$), and a deconvolved core size of 2.1$\times$1.3 kpc. This object has a strong star formation rate of about 89 M$_\odot$ yr$^{-1}$ and a high q22, 1.48, suggesting that the majority of the radio emission in such object comes from starburst. The brightness temperature of this object, 138 K, is one of the lowest in the sample, in agreement with the starburst-dominated radio emission. 
\subsection{J0806+7248}
RGB J0806+728 was classified as an S source given its steep broad-band index \citet{Berton15a}. However, both core and integrated spectral indexes we measured are flat. This might be interpreted as a sign of variability, moving the classification from S to F source. The source was indeed already observed at 5 GHz, and showed a flux of 20 mJy \citep{Doi07}, two times larger than our measurement. At 1.7 GHz on VLBA scales this source shows a diffuse morphology extended up to 100 pc from the core, while it appears as two sided for VLBA at 8.6 GHz \citep{Doi11}. Our map (Fig.~\ref{fig:J0806p7248}) do not show extended emission, which is clearly on smaller scales, but we find a significant diffuse emission around the unresolved core, whose size is approximately 250$\times$200 pc. The source was not detected at high frequency (22 and 37 GHz). 
\subsection{J0814+5609}
\begin{figure*}
\centering
\includegraphics[trim={0cm 2cm 0cm 0cm},clip,width=7cm]{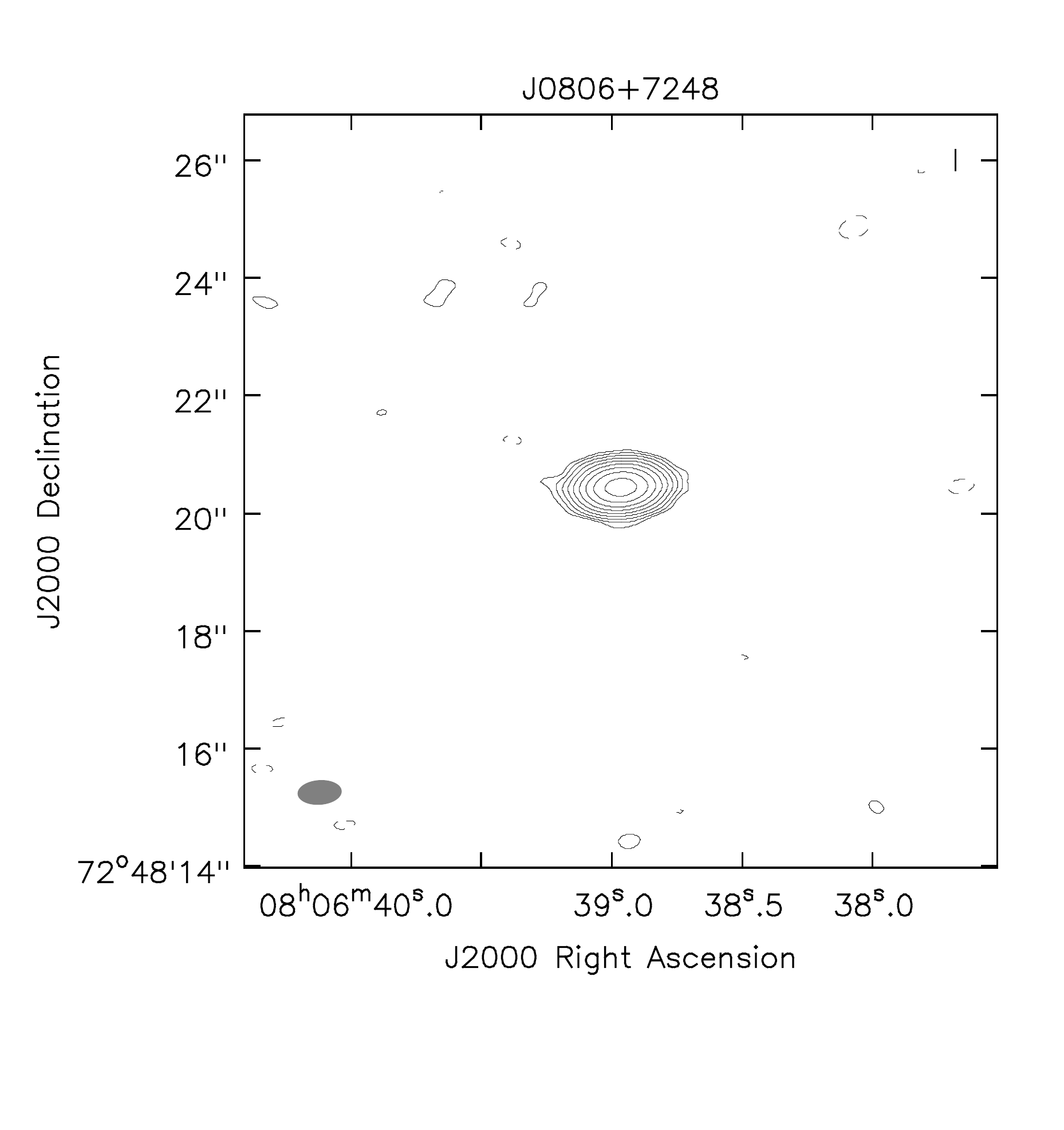} 
\includegraphics[trim={0cm 2cm 0cm 0cm},clip,width=7cm]{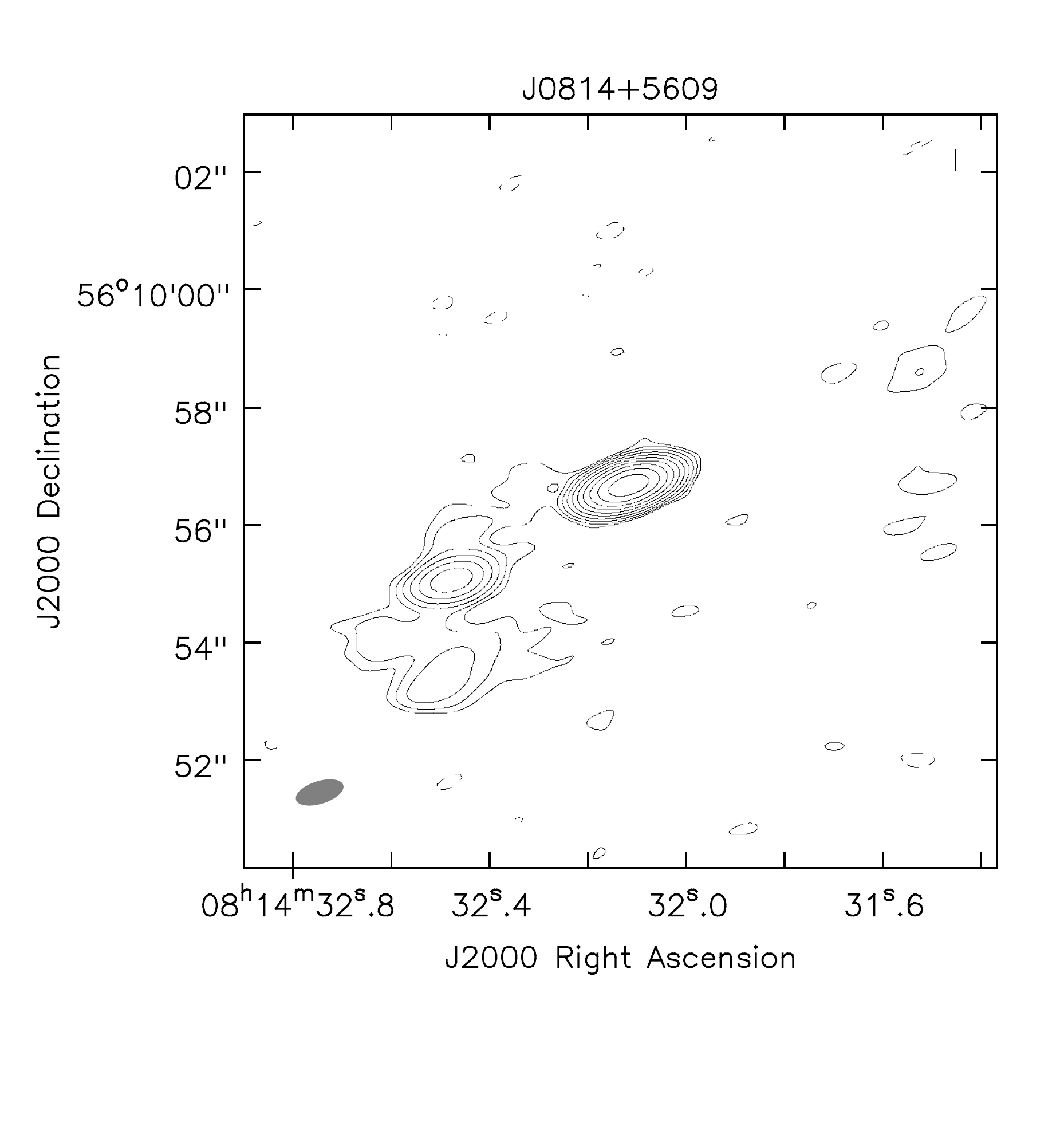} 
\caption{\textbf{Left panel:} J0806+7248, rms = 10 $\mu$Jy, contour levels at -3, 3$\times$2$^n$, n $\in$ [0,8], beam size 1.32$\times$0.72 kpc. \textbf{Right panel:} J0814+5609, rms = 10 $\mu$Jy, contour levels at -3, 3$\times$2$^n$, n $\in$ [0,9], beam size 5.06$\times$2.28 kpc.}
\label{fig:J0806p7248}
\label{fig:J0814p5609}
\end{figure*}

SDSS J081432.11+560956.6 was classified as a F source by \citet{Foschini15}, with a spectral index below 1.4 GHz of 0.38$\pm$0.01. The map (Fig.~\ref{fig:J0814p5609}) clearly shows that the source has a large scale extended emission directed toward South-East. The extended emission, accounting for $\sim$12\% of the total flux, shows a bright spot at PA$\sim$119$^\circ$, with a peak of 1.58$\pm$0.02 mJy, at 3.4 arcsec ($\sim$21 kpc), while the total emission on the map reaches a distance of 5.3 arcsec ($\sim$33 kpc), where it shows an edge-brightening at PA$\sim$137$^\circ$. The deconvolved size of the core is 0.5$\times$0.2 kpc. On VLBA scales \citep{Gu15} the structures we observed are resolved-out, but the source still shows an elongated structure at 2, 5 and 8 GHz, extended toward East just like on our map. The in-band spectral indexes are flat, showing that both the core and the secondary bright spot have a flat spectrum. The broad-band index instead is slightly steeper. This might be due both to the extended emission, but also to variability. \citet{Doi16} found a VLBI flux of 117 mJy at 22 GHz, sign of strong variability in the core. Conversely, despite multiple observations, the object was not detected at 37 GHz. 
\subsection{J0849+5108}
SBS J0846+513 is one of the brightest $\gamma$-ray emitting NLS1, that on VLBA scales at 5 GHz and higher frequencies shows a core-jet morphology and superluminal motion (up to 8c, \citealp{Dammando12}). It was also monitored at multiple frequencies, showing a rather strong variability both in flux and polarization \citep{Maune14, Angelakis15}. It is a flat spectrum source, and as expected it was detected both at 22 GHz (flux 454 mJy), and at 37 GHz (highest measured flux of 1.18 Jy). The source on kpc scales has a compact morphology (Fig.~\ref{fig:J0849p5108}), with a core size 320$\times$215 pc. The brightness temperature calculated from our data, although relatively low because of the large beam size of the VLA, is the second highest of our sample, with $\sim$1.3$\times$10$^7$ K. 
\subsection{J0850+4626}
\begin{figure*}
\centering
\includegraphics[trim={0cm 2cm 0cm 0cm},clip,width=7cm]{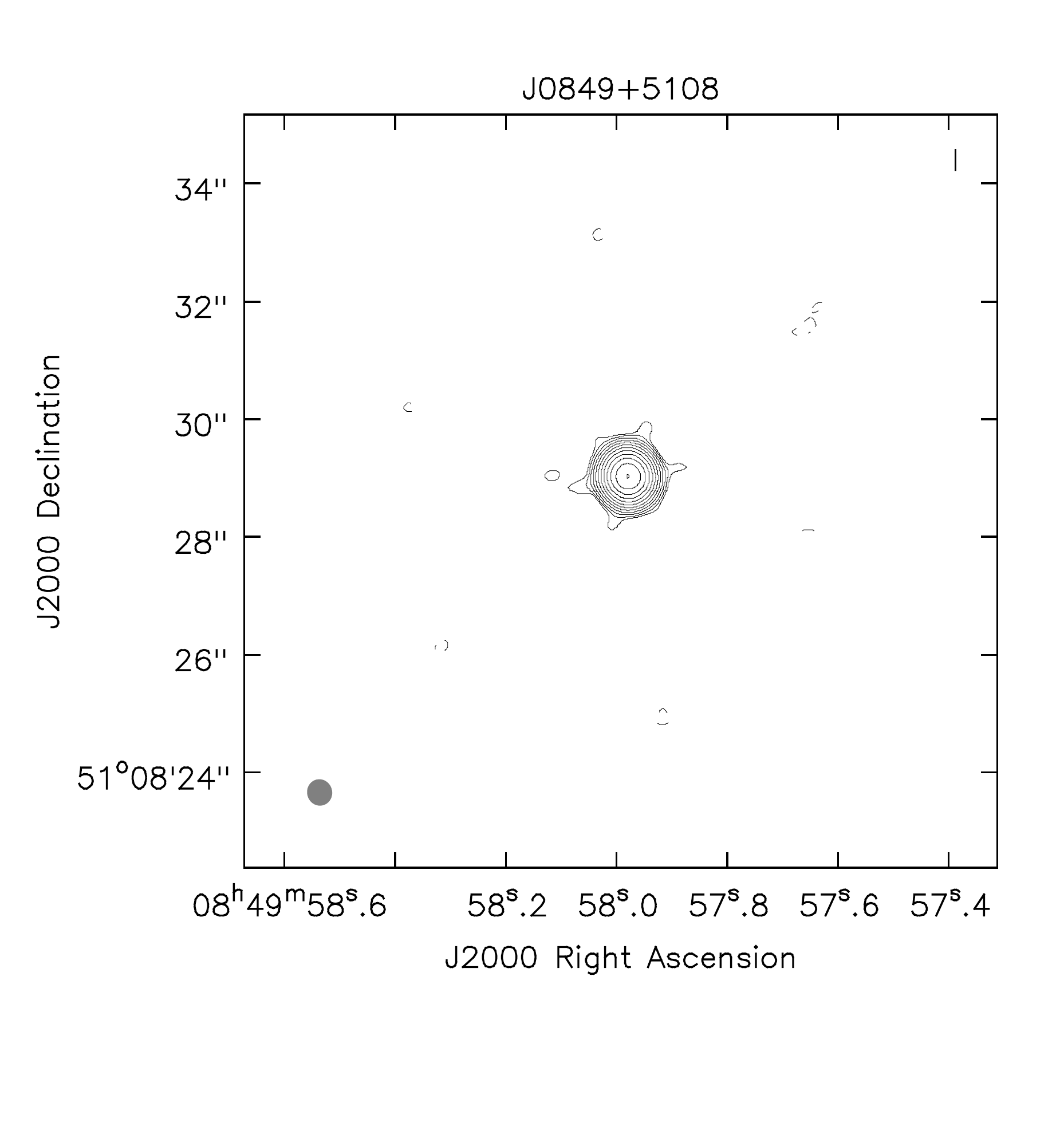} 
\includegraphics[trim={0cm 2cm 0cm 0cm},clip,width=7cm]{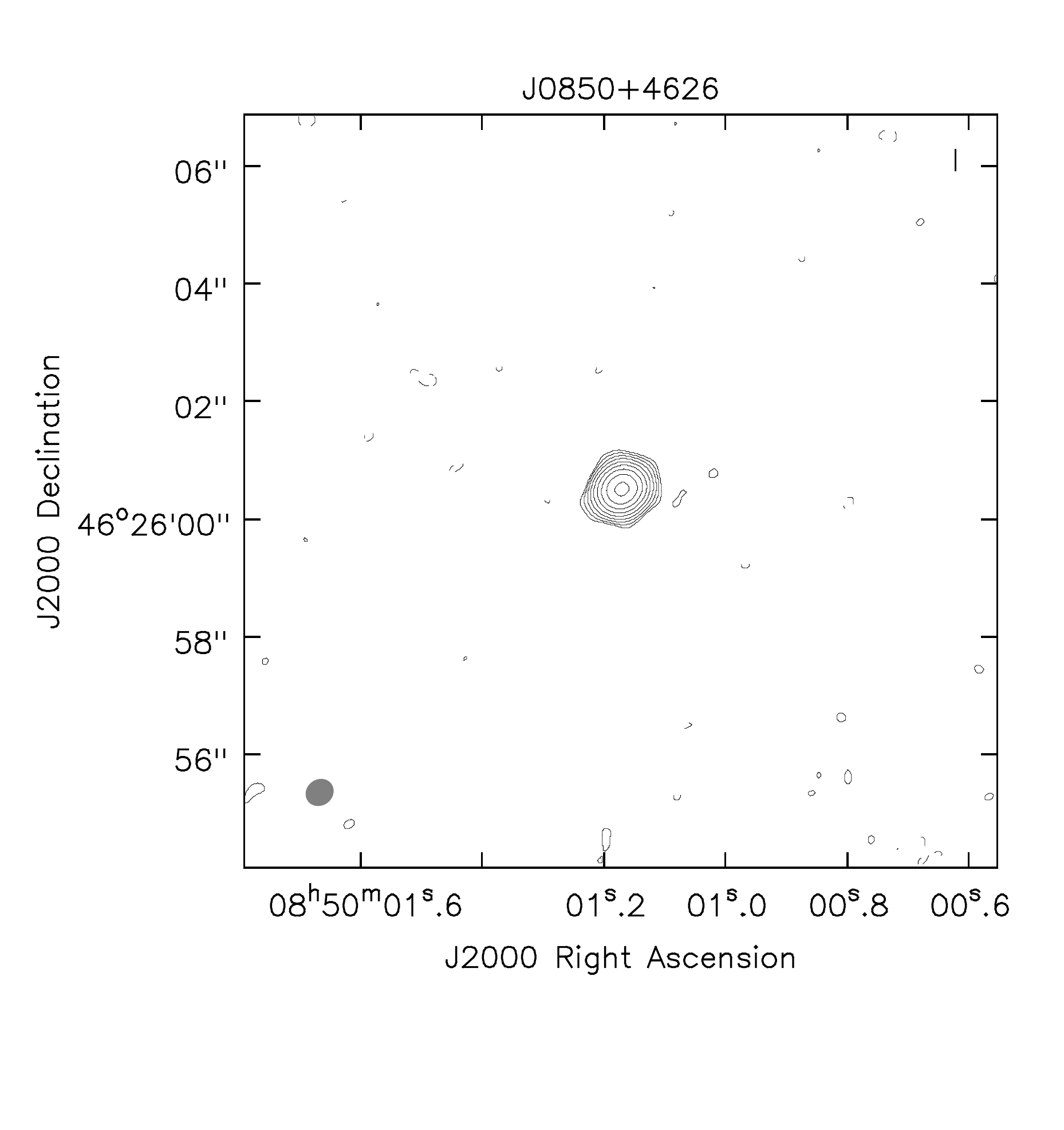} 
\caption{\textbf{Left panel:} J0849+5108, rms = 35 $\mu$Jy, contour levels at -3, 3$\times$2$^n$, n $\in$ [0,11], beam size 2.84$\times$2.71 kpc. \textbf{Right panel:} J0850+4626, rms = 10 $\mu$Jy, contour levels at -3, 3$\times$2$^n$, n $\in$ [0,8], beam size 3.00$\times$2.63 kpc.}
\label{fig:J0849p5108}
\label{fig:J0850p4626}
\end{figure*}

SDSS J085001.16+462600.5 was classified as S source according to low frequency ($\nu <$ 1.4 GHz) data. Our spectral indexes however seem to be consistent with a flat-spectrum source. The source, however, is detected neither at 22 GHz, nor at 37 GHz. The object appears as an unresolved core (Fig.~\ref{fig:J0850p4626}), with a deconvolved size of 375$\times$350 pc. This source was studied on VLBA scales by \citet{Gu15}, who found a compact morphology. The flux they measured at 5 GHz (11 mJy) is slightly larger than our value, possibly indicating a relatively small scale variability. 
\subsection{J0902+0443} %cateti misurati: W-E 0.637594, N-S 0.818045
\label{sec:J0902+0443}
SDSS J090227.16+044309.5 was included in the F sample by \citet{Foschini15}, showing a broad-band spectral index of 0.07$\pm$0.01. Our results confirm this classification. The source is one of the few showing a flux larger than 100 mJy, with a luminosity of $\sim$6$\times$10$^{42}$ \ergs. A significant flux variability ($\sim$10 mJy) was found by previous measurements at 5 GHz with the NRAO 91m telescope \citep{Becker91, Griffith95, Yuan08}, and is also consistent with our result. Our map, shown in Fig.~\ref{fig:J0902p0443} shows a partially resolved structure of $\sim$6.5 kpc at PA$\sim$142$^\circ$, in the South-East direction. An elongated structure was already visible on VLBA scales, although its direction appears to be slightly different \citep{Gu15}. The source was detected at 22 GHz, but not at 37 GHz. This source has a very high star formation rate ($\sim$500 M$_\odot$ yr$^{-1}$), which however is not strongly contributing to the radio emission, as suggesting by the q22 = -0.65. 
\subsection{J0913+3658}
\begin{figure*}
\centering
\includegraphics[trim={0cm 2cm 0cm 0cm},clip,width=7cm]{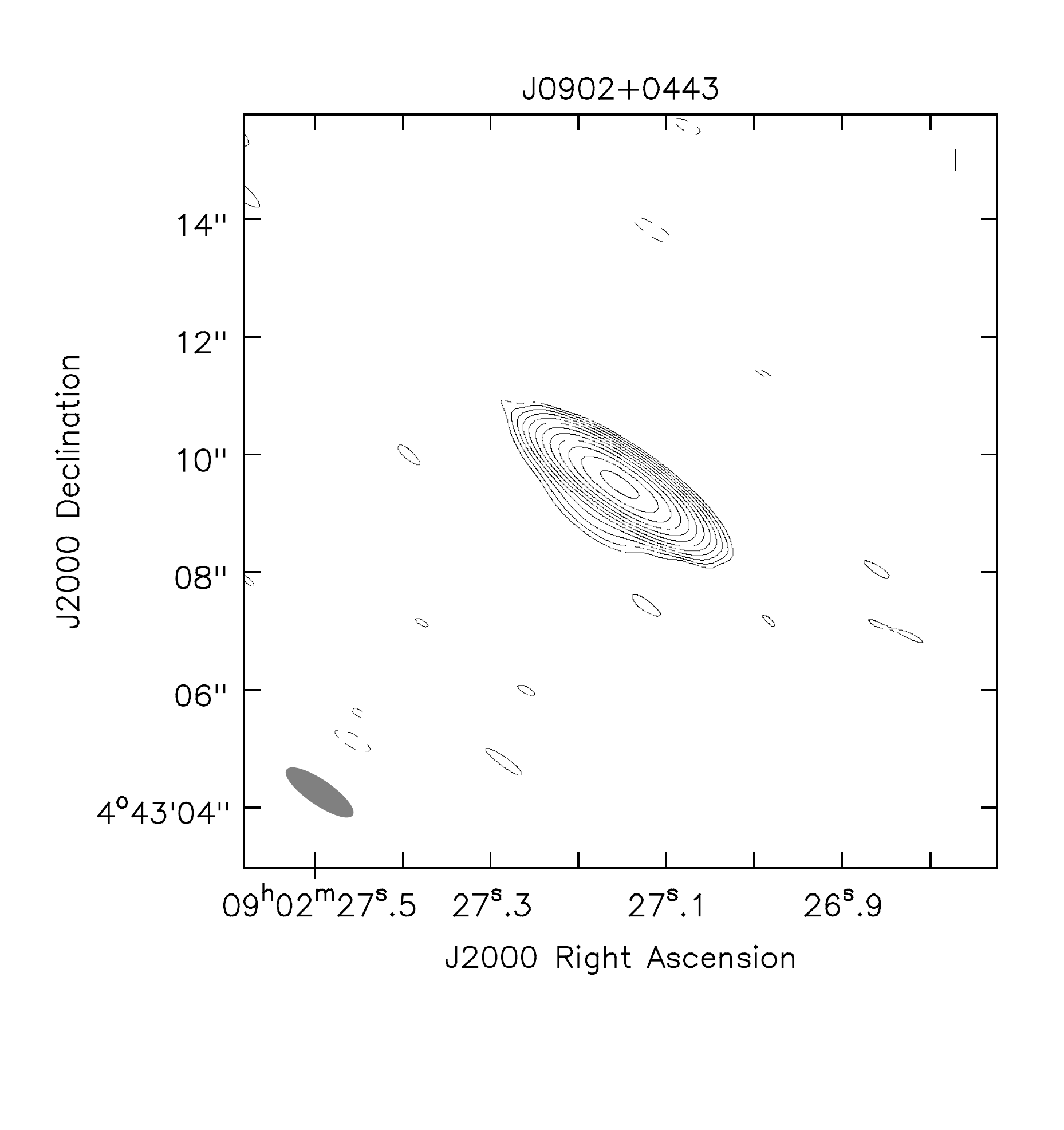} 
\includegraphics[trim={0cm 2cm 0cm 0cm},clip,width=7cm]{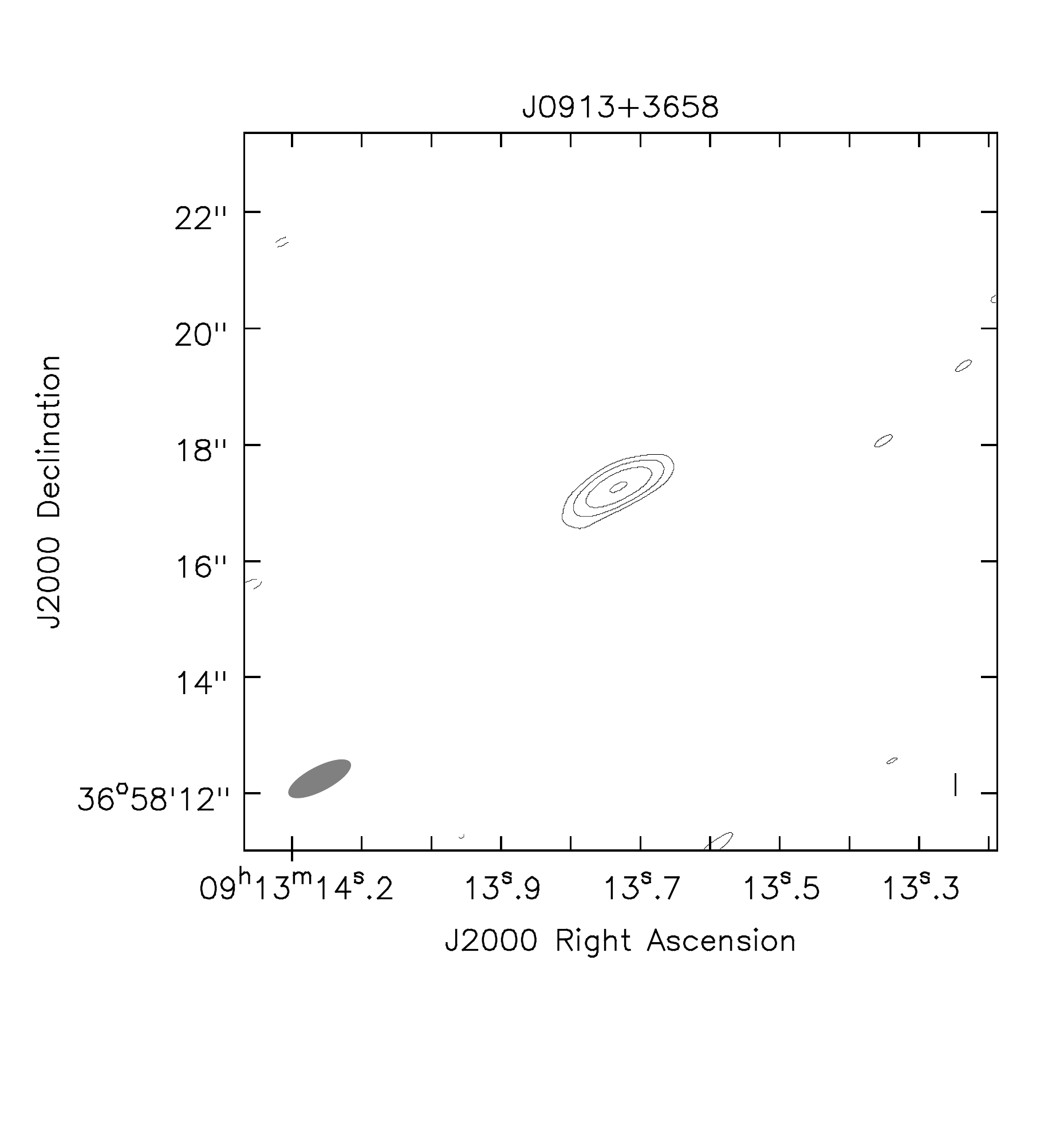} 
\caption{\textbf{Left panel:} J0902+0443, rms = 13 $\mu$Jy, contour levels at -3, 3$\times$2$^n$, n $\in$ [0,11], beam size 8.52$\times$2.59 kpc. \textbf{Right panel:} J0913+3658, rms = 11 $\mu$Jy, contour levels at -3, 3$\times$2$^n$, n $\in$ [0,3], beam size 2.31$\times$0.24 kpc.}
\label{fig:J0902p0443}
\label{fig:J0913p3658}
\end{figure*}

RX J0913.2+3658 is a radio-quiet object, already observed by the FIRST survey with a flux of approximately 1 mJy. Our observation is the first carried out at 5 GHz, and it revealed a rather steep spectrum and a low brightness temperature. In the map shown in Fig.~\ref{fig:J0913p3658} the source shows a significant diffuse emission outside of an unresolved core whose size is 680$\times$350 pc. 
\subsection{J0925+5217} %da qui uso l'est giusto. verso sud misuro 2.790998 x, 5.10075 y (5.81"), PA 209. verso nord struttura piu brillante, 0.625564 x, 2.18947 y (2.28"), PA 16. 
Mrk 110 is a well studied radio-quiet NLS1, detected by NVSS and often studied at 5 GHz, finding fluxes consistent with our measurements \citep{Nelson00, Wu05, Kataoka11}. Its core has a rather flat spectral index, while the total flux is steep. At VLBA scales, this object shows only a point-like core of 1.1 mJy \citep{Doi13}. Our map (Fig.~\ref{fig:J0925p5217}) shows an extended diffuse emission toward North, West and South. In particular, at PA 209$^\circ$ the emission reaches a distance of 5.8 arcsec (11.4 kpc), while a bright spot can be found at PA 16$^\circ$ at 2.3 arcsec (4.5 kpc). The emission in the North direction was already detected by \citet{Kukula98}. The diffuse flux represents approximately 51\% of the source total emission. The unresolved core in our image has a size of 200$\times$130 pc.
\subsection{J0926+1244} %S1, x 1.56391, y 0.661789. S2 x 0.676692, y 0.857317
\begin{figure*}
\centering
\includegraphics[trim={0cm 2cm 0cm 0cm},clip,width=7cm]{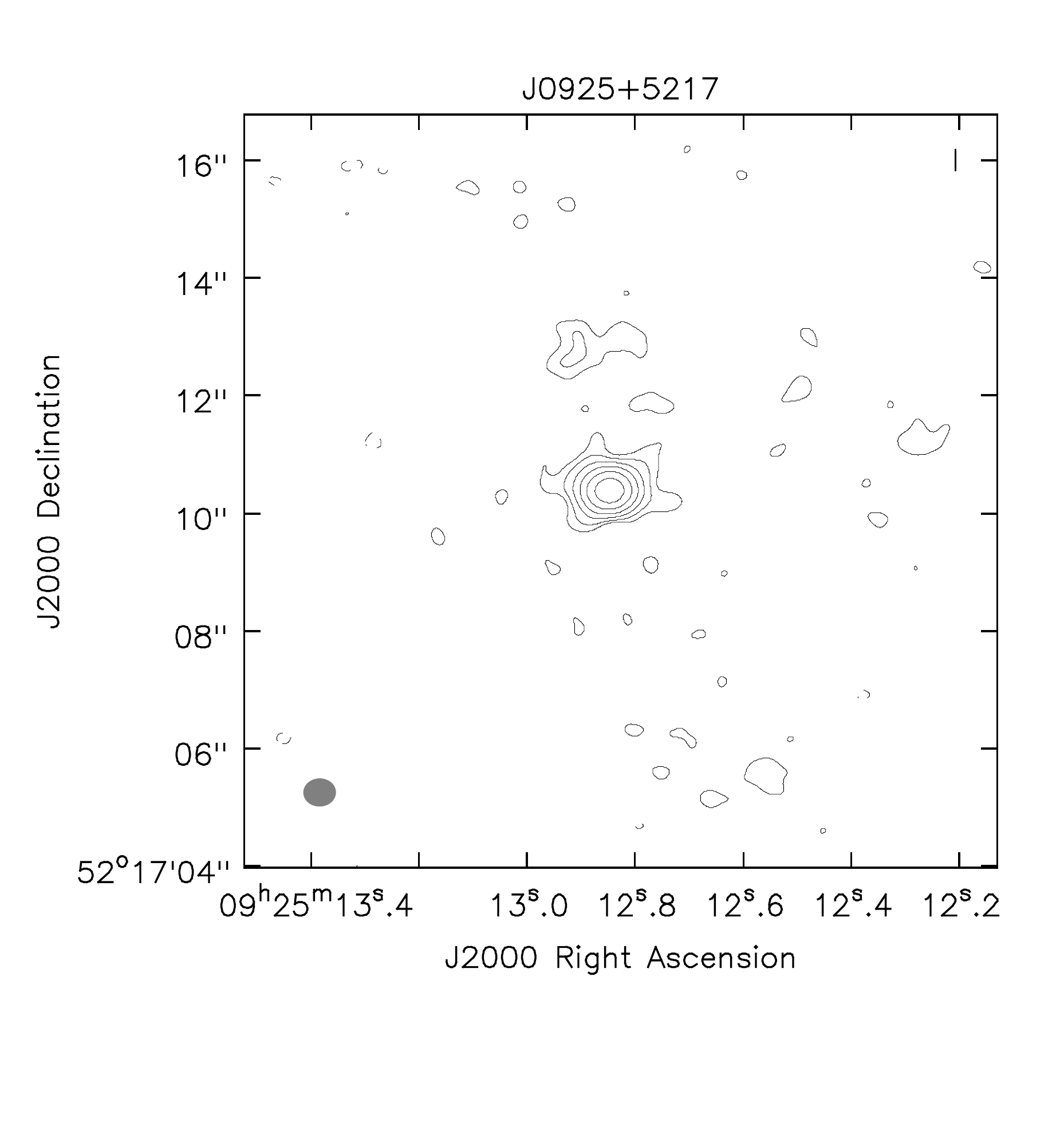} 
\includegraphics[trim={0cm 2cm 0cm 0cm},clip,width=7cm]{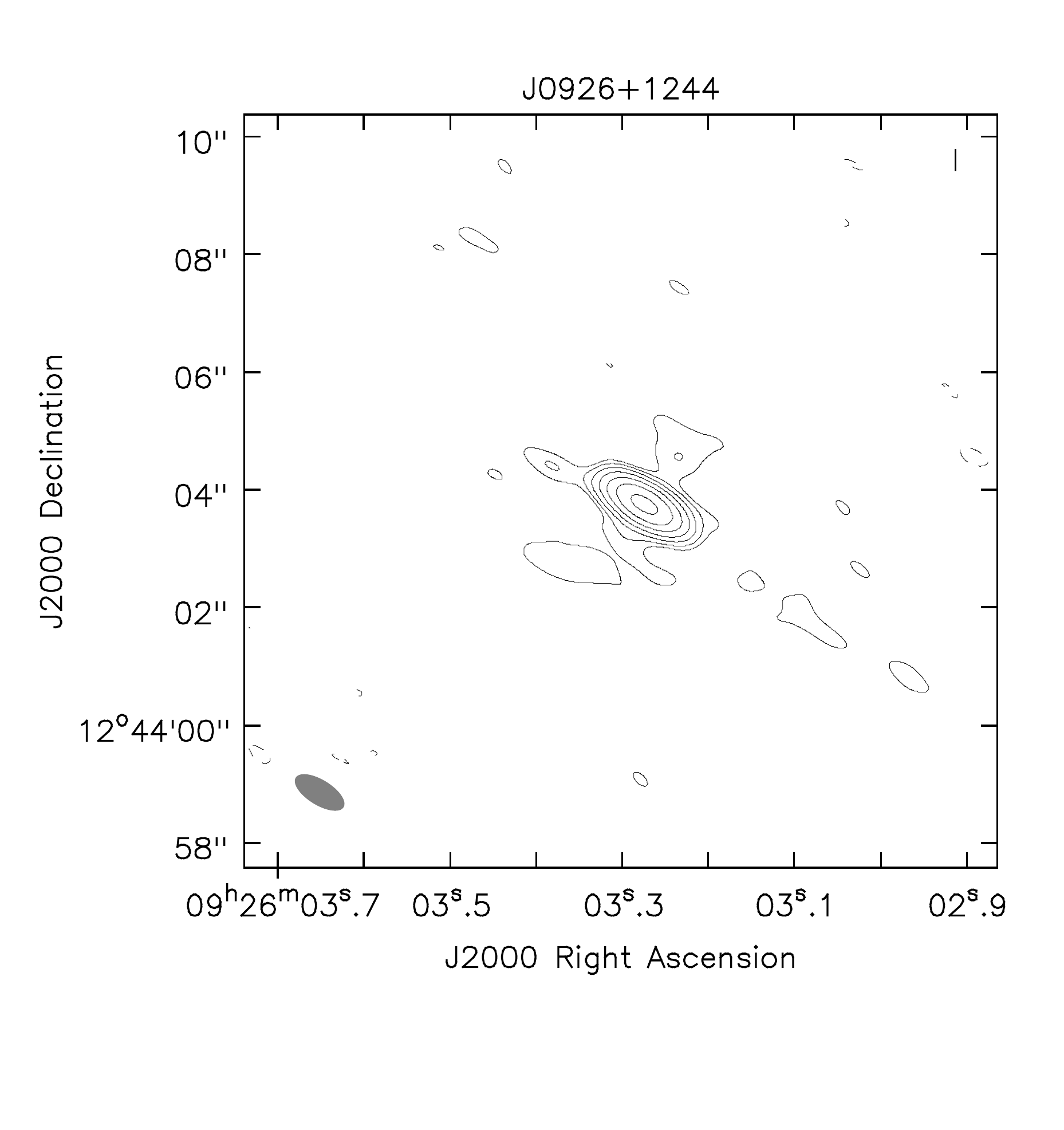} 
\caption{\textbf{Left panel:} J0925+5217, rms = 11 $\mu$Jy, contour levels at -3, 3$\times$2$^n$, n $\in$ [0,5], beam size 0.38$\times$0.32 kpc. \textbf{Right panel:} J0926+1244, rms = 12 $\mu$Jy, contour levels at -3, 3$\times$2$^n$, n $\in$ [0,6], beam size 0.55$\times$0.80 kpc.}
\label{fig:J0925p5217}
\label{fig:J0926p1244}
\end{figure*}

Mrk 705 is a radio-quiet NLS1, that in our analysis shows steep spectral indexes. There is a nearby radio source at 30 arcsec at PA 175$^\circ$, SDSS J092603.50+124333.6, with an FR II morphology, a flux of $\sim$10 mJy, and the optical spectrum of an elliptical galaxy. The main source on VLBA scales showed a possible linear structure of 26 pc extending toward East \citep{Doi13}, while previous VLA-A observations at 8.46 GHz showed only a compact source \citep{Schmitt01}. Fig.~\ref{fig:J0926p1244} also shows a relatively bright (76 $\mu$Jy) spot at 1.7 arcsec ($\sim$1 kpc) at PA 67$^\circ$, and a second one (73 $\mu$Jy) at 1.1 arcsec (0.6 kpc) at PA 38$^\circ$. The core has a size of 96$\times$60 pc. The flux outside the central beam represents approximately 15\% of the total emission, making it an intermediate morphology source according to our classification. 
\subsection{J0937+3615}
FBQS J0937+3615 is one of the \textit{bona-fide} F sources, which shows a fairly high star formation rate that can significantly contribute to the radio emission (83 M$_\odot$ yr$^{-1}$, \citealp{Caccianiga15}). Our measurements indicate that both the in- and broad-band spectral indexes are steep. This object, shown in Fig.~\ref{fig:J0937p3615}, has an unresolved core whose size is 1.4$\times$0.9 kpc, and its flux fairly is spreaded out, with peak flux representing 74\% of the total emission and a very low brightness temperature (436 K). It has not been detected at 37 GHz. These properties seem to suggest that the source radio emission is dominated by star formation. 
\subsection{J0945+1915} %extended up to x 0.769925, y 0.745865, primo quadrante, 1.1 arcsec, PA 314
\begin{figure*}
\centering
\includegraphics[trim={0cm 2cm 0cm 0cm},clip,width=7cm]{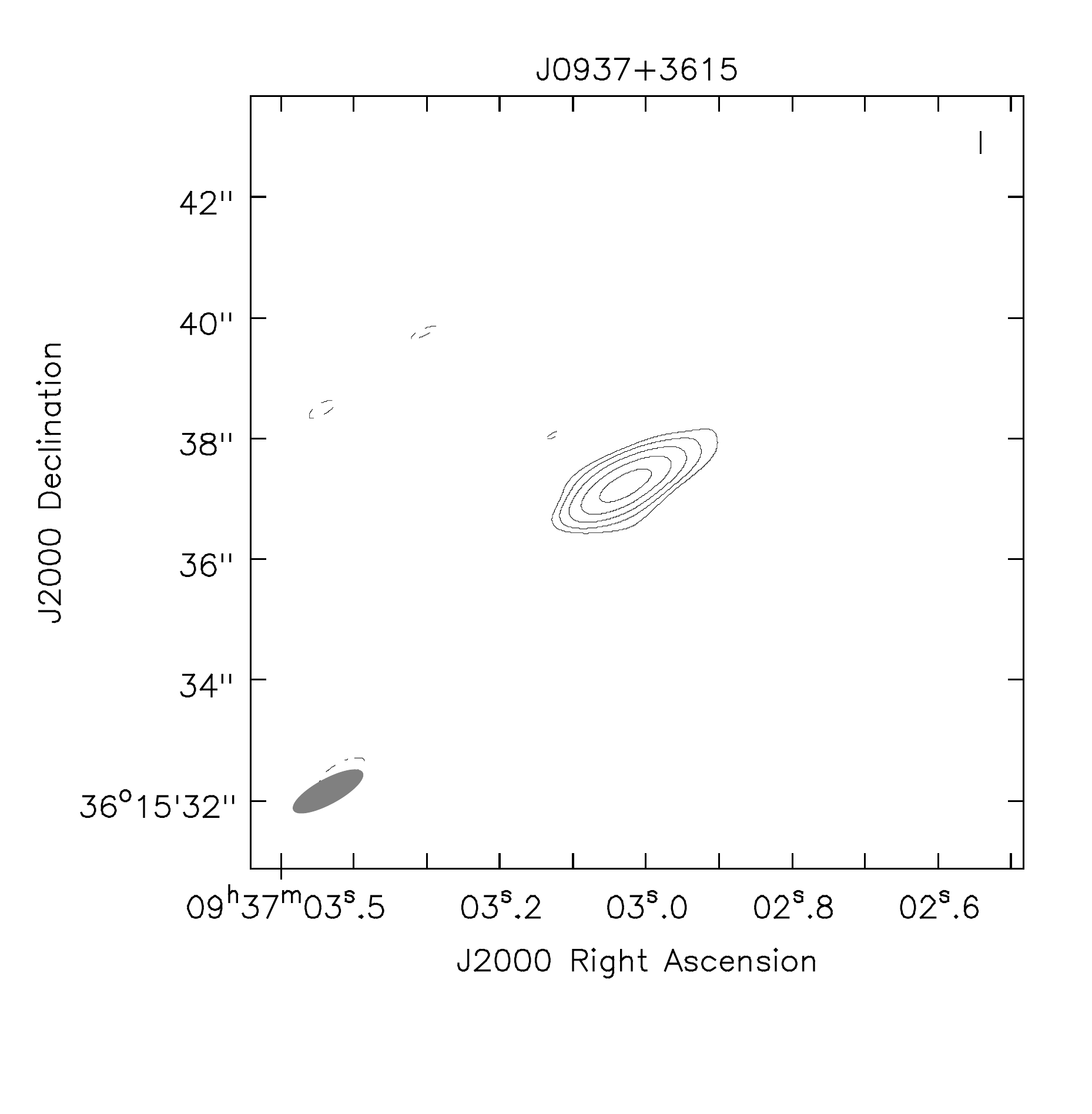} 
\includegraphics[trim={0cm 2cm 0cm 0cm},clip,width=7cm]{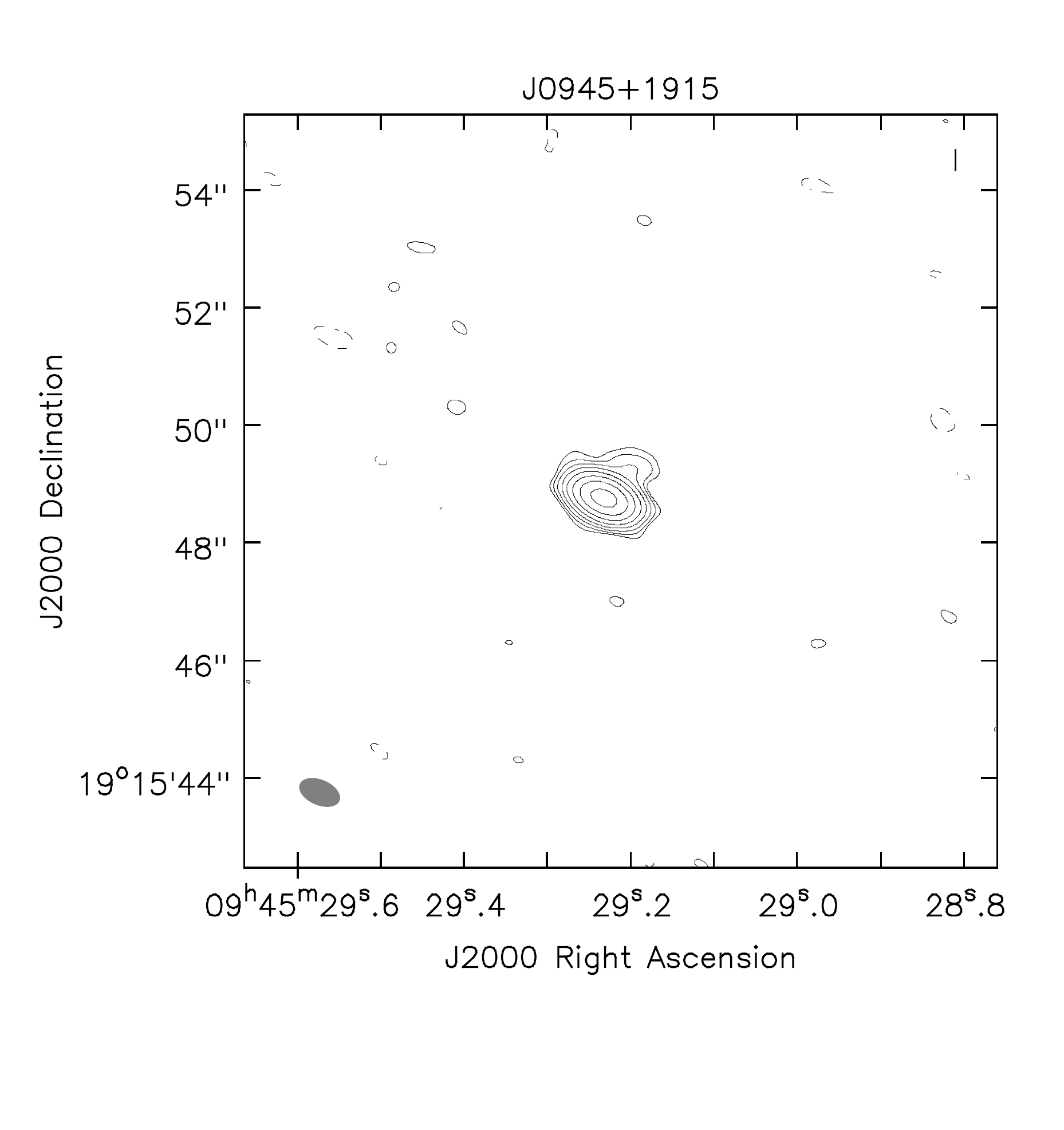} 
\caption{\textbf{Left panel:} J0937+3615, rms = 12 $\mu$Jy, contour levels at -3, 3$\times$2$^n$, n $\in$ [0,4], beam size 3.92$\times$1.24 kpc. \textbf{Right panel:} J0945+1915, rms = 11 $\mu$Jy, contour levels at -3, 3$\times$2$^n$, n $\in$ [0,7], beam size 3.09$\times$1.80 kpc.}
\label{fig:J0937p3615}
\label{fig:J0945p1915}
\end{figure*}

SDSS J094529.23+191548.8 is a \textit{bona-fide} F source, also showing a high star formation rate (138 M$_\odot$ yr$^{-1}$, \citealp{Caccianiga15}). The spectral indexes are all steep, therefore this source should be reclassified as S. Fig.~\ref{fig:J0945p1915} shows a partially resolved structure at PA 314$^\circ$ extended for $\sim$1.1 arcsec (4.7 kpc), but the diffuse emission is rather low. The core is unresolved, and its deconvolved size is 490$\times$330 pc. It was not detected at 37 GHz. Given its luminosity of $\sim$8$\times$10$^{41}$ \ergs, it might be classified as a classical CSS. 
\subsection{J0948+5029}
Mrk 124 is a radio-quiet NLS1 which was never observed before at 5 GHz. All the spectral indexes are consistent with a steep spectrum within the error bars. The source in the map (Fig.~\ref{fig:J0948p5029}) appears as an unresolved compact core, whose deconvolved size is 160$\times$50 pc. 
\subsection{J0952-0136} %muso dello squalo, x 1.97763, y 3.14807 terzo quadrante 3.7
\begin{figure*}
\centering
\includegraphics[trim={0cm 2cm 0cm 0cm},clip,width=7cm]{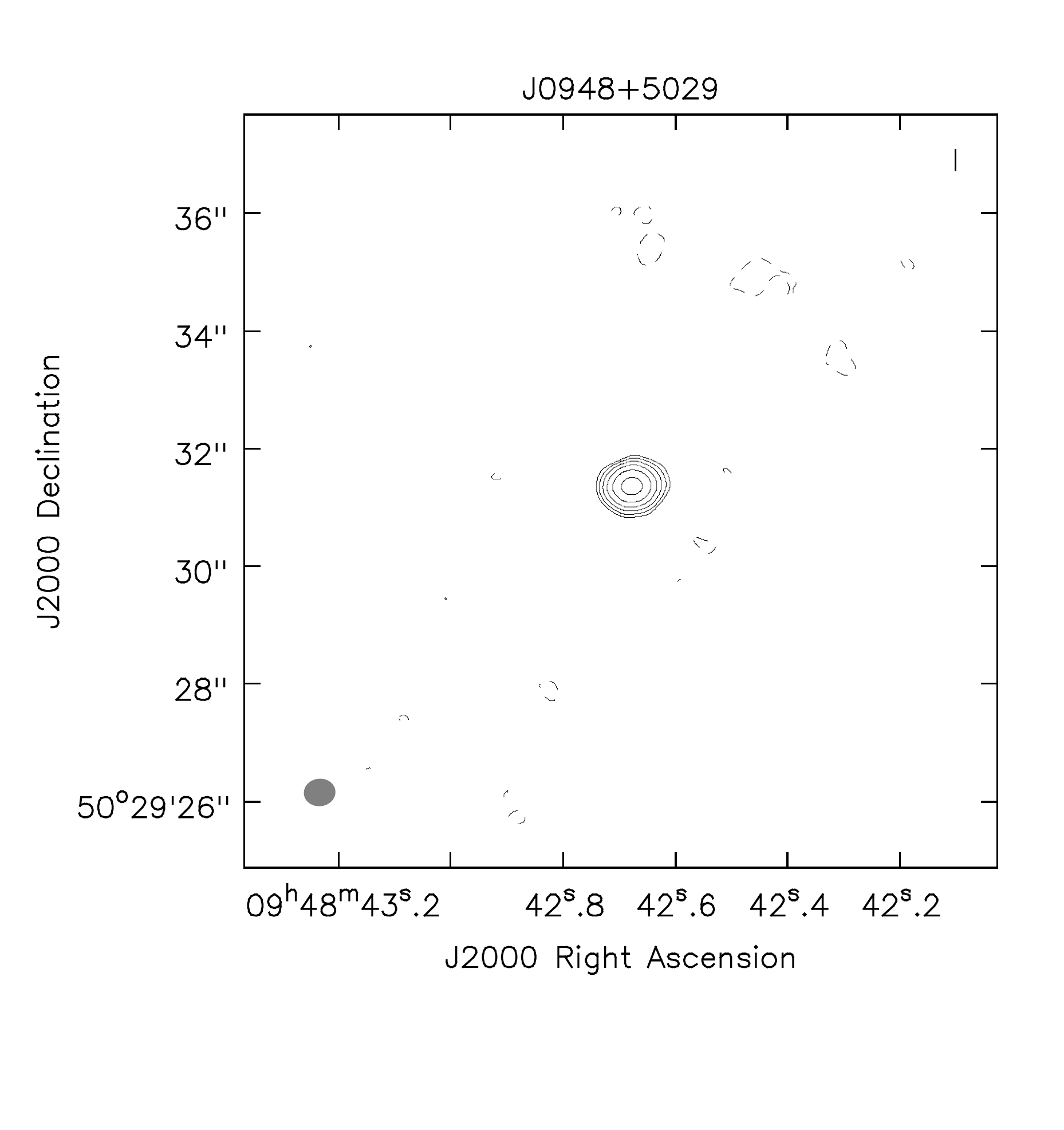} 
\includegraphics[trim={0cm 2cm 0cm 0cm},clip,width=7cm]{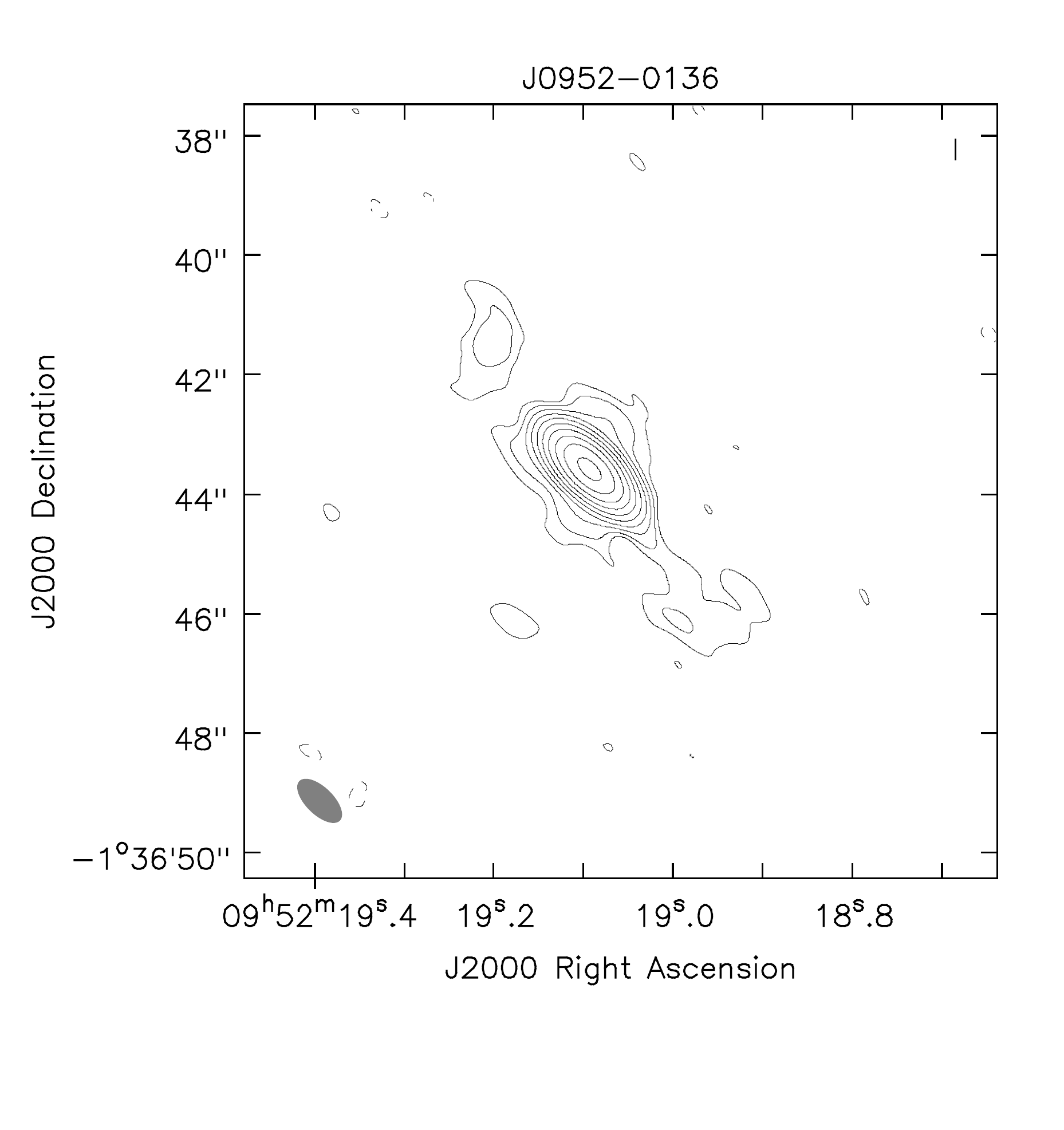} 
\caption{\textbf{Left panel:} J0948+5029, rms = 15 $\mu$Jy, contour levels at -3, 3$\times$2$^n$, n $\in$ [0,5], beam size 0.57$\times$0.49 kpc. \textbf{Right panel:} J0952-0136, rms = 11 $\mu$Jy, contour levels at -3, 3$\times$2$^n$, n $\in$ [0,9], beam size 0.38$\times$0.18 kpc.}
\label{fig:J0948p5029}
\label{fig:J0952m0136}
\end{figure*}

Mrk 1239 is a NLS1 which has been classified as radio-quiet \citep{Doi15} or barely radio-loud \citep{Berton15a}. This object was studied repeatedly in radio, finding extended emission reminiscent of an FR I radio galaxy \citep{Doi13, Doi15}. Comparing VLA and VLBA observations, \citet{Orienti10} found that almost 80\% of the VLA (at 15 GHz) flux was missing on VLBA scales, suggesting the presence of a very significant extended emission. Our map, shown in Fig.~\ref{fig:J0952m0136}, reveals indeed extended emission everywhere around the core, which was the only component visible in 15 GHz VLA images. The largest extension is measured at PA 212$^\circ$, at a projected distance of 3.7 arcsec (1.5 kpc). The unresolved core has a size of 84$\times$33 pc, and it shows a flat in-band spectrum. The total emission also has a flat spectrum, but it is dominated by the core, which represents the 90\% of the entire flux. The broad-band spectrum is instead steep between 1.4 and 5 GHz and, as expected from this slope, the source is not detected at 22 GHz. 
\subsection{J0957+2433}
RX J0957.1+2433 is a radio-quiet NLS1. All of its spectral indexes are steep, and in particular its broad-band index is the second steepest of our sample. The map of Fig.~\ref{fig:J0957p2433} does not show any structure with the exception of the unresolved core, whose size could not be calculated because of the source faintness. 
\subsection{J1031+4234}
\begin{figure*}
\centering
\includegraphics[trim={0cm 2cm 0cm 0cm},clip,width=7cm]{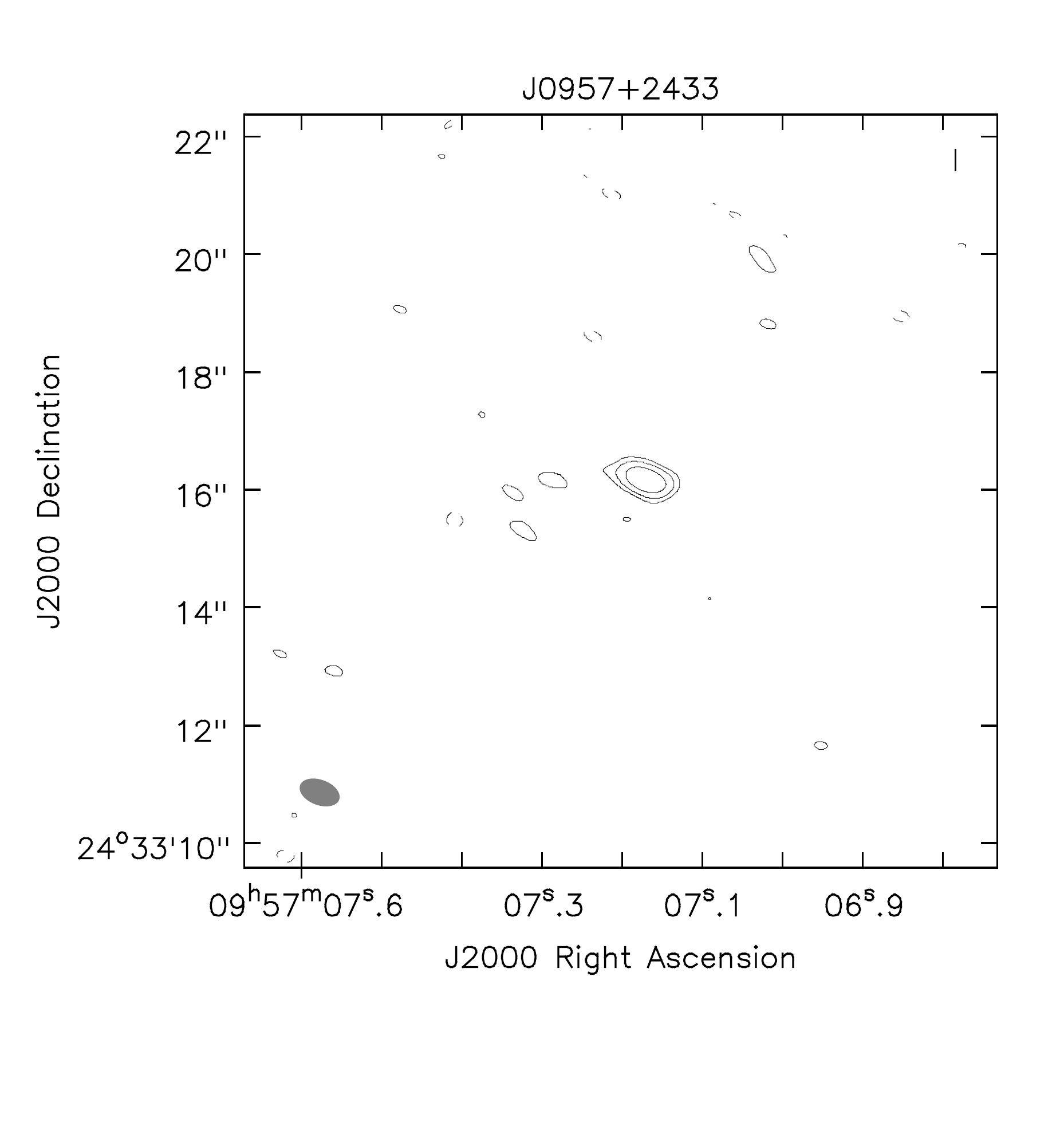} 
\includegraphics[trim={0cm 2cm 0cm 0cm},clip,width=7cm]{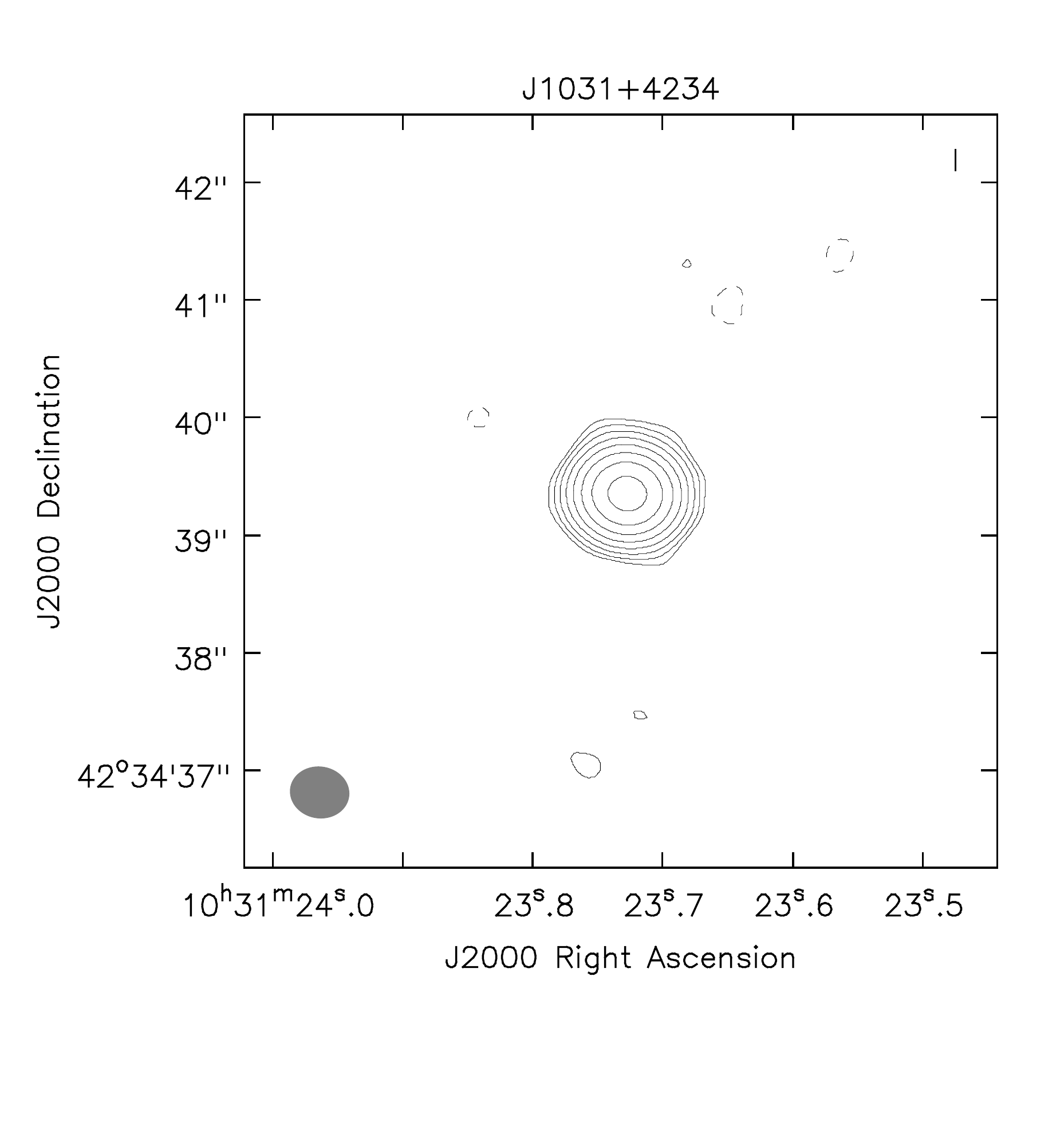} 
\caption{\textbf{Left panel:} J0957+2433, rms = 10 $\mu$Jy, contour levels at -3, 3$\times$2$^n$, n $\in$ [0,2], beam size 1.07$\times$0.63 kpc. \textbf{Right panel:} J1031+4234, rms = 10 $\mu$Jy, contour levels at -3, 3$\times$2$^n$, n $\in$ [0,7], beam size 2.59$\times$2.23 kpc.}
\label{fig:J0957p2433}
\label{fig:J1031p4234}
\end{figure*}

SDSS J103123.73+423439.3 is classified as an F source, with a spectral index below 1.4 GHz of -0.40$\pm$0.06 \citep{Yuan08, Foschini15}. Our spectral index measurements point in a different direction, showing a much steeper broad-band index, and relatively steep in-band indexes (although close to the 0.5 threshold). This might be an indication of a curved spectrum peaking at low frequencies, possibly in the MHz range, or of source variability. The map (Fig.~\ref{fig:J1031p4234}) clearly shows a compact source, with a deconvolved size lower than 0.6$\times$0.1 kpc and a luminosity higher than 10$^{41}$ \ergs. If the source spectrum is actually peaked at low frequencies, these properties might indicate that this object is classifiable as a classical CSS. 
\subsection{J1034+3928} 
KUG 1031+398 is a steep-spectrum radio-loud source. While the broad-band index is clearly in agreement with this classification, the in-band indexes are flat, particularly in the core. The extended emission represents approximately 13\% of the total flux, and it is visible for $\sim$5 arcsec (4.5 kpc) both East and West of the source. The unresolved core has a deconvolved size of 100$\times$65 kpc. The source was not detected at 22 GHz with the VLBA. 
\subsection{J1037+0036}
\begin{figure*}
\centering
\includegraphics[trim={0cm 2cm 0cm 0cm},clip,width=7cm]{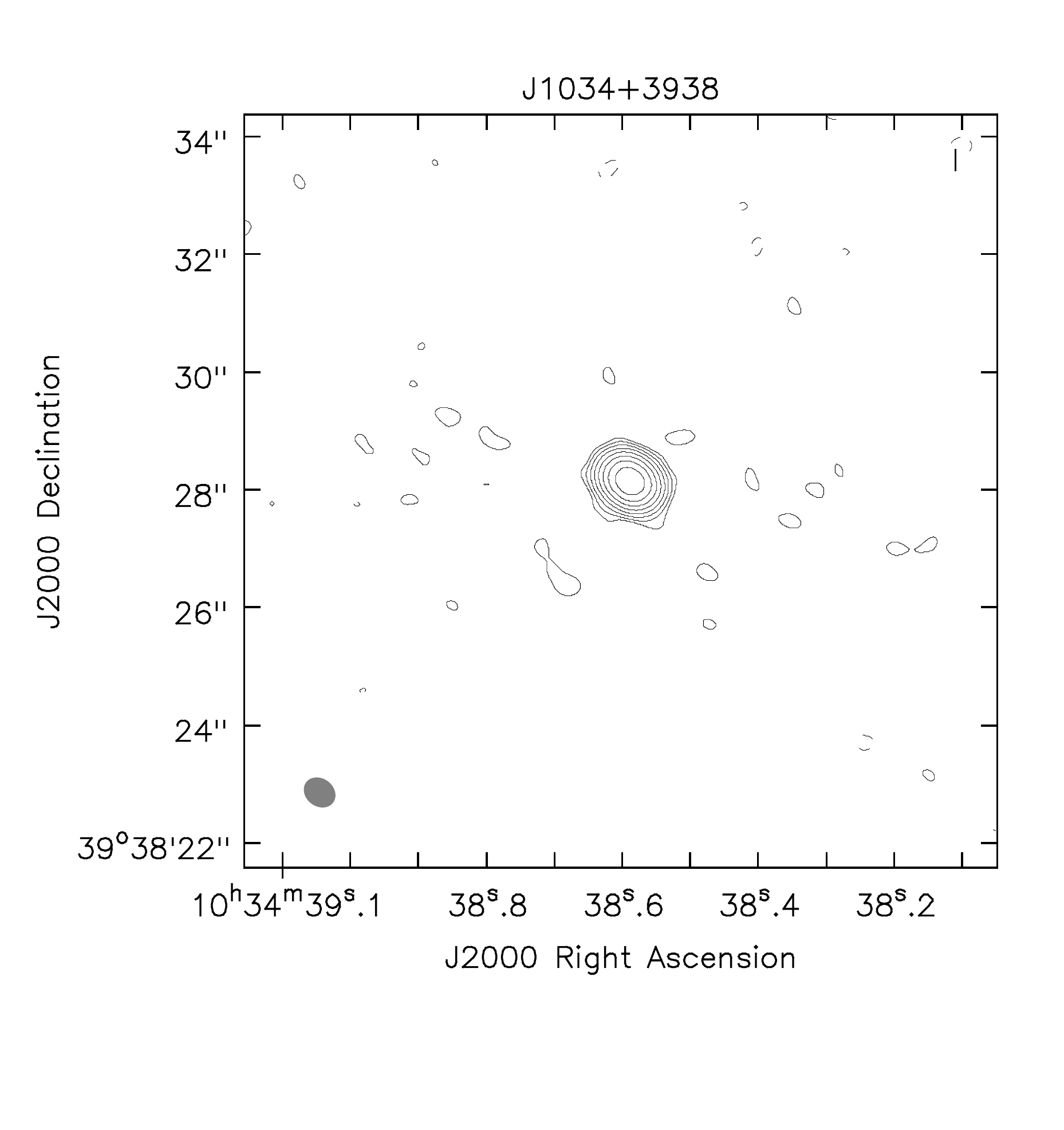} 
\includegraphics[trim={0cm 2cm 0cm 0cm},clip,width=7cm]{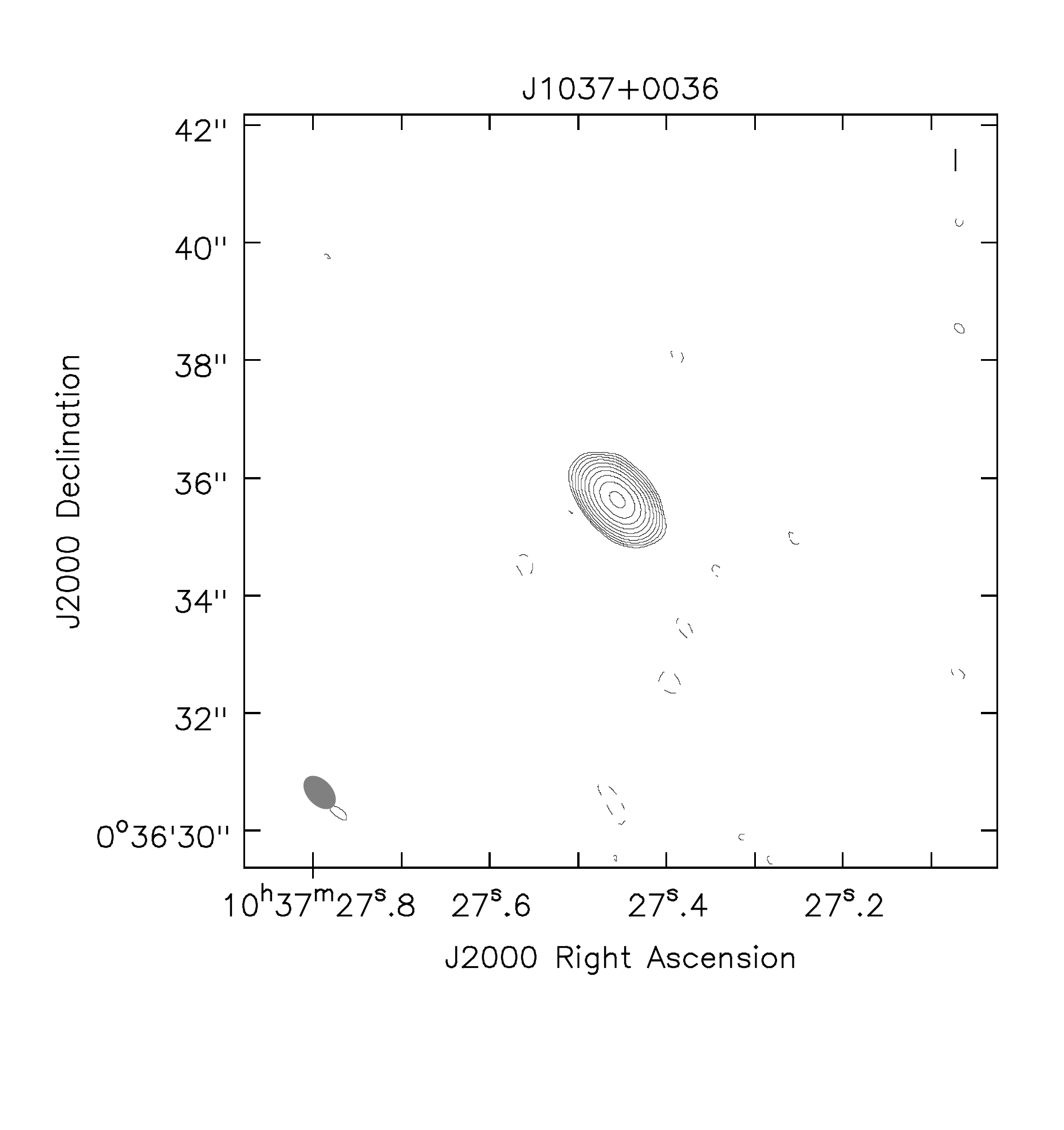} 
\caption{\textbf{Left panel:} J1034+3938, rms = 10 $\mu$Jy, contour levels at -3, 3$\times$2$^n$, n $\in$ [0,7], beam size 0.46$\times$0.37 kpc. \textbf{Right panel:} J1031+4234, rms = 11 $\mu$Jy, contour levels at -3, 3$\times$2$^n$, n $\in$ [0,9], beam size 4.32$\times$2.73 kpc.}
\label{fig:J1034p3938}
\label{fig:J1037p0036}
\end{figure*}

SDSS J103727.45+003635.6 is one of the \textit{bona-fide} F sources. Our spectral index measurements indicate that its assumption was correct. The map, shown in Fig.~\ref{fig:J1037p0036}, shows only an unresolved core, with a luminosity of 1.5$\times$10$^{42}$ \ergs and a deconvolved size of 370$\times$90 pc. The brightness temperature, above 10$^6$ K, is one of the highest of the sample. The source is however too weak to be detected at 22 and 37 GHz. 
\subsection{J1038+4227} %x 6.1594, y 5.72632, 1 quad
FBQS J1038+4227 is a \textit{bona-fide} flat-spectrum source. However, our measurements suggest that this source is instead a steep-spectrum object. The map of Fig.~\ref{fig:J1038p4227} shows an unresolved core of size 450$\times$410 pc, and a spreaded-out extended emission that accounts for 69\% of the total flux and is found as far as 8.4 arcsec from the radio core ($\sim$30 kpc) at PA 313$^\circ$. The source is not detected at 37 GHz.
\subsection{J1047+4725} %0.842105, 0.938346, 1 quad
\begin{figure*}
\centering
\includegraphics[trim={0cm 2cm 0cm 0cm},clip,width=7cm]{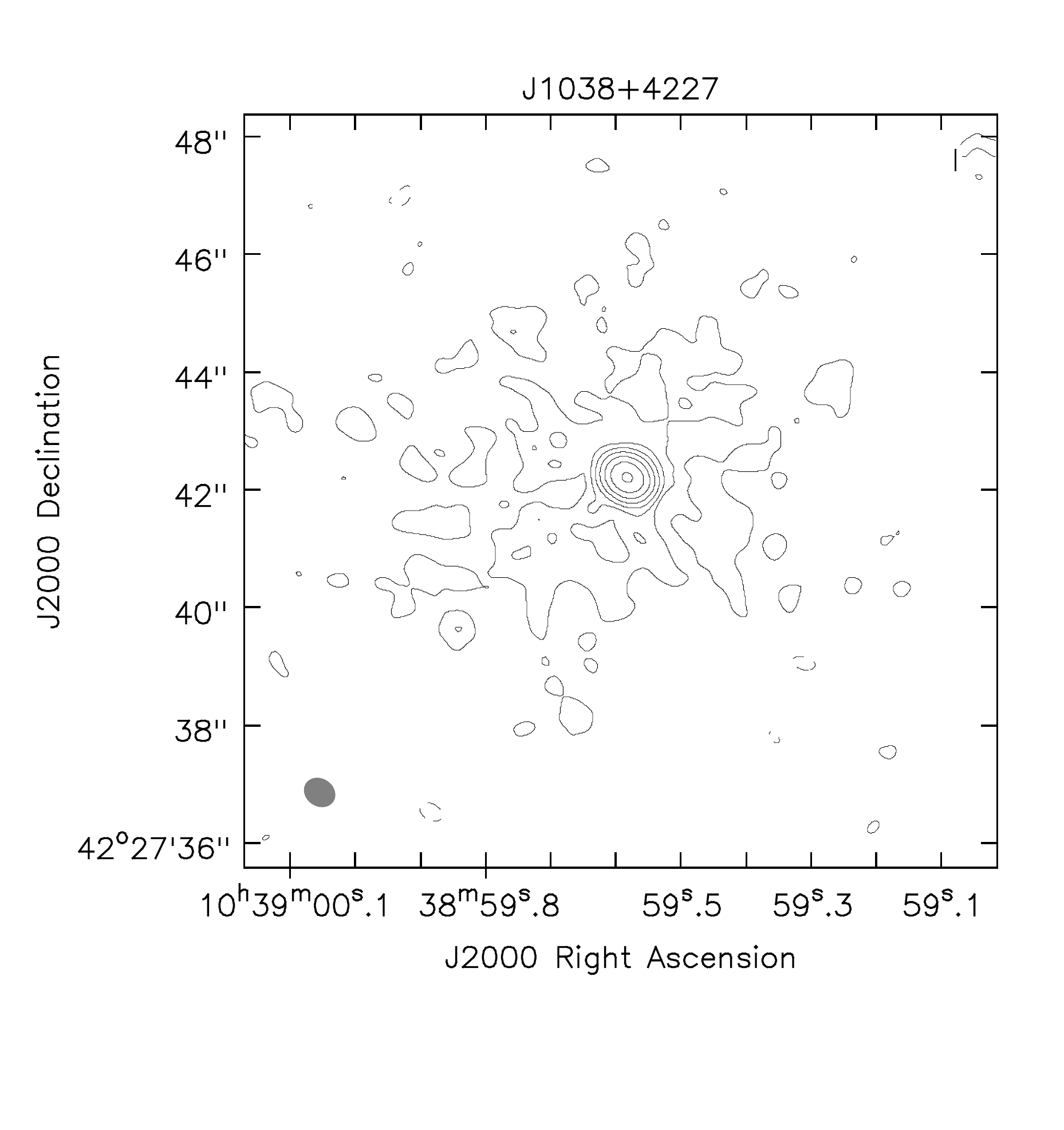} 
\includegraphics[trim={0cm 2cm 0cm 0cm},clip,width=7cm]{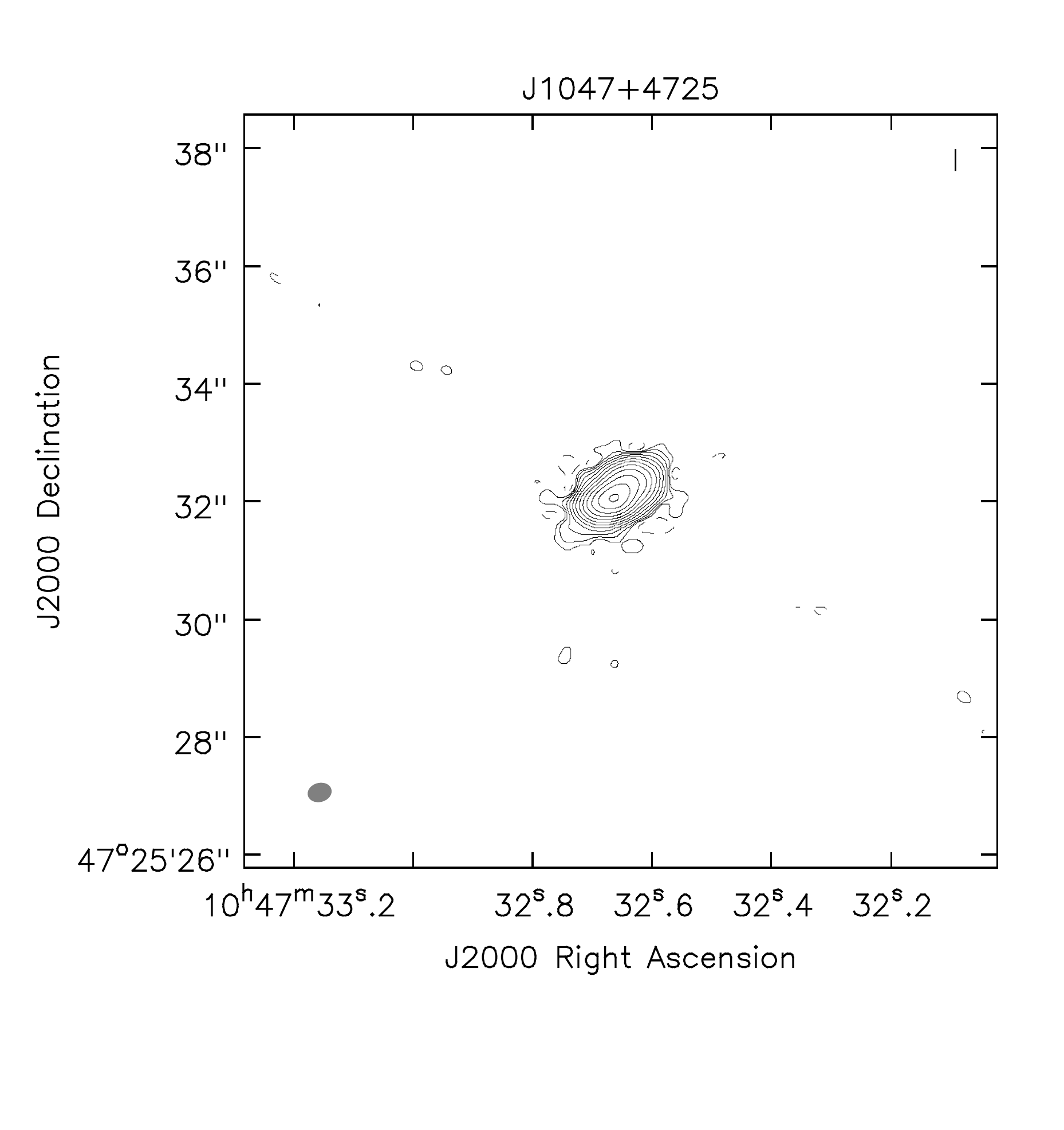} 
\caption{\textbf{Left panel:} J1038+4227, rms = 10 $\mu$Jy, contour levels at -3, 3$\times$2$^n$, n $\in$ [0,6], beam size 1.92$\times$1.60 kpc. \textbf{Right panel:} J1047+4725, rms = 37 $\mu$Jy, contour levels at -3, 3$\times$2$^n$, n $\in$ [0,11], beam size 3.00$\times$2.33 kpc.}
\label{fig:J1038p4227}
\label{fig:J1047p4725}
\end{figure*}

SDSS J104732.68+472532.0 is classified as an F source, and our in-band indexes confirm this classification. The broad-band index is steeper than that found by \citet{Foschini15} (0.33$\pm$0.01), likely as a consequence of variability. The source has the highest luminosity of our sample, 6$\times$10$^{43}$ \ergs, 55\% of which comes from an unresolved core of 3.6$\times$1.2 kpc. The rest of the flux, as shown in Fig.~\ref{fig:J1047p4725}, is originated in a partially resolved diffuse emission extended up to 1.3 arcsec (9.5 kpc) from the nucleus at PA 318$^\circ$. Such structure is resolved into a core-jet system in high resolution images with the VLBA at 5 GHz \citep{Gu15}. This object is the second farthest in our sample, at z = 0.799. The source was not detected at 22 and 37 GHz, indicating a possible steepening of the spectrum above 5 GHz. 
\subsection{J1048+2222}
SDSS J104816.57+222238.9 was classified as a F source \textit{bona-fide}, but our spectral indexes seem to indicate a steep-spectrum nature. The map in Fig.~\ref{fig:J1048p2222} does not show any particular structure, but an unresolved core whose size is lower than 1.2$\times$0.9 kpc. The source was not detected at 37 GHz.
\subsection{J1102+2239}
\begin{figure*}
\centering
\includegraphics[trim={0cm 2cm 0cm 0cm},clip,width=7cm]{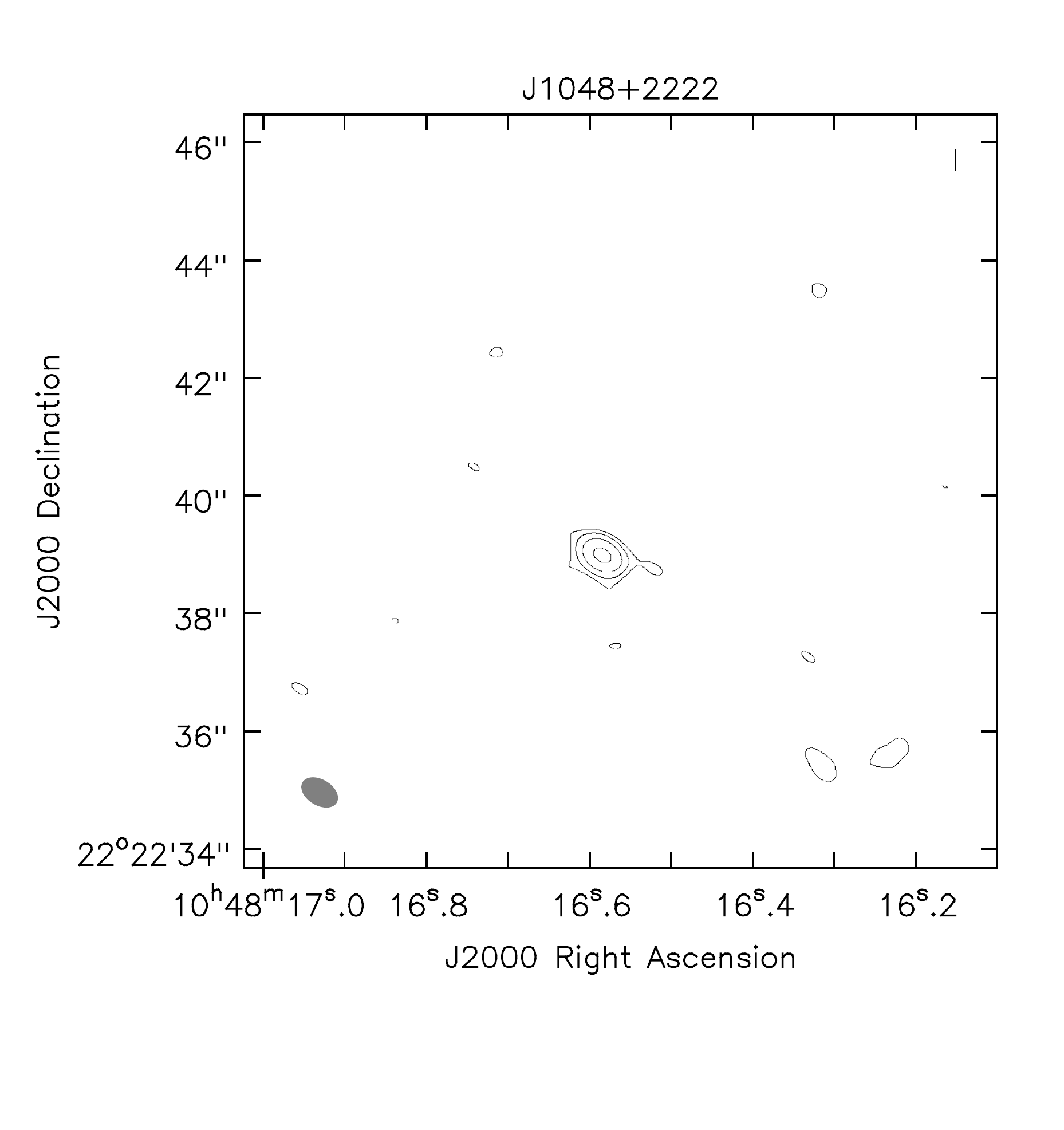} 
\includegraphics[trim={0cm 2cm 0cm 0cm},clip,width=7cm]{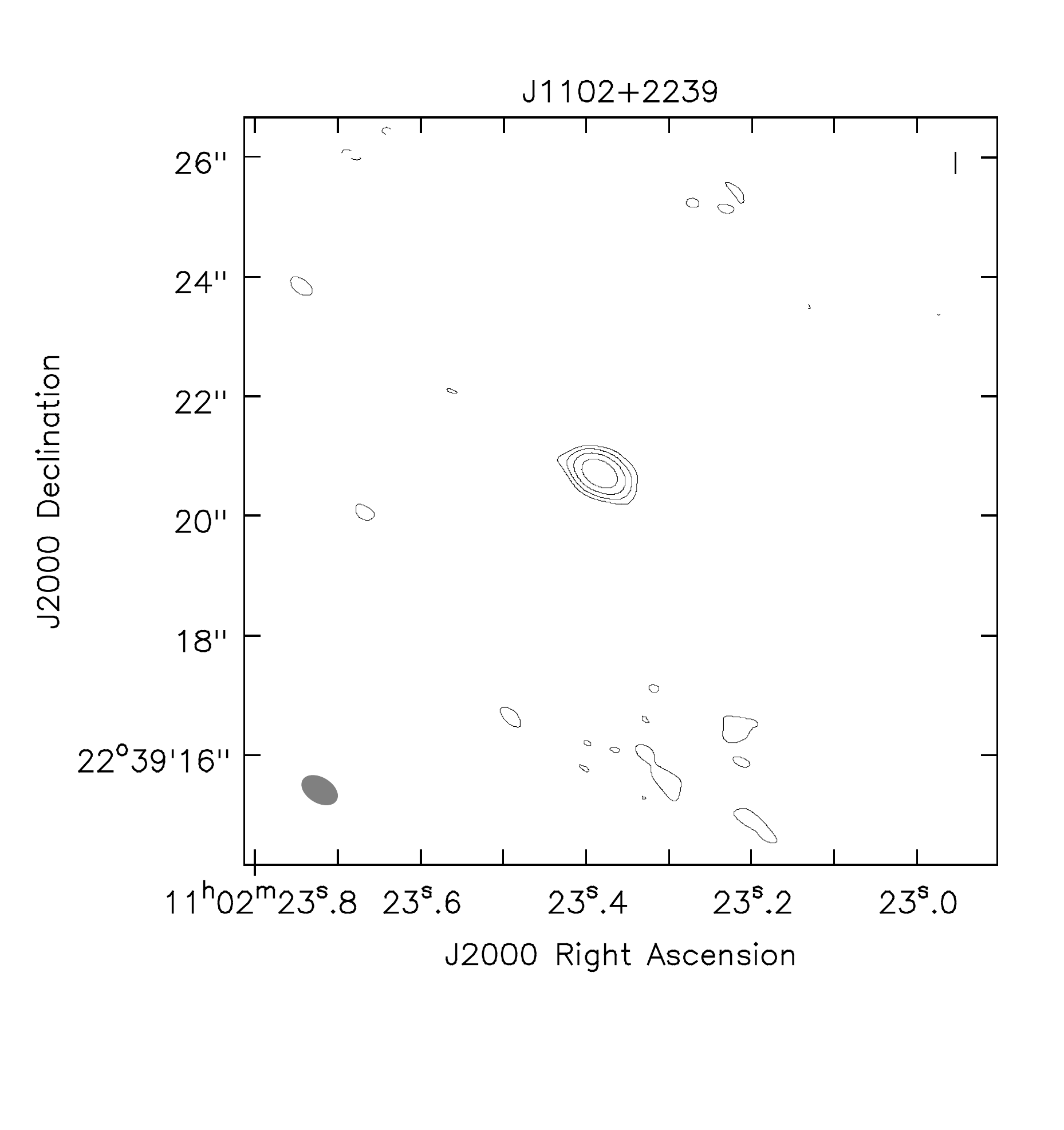} 
\caption{\textbf{Left panel:} J1048+2222, rms = 10 $\mu$Jy, contour levels at -3, 3$\times$2$^n$, n $\in$ [0,3], beam size 3.14$\times$2.00 kpc. \textbf{Right panel:} J1102+2239, rms = 15 $\mu$Jy, contour levels at -3, 3$\times$2$^n$, n $\in$ [0,3], beam size 3.76$\times$2.37 kpc.}
\label{fig:J1048p2222}
\label{fig:J1102p2239}
\end{figure*}

FBQS J1102+2239 was classified as one of the $\gamma$-ray emitters \citep{Foschini11}, although this property has not been confirmed \citep{Foschini15}. The map of Fig.~\ref{fig:J1102p2239} shows only the unresolved core with a size of 1.2$\times$0.5 kpc. The core spectral index, which has never been estimated before, unlike other G sources indicate a steep in-band spectrum, in agreement with the broad-band spectrum. The object was not detected at 37 GHz. The source is characterized by a very high level of star formation, 302 M$_\odot$ yr$^{-1}$, and it has a q22 parameter of 1.34, which is characteristic of starburst galaxies \citep{Caccianiga15}. If the $\gamma$-ray emission were to be confirmed, this would one of the few examples of $\gamma$-ray emitting S-NLS1 \citep{Liao15, Komossa15, Berton17a}. 
\subsection{J1110+3653} %3 quad, 1.28722, 2.8391
SDSS J111005.03+365336.3 is another \textit{bona-fide} F source, whose nature is confirmed by the flat in-band spectral indexes. The broad-band index is steeper, but still consistent with the flat classification within the error bars. The source (Fig.~\ref{fig:J1110p3653}) presents an unresolved core whose size is 500$\times$390 pc, and a diffuse emission extended up to 3.1 arcsec (21 kpc) from the nucleus at PA 156$^\circ$ that accounts for $\sim$13\% of the total flux. This diffuse emission was not detected with the VLBA at 5 GHz, where the core flux is 8.8 mJy \citep{Gu15}. The source was detected neither at 22 GHz, nor at 37 GHz. 
\subsection{J1114+3241}
\begin{figure*}
\centering
\includegraphics[trim={0cm 2cm 0cm 0cm},clip,width=7cm]{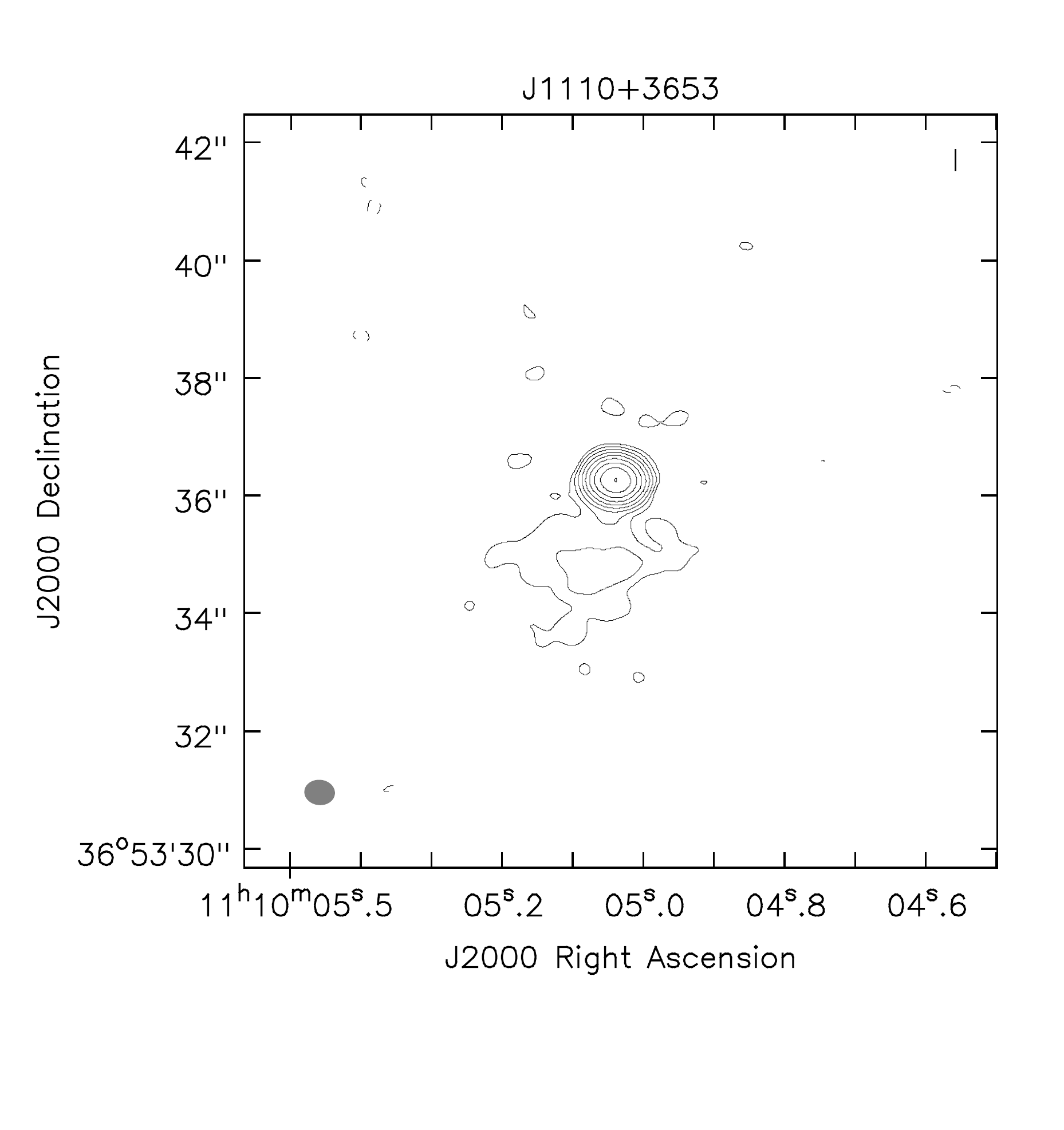} 
\includegraphics[trim={0cm 2cm 0cm 0cm},clip,width=7cm]{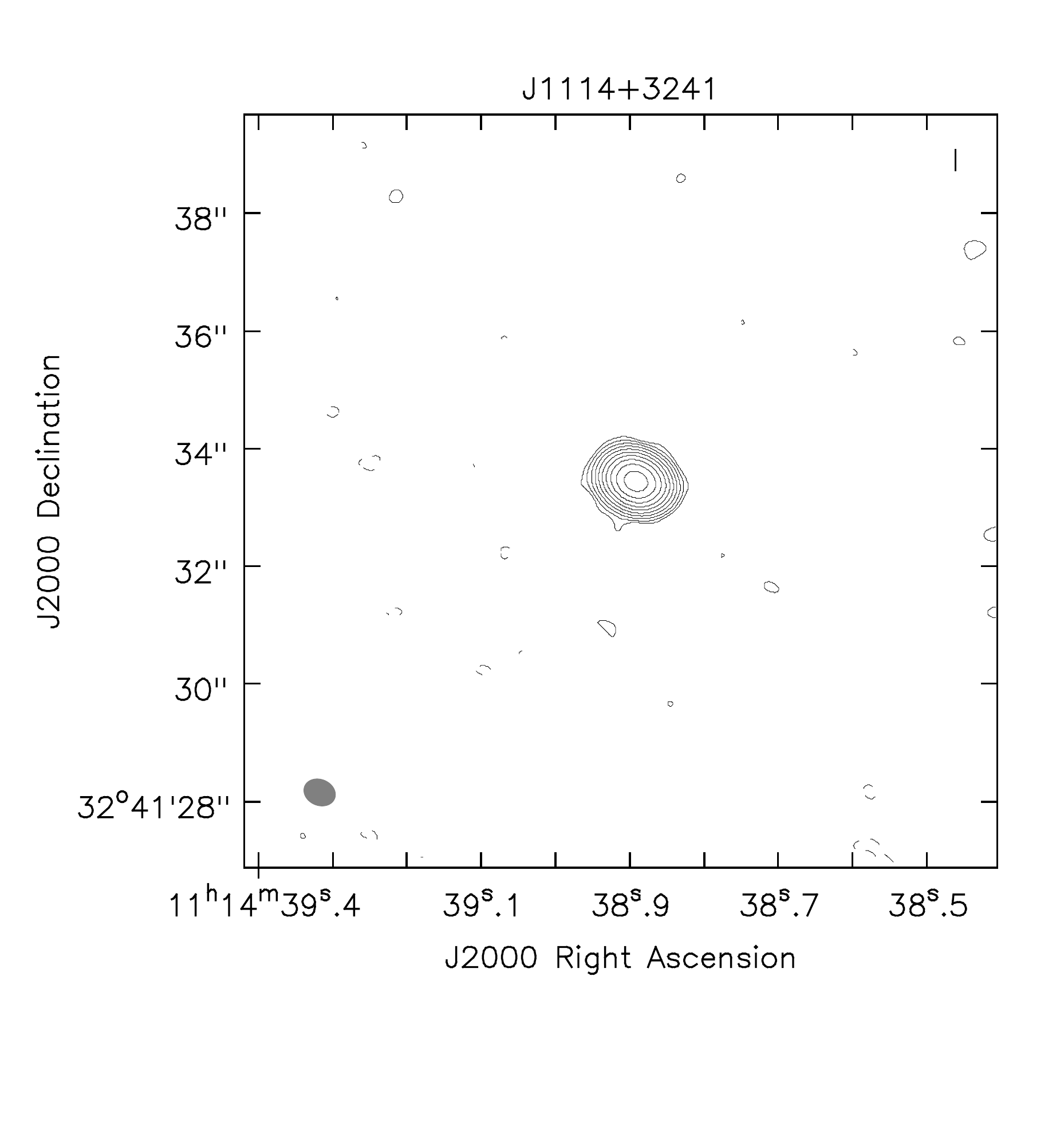} 
\caption{\textbf{Left panel:} J1110+3653, rms = 10 $\mu$Jy, contour levels at -3, 3$\times$2$^n$, n $\in$ [0,8], beam size 3.42$\times$2.80 kpc. \textbf{Right panel:} J1114+3241, rms = 9 $\mu$Jy, contour levels at -3, 3$\times$2$^n$, n $\in$ [0,9], beam size 1.74$\times$1.39 kpc.}
\label{fig:J1110p3653}
\label{fig:J1114p3241}
\end{figure*}

B2 1111+32 was classified as S source by \citet{Komossa06}. However, our spectral indexes indicate instead a flat spectrum. The broad-band index is also strongly inverted, a behavior observed only in seven of our sources. These indexes strongly support the presence of a powerful relativistic jet in this object, possibly connected with the generation of the powerful mildly relativistic outflows observed in this superluminous infrared galaxy \citep{Tombesi15}. The map of Fig.~\ref{fig:J1114p3241} shows only an unresolved core, whose deconvolved size is 230$\times$200 pc. 
\subsection{J1121+5351}
SBS 1118+541 is a radio-quiet NLS1. All of its spectral indexes might be steep, although they are consistent with a flat behavior within the error bars. The brightness temperature however is very low ($\sim$ 2000 K), which seems to favor the steep-spectrum hypothesis. The map (Fig.~\ref{fig:J1121p5351}) shows only an unresolved core of size 490$\times$270 pc. 
\subsection{J1138+3653}
\begin{figure*}
\centering
\includegraphics[trim={0cm 2cm 0cm 0cm},clip,width=7cm]{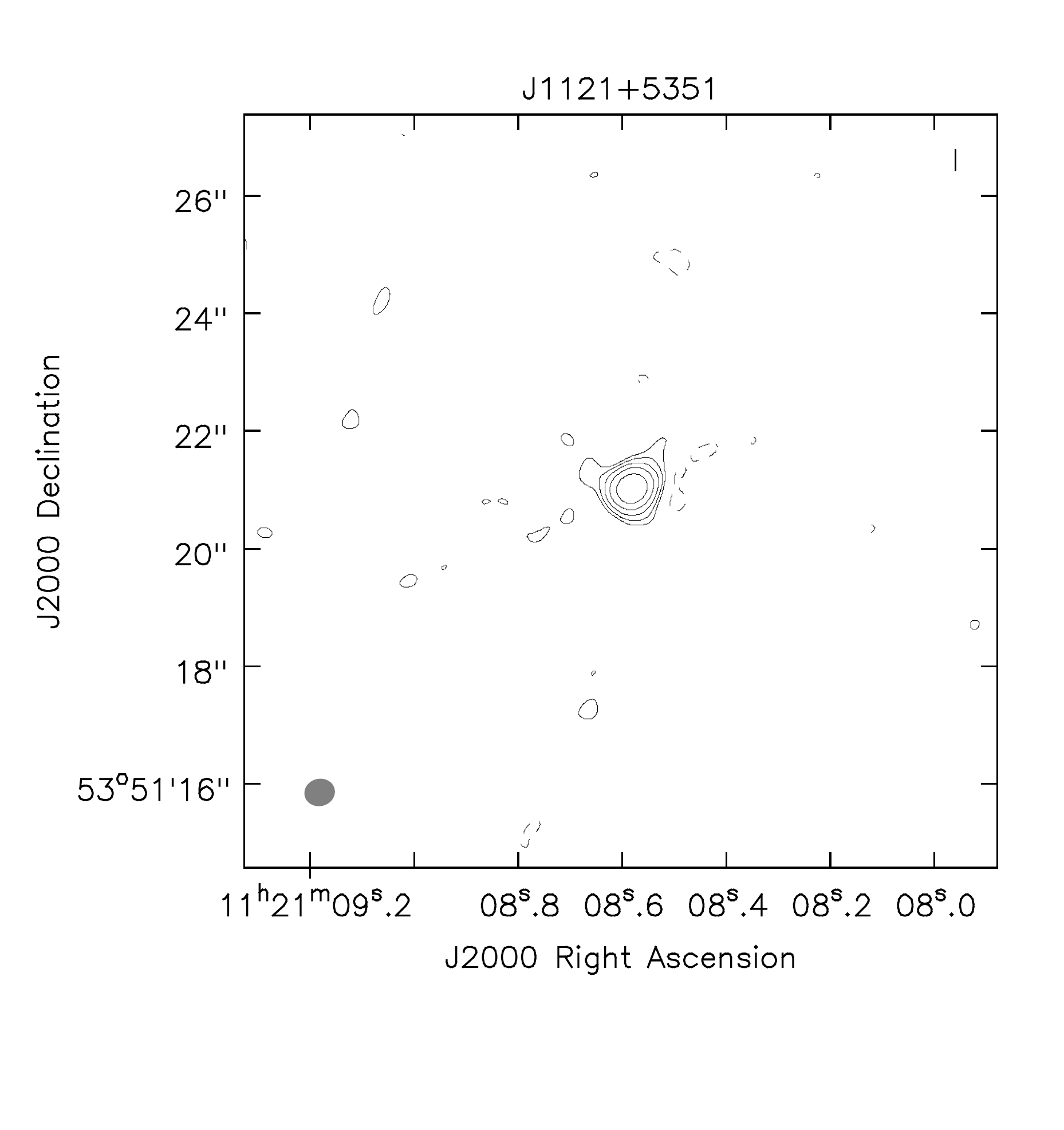} 
\includegraphics[trim={0cm 2cm 0cm 0cm},clip,width=7cm]{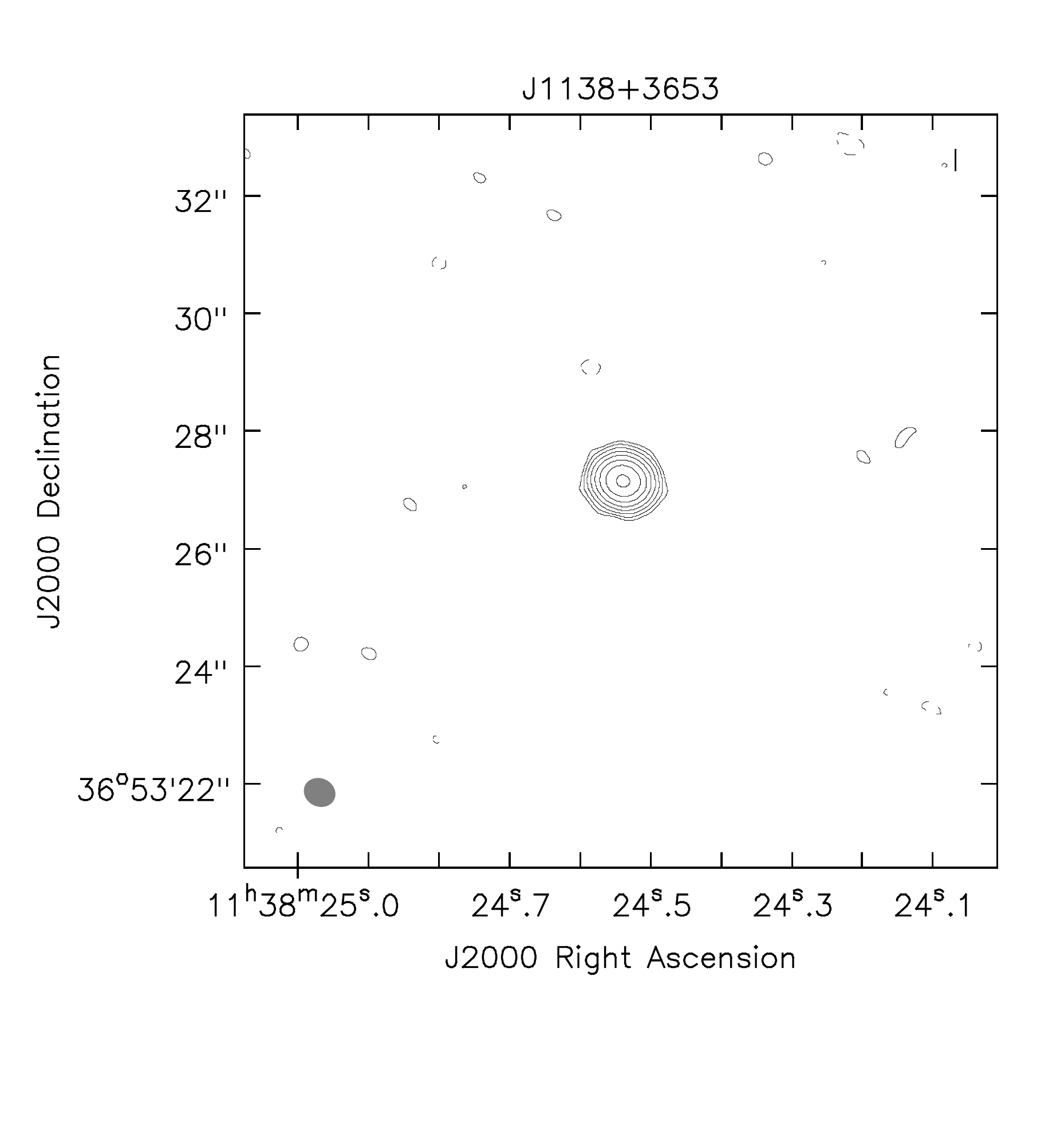} 
\caption{\textbf{Left panel:} J1121+5351, rms = 11 $\mu$Jy, contour levels at -3, 3$\times$2$^n$, n $\in$ [0,4], beam size 0.97$\times$0.85 kpc. \textbf{Right panel:} J1138+3653, rms = 10 $\mu$Jy, contour levels at -3, 3$\times$2$^n$, n $\in$ [0,8], beam size 2.65$\times$2.15 kpc.}
\label{fig:J1121p5351}
\label{fig:J1138p3653}
\end{figure*}

SDSS J113824.54+365327.1 was classified as a flat-spectrum source according to its spectral index 0.50 $\pm$ 0.09 below 1.4 GHz \citep{Foschini15}. Our in-band indexes confirm the classification, while the broad-band index indicates a steep-spectrum, likely because of variability. The unresolved core has a deconvolved size of 790$\times$245 pc, and it is the only structure visible in our map (Fig.~\ref{fig:J1138p3653}), and also in the VLBA map at 5 GHz by \citet{Gu15}. The latter observations found a core flux of 9 mJy, almost double of our measurement, an indication of significant variability of the source. Despite this, the source was detected neither at 22 GHz nor at 37 GHz. Finally, this source shows a significant amount of star formation (14 M$_\odot$ yr$^{-1}$), which however does not seem to be relevant when compared to the AGN activity \citep{Caccianiga15}.
\subsection{J1146+3236} %3 quad, 1.07669, 0.884211, second 3 quad, 1.96494, 4.06512, coord 11 46 54.447 +32 36 48.242
SDSS J114654.28+323652.3 is a flat-spectrum source which was detected at frequencies below 1.4 GHz \citep{Foschini15}. Our spectral indexes all confirm the classification, even suggesting a slightly inverted broad-band radio spectrum. The source appears essentially as an unresolved core whose deconvolved size is 360$\times$235 pc. However, some dim diffuse emission might be present in the map (see Fig.~\ref{fig:J1146p3236}). In particular there are two significantly brighter regions (6$\sigma$ detection). The closest one, at 1.4 arcsec (8.2 kpc) at P.A. 130$^\circ$, is relatively compact and has a peak of 68 $\mu$Jy, while the second one, more diffuse, has a peak of 97 $\mu$Jy and a total flux of 433 $\mu$Jy, and is located at P.A. 154$^\circ$ 4.5 arcsec (26.5 kpc) away from the source core. There are no nearby sources that can affect the map with their sidelobes. This source was detected at 22 GHz with a flux of 104 mJy, a sign of high variability, but not at 37 GHz. 
\subsection{J1159+2838}
\begin{figure*}
\centering
\includegraphics[trim={0cm 2cm 0cm 0cm},clip,width=7cm]{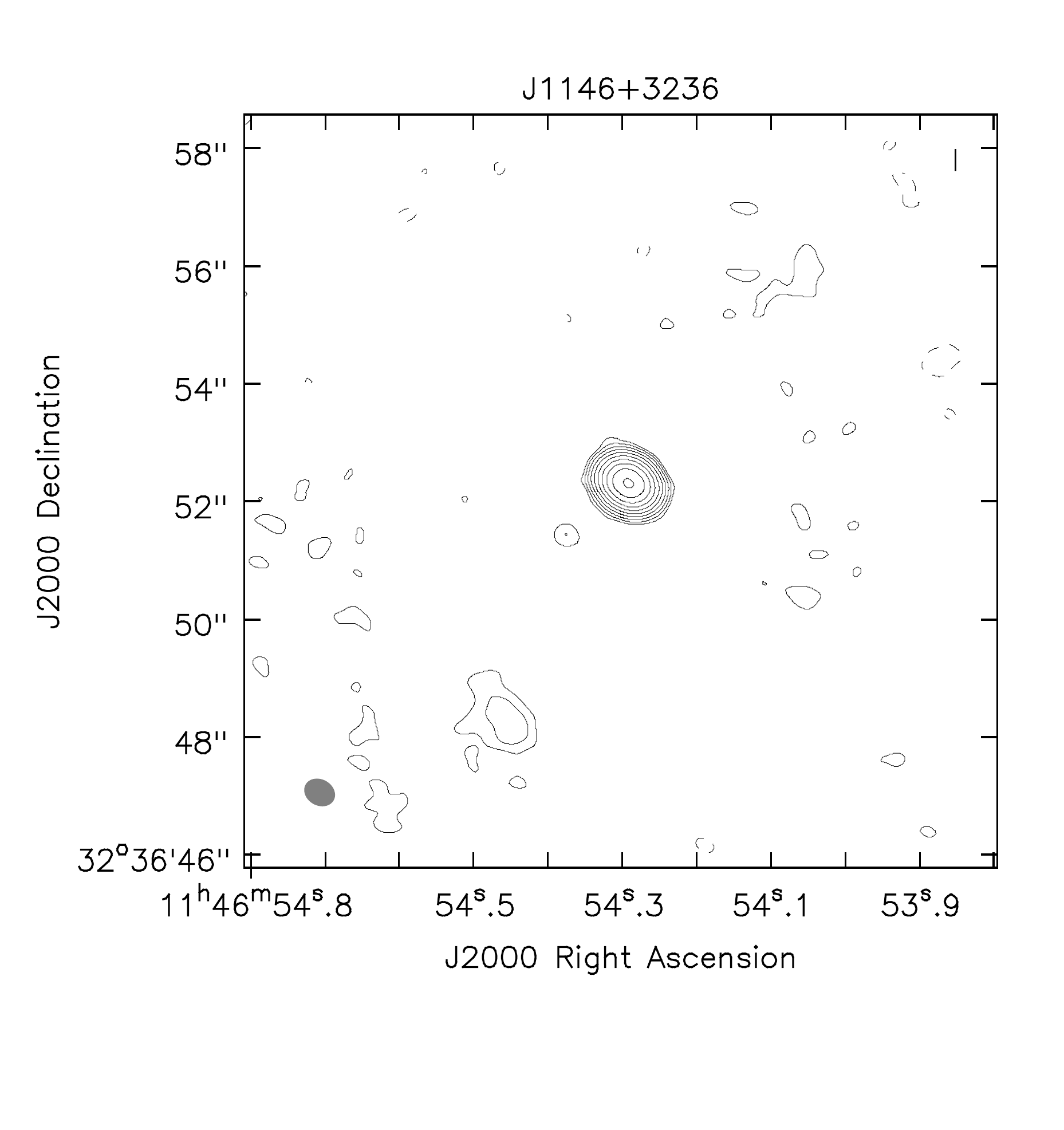} 
\includegraphics[trim={0cm 2cm 0cm 0cm},clip,width=7cm]{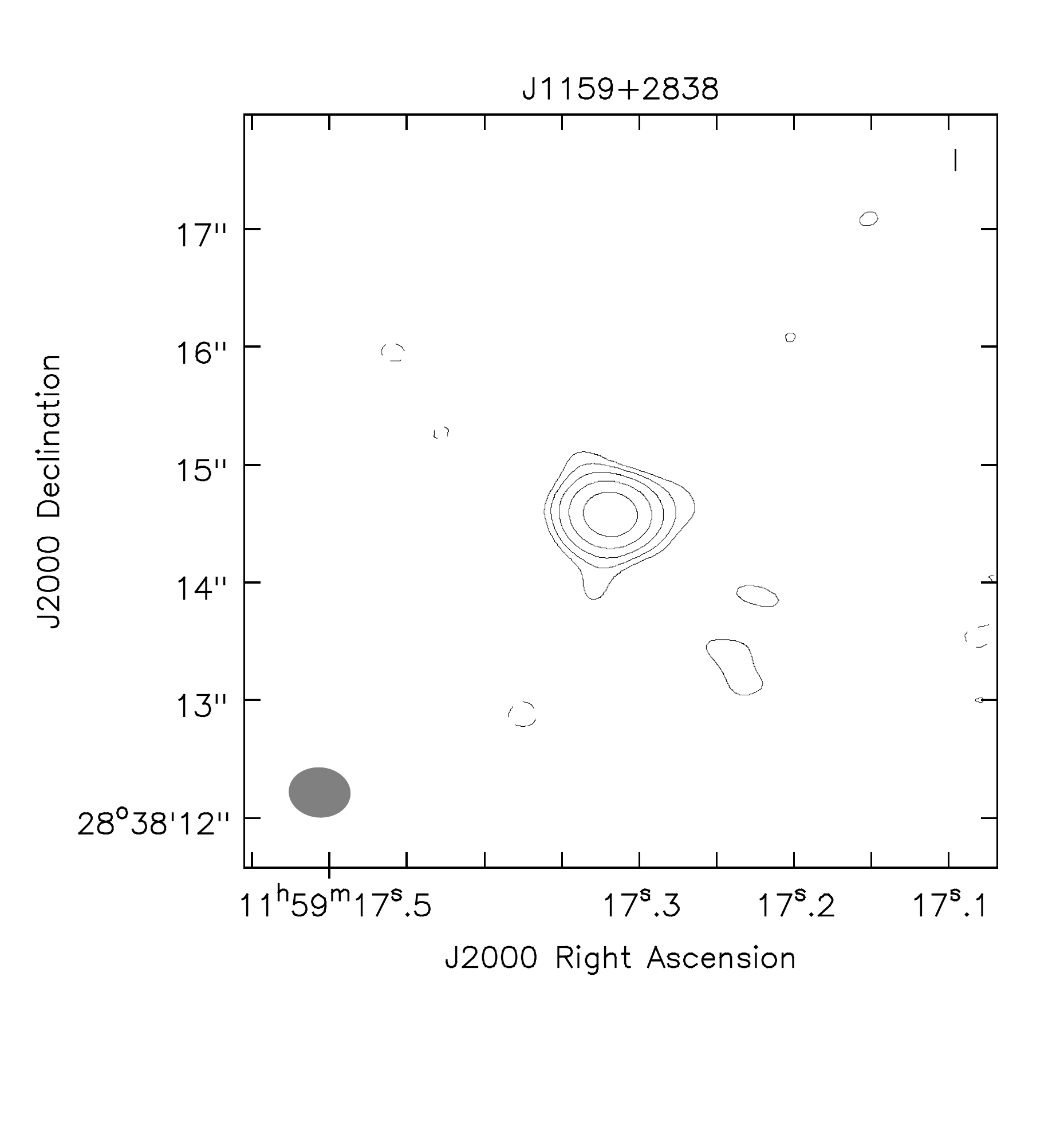} 
\caption{\textbf{Left panel:} J1146+3236, rms = 11 $\mu$Jy, contour levels at -3, 3$\times$2$^n$, n $\in$ [0,9], beam size 3.11$\times$2.52 kpc. \textbf{Right panel:} J1159+2838, rms = 10 $\mu$Jy, contour levels at -3, 3$\times$2$^n$, n $\in$ [0,4], beam size 1.78$\times$1.44 kpc.}
\label{fig:J1146p3236}
\label{fig:J1159p2838}
\end{figure*}

FBQS J1159+2838 is one of the \textit{bona-fide} F sources included in the \citet{Foschini15} sample. Our spectral indexes however led us to classify this object as a S-NLS1. The source appears in Fig.~\ref{fig:J1159p2838} as an unresolved core whose deconvolved size is approximately 430$\times$350 pc. With respect with other similar objects J1159+2838 seem to have a lower luminosity, 6$\times$10$^{39}$ \ergs, below the S average, while according to \citet{Caccianiga15} the star formation is fairly high, 128 M$_\odot$ yr$^{-1}$. This, along with the low brightness temperature, might indicate that starburst is a major source of radio emission in this object, as indicated also by its q22 ratio (1.39). This source was not detected at 37 GHz. Given these properties, this object might be classified as a low-luminosity CSS.
\subsection{J1203+4431} %4 quad, 1.36541, 0.82406
NGC 4051 is the closest NLS1, and a radio-quiet source. The core spectral index is significantly flatter than the integrated flux and broad-band indexes, possibly because of the presence of a non-thermal radio source in the nucleus, possibly a jet-base structure \citep{Giroletti09}. Its core revealed also the presence of a water-maser, consistent with a nuclear outflow \citep{Tarchi11}. Our map of Fig.~\ref{fig:J1203p4431} shows a large scale diffuse emission surrounding the unresolved core. The latter has a size of 33$\times$11 pc. The extended emission in the core is instead directed approximately at P.A. 240$^\circ$, but then it bends toward South-East forming a sort of horseshoe structure. More diffuse emission is visible on the other side of the nucleus, at P.A.$\sim$60$^\circ$. The diffuse emission accounts roughly for 61\% of the total luminosity of the source. NGC 4051 is detected only because of its vicinity: its luminosity is indeed the lowest of the sample, 3$\times$10$^{36}$ \ergs. 
\subsection{J1209+3217}
\begin{figure*}
\centering
\includegraphics[trim={0cm 2cm 0cm 0cm},clip,width=7cm]{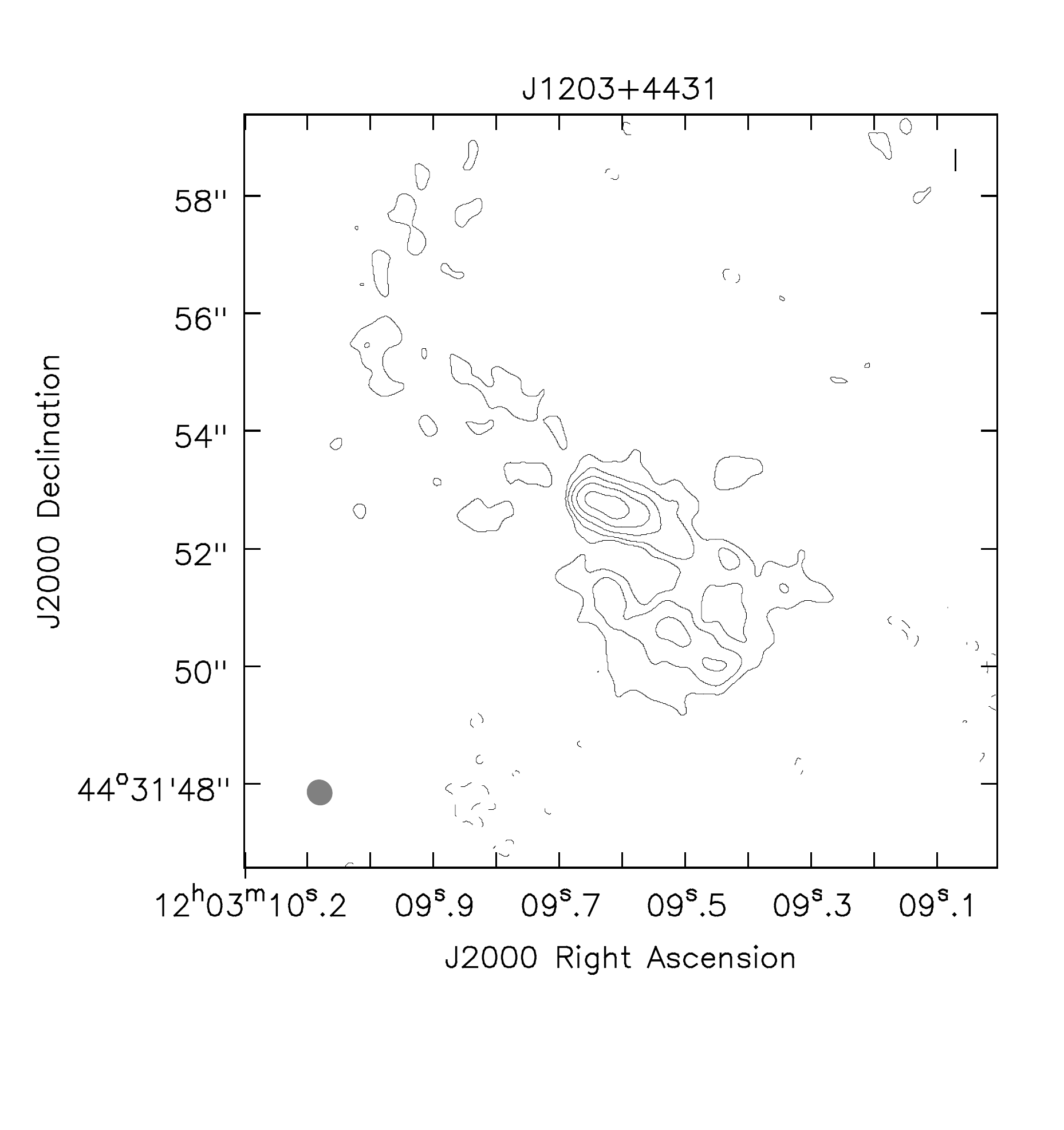} 
\includegraphics[trim={0cm 2cm 0cm 0cm},clip,width=7cm]{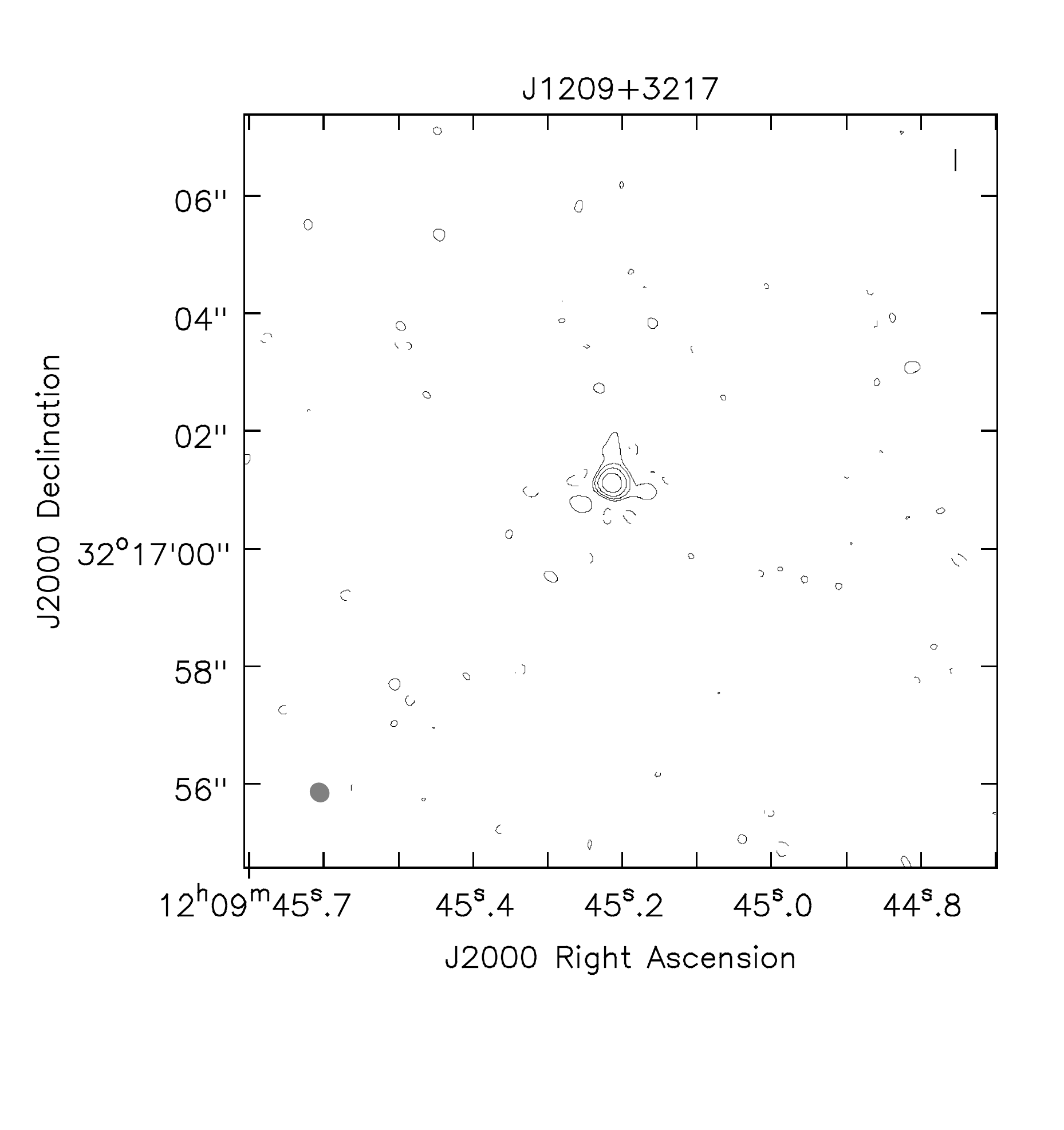} 
\caption{\textbf{Left panel:} J1203+4431, rms = 12 $\mu$Jy, contour levels at -3, 3$\times$2$^n$, n $\in$ [0,4], beam size 0.02$\times$0.02 kpc. \textbf{Right panel:} J1209+3217, rms = 13 $\mu$Jy, contour levels at -3, 3$\times$2$^n$, n $\in$ [0,3], beam size 0.83$\times$0.76 kpc.}
\label{fig:J1203p4431}
\label{fig:J1209p3217}
\end{figure*}

RX J1209.7+3217 is a radio-quiet source that appears in the map (Fig.~\ref{fig:J1209p3217}) with no particular features, and an unresolved core of 350$\times$170 pc. The steep spectral indexes and the low brightness temperature are typical of radio-quiet sources. 
\subsection{J1215+5442} %3 quad, 0.505263, 1.02256
SBS 1213+549A is a radio-quiet NLS1 with steep spectral indexes and a low brightness temperature. The map of Fig.~\ref{fig:J1215p5442} indicates the presence of some weak diffuse emission toward South-East, at P.A. 154$^\circ$, approximately extended for 1.1 arcsec (3 kpc). The unresolved core has a size of 480$\times$260 pc. 
\subsection{J1218+2948}
\begin{figure*}
\centering
\includegraphics[trim={0cm 2cm 0cm 0cm},clip,width=7cm]{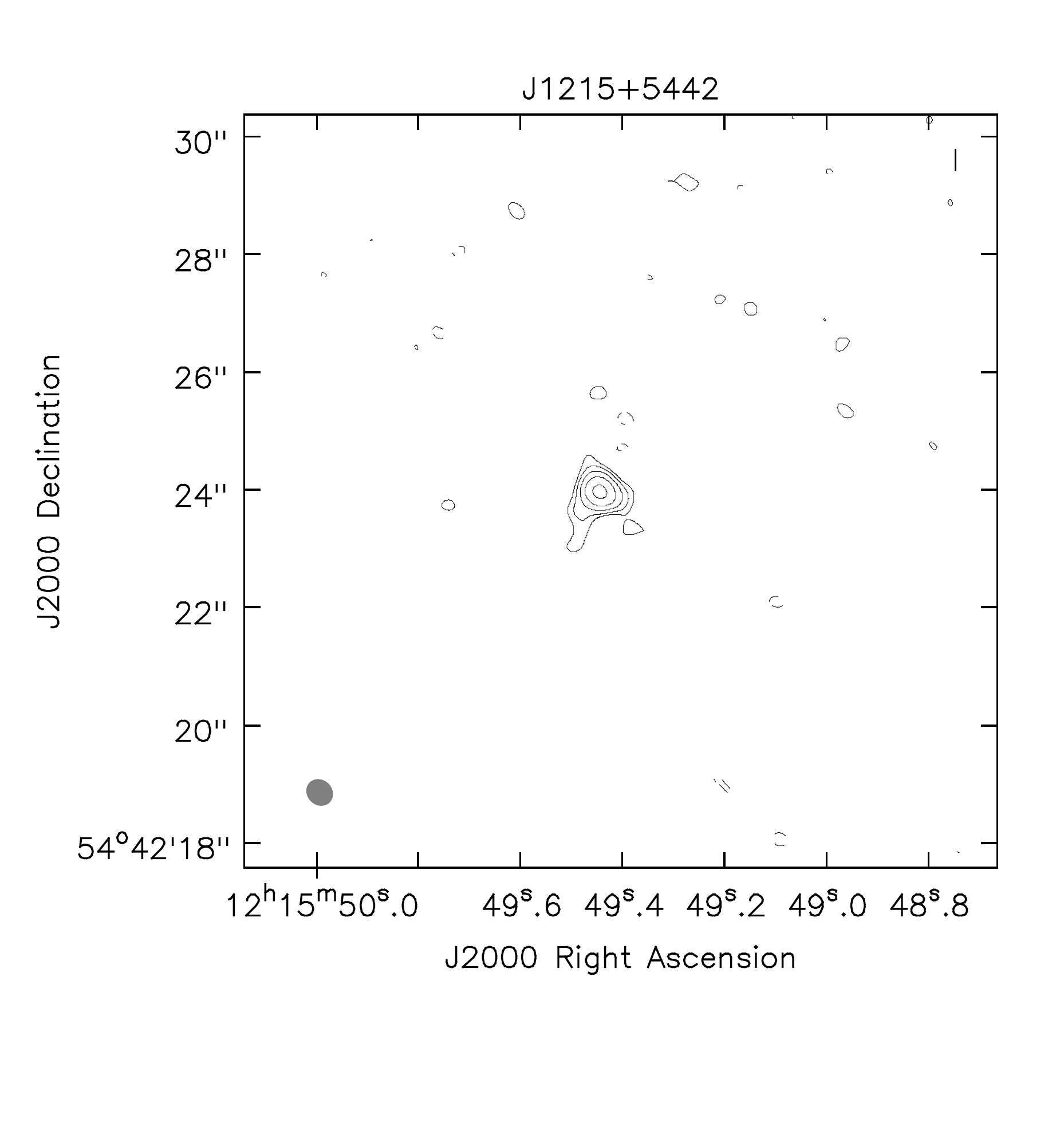} 
\includegraphics[trim={0cm 2cm 0cm 0cm},clip,width=7cm]{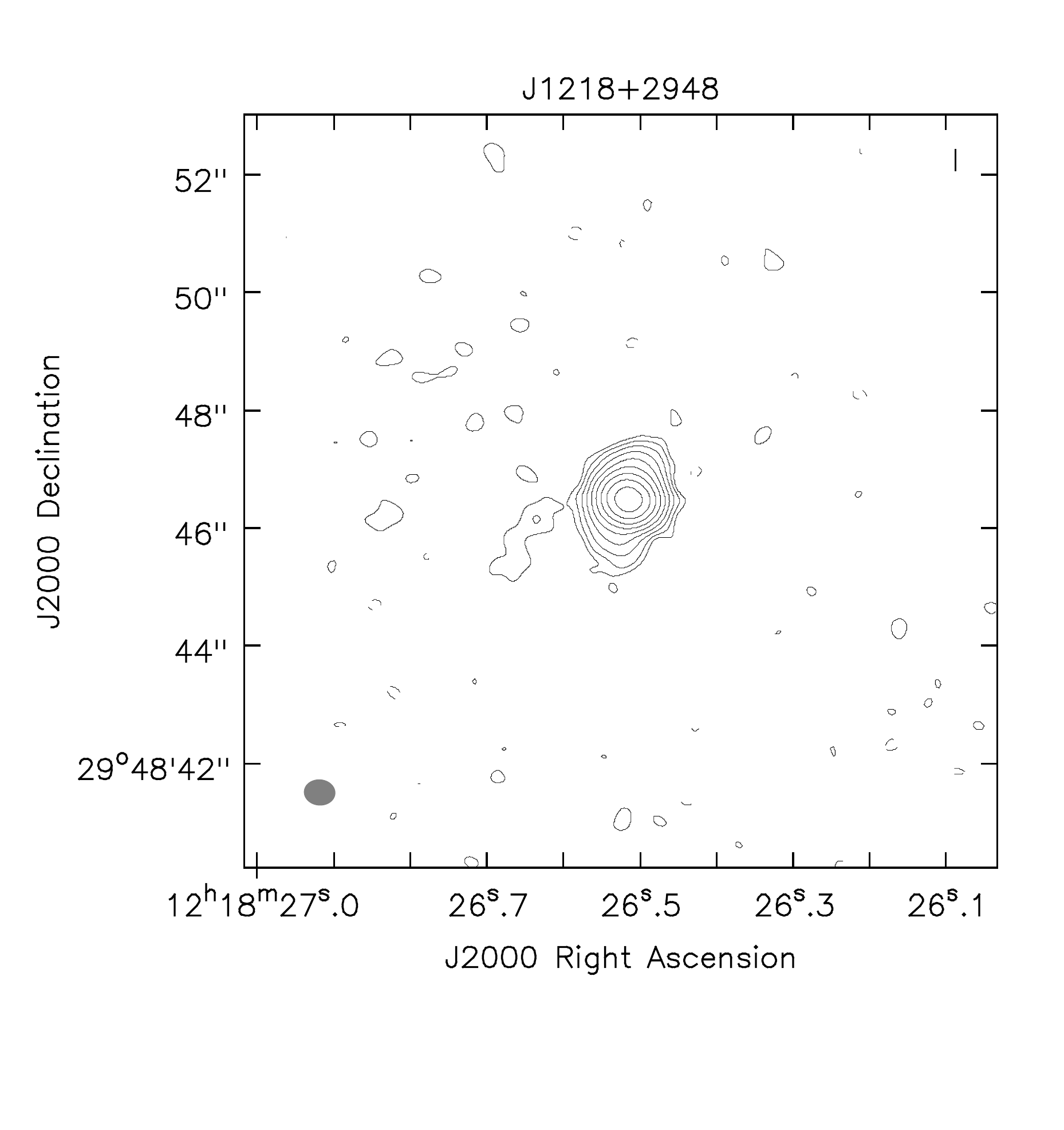} 
\caption{\textbf{Left panel:} J1215+5442, rms = 10 $\mu$Jy, contour levels at -3, 3$\times$2$^n$, n $\in$ [0,4], beam size 1.20$\times$1.07 kpc. \textbf{Right panel:} J1218+2948, rms = 10 $\mu$Jy, contour levels at -3, 3$\times$2$^n$, n $\in$ [0,8], beam size 0.14$\times$0.11 kpc.}
\label{fig:J1215p5442}
\label{fig:J1218p2948}
\end{figure*}

Mrk 766 is a well-known NLS1, harboring a strong water maser which seem to indicate the presence of a jet \citep{Tarchi11}. The unresolved core of the source has a size of 80$\times$60 pc, and it shows a flat spectrum, just like the in-band integrated index. Only the broad-band spectral index is consistent with a steep spectrum. Our estimated flux is in good agreement with previous observation at 5 GHz with the VLA-A (14.5 mJy, \citealp{Lal11}). The brightness temperature of $\sim$10000 K is higher with respect to other radio-quiet objects. The map of Fig.~\ref{fig:J1218p2948} shows a diffuse emission around the source particularly on the North-East side, and a brighter spot at P.A. 102$^\circ$, at 1.6 arcsec (430 pc) from the core. This diffuse emission accounts for $\sim$26\% of the total flux. This object appears compact on VLBI scale, and it was not detected at 22 GHz. 
\subsection{J1227+3214}
FBQS J1227+3214 was classified as a flat source according to its inverted spectral index (-1.04$\pm$0.07, \citealp{Foschini15}). Albeit we do not find the same strongly inverted spectrum, our classification remain consistent with the previous one. The source map in Fig.~\ref{fig:J1227p3214} reveals an unresolved core whose deconvolved size is 450$\times$225 pc, and possibly some diffuse emission in the North-East direction (PA$\sim$45$^\circ$). The star formation in this object is relatively high, 54 M$_\odot$ yr$^{-1}$ as reported by \citet{Caccianiga15}. The q22 ratio is also fairly high (1.29), possibly suggesting that a significant contribution to radio emission is provided by starburst. The source was not detected at 37 GHz.
\subsection{J1238+3942}
\begin{figure*}
\centering
\includegraphics[trim={0cm 2cm 0cm 0cm},clip,width=7cm]{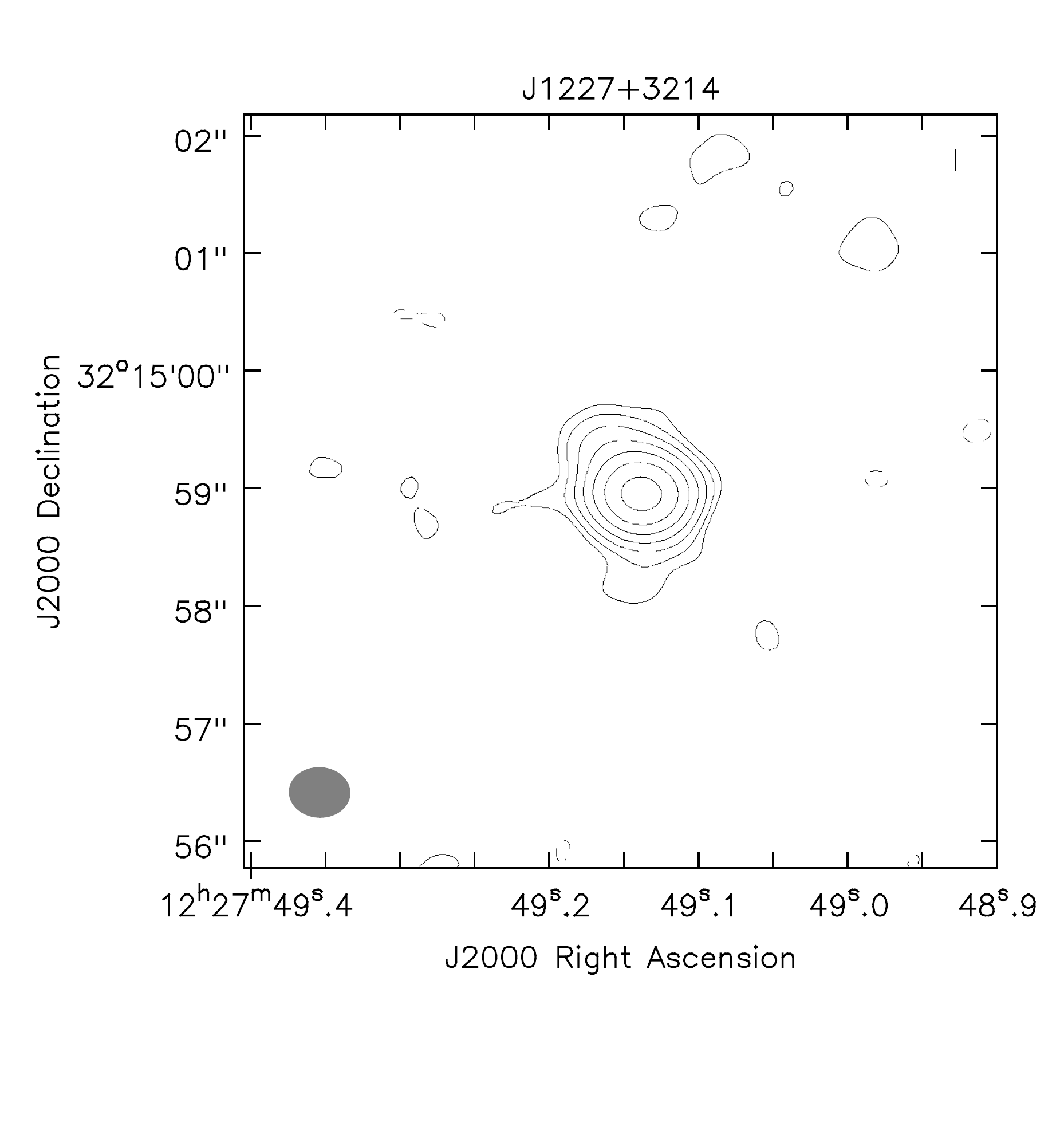} 
\includegraphics[trim={0cm 2cm 0cm 0cm},clip,width=7cm]{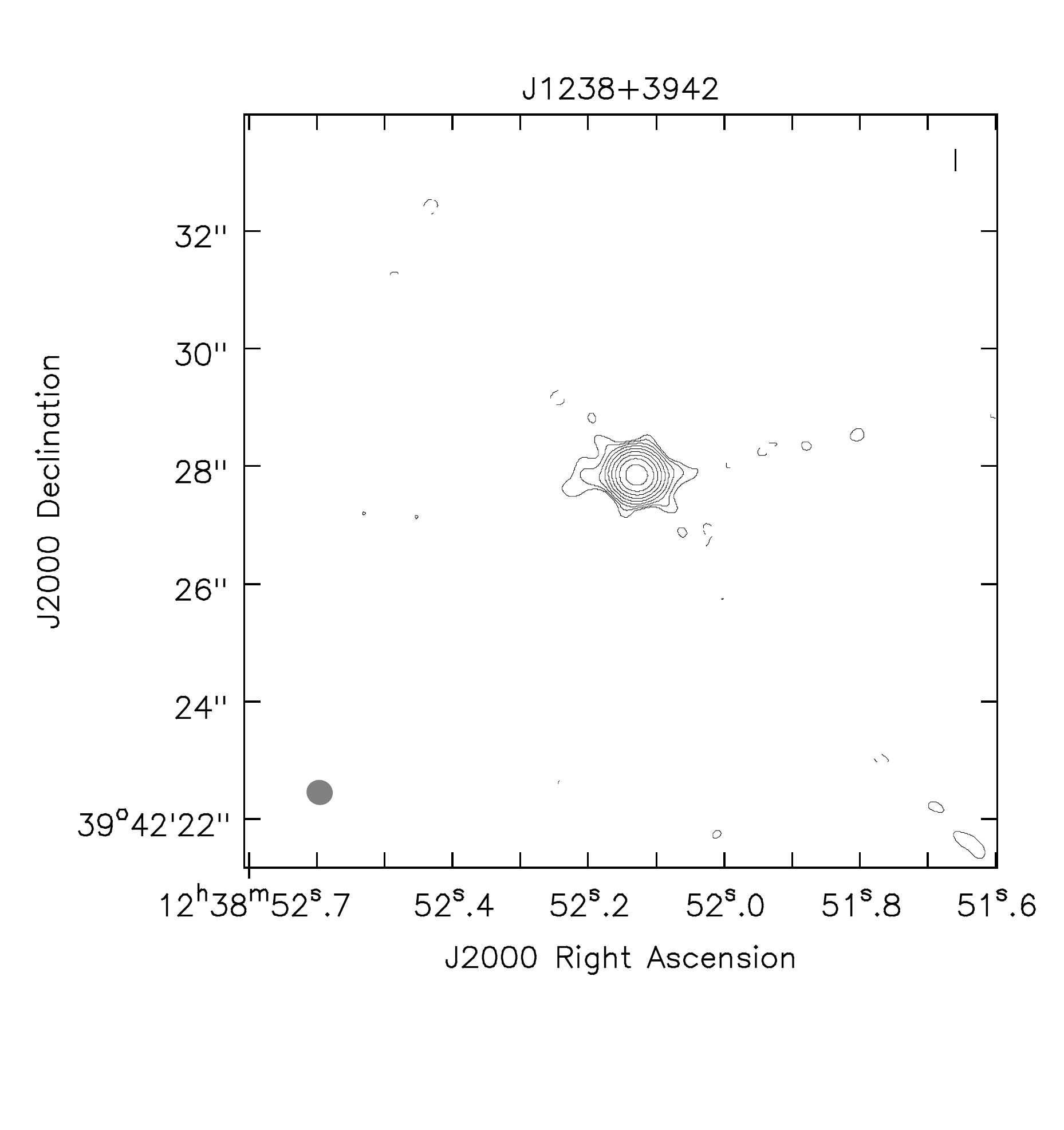} 
\caption{\textbf{Left panel:} J1227+3214, rms = 12 $\mu$Jy, contour levels at -3, 3$\times$2$^n$, n $\in$ [0,6], beam size 1.23$\times$1.01 kpc. \textbf{Right panel:} J1238+3942, rms = 13 $\mu$Jy, contour levels at -3, 3$\times$2$^n$, n $\in$ [0,8], beam size 2.92$\times$2.79 kpc.}
\label{fig:J1227p3214}
\label{fig:J1238p3942}
\end{figure*}

SDSS J123852.12+394227.8 is a \textit{bona-fide} F source, that in our measurements show a rather inverted radio spectrum both according to in- and broad-band spectral indexes. The map of Fig.~\ref{fig:J1238p3942} shows an unresolved core with a size lower than 400$\times$200 pc. Despite the inverted spectrum, the source was not detected neither at 22 GHz nor at 37 GHz, possibly indicating a slope change toward higher frequencies. 
\subsection{J1242+3317}
WAS 61 is a radio-quiet source with steep spectral indexes, not different from several others Q objects. The source, as seen in the map of Fig.~\ref{fig:J1242p3317}, is elongated in the East direction (P.A.$\sim$90$^\circ$). This extended emission accounts for approximately 40\% of the source total flux. The unresolved core size is 440$\times$140 pc. 
\subsection{J1246+0222}
\begin{figure*}
\centering
\includegraphics[trim={0cm 2cm 0cm 0cm},clip,width=7cm]{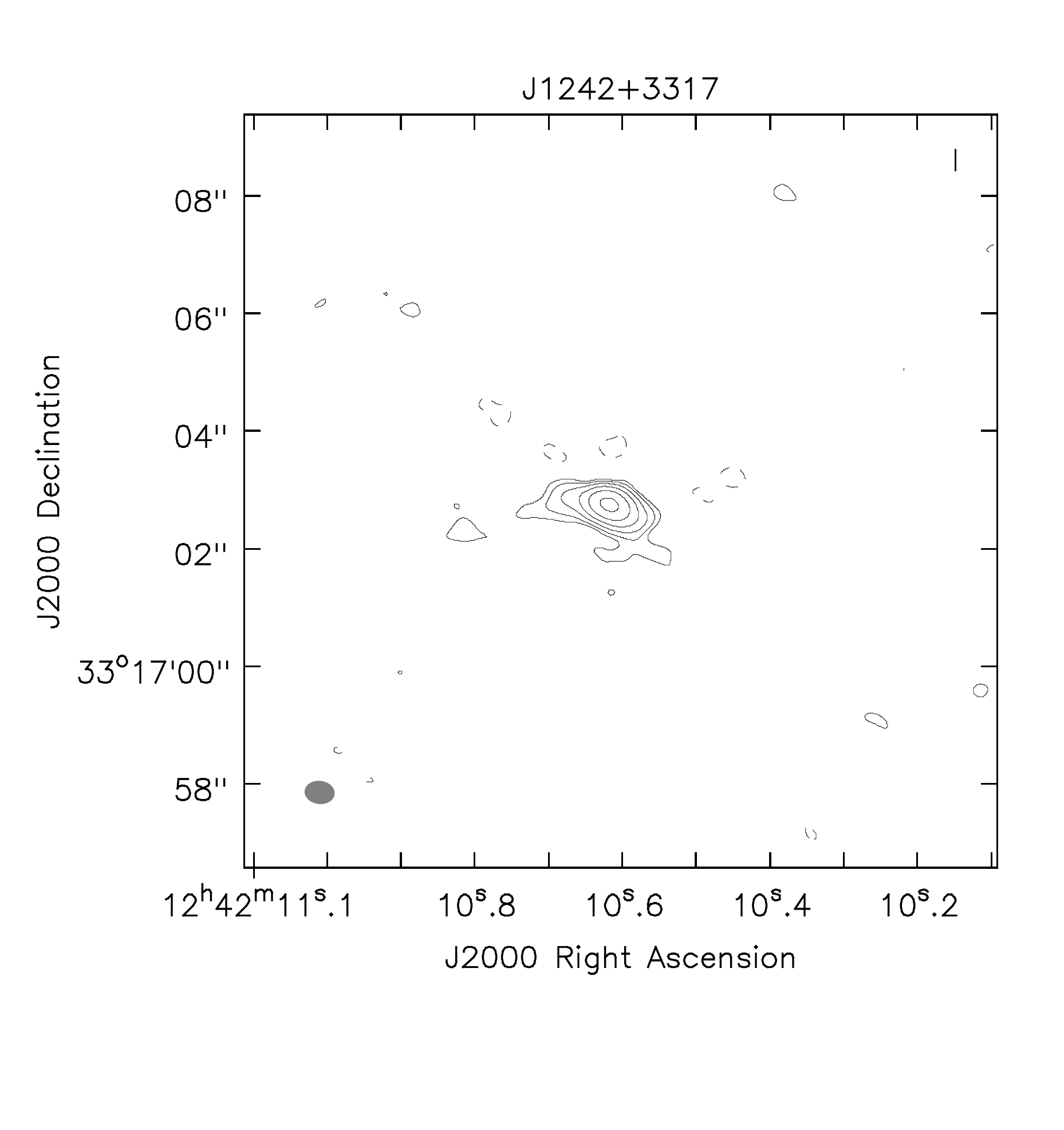} 
\includegraphics[trim={0cm 2cm 0cm 0cm},clip,width=7cm]{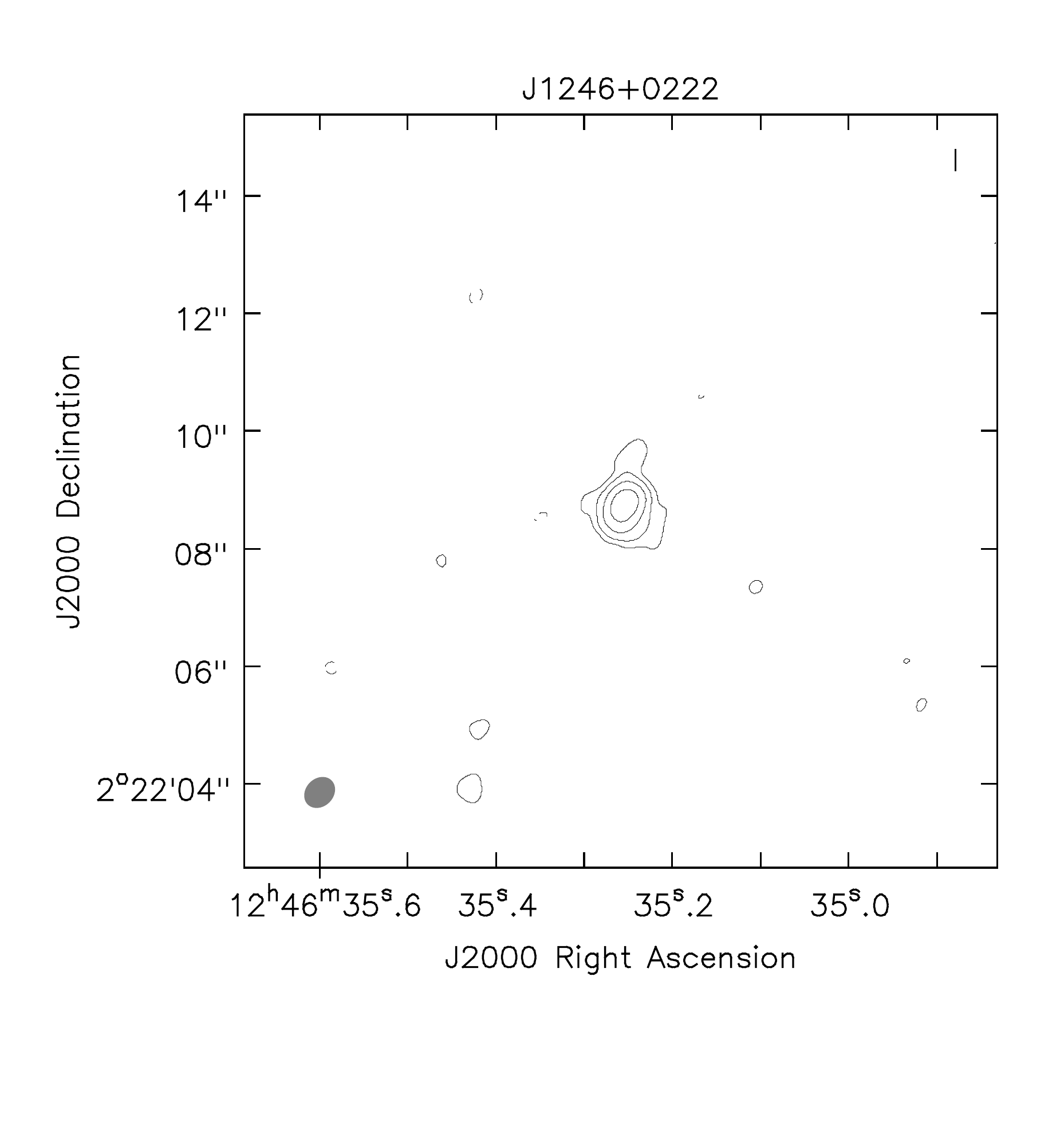} 
\caption{\textbf{Left panel:} J1242+3317, rms = 12 $\mu$Jy, contour levels at -3, 3$\times$2$^n$, n $\in$ [0,5], beam size 0.43$\times$0.33 kpc. \textbf{Right panel:} J1246+0222, rms = 11 $\mu$Jy, contour levels at -3, 3$\times$2$^n$, n $\in$ [0,3], beam size 0.52$\times$0.42 kpc.}
\label{fig:J1242p3317}
\label{fig:J1246p0222}
\end{figure*}

PG 1244+026 is a well-known radio-quiet NLS1, that in the map of Fig.~\ref{fig:J1246p0222} shows an unresolved core of 440$\times$240 pc that accounts only for 66\% of the total emission. Our flux measurement is rather similar to a previous measurement by \citet{Sikora07}, who found a slightly higher flux of 0.83 mJy. The core spectral index is uncertain, consistent both with flat and steep values, while the spectral index of the extended emission seems to be flat. Conversely, the broad-band index is definitely steep. However, the brightness temperature is the second lowest of the sample, 275 K, suggesting that the radio emission might be of thermal origin.
\subsection{J1246+0238}
According to \citet{Foschini11}, SDSS J124634.65+023809.0 is a $\gamma$-ray source. The spectral indexes are all flat, as expected from this kind of source, and in agreement with the index already measured by \citet{Foschini15}. The source, as shown in Fig.~\ref{fig:J1246p0238}, appears only as an unresolved core. The same compact core is reported by \citet{Gu15} at 5 GHz with VLBA observations. The flux they measured was less than half our value (8.7 mJy against 19.66 mJy in the core), suggesting that this source is strongly variable. Despite this, it was not detected neither at 22 GHz nor at 37 GHz. The star formation is not negligible (68 M$_\odot$ yr$^{-1}$), but the q22 parameter of -0.36 suggests that it is not the main contributor to the radio emission. 
\subsection{J1302+1624}
\begin{figure*}
\centering
\includegraphics[trim={0cm 2cm 0cm 0cm},clip,width=7cm]{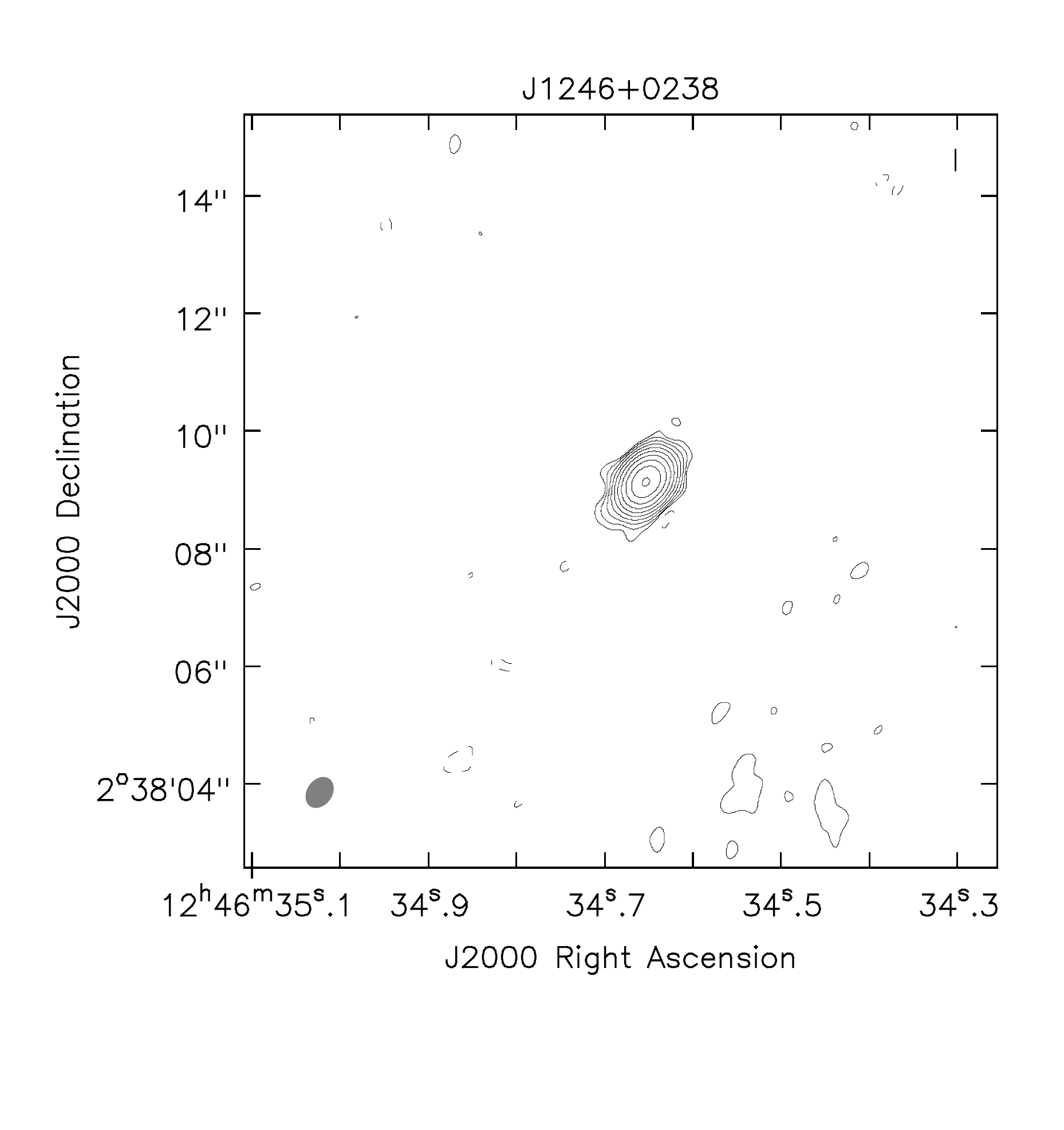} 
\includegraphics[trim={0cm 2cm 0cm 0cm},clip,width=7cm]{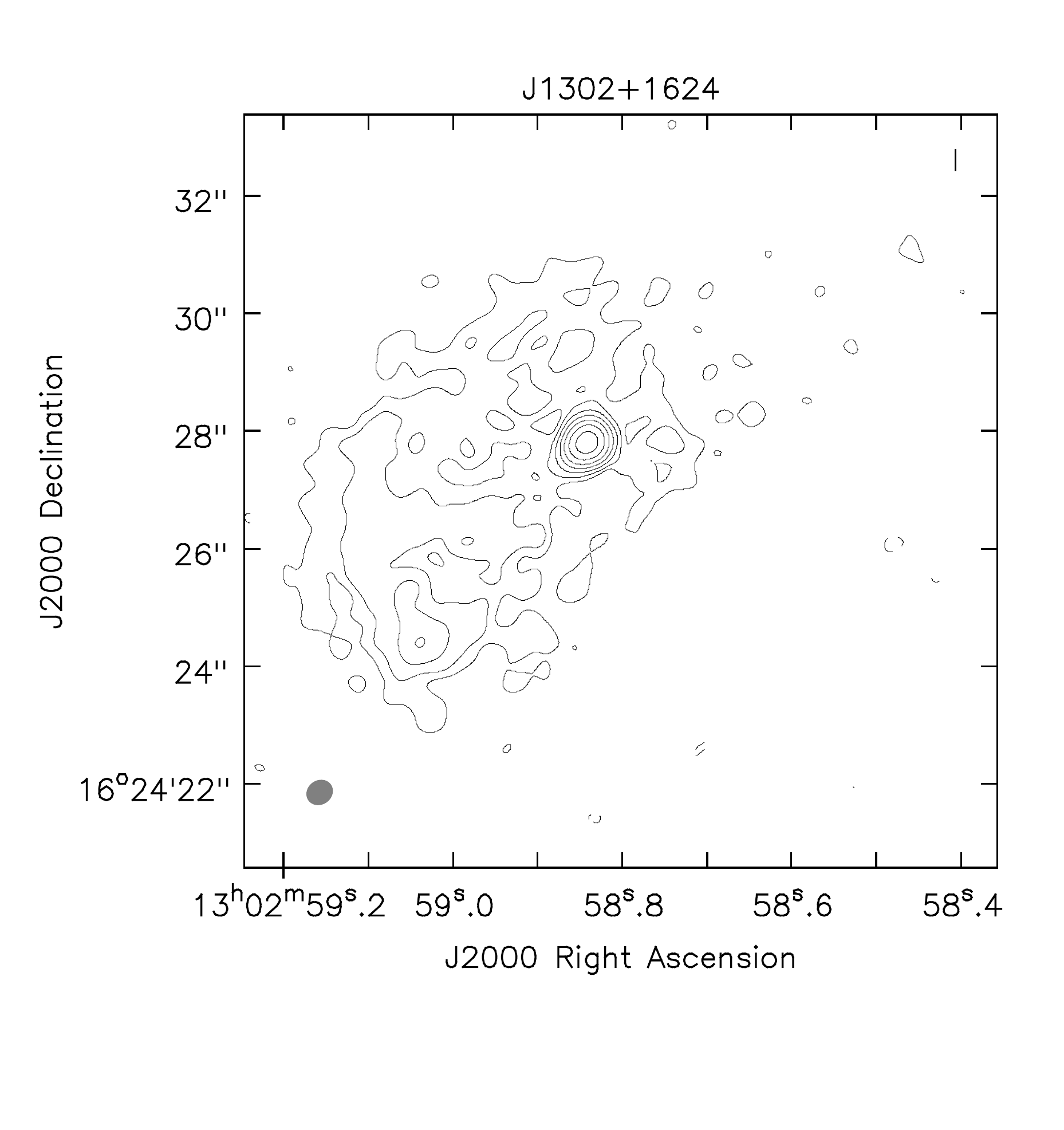} 
\caption{\textbf{Left panel:} J1246+0238, rms = 12 $\mu$Jy, contour levels at -3, 3$\times$2$^n$, n $\in$ [0,9], beam size 2.80$\times$2.09 kpc. \textbf{Right panel:} J1302+1624, rms = 11 $\mu$Jy, contour levels at -3, 3$\times$2$^n$, n $\in$ [0,6], beam size 0.58$\times$0.51 kpc.}
\label{fig:J1246p0238}
\label{fig:J1302p1624}
\end{figure*}

Mrk 783 is a NLS1 at the edge of the radio-quiet/radio-loud threshold. Already studied in detail by \citet{Congiu17}, we confirm the results of their analysis. Its in-band integrated spectral index is the steepest of our entire sample, in agreement with the hypothesis that the large extended emission visible in Fig.~\ref{fig:J1302p1624} is a relic. Interestingly, this faint extended emission is associated with an ionized region of gas located 35 kpc away from the nucleus, a unique case among NLS1s \citep{Congiu17b}. The source was not detected with the VLBA at 22 GHz. 
\subsection{J1305+5116} %3 quad, 0.228571, 1.05865
SDSS J130522.74+511640.2 was originally classified as an S source and analyzed by \citet{Berton15a} because of a detection at low frequency (150 MHz, 320 mJy, \citealp{Yuan08}) which allowed the spectral classification. The source was later discovered to be a $\gamma$-ray source \citep{Liao15}, and the in-band indexes allow us to change its classification in F-NLS1. In the map of Fig.~\ref{fig:J1305p5116} it clearly shows two components separated by 1.1 arcsec (8.1 kpc) at P.A. 168$^\circ$. It is not certain where the core is, since both of the components have a flat or inverted spectrum. However, the North (N) component core accounts for $\sim$64\% of the total flux, while the South (S) component has a peak of 18.3 mJy beam$^{-1}$. The NVSS coordinates point on the N source, but the uncertainty on the declination is fairly large (0.6 arcsec). \citet{Gu15} detected the same kind of structure at 2.2, 5, and 8.6 GHz with the VLBA. At 5 GHz they found however a flux of 15 mJy, which is less than half the flux we obtained for the N component only, indicating that there is a large fraction of diffuse emission which VLBA cannot resolve. The N component has a unresolved core whose size is 1.1$\times$0.4 kpc, while the S component core is 2.8$\times$1.8 kpc large. A possibility is that the N component is the core, while the S component is the hot-spot of a radio lobe. The source was not detected both at 22 GHz and 37 GHz. 
\subsection{J1317+6010}
\begin{figure*}
\centering
\includegraphics[trim={0cm 2cm 0cm 0cm},clip,width=7cm]{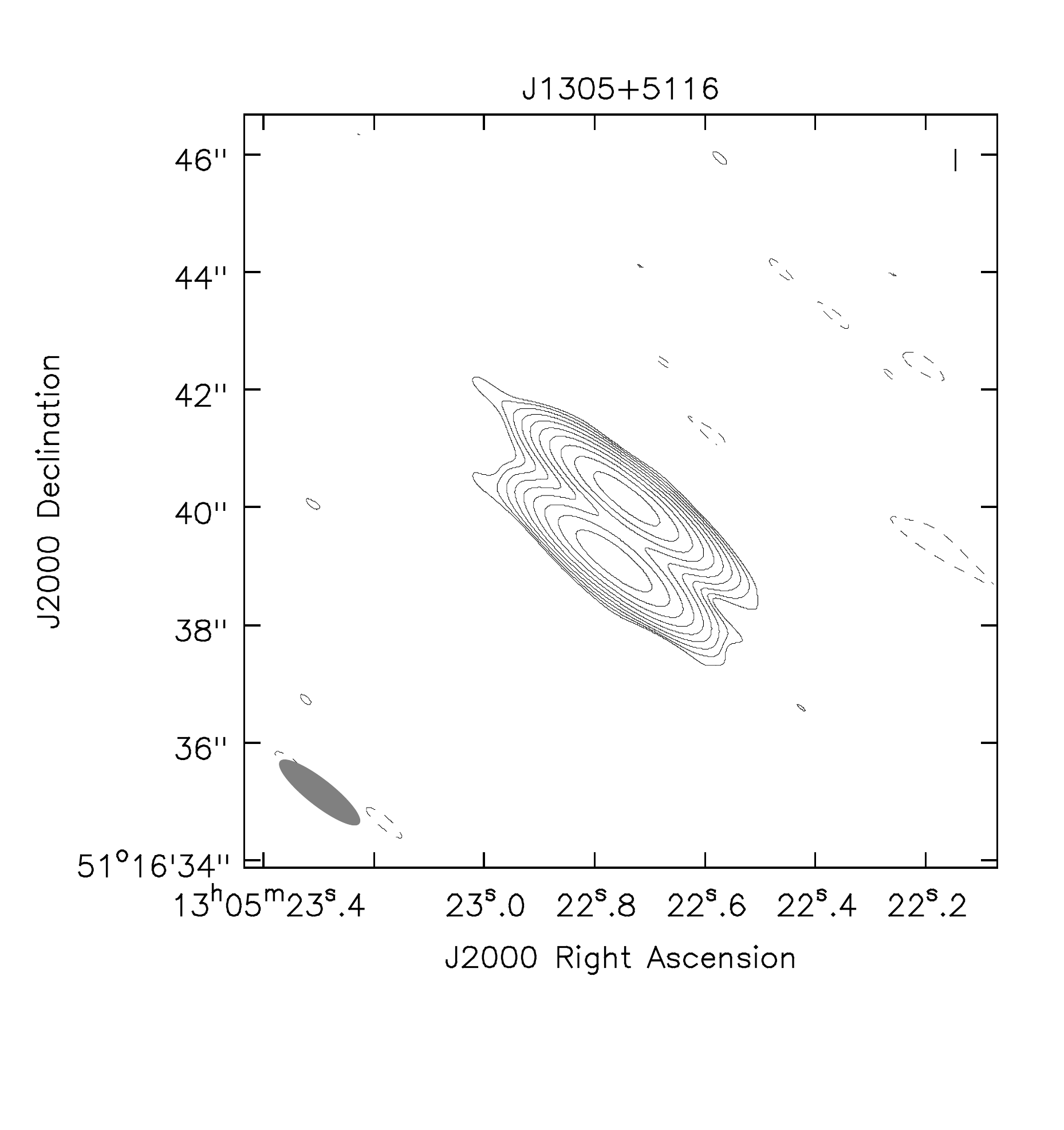} 
\includegraphics[trim={0cm 2cm 0cm 0cm},clip,width=7cm]{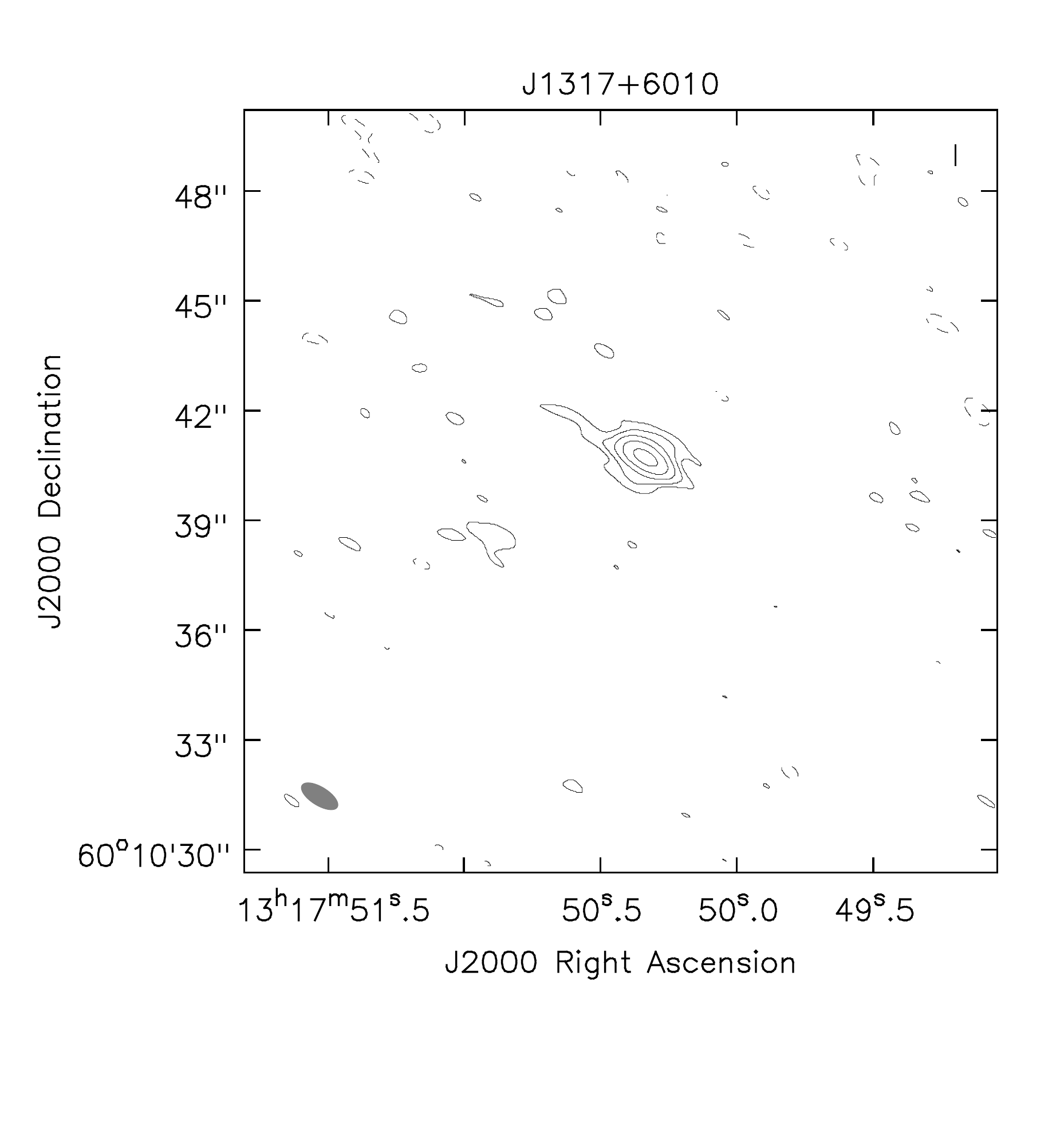} 
\caption{\textbf{Left panel:} J1305+5116, rms = 13 $\mu$Jy, contour levels at -3, 3$\times$2$^n$, n $\in$ [0,9], beam size 12.70$\times$3.29 kpc. \textbf{Right panel:} J1317+6010, rms = 11 $\mu$Jy, contour levels at -3, 3$\times$2$^n$, n $\in$ [0,4], beam size 2.74$\times$1.19 kpc.}
\label{fig:J1305p5116}
\label{fig:J1317p6010}
\end{figure*}

SBS 1315+604 is a radio-quiet NLS1. In the map of Fig.~\ref{fig:J1317p6010} the source shows no significant structure, and an unresolved core of 1.1$\times$0.9 kpc. The core has a flat spectrum, while the total emission is steeper, although still consistent with the flat classification. The broad-band index instead is steep. The brightness temperature is extremely low, 282 K, and suggests a possible thermal origin of the radiation. 
\subsection{J1333+4141}
SDSS J133345.47+414127.7 is one of the \textit{bona-fide} F sources, but according to our spectral indexes this source appears to be steep-spectrum. In Fig.~\ref{fig:J1333p4141} J1333+4141 is compact, with a deconvolved core size of 1.4$\times$0.2 kpc. The brightness temperature is low, $\sim$4500 K, and star formation appears to be very relevant in the radio. The q22 parameter is indeed 1.39, and the star formation rate is 148 M$_\odot$ yr$^{-1}$ \citep{Caccianiga15}. It was not detected at 37 GHz. Because of its properties, it can be considered as a low-luminosity CSS.
\subsection{J1337+2423} %2 quad, 1.9609, 0.132331
\begin{figure*}
\centering
\includegraphics[trim={0cm 2cm 0cm 0cm},clip,width=7cm]{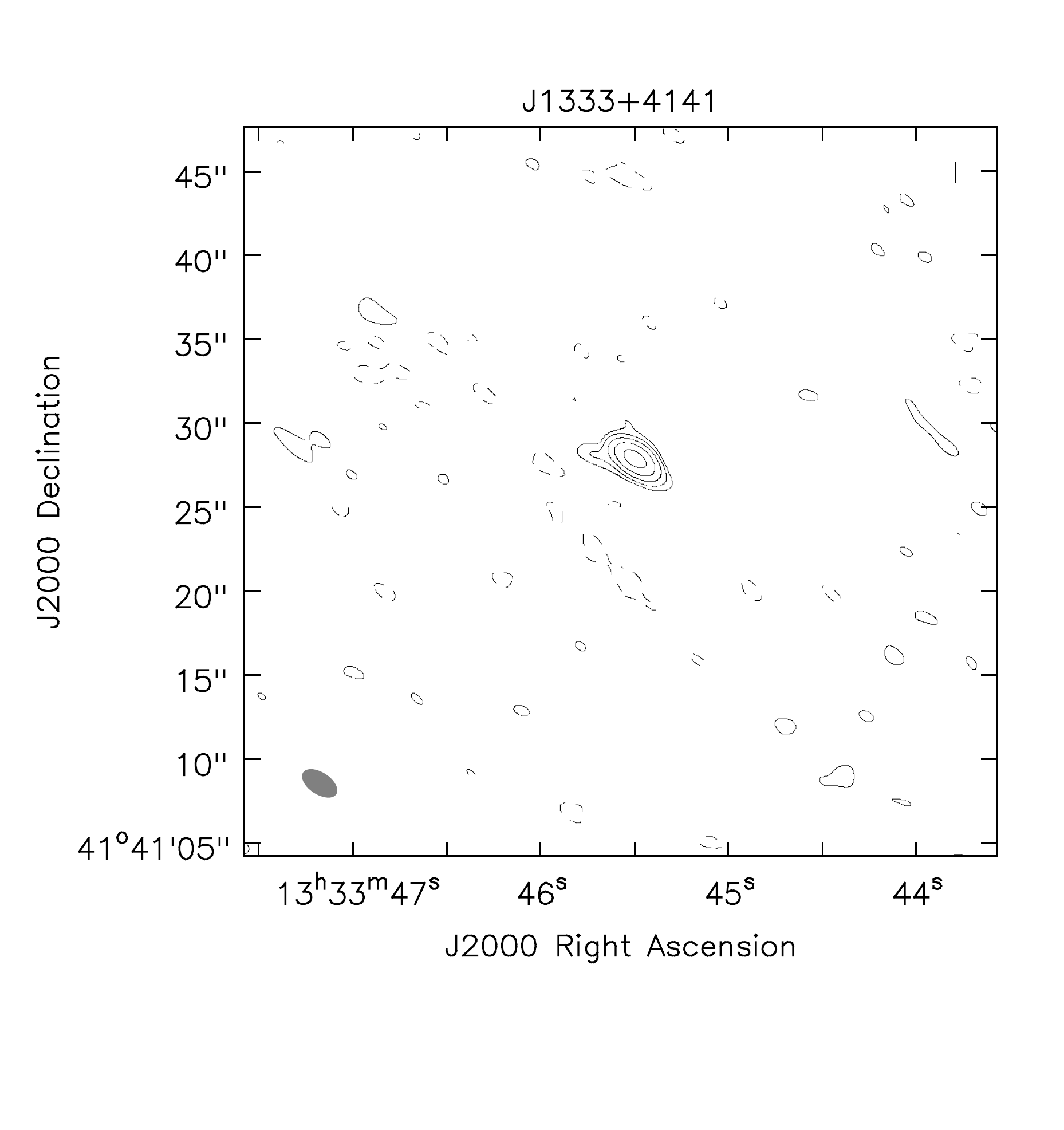} 
\includegraphics[trim={0cm 2cm 0cm 0cm},clip,width=7cm]{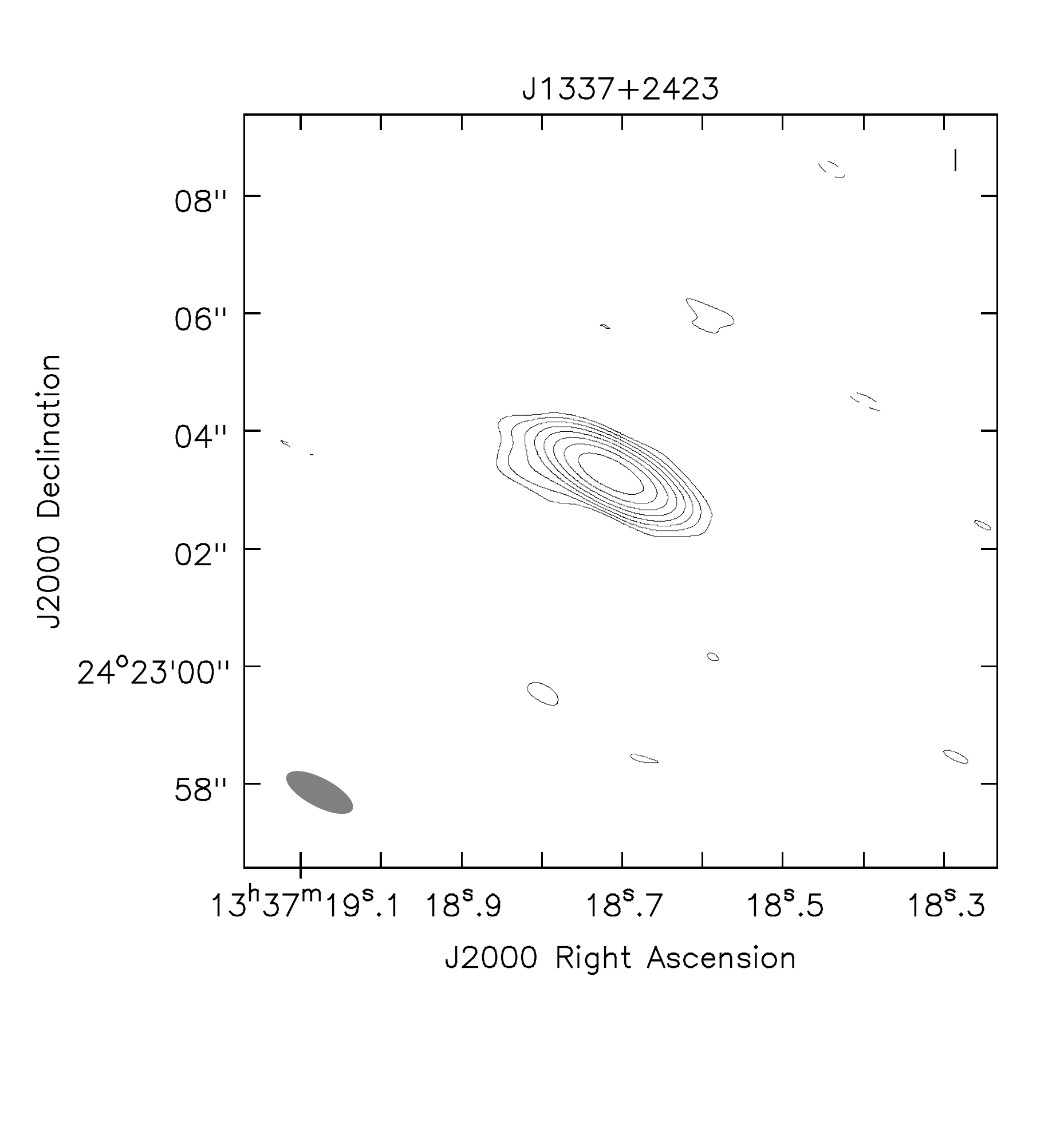} 
\caption{\textbf{Left panel:} J1333+4141, rms = 14 $\mu$Jy, contour levels at -3, 3$\times$2$^n$, n $\in$ [0,4], beam size 8.31$\times$4.52 kpc. \textbf{Right panel:} J1337+2423, rms = 13 $\mu$Jy, contour levels at -3, 3$\times$2$^n$, n $\in$ [0,7], beam size 2.45$\times$0.93 kpc.}
\label{fig:J1333p4141}
\label{fig:J1337p2423}
\end{figure*}

IRAS 13349+2438 is a well-studied radio-quiet NLS1. The map of Fig.~\ref{fig:J1337p2423} shows an intermediate morphology, with an extended emission at P.A. 86$^\circ$ reaching up to $\sim$2 arcsec (7.1 kpc). Both the in-band spectral indexes are flat, just like the broad-band spectral index. The unresolved core has a size of 560$\times$235 pc. The brightness temperature, with respect to other radio-quiet sources, is high, $\sim$72000 K. Given these properties, we cannot rule out that this source harbors a low-power relativistic beamed jet. 
\subsection{J1346+3121}
SDSS J134634.97+312133.7 is one of the \textit{bona-fide} F sources, and our measurements confirm this classification, since all the spectra are strongly inverted, more than any other source in the sample. Despite this, this object was not detected yet at 37 GHz. The map of Fig.~\ref{fig:J1346p3121} shows no particular features, with a core of size 720$\times$260 pc and a relatively low brightness temperature ($\sim$30000 K). 
\subsection{J1348+2622}
\begin{figure*}
\centering
\includegraphics[trim={0cm 2cm 0cm 0cm},clip,width=7cm]{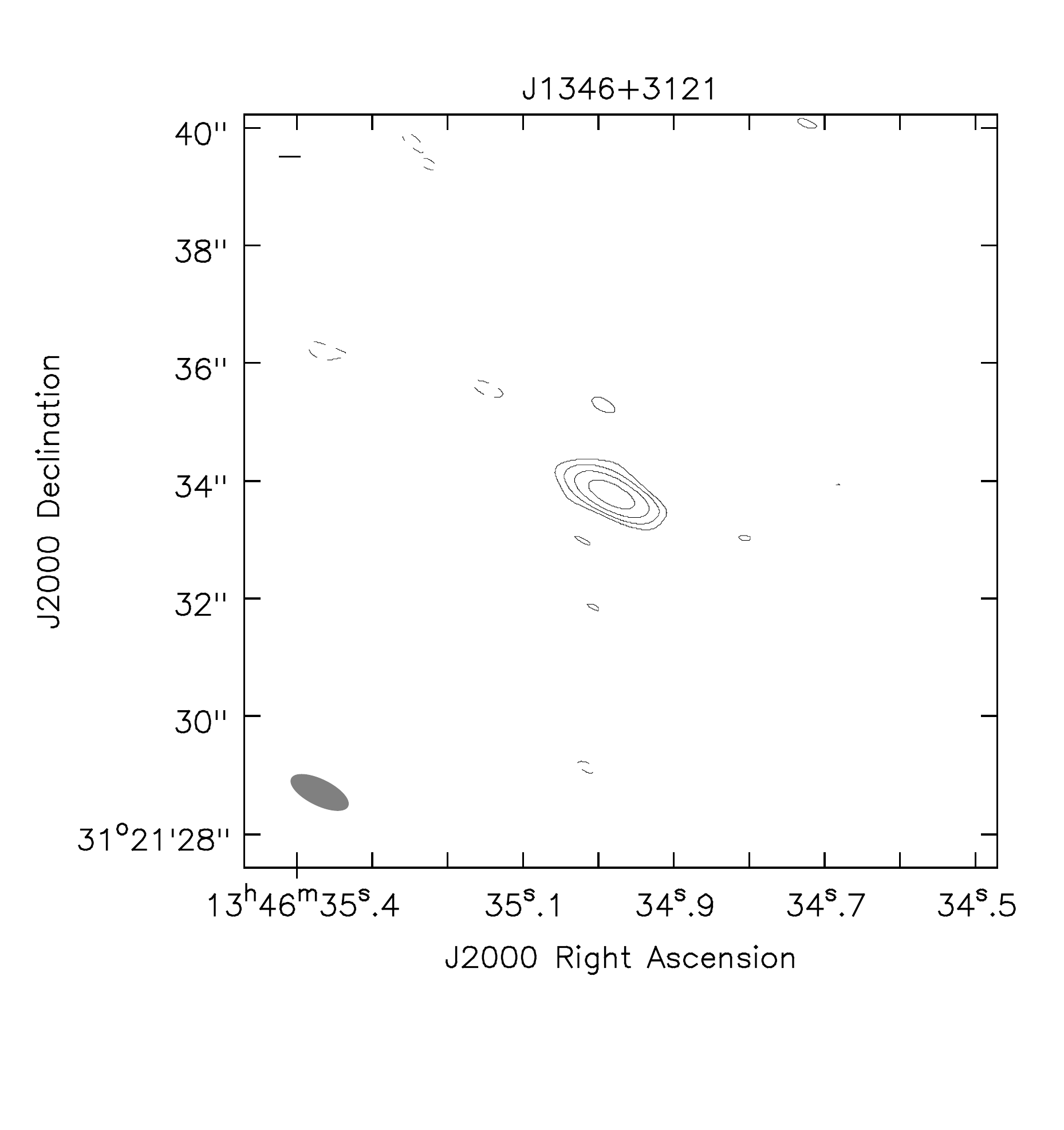} 
\includegraphics[trim={0cm 2cm 0cm 0cm},clip,width=7cm]{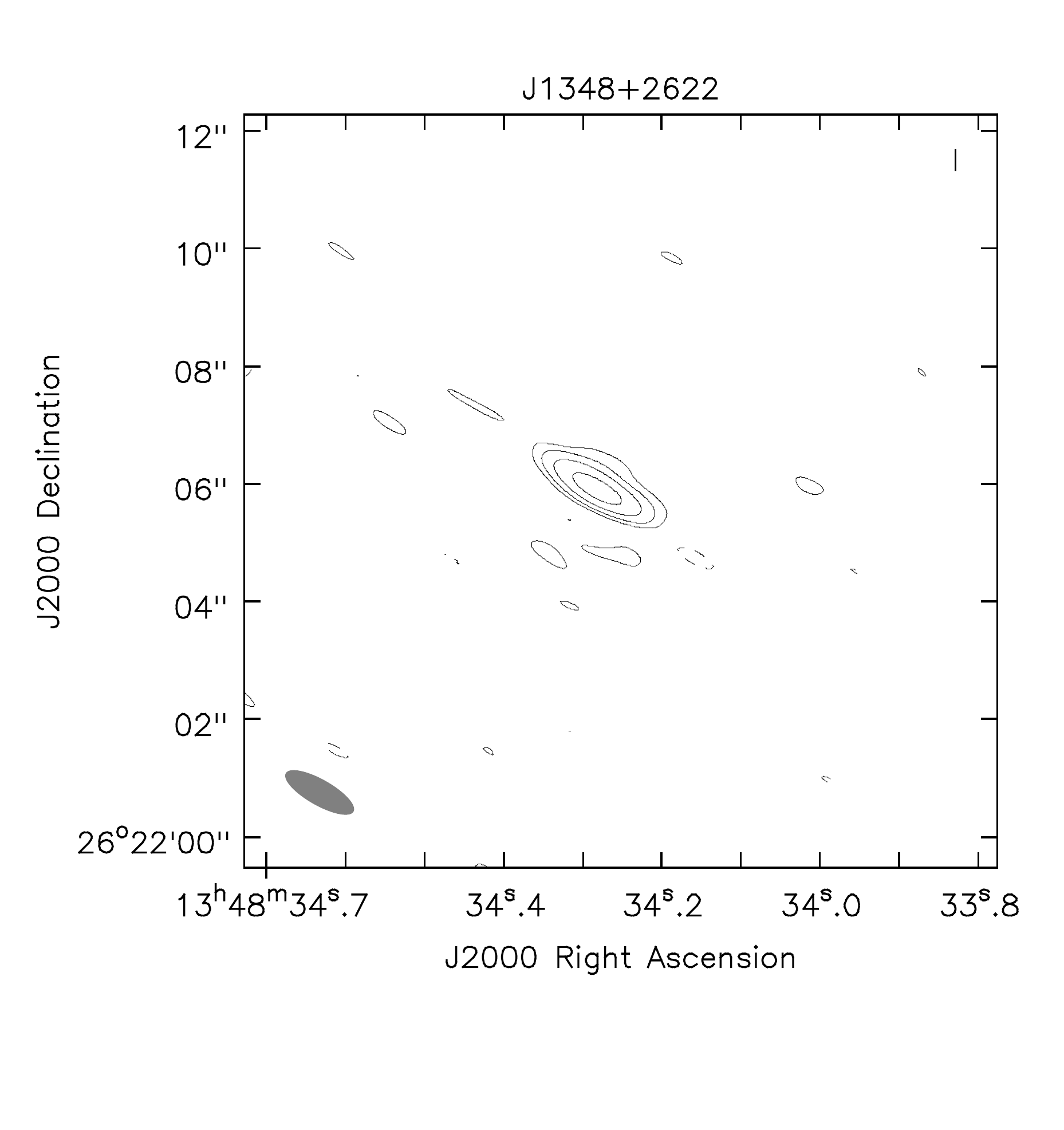} 
\caption{\textbf{Left panel:} J1346+3121, rms = 12 $\mu$Jy, contour levels at -3, 3$\times$2$^n$, n $\in$ [0,3], beam size 4.10$\times$1.66 kpc. \textbf{Right panel:} J1348+2622, rms = 12 $\mu$Jy, contour levels at -3, 3$\times$2$^n$, n $\in$ [0,3], beam size 10.26$\times$3.29 kpc.}
\label{fig:J1346p3121}
\label{fig:J1348p2622}
\end{figure*}

SDSS J134834.28+262205.9 is another \textit{bona-fide} F source, which we reclassify as steep-spectrum. It is the most distant source of the sample (z = 0.917), and in the map of Fig.~\ref{fig:J1348p2622} it appears as a compact source with a deconvolved core size of 3.4$\times$1.3 kpc and some faint diffuse emission toward South and East. The spectral indexes are all rather steep, of the order of unity, and the brightness temperature is low, $\sim$700 K. These characteristics seem to recall those of sources with high star formation rates. However, this is in contrast with the result of \citet{Caccianiga15}, who did not find any sign of strong starbursts using WISE colors. 
\subsection{J1355+5612}
SBS 1353+564 is a radio-quiet NLS1. The in-band spectral index of the core is flat, and the same is true for the in-band index of the integrated emission. The broad-band index is instead steep. We could provide only an upper limit to the deconvolved core size, which is smaller than 650$\times$265 pc. The map of Fig.~\ref{fig:J1355p5612} does not reveal any particular structure beside the core.
\subsection{J1358+2658}
\begin{figure*}
\centering
\includegraphics[trim={0cm 2cm 0cm 0cm},clip,width=7cm]{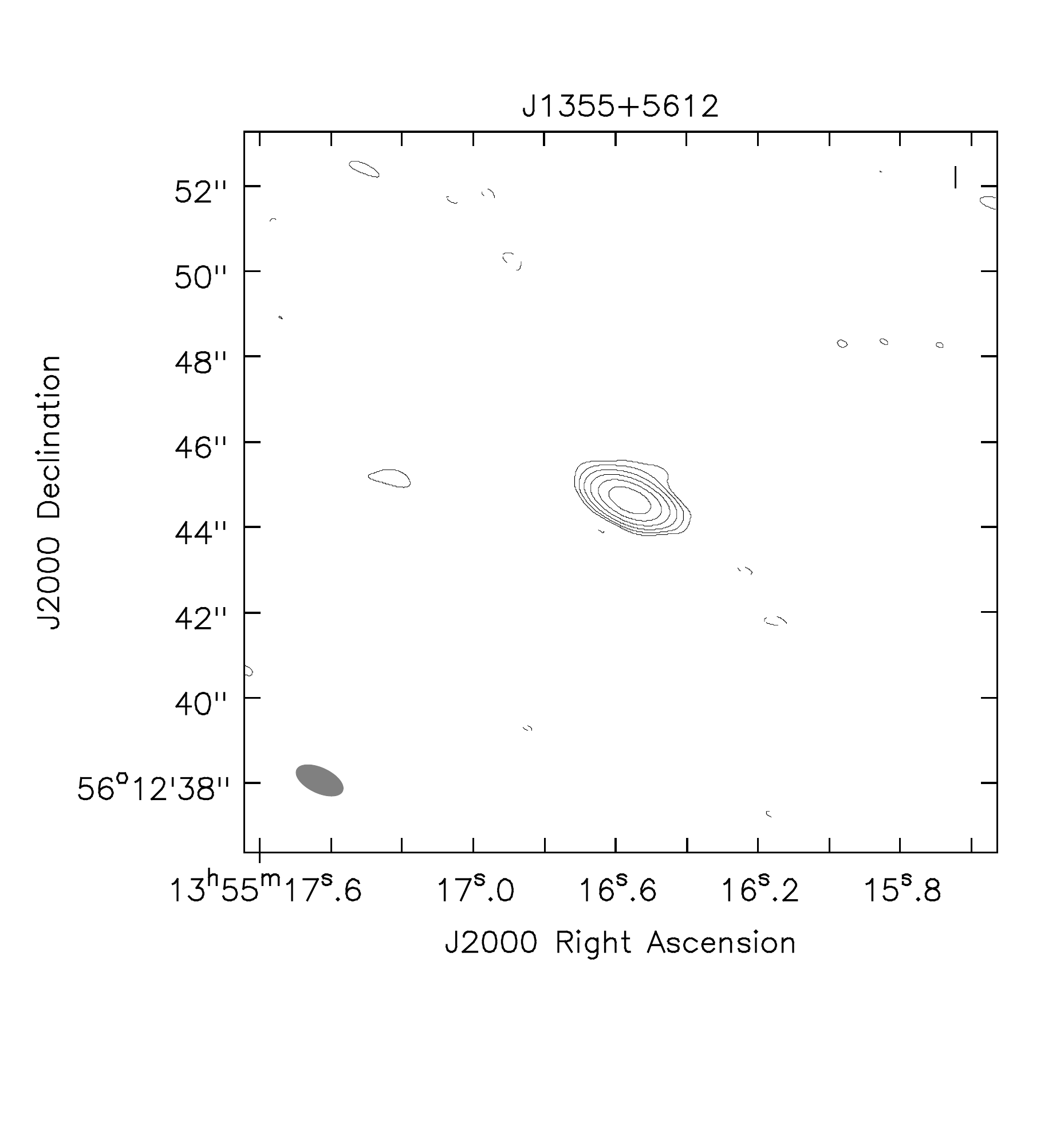} 
\includegraphics[trim={0cm 2cm 0cm 0cm},clip,width=7cm]{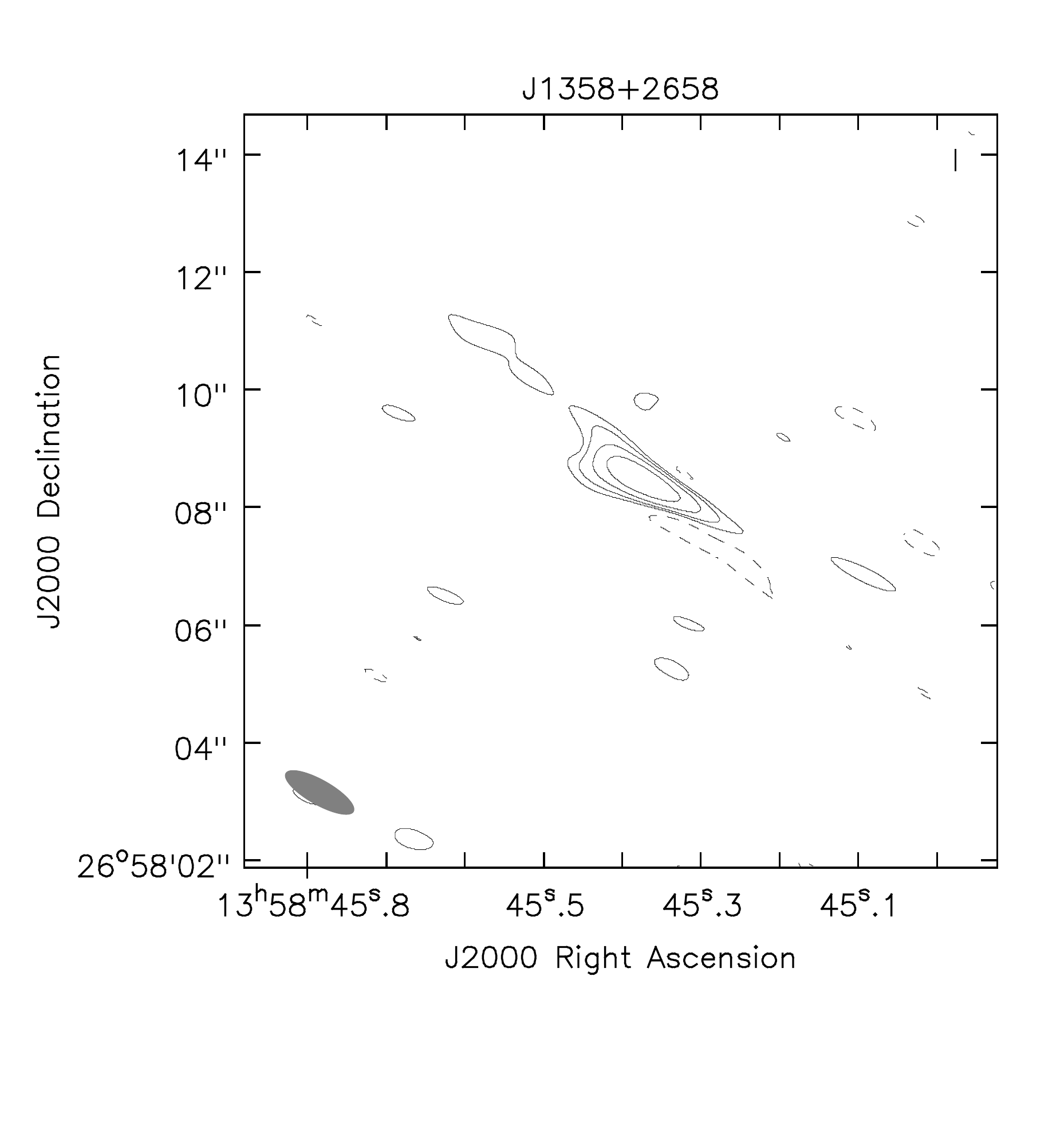} 
\caption{\textbf{Left panel:} J1355+5612, rms = 12 $\mu$Jy, contour levels at -3, 3$\times$2$^n$, n $\in$ [0,5], beam size 2.59$\times$1.27 kpc. \textbf{Right panel:} J1358+2658, rms = 12 $\mu$Jy, contour levels at -3, 3$\times$2$^n$, n $\in$ [0,3], beam size 6.24$\times$1.95 kpc.}
\label{fig:J1355p5612}
\label{fig:J1358p2658}
\end{figure*}

SDSS J135845.38+265808.4 was classified \textit{bona-fide} as F source by \citet{Foschini15}, but given our measurements we classified this source as steep-spectrum. Our map in Fig.~\ref{fig:J1358p2658} reveals an unresolved core, whose size is smaller than 3.6$\times$0.5 kpc, surrounded by a faint diffuse emission that accounts for 14\% of the total flux. This object has a relatively high star formation rate (130 M$_\odot$ yr$^{-1}$) and a q22 = 1.30, suggesting that the contribution of starbursts is strong in radio.
\subsection{J1402+2159} %2 quad, 3.72932, 1.70245
RX J1402.5+2159 is a radio-quiet NLS1. The spectral indexes indicate a steep-spectrum source, although the in-band index of the integrated flux is, within the error bars, consistent with a flat spectrum. The map (Fig.~\ref{fig:J1402p2159}), despite the very elongated beam, reveals a diffuse emission directed at P.A. 65$^\circ$ extended for 4.1 arcsec (5.2 kpc). The diffuse emission accounts for 20\% of the total flux. It is worth noting that within 400 arcsec from the source there are four other neighbors. One of them is a NVSS radio source, J140250+220154, with a flux of 24 mJy. 
\subsection{J1421+2824}% 2 quad, 2.81504, 0.409023
\begin{figure*}
\centering
\includegraphics[trim={0cm 2cm 0cm 0cm},clip,width=7cm]{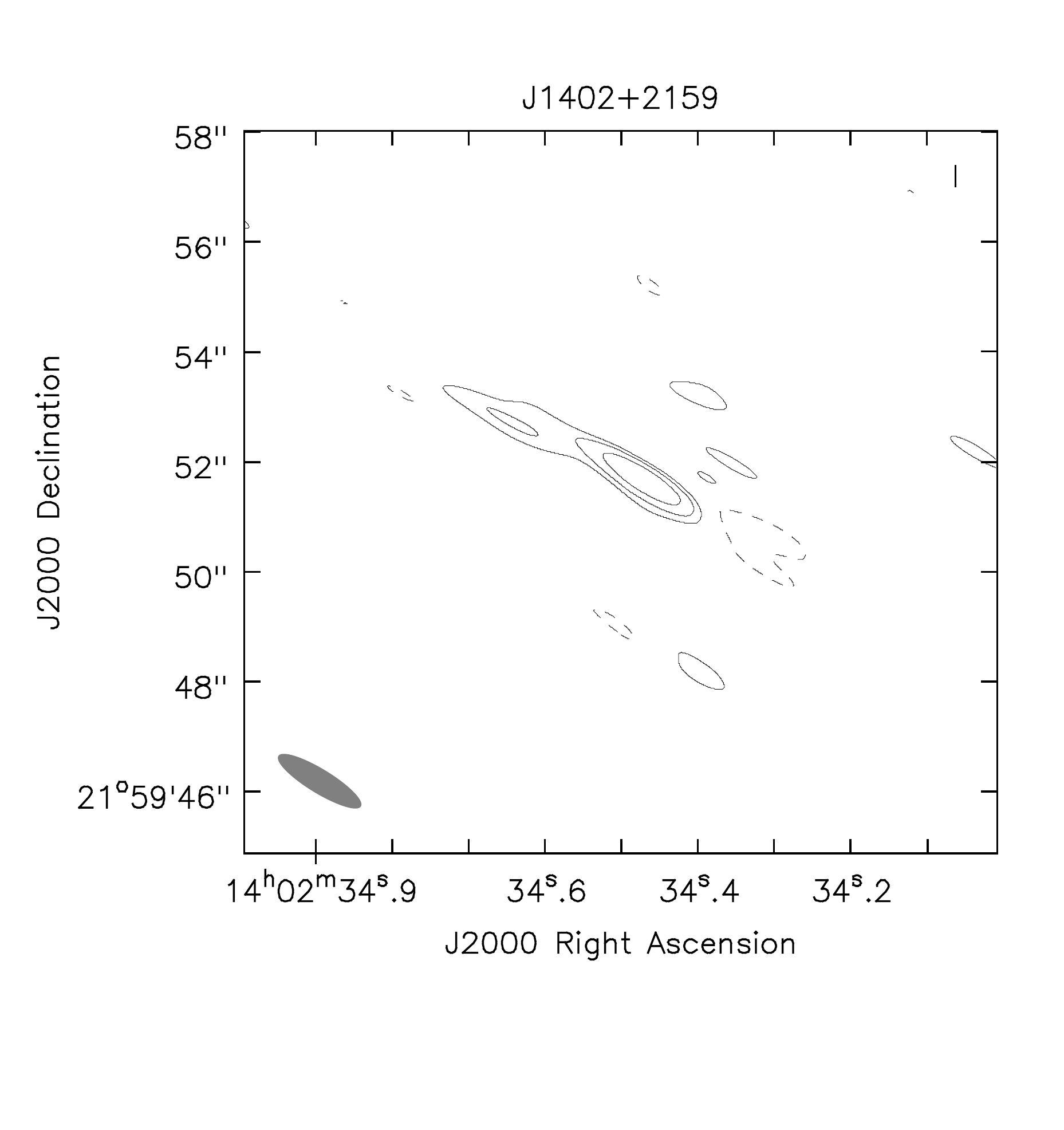} 
\includegraphics[trim={0cm 2cm 0cm 0cm},clip,width=7cm]{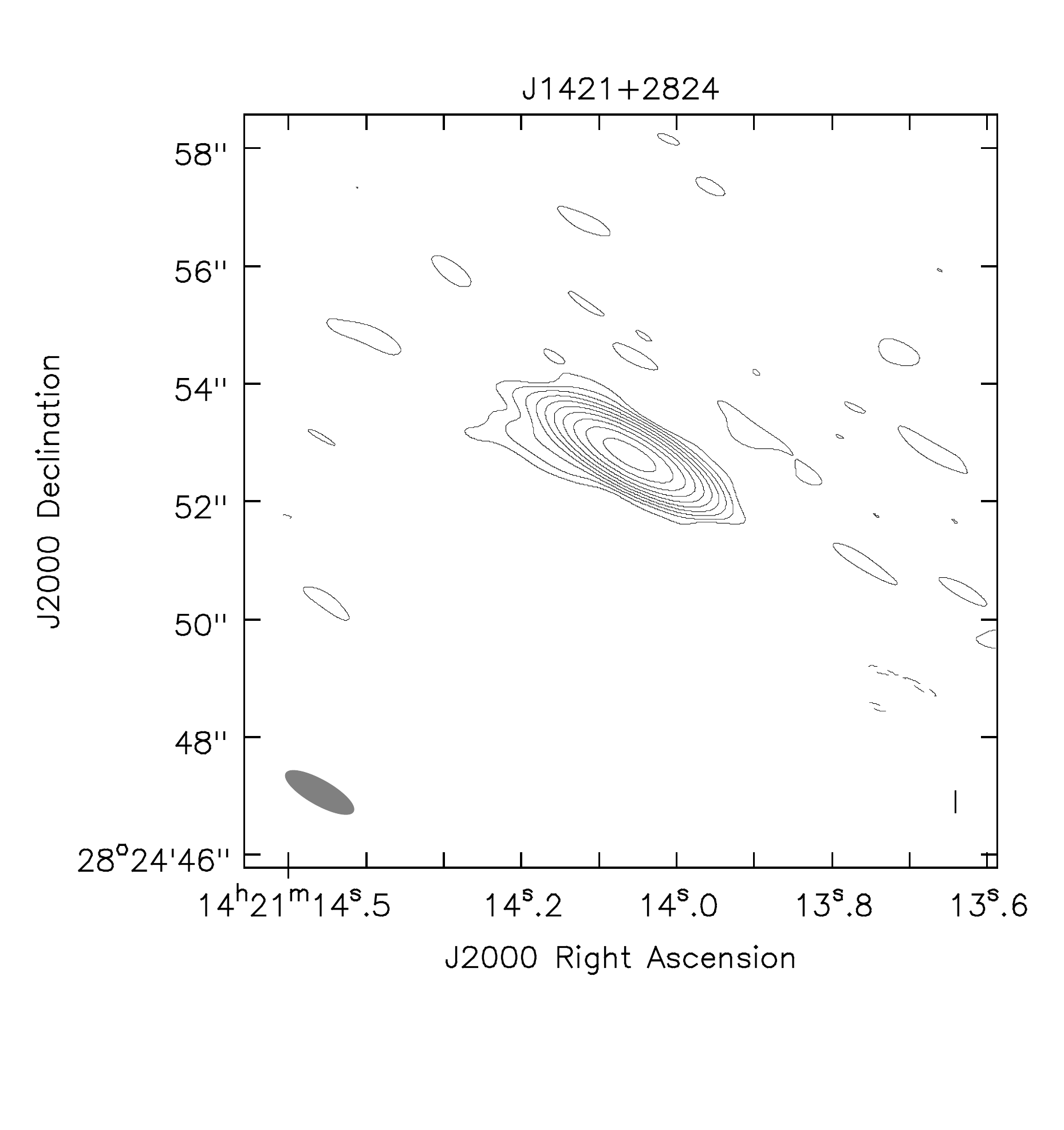} 
\caption{\textbf{Left panel:} J1402+2159, rms = 12 $\mu$Jy, contour levels at -3, 3$\times$2$^n$, n $\in$ [0,2], beam size 2.20$\times$0.53 kpc. \textbf{Right panel:} J1421+2824, rms = 11 $\mu$Jy, contour levels at -3, 3$\times$2$^n$, n $\in$ [0,9], beam size 8.31$\times$2.67 kpc.}
\label{fig:J1402p2159}
\label{fig:J1421p2824}
\end{figure*}

SDSS J142114.05+282452.8 was classified as F source according to its inverted radio spectrum (-0.20$\pm$0.01, \citealp{Foschini15}). Our spectral indexes all confirm the same classification, although none of them is inverted, likely because of the source variability. The map of Fig.~\ref{fig:J1421p2824} shows, outside the core, an elongated structure extended for 2.8 arcsec (18 kpc) at P.A. 82$^\circ$. The deconvolved core size is 1.5$\times$0.4 kpc. VLBA observations of this object revealed a partially resolved core-jet structure, with a flux in very good agreement with our estimate (27 mJy against 26.35), and confirm the flat-spectrum of the source \citep{Gu15}. Finally, this object was detected at 22 GHz by VLBA with a flux of 117 mJy, but not at 37 GHz. 
\subsection{J1443+4725} %3 quad, 0.607519, 0.90256
B3 1441+476 was classified as a S source according to data found in the literature \citep[e.g.][]{Gregory91, Douglas96, Condon98}, which suggested a spectral index between 151 MHz and 5 GHz of $\sim$0.65. Given the compact spectrum, its relatively small size and its luminosity (9.5$\times$10$^{42}$ \ergs), the source was classified as a CSS. However, our in-band measurements indicate that the radio spectrum is flat, a result in good agreement with the detection of $\gamma$-ray emission found in this object \citep{Liao15}. This might suggest that the apparent steep spectrum measured in the past was due to non-simultaneous observations and source variability. In support of the variability hypothesis, the flux at 5 GHz measured in the 87GB catalog \citep{Gregory91} was 59 mJy, while we obtained a flux of $\sim$83 mJy. The different instrument they used (Green Bank telescope) is not enough to explain this different flux. On the VLBA scales, this object has a clear one-sided core-jet morphology \citep{Gu15}, while in our map (Fig.~\ref{fig:J1443p4725}) the source is only partially resolved, with an elongated structure at P.A. 146$^\circ$ extended for 1.1 arcsec (7.8 kpc). The unresolved core has instead a deconvolved size of 730$\times$515 pc. 
\subsection{J1505+0326} %4 quad, 2.30977, 0.733835, secondary 301 muJy 4.7, 236 at 5.7 
\begin{figure*}
\centering
\includegraphics[trim={0cm 2cm 0cm 0cm},clip,width=7cm]{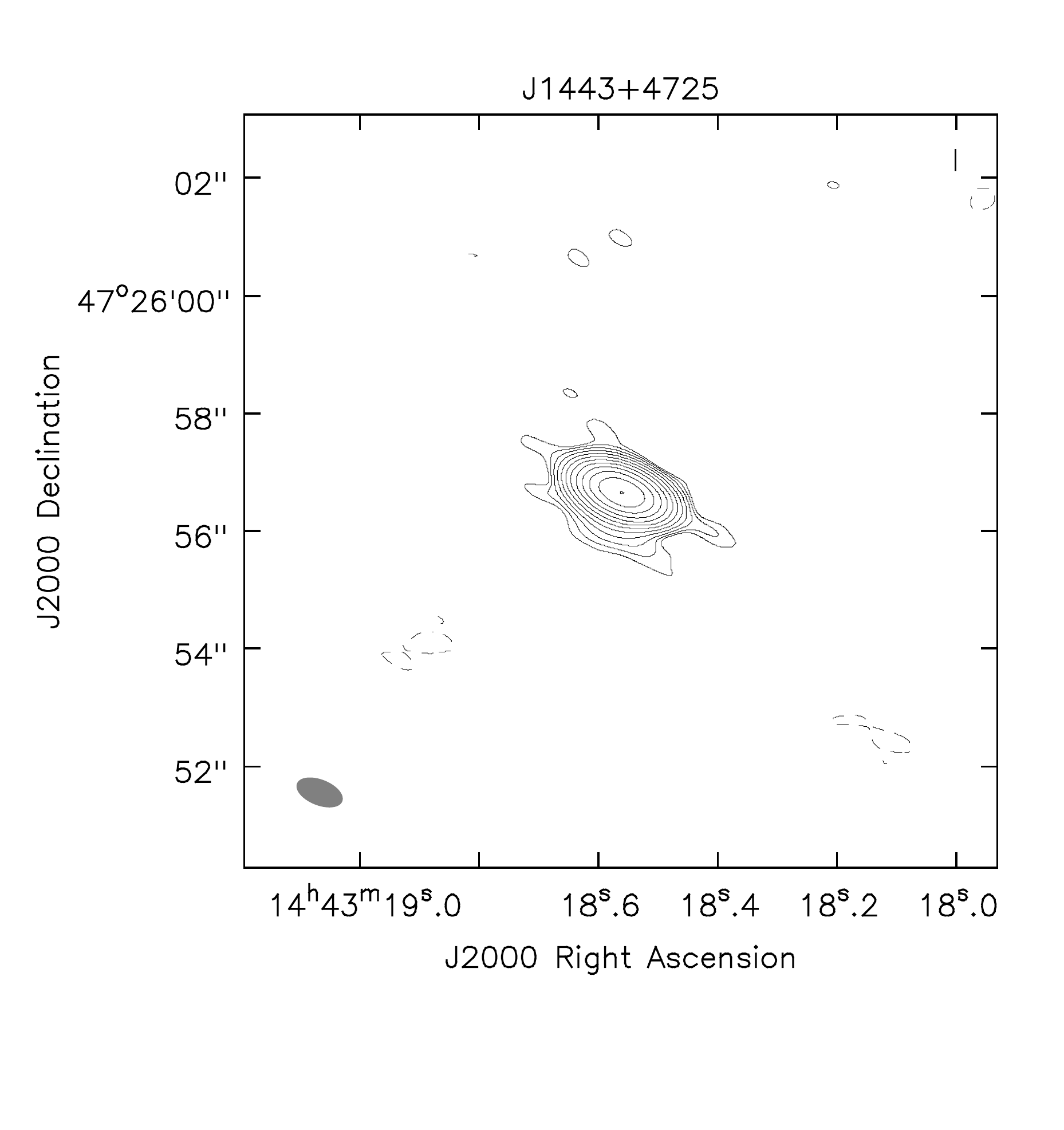} 
\includegraphics[trim={0cm 2cm 0cm 0cm},clip,width=7cm]{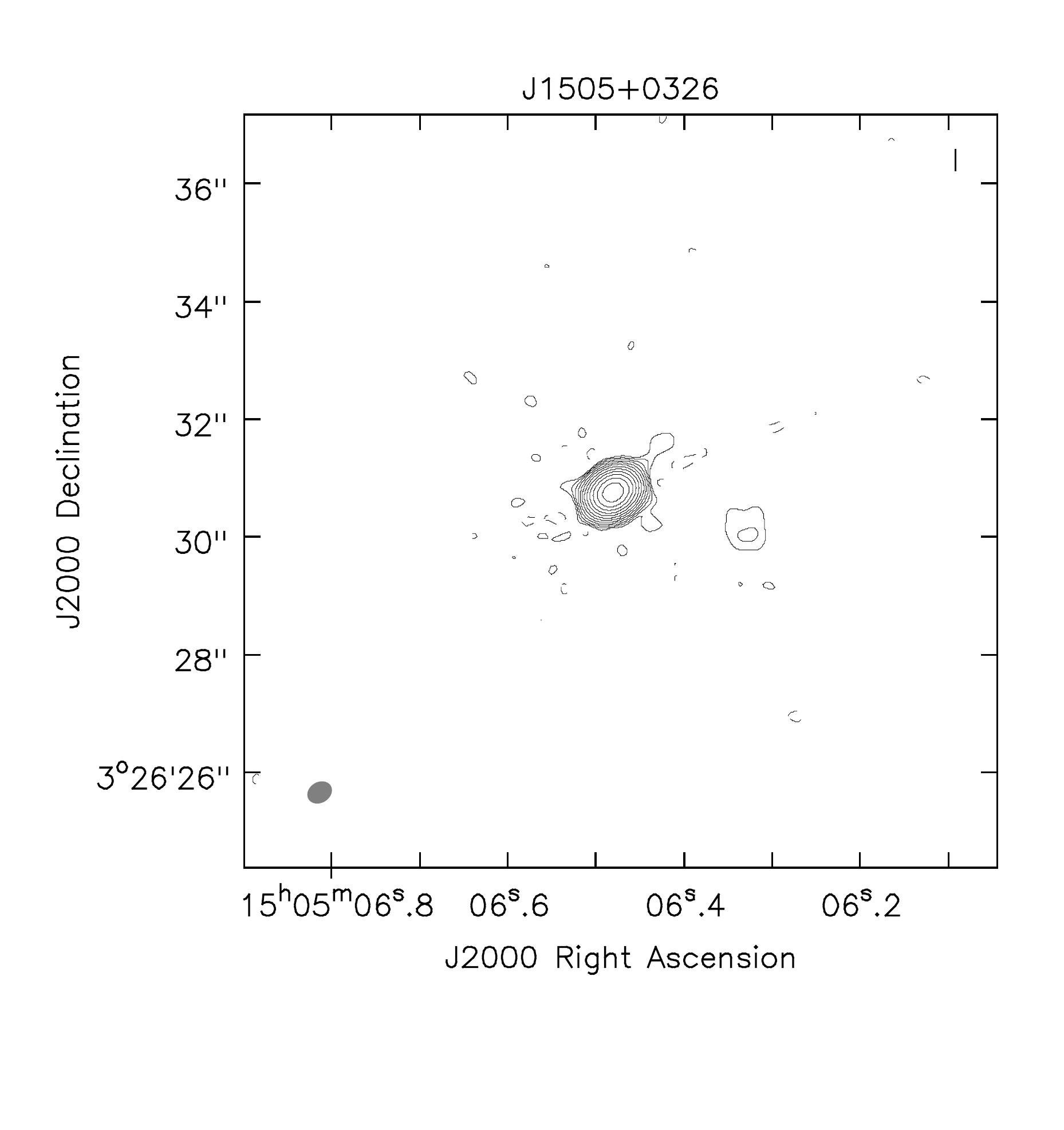} 
\caption{\textbf{Left panel:} J1443+4725, rms = 11 $\mu$Jy, contour levels at -3, 3$\times$2$^n$, n $\in$ [0,11], beam size 5.80$\times$3.01 kpc. \textbf{Right panel:} J1505+0326, rms = 38 $\mu$Jy, contour levels at -3, 3$\times$2$^n$, n $\in$ [0,12], beam size 2.28$\times$1.79 kpc.}
\label{fig:J1443p4725}
\label{fig:J1505p0326}
\end{figure*}

SDSS J150506.47+032630.8 is one of the $\gamma$-ray emitters, associated with 3FGL J1505.1+0326 by the Fermi collaboration \citep{Ackermann15}. All of its spectral indexes are flat, and its luminosity is the second highest of the sample, 1.2$\times$10$^{43}$ \ergs. The map of Fig.~\ref{fig:J1505p0326} shows an unresolved core with a deconvolved size of 230$\times$100 pc and a brightness temperature 4.8$\times$10$^7$ K, the highest in the sample, but also a second source at P.A. 252$^\circ$ and 2.4 arcsec (13.2 kpc) away from the source core. This secondary source has a peak flux of 271 $\mu$Jy and an integrated flux of 621 $\mu$Jy, and a rather steep spectral index in the core, $\sim$1.2. Given the brightness of the main source, the noise level is higher than in other maps, therefore we could not establish whether this secondary source is associated with J1505+0326, or it is a background object unconnected with it. However, it is worth noting that a jet-core structure with the same position angle as the secondary soure was observed at 2.3 GHz with the VLBA \citep{Dallacasa98}. J1505+0326 was detected both at 22 GHz and at 37 GHz. Flaring events are not uncommon in this object \citep[e.g.][]{Dammando16}, as proven by past flux VLA-A measurement at 5 GHz that provided a flux of 926 mJy, twice as much with respect to our present estimate \citep{Dallacasa00}. 
\subsection{J1536+5433}
Mrk 486 is a radio-quiet NLS1, detected only by FIRST. The map of Fig.~\ref{fig:J1536p5433} shows only an unresolved core with a deconvolved size of 270$\times$90 pc. The spectral indexes are all steep, even if the error bars of the in-band indexes do not exclude the possibility of a flat spectrum. The low brightness temperature (759 K), however, seems to be more consistent with a steep spectrum of starburst origin. 
\subsection{J1537+4942}
\begin{figure*}
\centering
\includegraphics[trim={0cm 2cm 0cm 0cm},clip,width=7cm]{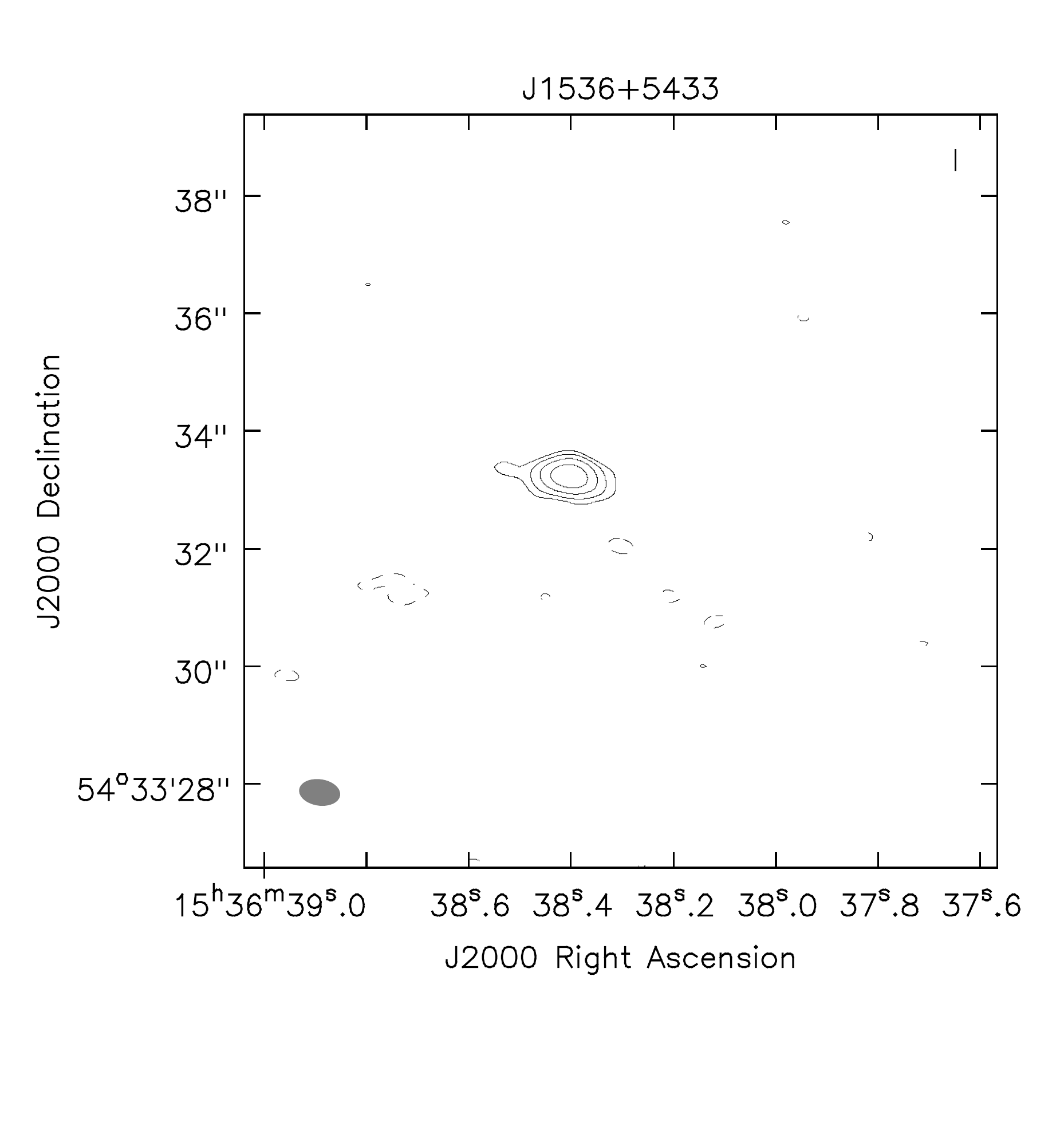} 
\includegraphics[trim={0cm 2cm 0cm 0cm},clip,width=7cm]{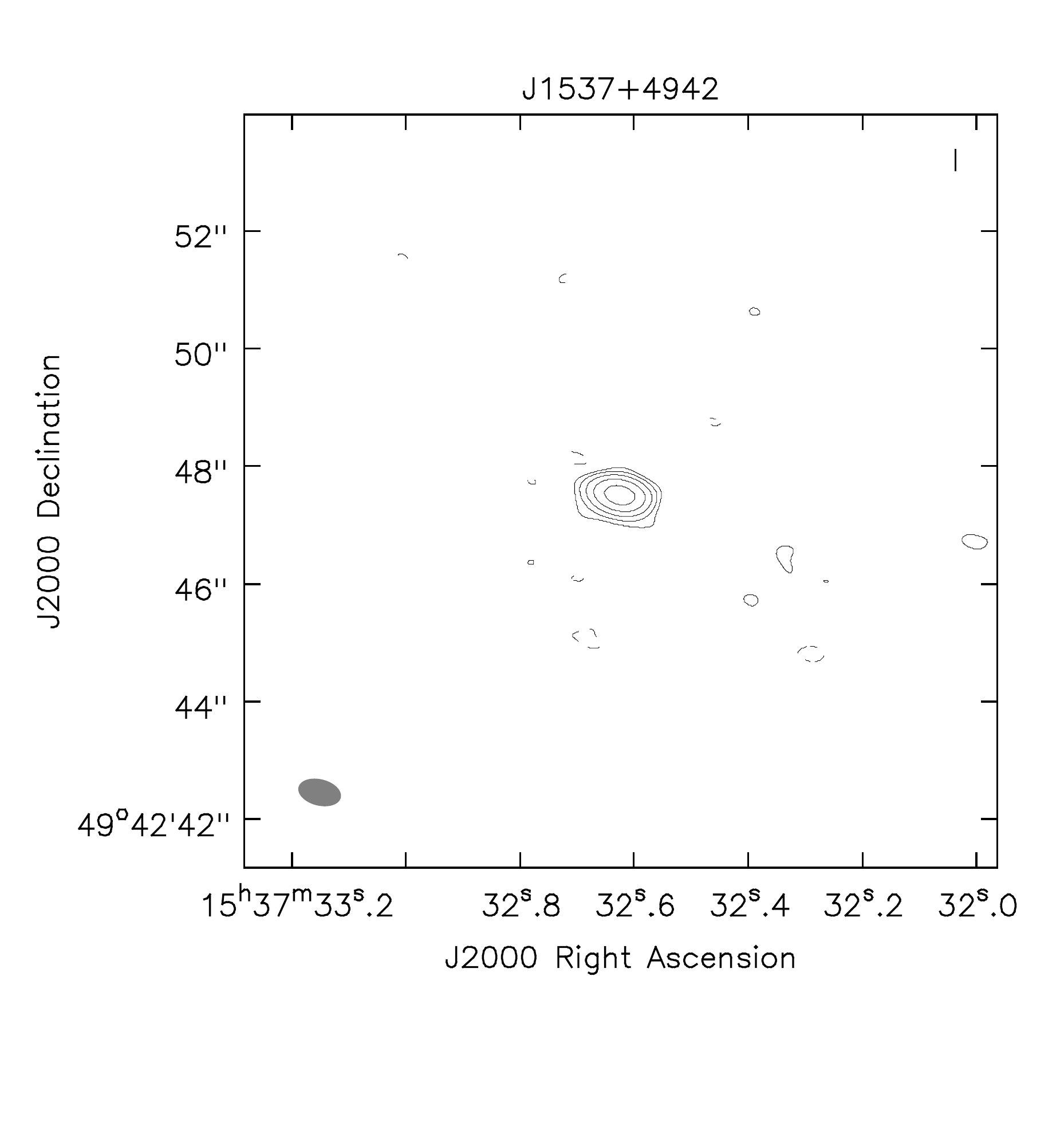} 
\caption{\textbf{Left panel:} J1536+5433, rms = 11 $\mu$Jy, contour levels at -3, 3$\times$2$^n$, n $\in$ [0,3], beam size 0.53$\times$0.33 kpc. \textbf{Right panel:} J1537+4942, rms = 10 $\mu$Jy, contour levels at -3, 3$\times$2$^n$, n $\in$ [0,4], beam size 3.06$\times$1.82 kpc.}
\label{fig:J1536p5433}
\label{fig:J1537p4942}
\end{figure*}

SBS 1536+498 is the most distant radio-quiet NLS1 of the sample (z = 0.280). Fig.~\ref{fig:J1537p4942} shows only a compact source with spectral indexes that are consistent with both a flat- and a steep-spectrum. The broad-band index instead seems to rule out the steep spectrum, since it is lower than 0.5 even considering the error bars. 
\subsection{J1548+3511}
SDSS J154817.92+351128.0 was classified as a flat-spectrum source, and detected below 1.4 GHz. Our in-band indexes confirm the classification, while the broad-band index is steep likely because of variability. Our flux measurement is in agreement with that already carried out by the VLA-A by \citet{Laurent97}, who found 60 mJy against our $\sim$64 mJy. The map of Fig.~\ref{fig:J1548p3511} shows only a compact source, whose unresolved core has a deconvolved size of 760$\times$320 pc. On VLBA scales, instead, this object has a core-jet morphology, extended in the North-West direction and an inverted-spectrum bright core \citep{Gu15}. The source has a flux of 22 mJy at 22 GHz, and it was detected also at 37 GHz with a flux of 340 mJy. Finally, J1548+3511 has a fairly high star formation rate of 151 M$_\odot$ yr$^{-1}$, and a q22 = -0.77. This latter value suggests that the radio is indeed originated in the relativistic jet.

\subsection{J1555+1911}
\begin{figure*}
\centering
\includegraphics[trim={0cm 2cm 0cm 0cm},clip,width=7cm]{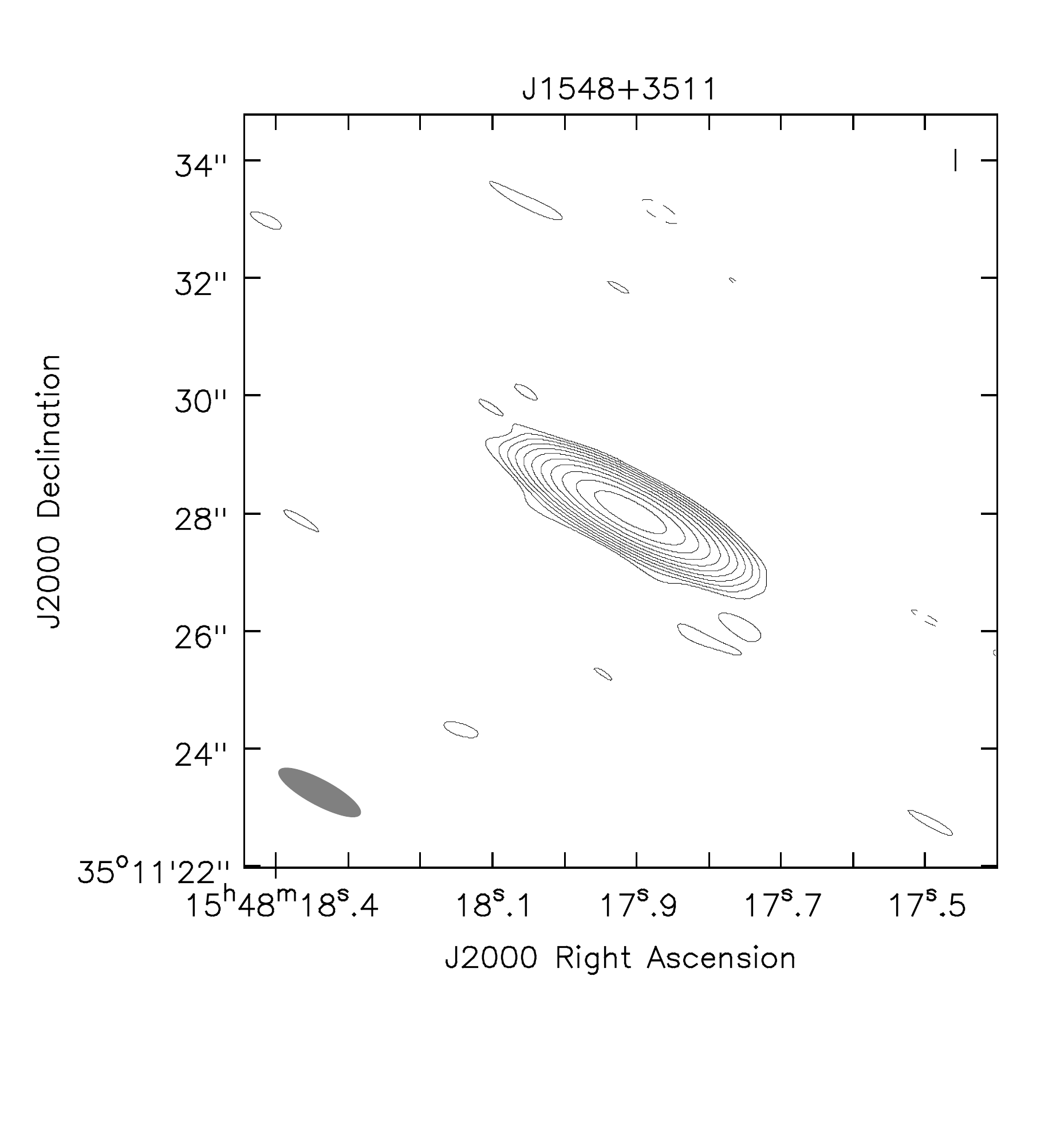} 
\includegraphics[trim={0cm 2cm 0cm 0cm},clip,width=7cm]{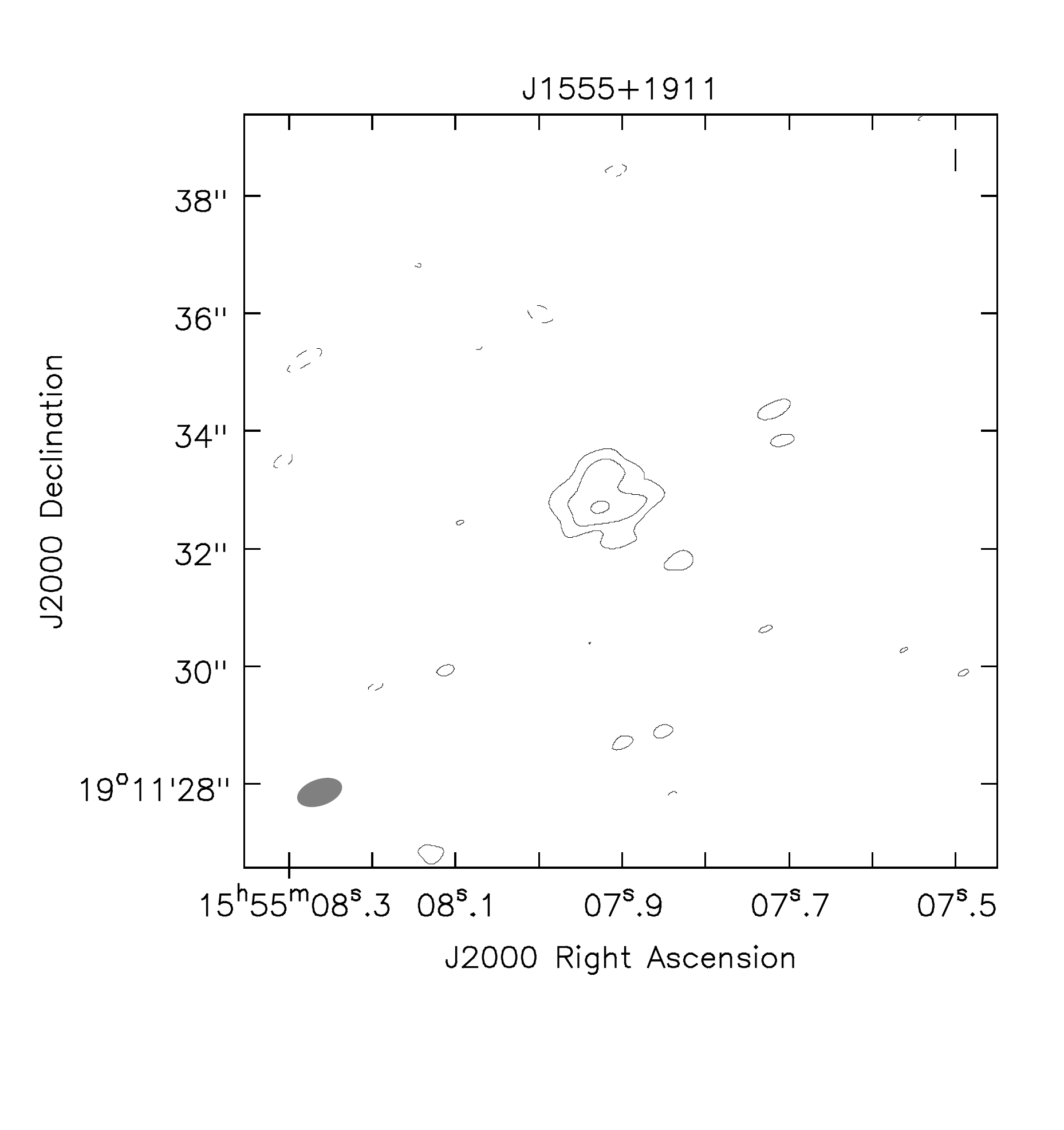} 
\caption{\textbf{Left panel:} J1548+3511, rms = 12 $\mu$Jy, contour levels at -3, 3$\times$2$^n$, n $\in$ [0,10], beam size 9.30$\times$2.56 kpc. \textbf{Right panel:} J1555+1911, rms = 11 $\mu$Jy, contour levels at -3, 3$\times$2$^n$, n $\in$ [0,2], beam size 0.54$\times$0.30 kpc.}
\label{fig:J1548p3511}
\label{fig:J1555p1911}
\end{figure*}

Mrk 291 is a radio-quiet source which in Fig.~\ref{fig:J1555p1911} shows only a spreaded-out emission extended for $\sim$2 arcsec (1.4 kpc) with a steep spectrum and a very low brightness temperature of $\sim$8 K. These properties might indicate that its radio emission has a thermal origin. 

\subsection{J1559+3501} 
Mrk 493 is a radio-quiet NLS1 which shows steep spectral indexes and a low brightness temperature (63 K). The map (Fig.~\ref{fig:J1559p3501}) shows a diffuse emission surrounding in every direction the unresolved core, which has a deconvolved size of 610$\times$335 pc. 

\subsection{J1612+4219}
\begin{figure*}
\centering
\includegraphics[trim={0cm 2cm 0cm 0cm},clip,width=7cm]{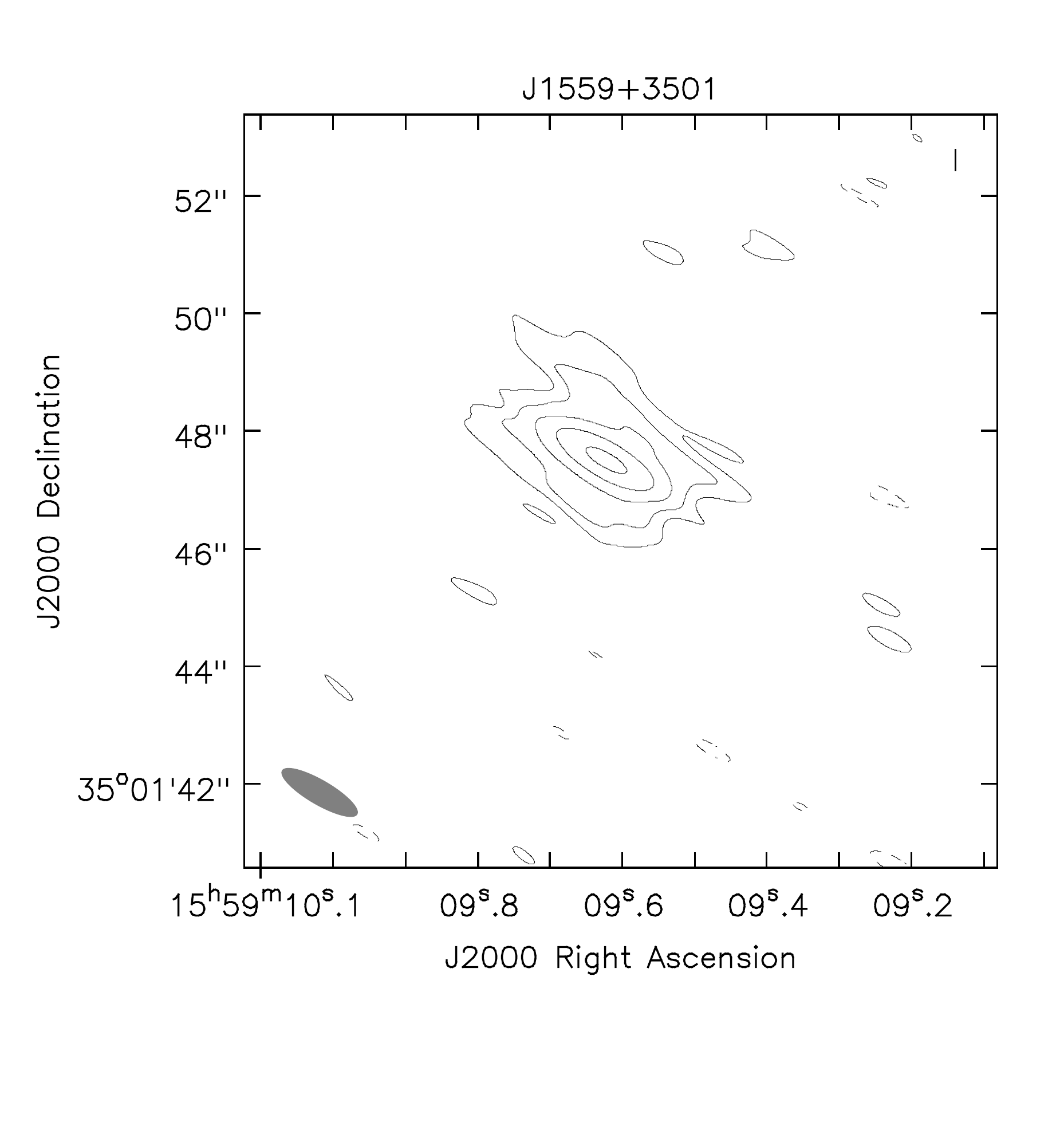} 
\includegraphics[trim={0cm 2cm 0cm 0cm},clip,width=7cm]{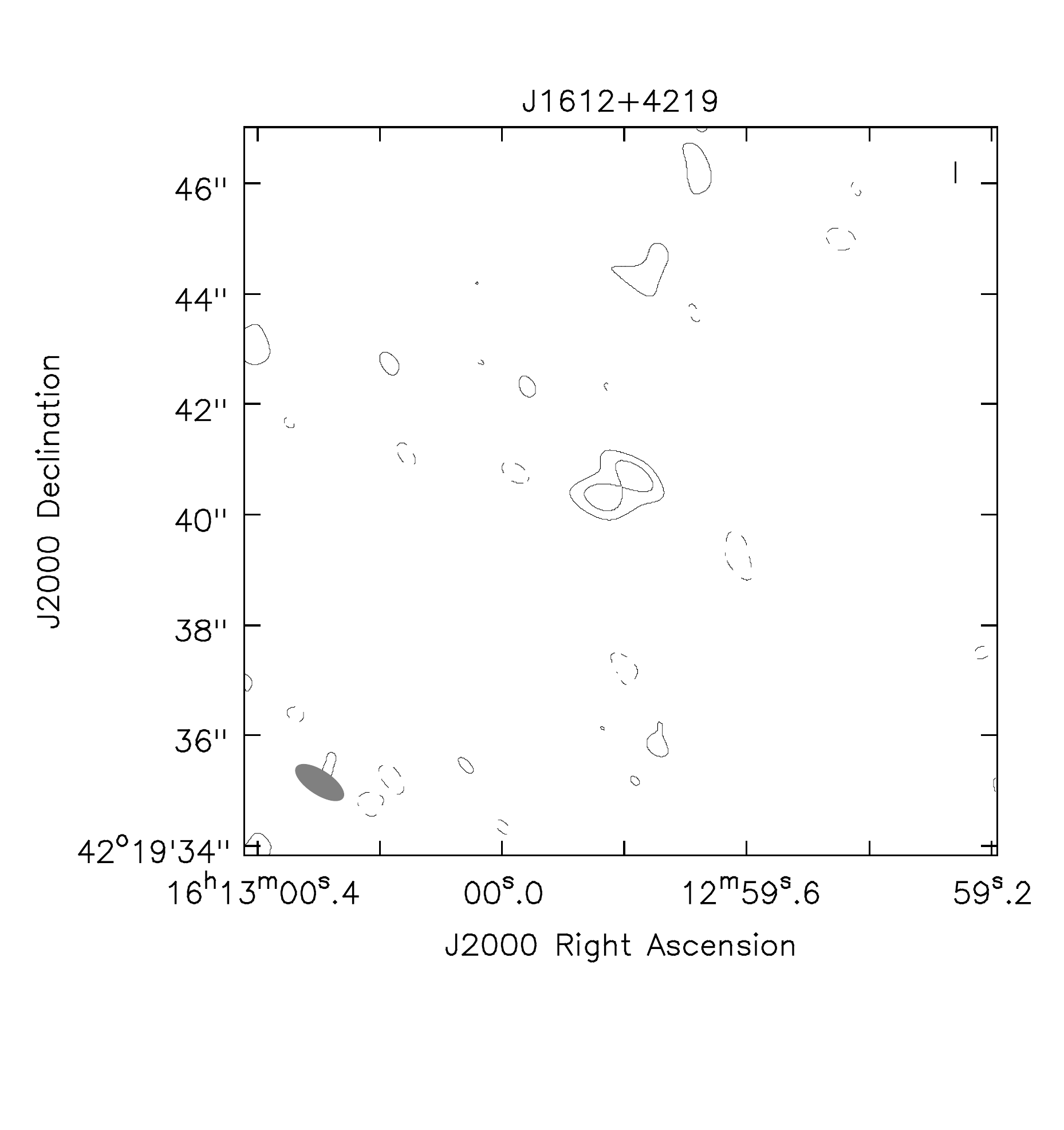} 
\caption{\textbf{Left panel:} J1559+3501, rms = 11 $\mu$Jy, contour levels at -3, 3$\times$2$^n$, n $\in$ [0,4], beam size 0.91$\times$0.25 kpc. \textbf{Right panel:} J1612+4219, rms = 39 $\mu$Jy, contour levels at -3, 3$\times$2$^n$, n $\in$ [0,2], beam size 3.71$\times$1.56 kpc.}
\label{fig:J1559p3501}
\label{fig:J1612p4219}
\end{figure*}

SDSS J161259.83+421940.3 is one of the \textit{bona-fide} NLS1 classified as F. Our spectral indexes seem to be in agreement with this classification, but the presence of a strong nearby source in the visibilities hampered our analysis, introducing large errors in our measurements. Indeed, in the map of Fig.~\ref{fig:J1612p4219}, the source morphology cannot be clearly understood, even if it might be somewhat extended. This object was detected at 37 GHz with a flux of 460 mJy. The star formation rate of this object was estimated as 224 M$_\odot$ yr$^{-1}$, with an associated q22 = 1.32. Starburst therefore might strongly contribute to the radio emission of this object. 

\subsection{J1629+4007} %3 quad, 2.59868, 0.11895
\begin{figure*}
\centering
\includegraphics[trim={0cm 2cm 0cm 0cm},clip,width=7cm]{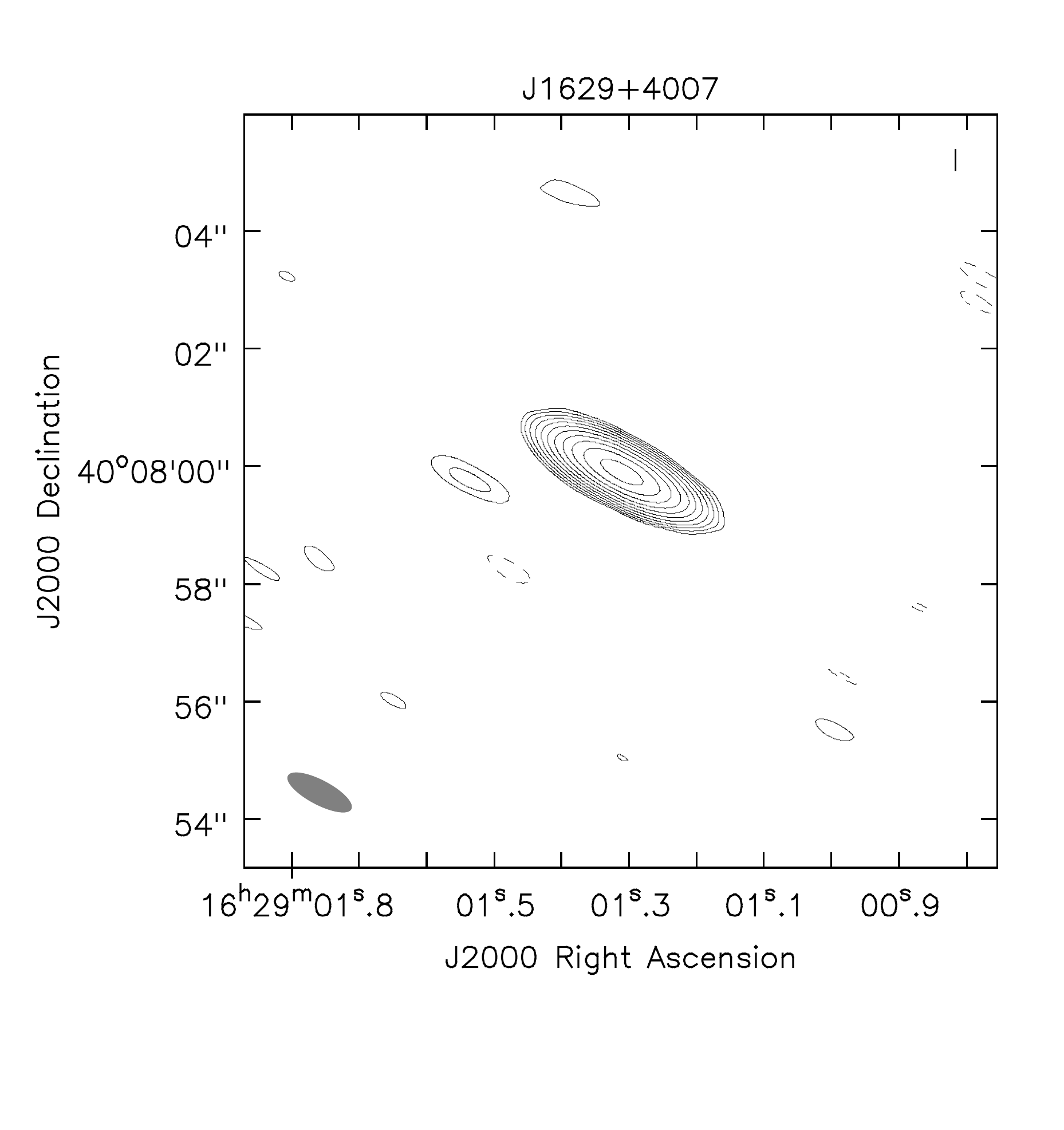} 
\includegraphics[trim={0cm 2cm 0cm 0cm},clip,width=7cm]{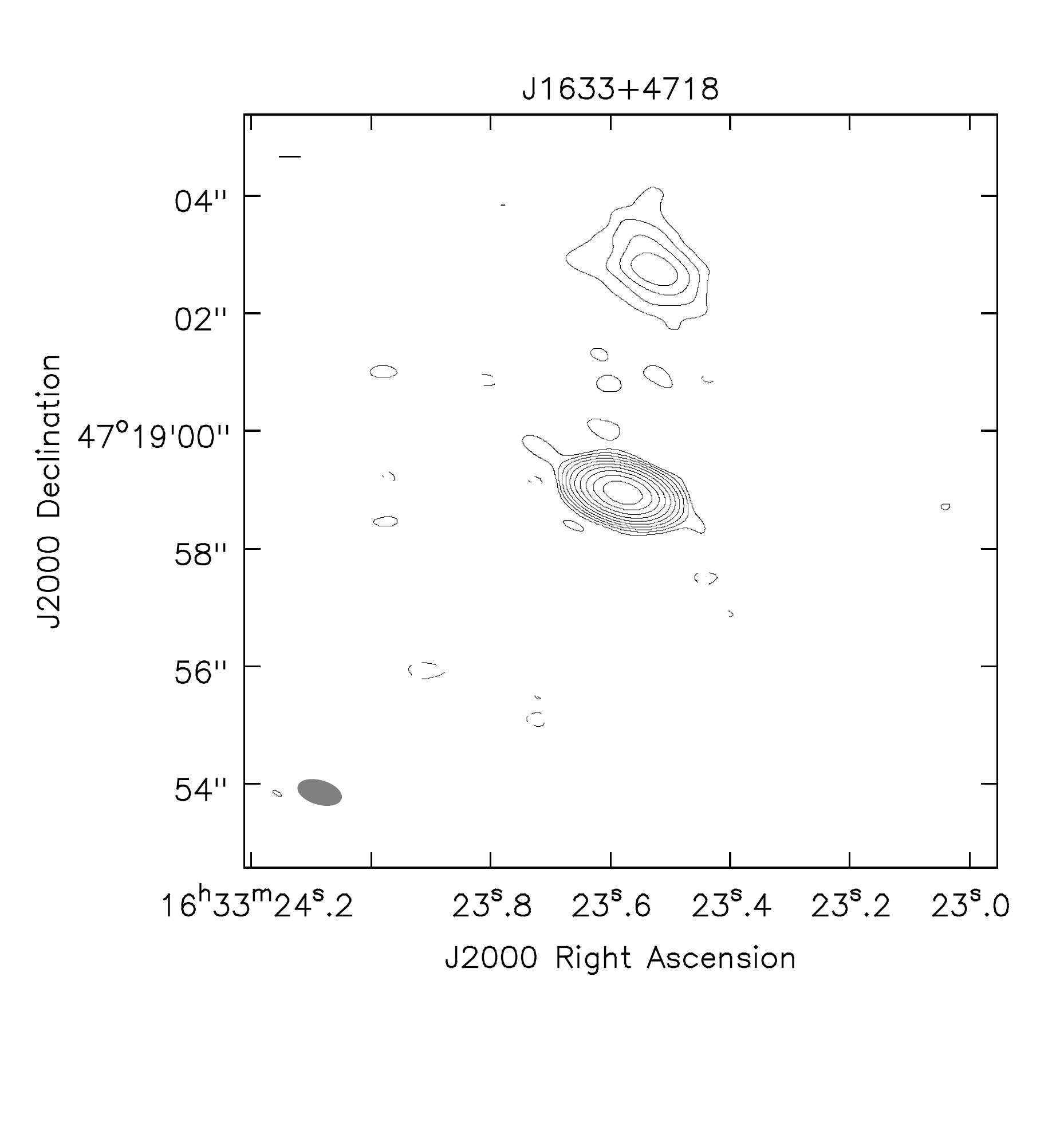} 
\caption{\textbf{Left panel:} J1629+4007, rms = 10 $\mu$Jy, contour levels at -3, 3$\times$2$^n$, n $\in$ [0,10], beam size 4.99$\times$1.66 kpc. \textbf{Right panel:} J1633+4718, rms = 9 $\mu$Jy, contour levels at -3, 3$\times$2$^n$, n $\in$ [0,9], beam size 1.60$\times$0.84 kpc.}
\label{fig:J1629p4007}
\label{fig:J1633p4718}
\end{figure*}

SDSS J162901.30+400759.9 was classified as F source, having a strongly inverted spectrum between 1.4 and 5 GHz (-0.68$\pm$0.02, \citealp{Foschini15}). Our spectral indexes confirmed this classification. The map of this source in Fig.~\ref{fig:J1629p4007} shows a compact structure with an unresolved core of 260$\times$70 pc, and a secondary, faint source located 2.6 arcsec (10.8 kpc) away from the core at P.A. 93$^\circ$. This second source has a flux density peak of 78 mJy beam$^{-1}$, and it is not clear whether it is related with J1629+4007 or not. The main source was detected both at 22 GHz with VLBI with 145 mJy, and at 37 GHz with 350 mJy. According to \citet{Caccianiga15}, J1629+4007 has a star formation rate of 25 M$_\odot$ yr$^{-1}$ and a q22 index of 0.29, suggesting that the bulk of the radio emission is of non-thermal origin. 

\subsection{J1633+4718} %1 quad, 0.547052, 3.78538
SDSS J163323.58+471858.9 was classified as F source, and our in-band indexes are in agreement with such classification. The broad-band index is instead steep, but variability is definitely present, since while our flux estimate is 26 mJy, \citet{Laurent97} report instead 30 mJy at 5 GHz. The map shows an unresolved core with a deconvolved size of 190$\times$120 pc, and a second fainter source located at P.A. 352$^\circ$ 3.8 arcsec (8 kpc) away from the core. Diffuse emission is present between these two bright spots, suggesting that they belong to the same structure, and indicating a core-jet morphology. The second structure has an integrated flux of 780 $\mu$Jy, and an in-band spectral index (integrated) of 0.4. This secondary component accounts for the 6\% of the total flux, and it was not seen with the VLBA either by \citet{Doi11} at 1.7 GHz or by \citet{Gu10} at 5 GHz, likely because it was resolved out. The star formation rate of this object is 68 M$_\odot$ yr$^{-1}$, with q22 = 0.23. These parameters seem to indicate that, although strong, starburst are not decisive in the radio emission. At 22 GHz the source was detected showing a flux of 163 mJy, while it was not detected at 37 GHz.  

\subsection{J1634+4809}
SDSS J163401.94+480940.2 is one of the \textit{bona-fide} F sources, and we can confirm this classification. In the VLBA image the source shows only an unresolved core \citep{Gu15}, and the same is true for our map (Fig.~\ref{fig:J1634p4809}), in which only an unresolved core of 900$\times$340 pc can be seen. However, there might be an elongated structure in the West direction, at P.A.$\sim$270$^\circ$. The star formation rate is 117 M$_\odot$ yr$^{-1}$, and q22 = 0.15 \citep{Caccianiga15}, suggesting that the radio might be dominated by the non-thermal emission of the relativistic jet.

\subsection{J1703+4540} %3 quad, 0.180451, 1.08872
\begin{figure*}
\centering
\includegraphics[trim={0cm 2cm 0cm 0cm},clip,width=7cm]{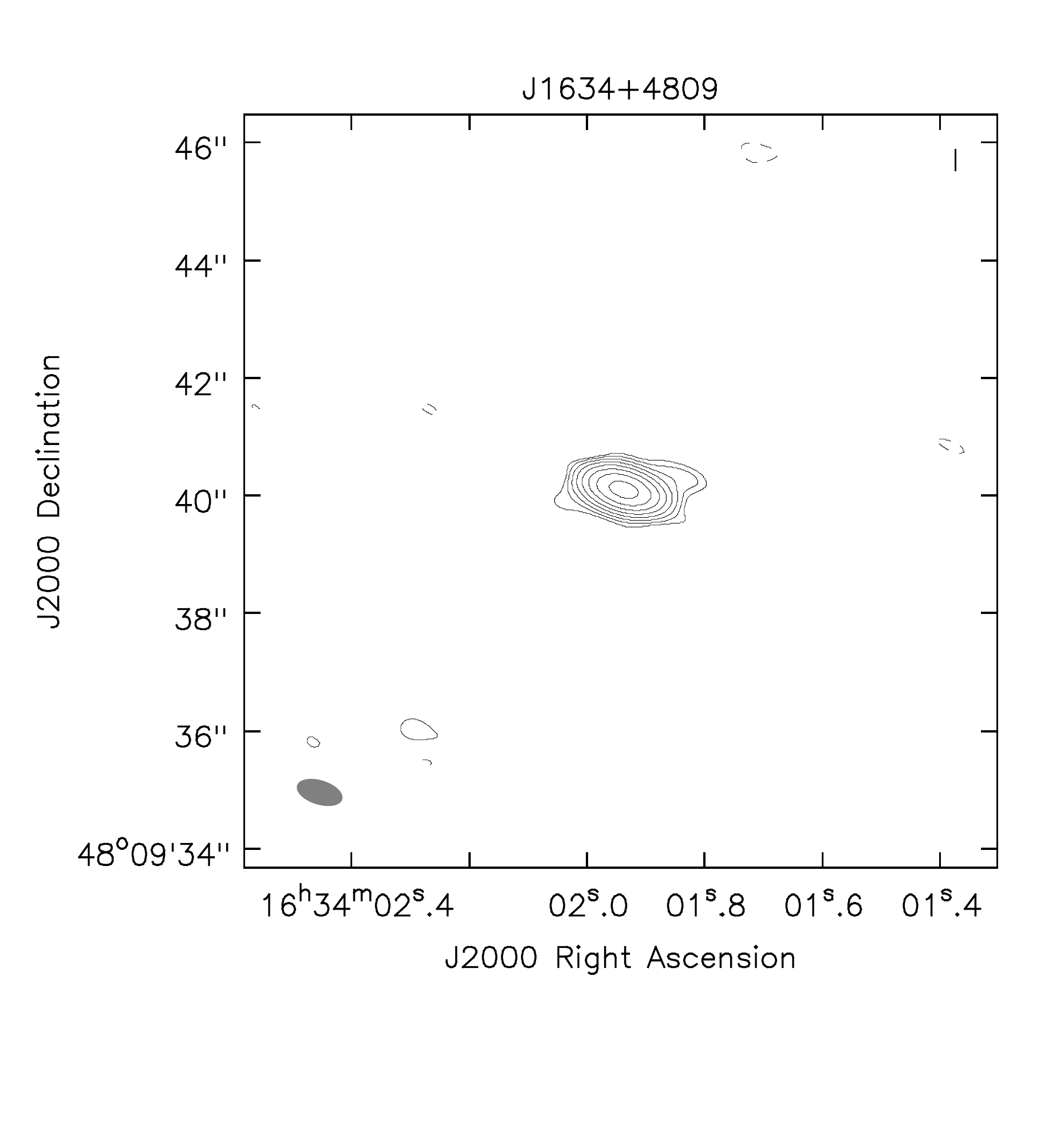} 
\includegraphics[trim={0cm 2cm 0cm 0cm},clip,width=7cm]{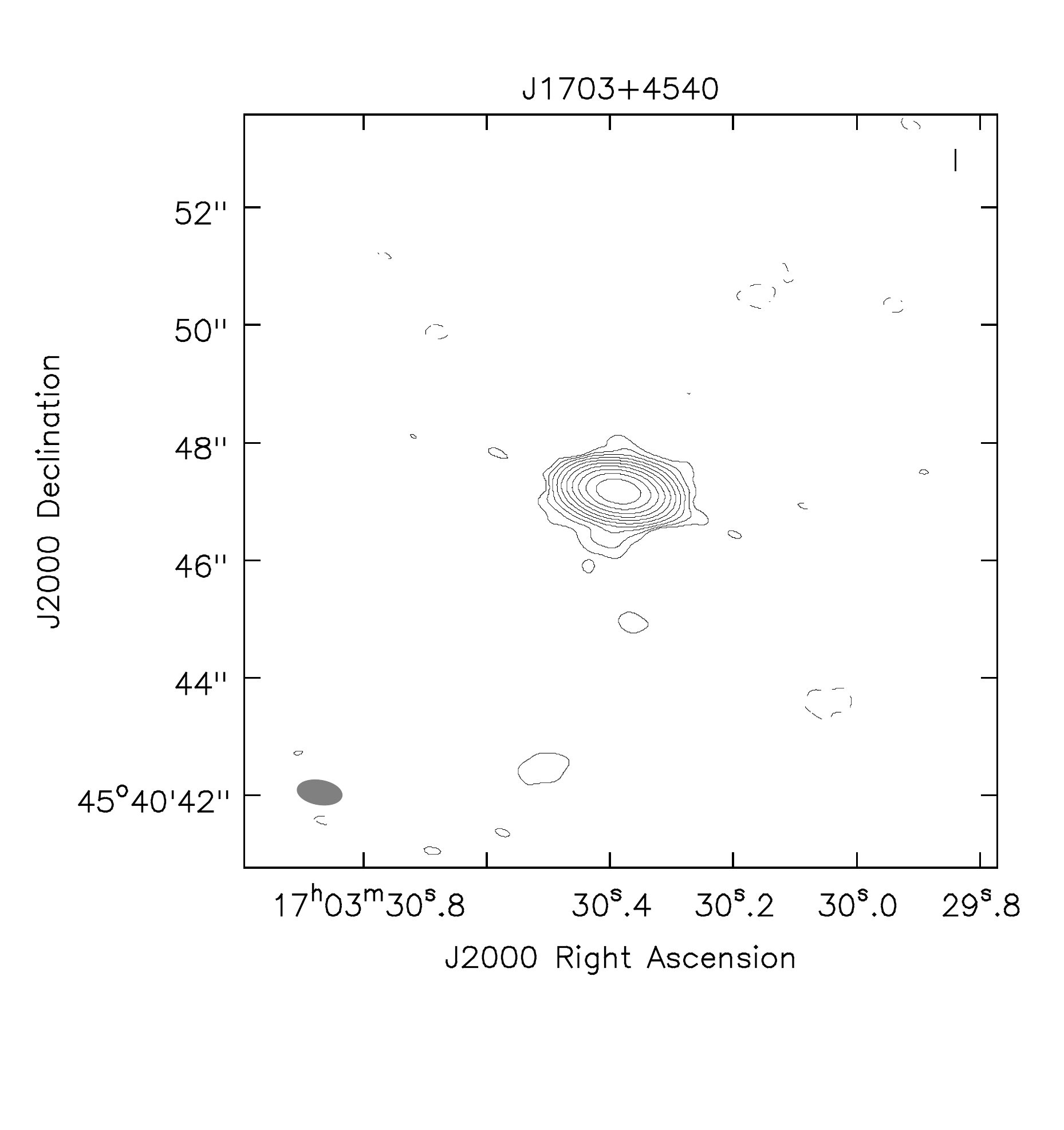} 
\caption{\textbf{Left panel:} J1634+4809, rms = 10 $\mu$Jy, contour levels at -3, 3$\times$2$^n$, n $\in$ [0,7], beam size 4.80$\times$2.43 kpc. \textbf{Right panel:} J1703+4540, rms = 11 $\mu$Jy, contour levels at -3, 3$\times$2$^n$, n $\in$ [0,9], beam size 0.89$\times$0.48 kpc.}
\label{fig:J1634p4809}
\label{fig:J1703p4540}
\end{figure*}

SDSS J170330.38+454047.1 is a S source, which was classified as a CSS with a turnover frequency below 150 MHz \citep{Snellen04}. Our in-band spectral indexes instead are flat, and as already pointed out by \citet{Gu10} the previously measured steep indexes might be affected by variability. Indeed, our broad-band index is also steep. The map of Fig.~\ref{fig:J1703p4540} reveals that the source is slightly extended toward South, at P.A. 170$^\circ$, while the unresolved core has a deconvolved size of 265$\times$120 pc. VLBA observations at 1.7 GHz revealed that the source has a core-jet structure, with the one-sided relativistic jet directed to South-West for 35 pc \citep{Doi11}. The source has not been detected at 37 GHz. 

\subsection{J1709+2348}
\begin{figure*}
\centering
\includegraphics[trim={0cm 2cm 0cm 0cm},clip,width=7cm]{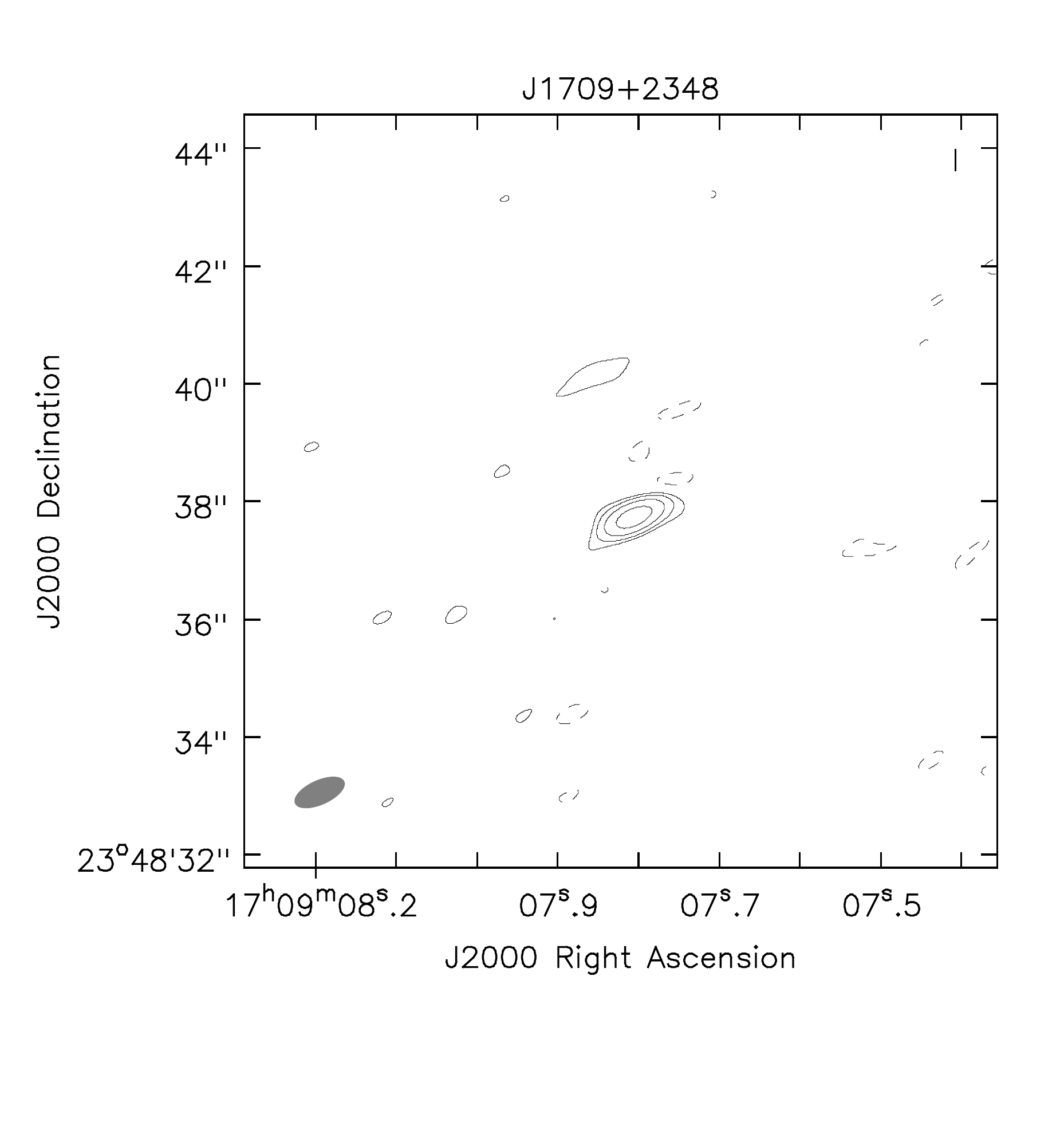} 
\includegraphics[trim={0cm 2cm 0cm 0cm},clip,width=7cm]{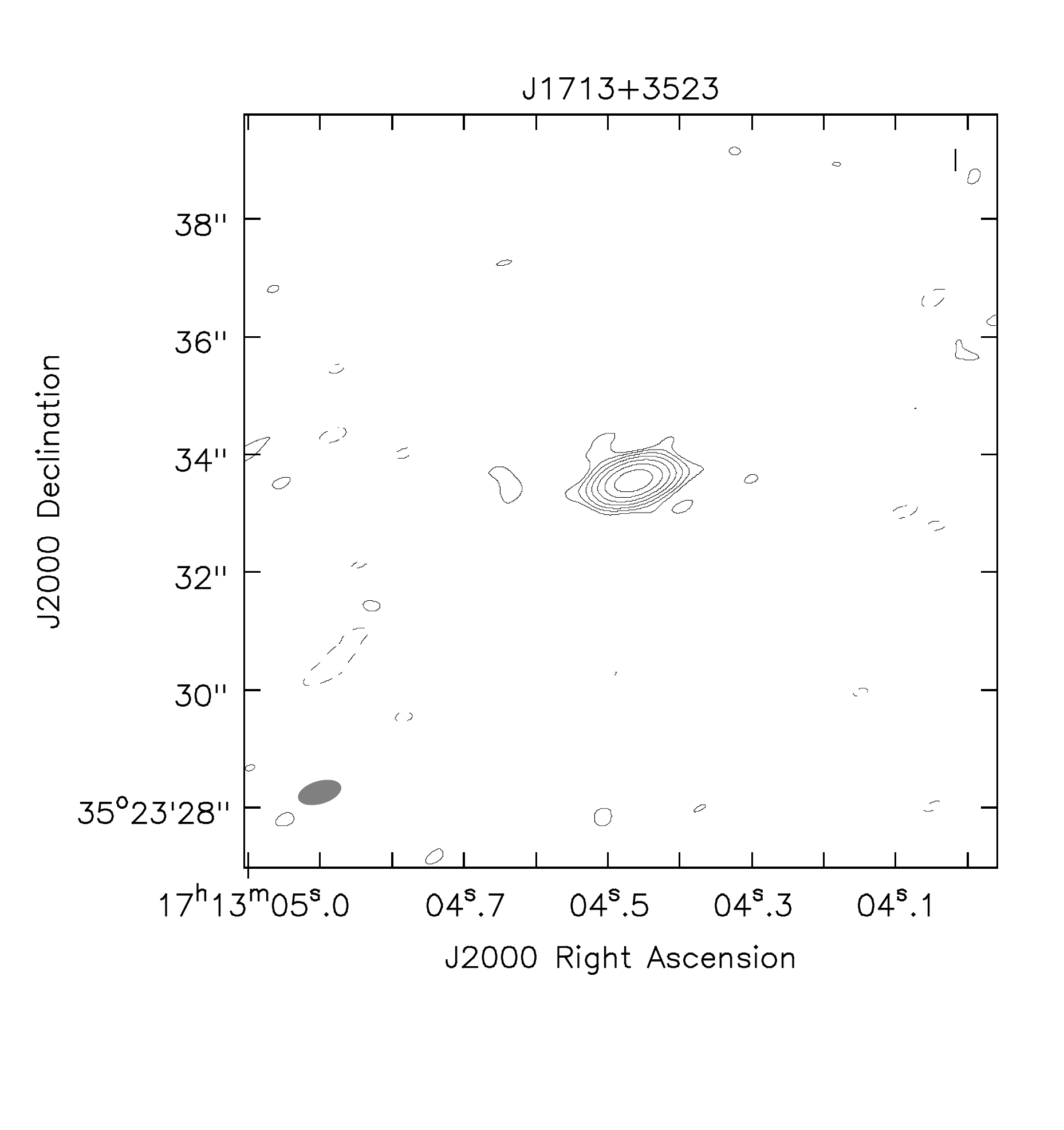} 
\caption{\textbf{Left panel:} J1709+2348, rms = 11 $\mu$Jy, contour levels at -3, 3$\times$2$^n$, n $\in$ [0,3], beam size 3.56$\times$1.58 kpc. \textbf{Right panel:} J1713+3523, rms = 11 $\mu$Jy, contour levels at -3, 3$\times$2$^n$, n $\in$ [0,6], beam size 1.16$\times$0.58 kpc.}
\label{fig:J1709p2348}
\label{fig:J1713p3523}
\end{figure*}

SDSS J170907.80+234837.7 was classified \textit{bona-fide} as a flat-spectrum source, but our spectral indexes do not agree with this hypothesis, suggesting instead a steep-spectrum. The map of Fig.~\ref{fig:J1709p2348} shows that the source is compact. The brightness temperature is rather low, $\sim$545 K. In this source the star formation rate is 23 M$_\odot$ yr$^{-1}$, with a q22 = 0.96. This might indicate that the radio emission of J1709+2348 might not be due only to the non-thermal emission of a jet, but also to the star formation, in agreement with the rather steep spectral indexes. Finally, spectral index, morphology, and luminosity are consistent with those of a low-luminosity CSS.

\subsection{J1713+3523}
FBQS J1713+3523 was classified as steep-spectrum below 1.4 GHz. This result is confirmed by all of our measurements. The map shows a compact structure with a deconvolved core size of 235$\times$165 pc. The brightness temperature is approximately 1.7$\times$10$^4$ K, a value not different from those observed in many of our radio-loud objects. It is worth noting that the source was detected with VLBI at 22 GHz, and showed a flux of 138 mJy, while it was not detected at 37 GHz. This huge flare at VLBI scales seems to suggest that the core is highly non-thermal. 

\subsection{J2242+2943}
Ark 564 is one of the most studied NLS1s, and in radio it always showed, at all scales, an elongated emission directed toward North, at P.A.$\sim$0$^\circ$, associated with a mildly relativistic outflow \citep[e.g.][]{Moran00, Gupta13}. Our map in Fig.~\ref{fig:J2242p2943} shows the same structure along with an unresolved core of 330$\times$110 pc. All the spectral indexes are rather steep, and along wiht the fairly low brightness temperature (2750 K), they suggest that star formation might contribute significantly to the radio emission. 

\subsection{J2314+2243}
\begin{figure*}
\centering
\includegraphics[trim={0cm 2cm 0cm 0cm},clip,width=7cm]{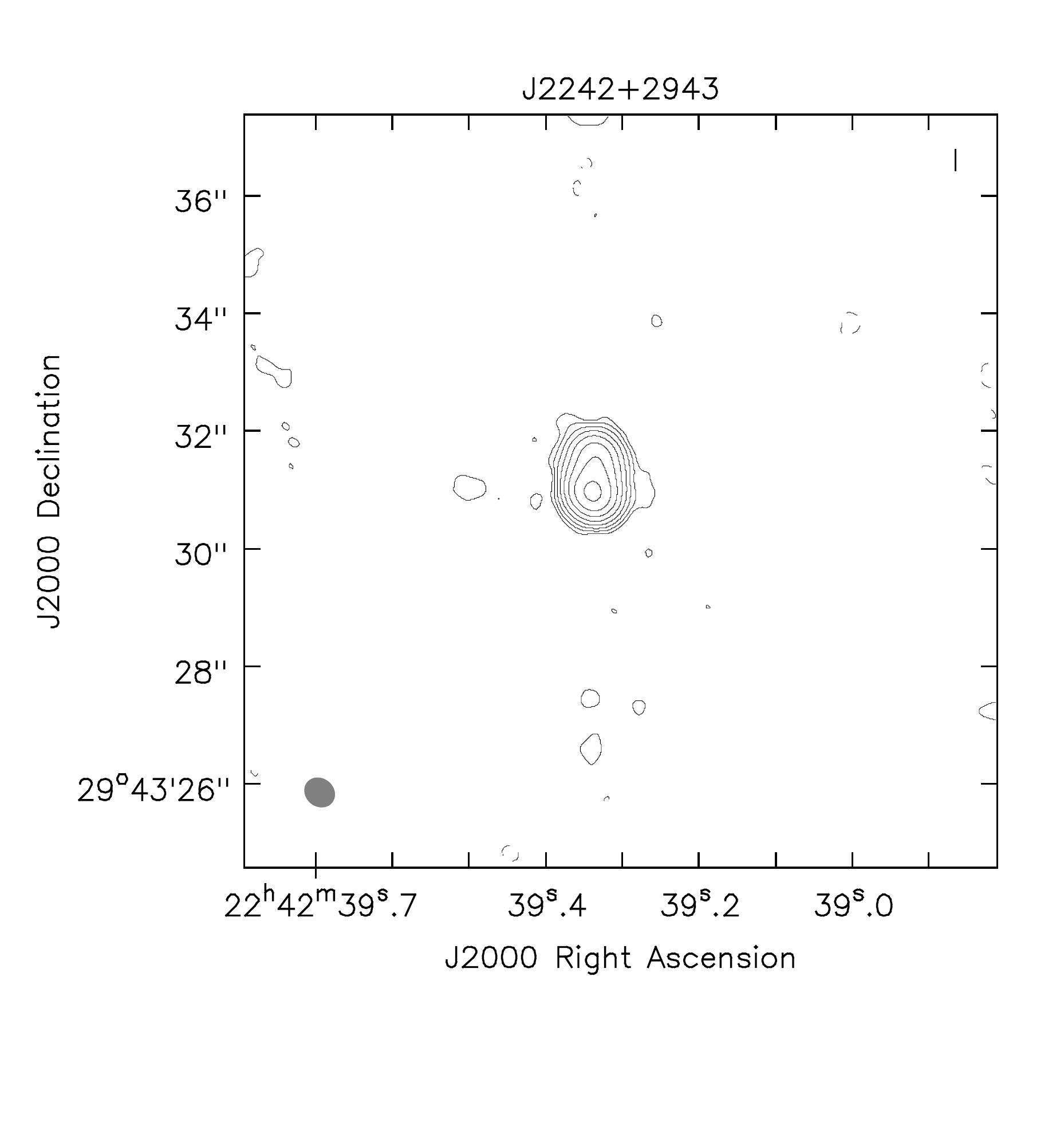} 
\includegraphics[trim={0cm 2cm 0cm 0cm},clip,width=7cm]{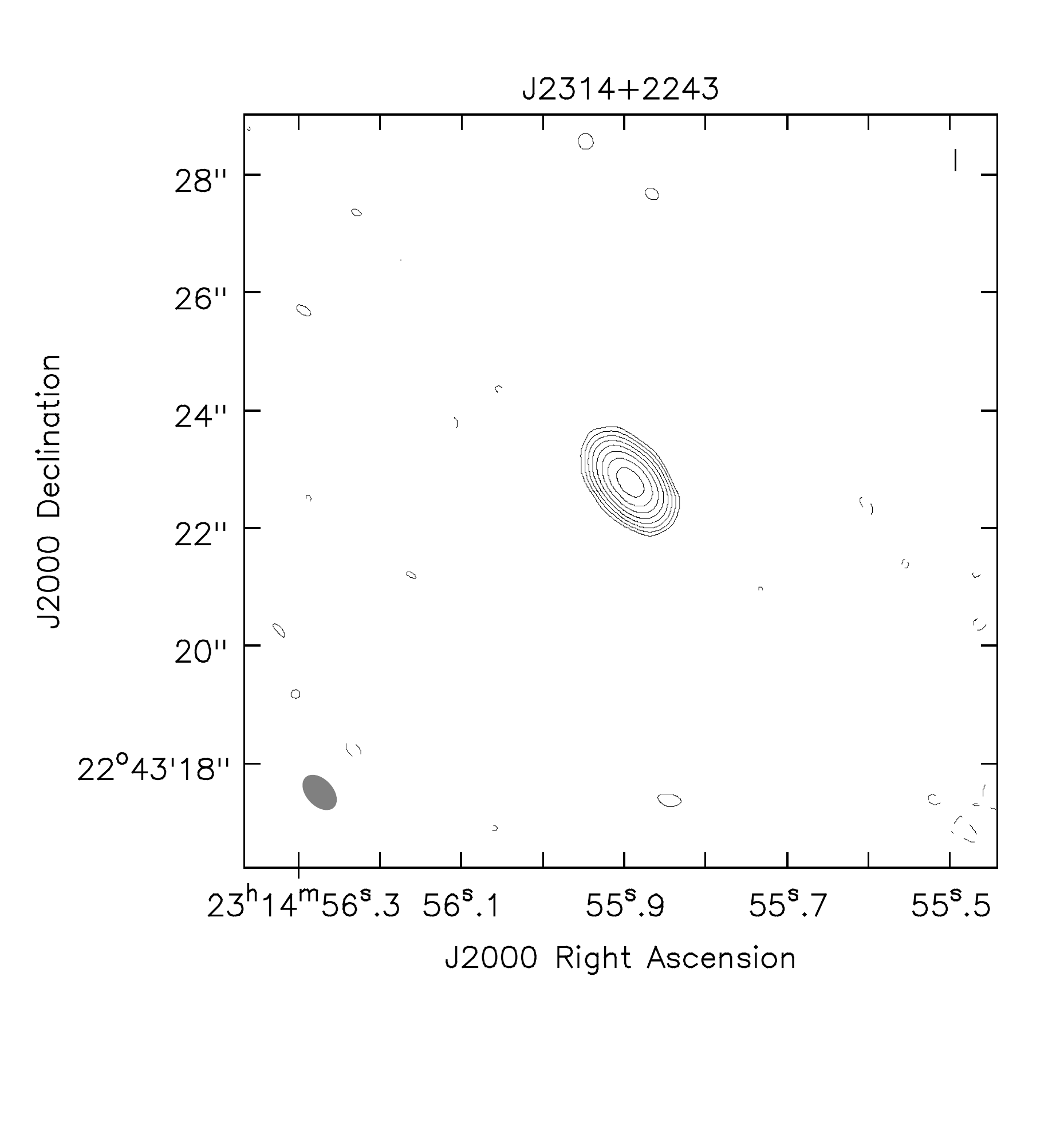} 
\caption{\textbf{Left panel:} J2242+2943, rms = 13 $\mu$Jy, contour levels at -3, 3$\times$2$^n$, n $\in$ [0,7], beam size 0.27$\times$0.23 kpc. \textbf{Right panel:} J2314+2243, rms = 11 $\mu$Jy, contour levels at -3, 3$\times$2$^n$, n $\in$ [0,7], beam size 1.96$\times$1.27 kpc.}
\label{fig:J2242p2943}
\label{fig:J2314p2243}
\end{figure*}

RX J2314.9+2243 was classified as a steep-spectrum source, but our in-band indexes seem to indicate a F classification. This objects had a tentative $\gamma$-ray detection \citep{Komossa15}, which however was not confirmed by other studies \citep[e.g.][]{Liao15}. The source in the map of Fig.~\ref{fig:J2314p2243} do not show any particular structure. The unresolved core with a size of 940$\times$280 pc. The brightness temperature, $\sim$1.6$\times$10$^4$ K, is similar to that of other radio-loud NLS1s. The source was not detected at 37 GHz.

\section{Radio maps}



\end{appendix}

\end{document}